\documentclass[11pt]{article}
\usepackage{amsmath}
\usepackage{graphicx}
\usepackage{subcaption}
\usepackage{amsfonts}
\usepackage{amssymb}
\usepackage{epsfig}
\usepackage{color}
\usepackage{psfrag}
\usepackage{epstopdf}
\usepackage{mathrsfs}
\usepackage{stmaryrd}
\usepackage{empheq}
\usepackage{ifsym}
\usepackage{bbding}
\usepackage{stmaryrd}
\usepackage[theorems,skins]{tcolorbox}
\usepackage{yhmath}
\usepackage{caption}
\usepackage{esint}
\usepackage{wasysym}
\usepackage{harmony}
\usepackage[title,titletoc]{appendix}

\newcommand{\1}[1]{\textbf{\textit{#1}}}						

\newcommand{\EQ}{\begin{equation}}
\newcommand{\EN}{\end{equation}}
\newcommand{\be}{\begin{equation}}
\newcommand{\ee}{\end{equation}}
\newcommand{\bea}{\begin{eqnarray}}
\newcommand{\eea}{\end{eqnarray}}

\usepackage{bm}

\usepackage{caption}

\begin{document} 


\topmargin 0pt
\oddsidemargin 5mm
\renewcommand{\thefootnote}{\arabic{footnote}}
\newpage
\topmargin 0pt
\oddsidemargin 5mm
\renewcommand{\thefootnote}{\arabic{footnote}}

\definecolor{cayenne}{RGB}{128, 0, 0}
\definecolor{1}{RGB}{98, 64, 27}
\definecolor{bordeaux}{RGB}{148, 23, 81}
\definecolor{2}{RGB}{128,0,128}
\definecolor{3}{RGB}{0, 64, 120}
\definecolor{4}{RGB}{0, 88, 254}
\definecolor{5}{RGB}{0, 228, 64}
\definecolor{6}{RGB}{0, 128, 0}
\definecolor{moka}{RGB}{128, 64, 0}
\definecolor{orange}{RGB}{255, 128, 0}
\definecolor{azzurro1}{RGB}{30, 82, 255}
\definecolor{azzurro2}{RGB}{15, 127, 255}
\definecolor{blu1}{RGB}{0, 40, 128}
\definecolor{blu2}{RGB}{30, 88, 100}
\definecolor{bluapple}{RGB}{0, 0, 255}
\definecolor{marroneapple}{RGB}{153, 102, 51}
\definecolor{cianoapple}{RGB}{0, 255, 255}
\definecolor{verdeapple}{RGB}{0, 255, 0}
\definecolor{magentaapple}{RGB}{255, 0, 255}
\definecolor{arancioapple}{RGB}{255, 128, 0}
\definecolor{violaapple}{RGB}{127, 0, 127}
\definecolor{rossoapple}{RGB}{255, 0, 0}
\definecolor{gialloapple}{RGB}{255, 255, 0}
\definecolor{aqua}{RGB}{0, 128, 255}
\definecolor{verde}{RGB}{1, 165, 1}

\numberwithin{equation}{section}



\begin{titlepage}

\vspace{-10mm}

\begin{center}
{\large {\bf Critical Casimir interaction between colloidal Janus-type particles in two spatial dimensions}}\\
\vspace{2mm}
{\large
A.~Squarcini$^{\eighthnote, \twonotes, {\small{\ViPa}} }$,
A.~Macio\l ek$^{\eighthnote, \twonotes, \AcPa}$,
E.~Eisenriegler${}^{ \SePa}$,
and S.~Dietrich$^{ \eighthnote, \twonotes   }$
}\\
%
%
${}^{\eighthnote}${\em Max-Planck-Institut f$\ddot{u}$r Intelligente Systeme,\\
Heisenbergstr. 3, D-70569 Stuttgart, Germany}\\
${}^{\twonotes}${\em IV. Institut f$\ddot{u}$r Theoretische Physik, Universit$\ddot{a}$t Stuttgart,\\
Pfaffenwaldring 57, D-70569 Stuttgart, Germany}\\
${}^{{\small{\AcPa}}}${\em Institute of Physical Chemistry, Polish Academy of Sciences, Kasprzaka 44/52, PL-01-224 Warsaw, Poland}\\
${}^{{\small{\SePa}}}${\em Theoretical Soft Matter and Biophysics, Institute of Complex Systems,\\
Forschungszentrum J\"{u}lich, D-52425 J\"{u}lich, Germany}

\end{center}
\vspace{-2mm}

\renewcommand{\thefootnote}{\arabic{footnote}}
\setcounter{footnote}{0}

\begin{abstract}
We study colloidal particles with chemically inhomogeneous surfaces suspended in a critical binary liquid mixture. The inhomogeneous particle surface is composed of patches with alternating adsorption preferences for the two components of the binary solvent. By describing the binary liquid mixture \emph{at} its consolute point in terms of the critical Ising model we exploit its conformal invariance in two spatial dimension. This allows us to determine exactly the universal profiles of the order parameter, the energy density, and the stress tensor as well as some of their correlation functions around a single particle for various shapes and configurations of the surface patches. The formalism encompasses several interesting configurations, including Janus particles of circular and needle shapes with dipolar symmetry and a circular particle with quadrupolar symmetry. From these single-particle properties we construct the so-called small particle operator expansion (SPOE), which enables us to obtain asymptotically exact expressions for the position- and orientation-dependent critical Casimir interactions of the particles with distant objects, such as another particle or the confining walls of a half plane, strip, or wedge, with various boundary conditions for the order parameter. In several cases we compare the interactions at large distances with the ones at close distance (but still large on the molecular scale). We also compare our analytical results for two Janus particles with recent simulation data.
\end{abstract}


\vfill

\rule[0.2cm]{35.1mm}{0.2mm}

${}^{{\small{\ViPa}}}$ squarcio@is.mpg.de
\end{titlepage}

\tableofcontents

\section{Introduction}
\label{intro}
Remarkable progress in colloidal synthesis has allowed one to fabricate particles with anisotropic shapes and interactions \cite{Weitz:2006,Kraft:2011,Kohler:2018}. Using such particles one can generate a large variety of nanostructured materials via self-assembly. Particularly promising for the buildup of complex colloidal structures are particles with surface patch properties, which act as analogues of molecular valences. Modern technologies are able to produce multivalent patchy particles with geometrically well-defined patterns, such as dimers (e.g., Janus particles), trimers, or even tetramers \cite{sacanna2013shaping,nguyen2017switching,ma10111265,Chen199}. In addition, one can prescribe a specific solvent affinity of the patches by suitable chemical treatments~\cite{nguyen2017switching,ma10111265,Iwashita_1,Iwashita_2}. This specific solvent affinity in combination with binary solvents provides the way to directed self-assembly controlled by fluid-mediated interactions between the particles or between the patches of different particles \cite{Beysens}. Recent experimental studies \cite{nguyen2017switching,ma10111265,Iwashita_1,Iwashita_2} have demonstrated that the fluid-mediated interactions, which involve the critical Casimir effect, are particularly useful for manipulating colloids (see also recent reviews \cite{Schall_review,review} and references therein). This is the case because their range and strength are set by the solvent correlation length, which can be finely tuned by temperature in a reversible and universal manner. The critical Casimir forces (CCFs) result from the spatial confinement of the fluctuating composition of the binary solvent close to its consolute point \cite{FdG}. Since the properties of CCFs, including their sign, depend sensitively on the adsorption preferences of a colloid surface  \cite{GMHNHBD_09,HHGDB_08}, one can generate a selective bonding between particle patches, e.g., among hydrophobic or hydrophilic patches in aqueous solutions. Non-spherical shapes of particles allow the assembly of more complex structures \cite{nguyen2017switching}. Due to the anisotropy of the particles, there is not only a force but also a critical Casimir torque acting on the particles \cite{KHD}. This may lead to additional interesting effects such as orientational ordering.

In order to investigate  the  assembly behavior of colloids one successfully uses computer simulations based, in a first step, on effective pair potentials \cite{nguyen2017switching, bianchi2012predicting, kern2003, Newton-et:2017}. It is therefore crucial to have a detailed knowledge of the critical Casimir pair potentials (CCPs) between two patchy particles. An accurate determination of the CCP is, however, a challenging task. Apart from a few exceptions and limiting cases, the presently available analytical results for CCFs and their potentials are of approximate character. These challenges have motivated the present study.

Here, we investigate the effect of chemical inhomogeneities at the surface of colloidal particles which are suspended in a binary liquid mixture at its consolute point with the latter belonging to the Ising universality class. The surfaces of the particles suspended in the binary solvent generically attract one of the two components of the mixture preferentially. In the corresponding Ising lattice model this amounts to fix surface spins in the $+$ or $-$ direction. A surface with no preference corresponds to free-spin boundary conditions \cite{NHB}. In the terminology of surface critical phenomena these two surface universality classes \cite{Binder, HWD_review, Diehl} are known as ``normal'' ($+$ or $-$) and ``ordinary'' ($O$), respectively. We focus on the case that the inhomogeneous particle surface is composed of patches with alternating preferences $+$ and $-$. Considering the system right at the bulk critical point of the solvent --- apart from being an important benchmark --- allows one to exploit conformal invariance \cite{Cardy_review} in order to map conformally the actual particle geometry to a simpler one. We study a system in two spatial dimensions ($d=2$) for which this is particularly effective due to the abundance of possible conformal mappings\footnote{In more than two dimensions it is only the invariance under inversion which, in addition to the dilatation, one has at ones disposal. Still, this yields interesting results such as the exact form of the profiles of the order parameter and of the energy around a single sphere (see Ref. \cite{BE_85}).}.

We introduce CCFs for the simple case of two macroscopically extended parallel plates with uniform boundary conditions $a$ and $b$, respectively, in $d$ spatial dimensions. In the universal scaling region close to the critical point of the bulk system, such that their mutual distance $\mathcal{W}$ and the bulk correlation length $\xi$ are much larger than microscopic lengths, the CCF (per cross-sectional area and in units of $k_{B}T$) $F_{ab}$ between the plates is given by \cite{FdG}
\begin{equation}
\label{Casimir_plates}
F_{ab} = \Theta_{ab}(\mathcal{W}/\xi) \mathcal{W}^{-d} \, ,
\end{equation} 
where $\Theta_{ab}(x)$ is a universal scaling function of the dimensionless ratio $\mathcal{W}/\xi$.  At the bulk critical point the scaling function reduces to the universal critical Casimir amplitude $\Delta_{ab}=\Theta_{ab}(x=0)$. Equation (\ref{Casimir_plates}) follows from scaling arguments and renormalization group analyses \cite{KD} which, in general, leave the explicit form of the function $\Theta_{ab}$ determined in the form of an $\epsilon$-expansion \cite{Krech}.

In $d=2$, where the two-plate system reduces to a strip, a complete understanding of the Casimir forces has been achieved. For the Ising strip on the square lattice, the corresponding scaling function $\Theta_{ab}(\mathcal{W}/\xi)$ in Eq.  (\ref{Casimir_plates}) is known exactly at any temperature and for arbitrary uniform boundary conditions \cite{AM,RZSA}. Moreover, its asymmetric behavior in terms of the scaling variable $\mathcal{W}/\xi$ has been explained in Ref. \cite{AM_13}. Mirror symmetric boundaries attract each other (i.e., $\Theta_{aa}<0$). This is a consequence of reflection positivity \cite{KK,Bachas}. This feature holds not only for the Ising universality class. Another important feature is that in the scaling regime off criticality, symmetry-breaking ($\pm$) and symmetry-preserving ($O$) boundary conditions exchange their role upon swapping the low and high temperature phases, as reported in Refs. \cite{AM,ES_94} for the Ising model and in Ref. \cite{DS_15} for a wider range of universality classes in two dimensions.

By applying conformal invariance methods \cite{BPZ_1,BPZ_2} the values $\Delta_{ab}$ have been calculated exactly for all combinations $(a,b)$ of the conformally invariant boundary conditions $(+,-,O)$ on its two edges \cite{Cardy_review,BCN_86,Cardy_BCFT,Cardy_annulus}:
\begin{equation}
\label{Casimir_plates'}
\Delta_{ab}=\pi[-1,-1,23,2]/48 \quad {\rm for} \quad ab=[OO,++,+-,+O] \, ; \quad \Delta_{ab}=\Delta_{ba} \, .
\end{equation}

As mentioned above, conformal invariance in $d=2$ allows one to relate the critical behavior in the presence of inclusions with distinct geometries. This refers not only to the order parameter and energy density profiles \cite{BE_85} and correlation functions \cite{CR_84}, but also to the CCFs between two immersed objects. As an illustration we consider the CCF between two arbitrarily arranged (non-crossing) semi-infinite and straight needles \cite{EB_16}. Like the interior of the infinite strip, the region outside the two needles is simply connected and can be conformally mapped onto the upper half plane (which can be considered as the simplest representative of the simply connected confined regions). From the ensuing mapping from the strip to the two-needle system, the CCFs between the needles can --- via a transformation \cite{Cardy_review} of the stress tensor density --- be determined from those of the strip. From a strip with $a=b$ the force between the two needles follows if they both have uniform boundary conditions $a$. If $a \neq b$, there are two switches between $a$ and $b$, one at each of the two infinitely far ends of the strip, and the mapping can be designed to put the switches anywhere on the boundaries of the two needles. This allows one, in particular, to determine the CCF for the case of a homogeneous boundary condition $a$ on one of the two needles and $b$ on the other. But this holds also for the case in which one needle has the boundary condition $a$ on one of its two sides and $b$ on the other, while the other needle has the same condition $a$ (or $b$) on both of its sides. Likewise, for the situation of an infinite needle forming the boundary of a half plane in which a semi-infinite needle is embedded, a corresponding mapping from the strip with $a \neq b$ allows one to determine the CCF when there is a ``chemical step'', i.e., a switch between $a$ and $b$, in the half plane boundary while the semi-infinite needle has the same condition $a$ (or $b$) on both of its sides. 

The case of the upper half plane, with the real axis consisting of an arbitrary number of successive intervals with arbitrarily distributed lengths on which the boundary condition alternates between $+$ and $-$, was analyzed in Ref. \cite{BG_93}. Using the corresponding stress tensor density \cite{EB_16} one can determine the interactions in a set of two or more inclusions with a simply connected outside region if the total number of $+/-$ switches in their boundaries is larger than two. A strip with alternating boundary conditions has been studied in Ref. \cite{DSE}. If it contains only + and $-$ segments in its two boundaries, it also belongs to the above type of systems.

An infinite plane containing two particles with at least one of them of {\it finite} extent represents a doubly connected region which can always be mapped conformally to the interior of an annulus (or, equivalently, to the surface of a finite cylinder). For the case of homogeneous boundary conditions $a$ and $b$ on the two boundary circles of the annulus --- with $(a,b)$ being any combination of the three aforementioned boundary universality classes --- in Ref. \cite{Cardy_annulus} the partition function of the critical system inside the annulus has been calculated for an arbitrary size ratio of the two circles. From these results one can derive the CCF between two particles, each with homogeneous boundary conditions, such as two circular particles with arbitrary size- and distance-to-size ratios \cite{BE_95,MVS}, between two finite needles on a line \cite{BEK1}, between the boundaries of a half plane or a strip and an embedded needle in some  special configurations \cite{VED}, and between two arbitrarily configured needles \cite{EB_16}.

\begin{figure*}[htbp]
\centering
        \begin{subfigure}[b]{0.21\textwidth}
            \centering
            \includegraphics[width=\textwidth]{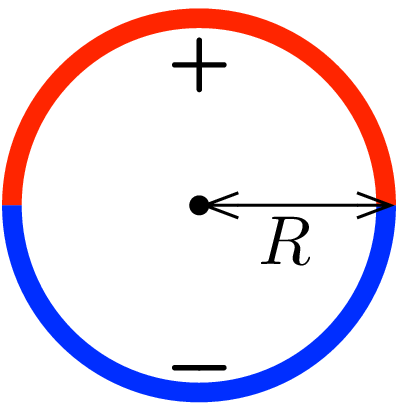}
            \caption[]%
            {{\small Janus (J)}}    
            \label{fig02_01a}
        \end{subfigure}
\hfill
        \begin{subfigure}[b]{0.21\textwidth}  
            \centering 
            \includegraphics[width=\textwidth]{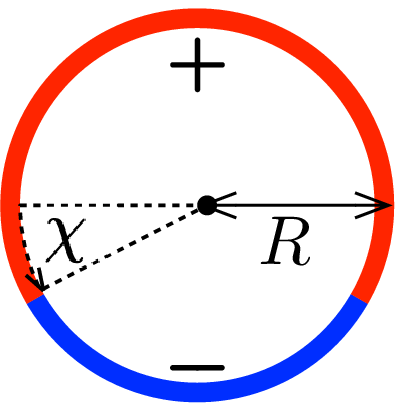}
            \caption[]%
            {{\small Janus $\chi$ ($\textrm{J}_{\chi}$)}}    
            \label{fig02_01b}
        \end{subfigure}
\hfill
        \begin{subfigure}[b]{0.21\textwidth}  
            \centering 
            \includegraphics[width=\textwidth]{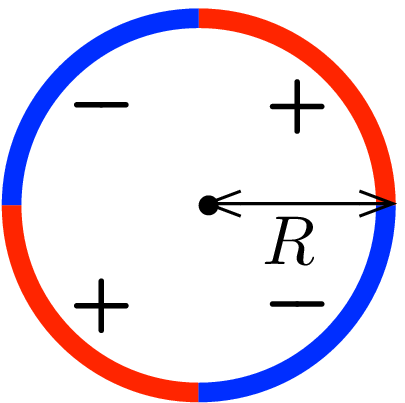}
            \caption[]%
            {{\small quadrupole (Q)}}    
            \label{fig02_01c}
        \end{subfigure}
\hfill
        \begin{subfigure}[b]{0.21\textwidth}  
            \centering 
            \includegraphics[width=\textwidth]{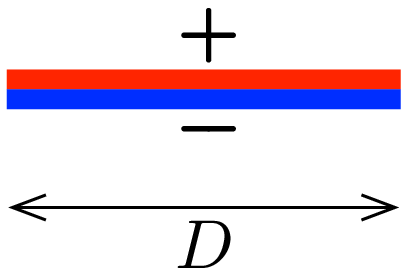}
            \caption[]%
            {{\small Janus needle (Jn) }}    
            \label{fig02_01d}
        \end{subfigure}
\vskip\baselineskip
\caption[]
{\small Circular and rod-like colloids with various patterns of chemical inhomogeneity. A symmetric Janus particle (a) and its generalized version (b) specified by the angle $\chi$. (c) A ``double'' Janus particle, corresponding to a quadrupole. (d) A needle-shaped Janus colloid.}
\label{fig02_01}
\end{figure*}

However, investigating the critical force and torque acting on a particle with an {\it inhomogeneous} surface (such as the Janus-type particles shown in Fig. \ref{fig02_01}), which interacts either with another such particle or is confined in a half plane, strip, or wedge, would require results for an annulus with an inhomogeneous boundary condition on at least one of the bounding circles.
These are presently not available and therefore we consider limiting cases.

Our main concern is to investigate the situation in which the size of the particle is much smaller than its closest distance to other inclusions or boundaries by using the efficient analytical tool of the ``small particle operator expansion'' (SPOE). Like in the ``operator product expansion'' \cite{OPE_1,OPE_2,OPE_3,BPZ_1,BPZ_2},  the large distance effects of a ``small'' object are encoded in a series of operators\footnote{In the literature the operators are sometimes addressed as ``fields''.} located at the position of the object which is reminiscent of the multipole expansion in electrostatics. Like the operator product expansion,  the SPOE is not limited to two dimensions and to the critical point. Up to now it has been established mainly for particles of spherical \cite{BE_95,ER_95} and anisotropic \cite{Eisenriegler_04,Eisenriegler_06} shape with a homogeneous boundary. An exception is the dumb-bell of two touching spheres, one with the $+$ and the other with the $-$ boundary condition \cite{Eisenriegler_04}.

\begin{figure*}[t!]
    \centering
    \begin{subfigure}[b]{0.23333\textwidth}
        \centering
        \includegraphics[height=37mm]{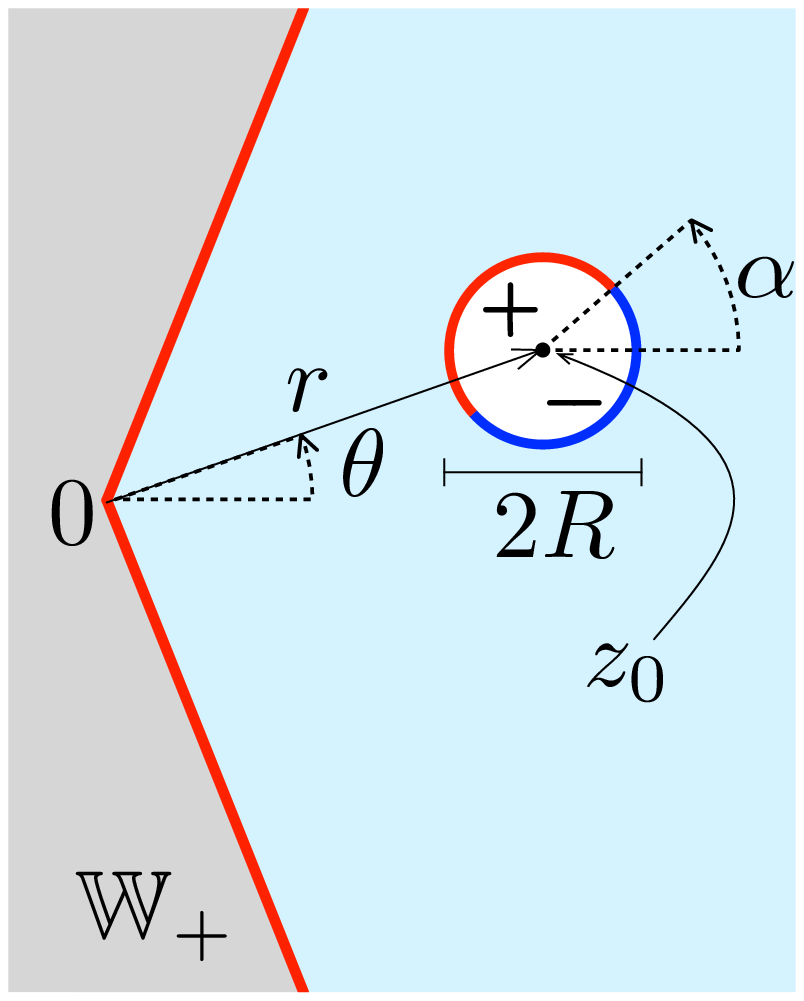}
        \caption{}
    \end{subfigure}%
    ~ 
    \begin{subfigure}[b]{0.35\textwidth}
        \centering
        \includegraphics[height=37mm]{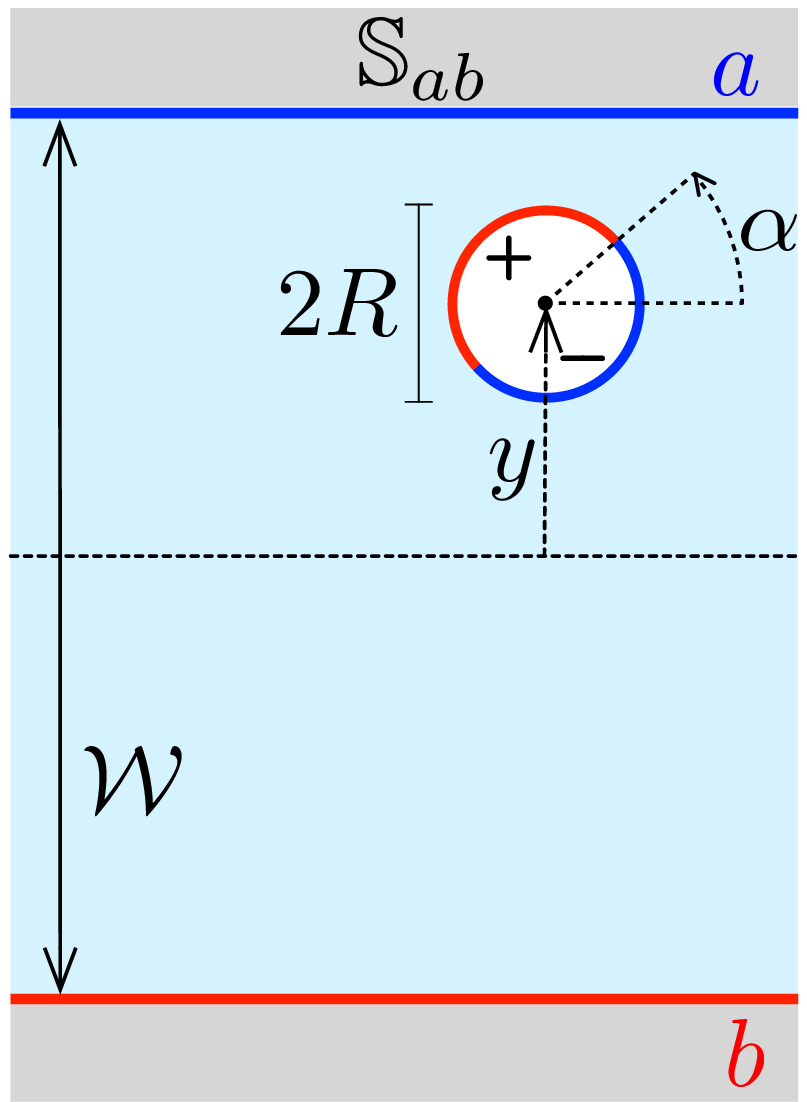}
        \caption{}
    \end{subfigure}
    ~ 
    \begin{subfigure}[b]{0.35\textwidth}
        \centering
        \includegraphics[height=37mm]{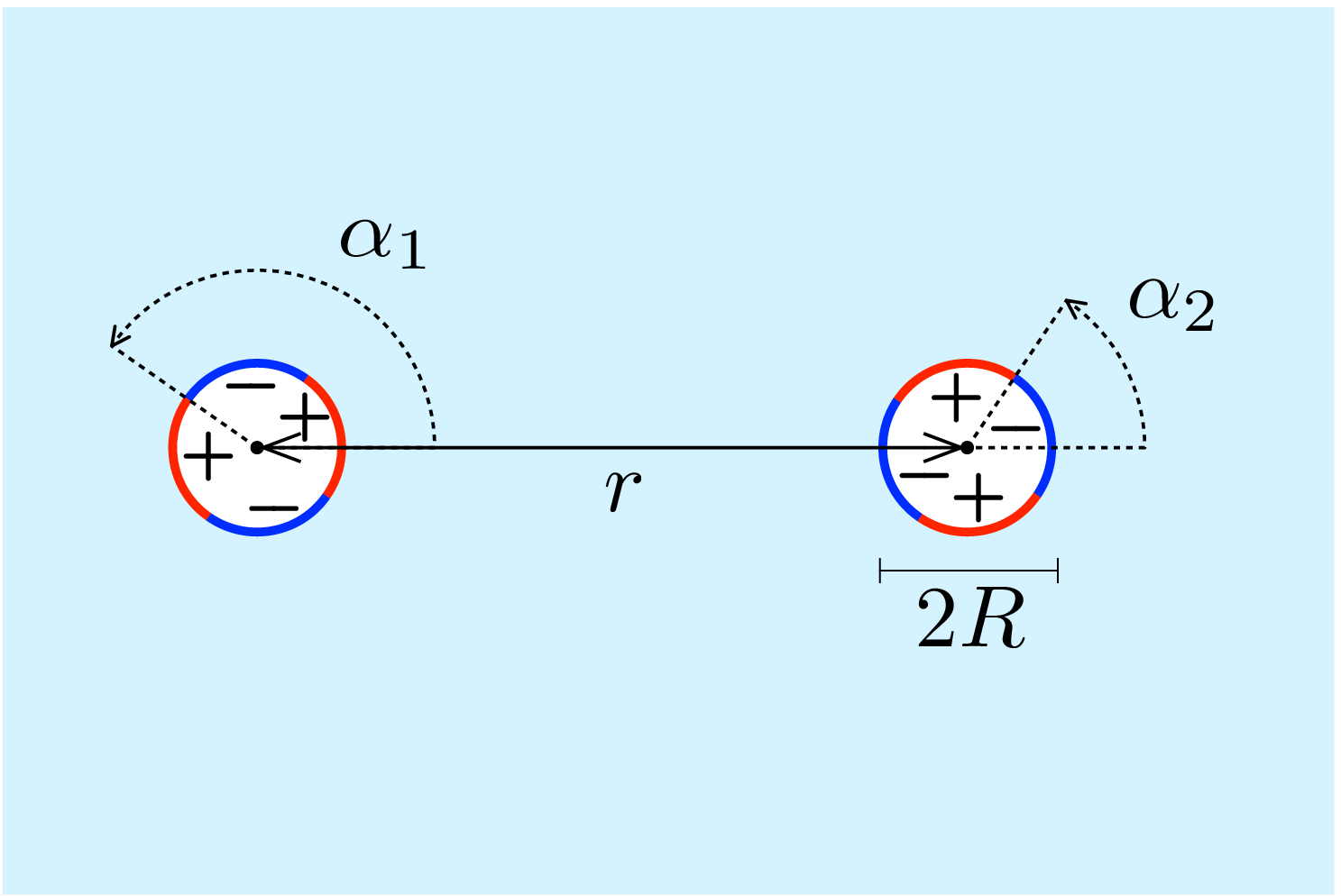}
        \caption{}
    \end{subfigure}
    \caption{Typical configurations of Janus and quadrupolar particles. (a) A Janus particle facing a wedge $\mathbb{W}_{+}$ with $+$ boundary condition. (b) A Janus particle in a strip $\mathbb{S}_{ab}$ with $(a,b)$ boundary conditions and width $\mathcal{W}$. (c) Two quadrupolar particles in the bulk.}
    \label{fig_wedge_bulk_strip}
\end{figure*}

Here, we provide the SPOEs for the particles shown in Fig. \ref{fig02_01}. For such particles the operators and their prefactors in the SPOE can be inferred from the profiles and correlation functions the particle induces when being solitary in the bulk. In this case the outside region is simply connected and the properties can be obtained by means of a conformal transformation\footnote{Like in the operator product expansion, in the vicinity of the critical point there are additional operator contributions with prefactors which vanish at the critical point and cannot be determined as described. (For the simple case of a spherical particle with an ordinary surface in the Gaussian model in $d > 2$ compare Ref. \cite{EBM_03}.) However, they are expected to yield only corrections of higher order to the leading distance- and orientation-dependent behaviors which are still determined by the corresponding operator terms at the critical point.} from those \cite{BX,BX_91,BG_93} in the empty half plane with an appropriate inhomogeneous boundary.

As mentioned above the SPOE enables one to obtain asymptotically exact analytical results for the free energy of interaction between the  particle and {\it distant} objects. It is interesting to compare these results with corresponding ones valid for {\it close} separations. In this context we present a discussion of the corresponding Derjaguin approximations. Finally, our analytical findings are compared with the numerical results for the interaction between Janus particles as obtained from Monte Carlo simulations in Ref. \cite{BPK}.

The paper is structured as follows. In Secs. \ref{symJanus}-\ref{quadrupoles} we determine the profiles of the order parameter, the energy density, and the stress tensor for all the particles shown in Fig.\ref{fig02_01}. For the generalized circular Janus particle in Fig. \ref{fig02_01}(b) details of the profiles are presented in Appendix \ref{appendix_C}. Using this information and the results for two-point correlation functions, as obtained in Appendices \ref{correlations} and \ref{appendix_D}, SPOEs for these particles are derived in Sec. \ref{SPOE}. In Sec. \ref{wall} it is explained how to use this operator expansion in order to calculate the critical Casimir free energy of interaction between a small particle and distant objects. We apply this method in order to calculate in Secs. \ref{sec_J_half} and \ref{wall_section2} the free energy necessary to transfer Janus and quadrupolar particles from the bulk into a half plane, a strip (as in Fig. \ref{fig_wedge_bulk_strip} (b)), or a wedge (as in Fig. \ref{fig_wedge_bulk_strip} (a)). In Sec. \ref{interaction_JJ} we discuss the critical Casimir interaction free energy between two distant Janus or quadrupolar particles as in Fig. \ref{fig_wedge_bulk_strip} (c) and between a Janus particle and a needle with ordinary boundary conditions. In several cases we compare the critical Casimir interactions at large distances with those at small distance. For the circular particles in Fig.\ref{fig02_01}, in Appendix \ref{appendix_F} we discuss the validity of the Derjaguin approximation and calculate the first correction to it. In Appendix \ref{discussion_degeneracy} we comment on level two degeneracy in two-dimensional conformal field theories. A glossary and a list of mathematical symbols used in the paper are provided in Appendix \ref{glossary}. Section \ref{conclusions} provides a summary and conclusions.

\section{Profiles induced by a single particle}
\label{single}
The four particles P in Figs. \ref{fig02_01} [(a), (b), (c), (d)], which we denote by ${\rm P}=[{\rm J}, \, {\rm J}_\chi , \, {\rm Q} , \, {\rm Jn}] = [ \textrm{Janus}, \textrm{Janus } \chi, \textrm{quadrupole}, \textrm{Janus needle} ]$, affect the critical system in the complemental, outside part of the plane via their boundaries, which are composed of alternating segments with the surface universality classes $+$ and $-$. These segments meet at switching points (sp) where the sign of the surface field flips. Burkhardt, Guim, and Xue \cite{BG_93,BX,BX_91} have investigated in detail the corresponding effect in the upper half plane with an arbitrary pattern $\mathcal{P}$ of consecutive boundary segments $+$ and $-$ on the real axis. They have calculated most of the resulting profiles and (multipoint) correlation functions of the order parameter $\Phi$ and the energy density $\varepsilon$. Here we use conformal invariance in order to determine the corresponding quantities for the particles from these results in the half plane. This approach proceeds in three steps

(i) One constructs a conformal transformation $w(z)$ which maps the region outside the particle in the plane $z=x+iy=r \exp(i\theta)$ to the upper half plane $\mathbb{H}$, with $w =u+iv= r_{w} \exp(i\theta_{w})$. Here one has $0<\theta<2\pi$ and $0< \theta_w < \pi$. According to Riemann's mapping theorem, such mappings exist because the outside region is simply connected.

(ii) For the segment composition of a given particle ${\rm P}$ the transformation $w(z)$ implies a corresponding segment pattern ${\cal P}$ of the half-plane boundary. For this one finds the resulting half-plane correlation functions $\langle ... \rangle_{\mathbb{H}_{\cal P}}$.

(iii) Finally, conformal invariance at the critical point enables one to obtain the correlation functions outside the particle because they are related to the ones in the half plane by a local scale transformation \cite{Cardy_review}. For $N$-point correlation functions of the primary, scalar operators $\Phi$ and $\varepsilon$, this relationship is given by \cite{Cardy_review,BPZ_1,BPZ_2,CFT_books_1,CFT_books_2}
\begin{eqnarray}
\label{03_08}
\langle \mathcal{O}_1 (z_1 ,\bar{z}_1 ) \mathcal{O}_2 (z_2 ,\bar{z}_2 ) ... \mathcal{O}_N (z_N ,\bar{z}_N)    \rangle_{\rm P} & = & \Big\vert \frac{\textrm{d}w}{\textrm{d}z_1} \Big\vert^{x_{\mathcal{O}_1}} \Big\vert \frac{\textrm{d}w}{\textrm{d}z_2} \Big\vert^{x_{\mathcal{O}_2}} \cdots \Big\vert \frac{\textrm{d}w}{\textrm{d}z_N} \Big\vert^{x_{\mathcal{O}_N}}  \times \\
&& \times  \langle \mathcal{O}_1 (w_1,\bar{w}_1)  \mathcal{O}_2 (w_2,\bar{w}_2) \cdots \mathcal{O}_N (w_N,\bar{w}_N)   \rangle_{\mathbb{H}_{\cal P}} \, ,\nonumber
\end{eqnarray}
$\langle \dots \rangle_{\rm P}$ is the thermal average in the presence of particle $\textrm{P}$,
\begin{equation}
w_j \equiv [w(z)]_{z=z_j} \, , \qquad \textrm{d} w/ \textrm{d}z_j \equiv [ \textrm{d} w(z)/\textrm{d} z]_{z=z_j} \, ,
\end{equation}
and each operator $\mathcal{O}_j$ can be either $\Phi$ or $\varepsilon$ with their bulk exponents $x_{\Phi} =1/8$ and $x_{\varepsilon} =1$, respectively. The operator $\varepsilon$ describes the deviation of the fluctuating energy density from its average in the bulk, i.e., the bulk average $\langle\varepsilon\rangle$ of $\varepsilon$ vanishes\footnote{Also for systems in more than two spatial dimensions $d$ this subtracted operator $\varepsilon$ is the most convenient quantity to work with when discussing energy density profiles by means of the field theoretic renormalization group, see, e.g., Eqs. (3.343) and (3.344) in Ref. \cite{HWD_review} as well as Ref. \cite{DD}, where a half space with homogeneous boundary condition is considered. For an ordinary boundary $O$ and at bulk $T_{c}$, Eq. (3.344) predicts a simple power law decay upon approaching the bulk, like in Eq. (\ref{03_03}) below, but with the value the exponent $x_{\varepsilon} \equiv (1-\alpha )/\nu$ adopts in $d$ dimensions; see also Ref. \cite{DD}.}. Here we follow common usage \cite{BPZ_1} to parametrize a point in the $z$ plane instead by its Cartesian coordinates $x, y$ by the complex coordinates $z=x+iy,\bar{z}=x-iy$, which are regarded to be independent. A corresponding notation is used for the $w$ plane.

In the present section we determine one-point correlation functions $\langle \mathcal{O} (z ,\bar{z} ) \rangle_{\rm P}$ and we shall use Eq. (\ref{03_08}) for $N=1$ only. Two-point correlation functions ($N=2$) will be considered in the appendices.

The simplest type of the particles ${\rm P}$ is the one with a {\it homogeneous} boundary. For this type both the particle boundary and the pattern ${\cal P}$ of the upper half plane consist of a {\it single} segment with surface universality class $a \in \{ +, \, -, O\}$. Denoting ${\mathbb{H}_{\cal P}}$ in this case by ${\mathbb{H}_{a}}$, the corresponding profiles of $\mathcal{O} =\Phi$, $\varepsilon$ are given by
\begin{equation}
\label{03_03}
\langle {\cal O}(w,\bar{w})\rangle_{\mathbb{H}_{a}} = \mathcal{A}_{{\cal O}}^{(a)} v^{-x_{ {\cal O}}}
\end{equation}
where $v=\textrm{Im} \, w$ and where within the convenient normalization of primary operators via their bulk two-point correlation function,
\begin{equation}
\label{03_04bis}
\langle \mathcal{O}(\1{r}_{1})\mathcal{O}(\1{r}_{2}) \rangle = |\1{r}_{1}-\1{r}_{2}|^{-2x_{\mathcal{O}}} \, ,
\end{equation}
the values of the amplitudes ${\cal A}_{{\cal O}}^{(a)}$ are \cite{Cardy_review}
\begin{eqnarray}
\label{03_04_01}
\mathcal{A}_{\Phi}^{(+)}  =  - \mathcal{A}_{\Phi}^{(-)} = 2^{1/8} \, , \quad \mathcal{A}_{\Phi}^{(O)}=0 \, , \quad  
\mathcal{A}_{\varepsilon}^{(+)} = \mathcal{A}_{\varepsilon}^{(-)} = - \mathcal{A}_{\varepsilon}^{(O)} = -\frac{1}{2}  \, .
\end{eqnarray}

\begin{figure*}[htbp]
\centering
        \begin{subfigure}[htbp]{0.48\textwidth}
            \centering
            \includegraphics[width=\textwidth]{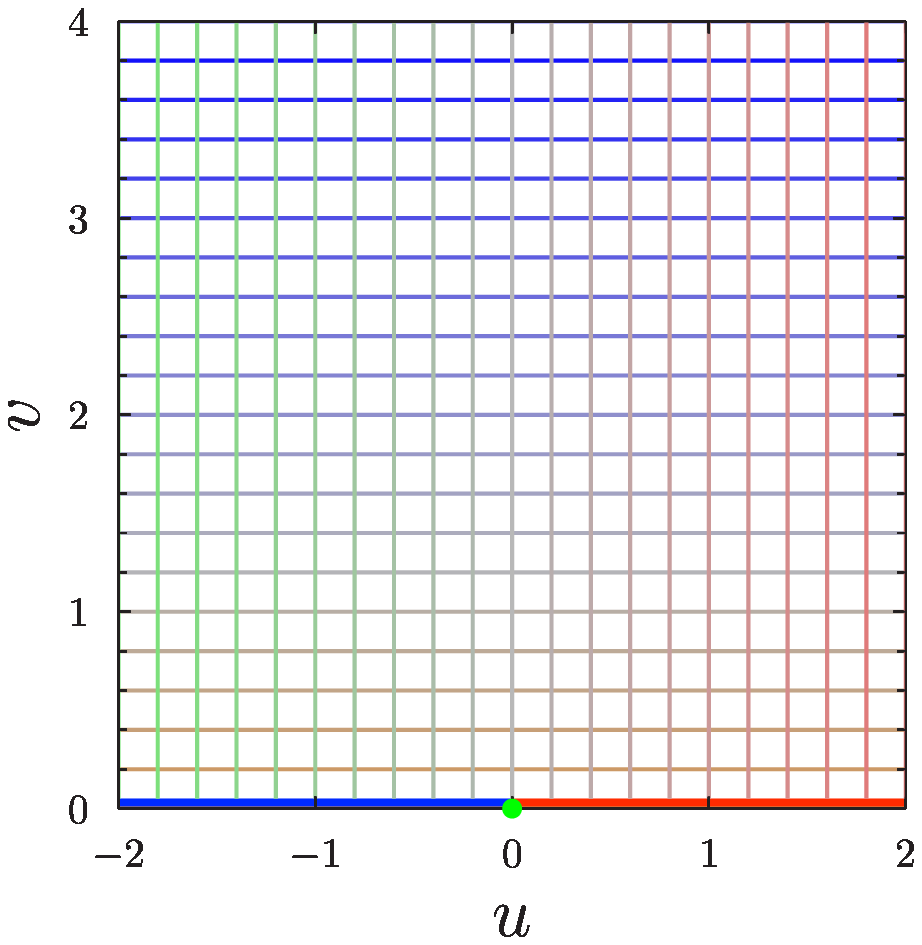}
            \caption[]%
            {{\small }}    
            \label{fig03_01a}
        \end{subfigure}
\hfill
        \begin{subfigure}[htbp]{0.48\textwidth}  
            \centering 
            \includegraphics[width=\textwidth]{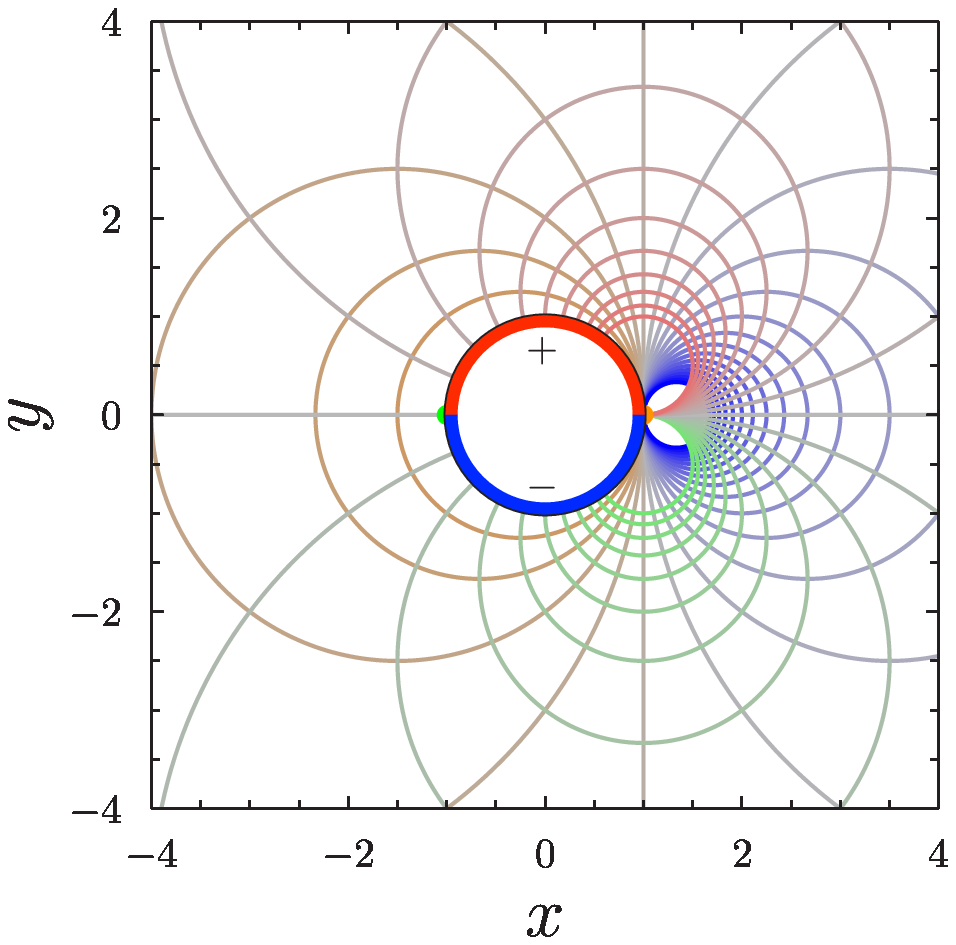}
            \caption[]%
            {{\small }}    
            \label{fig03_01b}
        \end{subfigure}
\caption[]
{\small The domain outside the circular particle centered at the origin in the $z$-plane and its boundary depicted in (b) are conformally mapped by the transformation $w(z)$ in Eq. (\ref{03_02}) to the upper half $w$-plane and its boundary depicted in (a). In particular, the  switching point $z=-1$ on the particle boundary is mapped to its counterpart at the origin of the $w$-plane; a green dot marks its location. The circles shown in (b) are the pre-images of the lines of constant $u$ and $v$ in (a). They shrink and accumulate towards the point $z=1$ (marked by an orange dot) which is the pre-image of $w=\infty$ in (a).  The two small white regions anchored at $z=1$ in (b) emerge due to mapping only to a finite number of lines in the $w$-plane in (a).}
\label{fig03_01}
\end{figure*}

For reasons of simplicity, the three circular particles ${\rm P}=[{\rm J}, \, {\rm J}_\chi , \, {\rm Q}]$ in Figs. \ref{fig02_01} (a)-(c) are taken to have radius $1$ and to be centered at the origin\footnote{The later generalization to particles with an arbitrary radius and position of the center trivially follows from a dilatation and translation.} $z=0$. In this case the mapping $w(z)$ from their outside region to $\mathbb{H}$ is provided by a  M$\ddot{\textrm{o}}$bius transformation which together with its derivative is given by\footnote{See for instance Ref. \cite{Kober}.}
\begin{equation}
\label{03_02}
w(z) = i\frac{z+1}{z-1} \, , \qquad  \frac{\textrm{d}w(z)}{\textrm{d}z} = -\frac{2i}{(z-1)^{2}} \, .
\end{equation}
This maps circles onto circles and, in particular, the points $z=\exp{(i\theta)}$ on the particle surface with $0<\theta<2\pi$ to the points
\begin{equation}
\label{circby}
w(\textrm{e}^{i\theta})=\cot(\theta/2) \, , \quad 0<\theta<2\pi
\end{equation}
on the boundary of $\mathbb{H}$. The action of the map given in Eq. (\ref{03_02}) is depicted in Fig. \ref{fig03_01}. 

Besides studying $\Phi$ and $\varepsilon$, we are also interested in how the particles $\textrm{P}$ affect correlation functions containing the symmetric and traceless Cartesian stress tensor $T_{kl}(x,y)$ or its complex counterpart \cite{CFT_books_1, CFT_books_2} 
\begin{equation}
\label{cartcompstress}
T(z) \equiv -\pi [T_{xx}(x,y)-iT_{xy}(x,y)] \, .
\end{equation}
The average $\langle T(z) \rangle_{\rm P}$ follows from its transformation formula \cite{BPZ_1} 
\begin{equation}
\label{stresstrafo}
\langle T(z) \rangle_{\rm P}=\Biggl( {\textrm{d}w \over \textrm{d}z} \Biggr)^2 \langle T(w) \rangle_{\mathbb{H}_{\cal P}} + {1 \over 24} S(w(z)) \, , \quad S(w(z)) \equiv {\textrm{d}^3 w/\textrm{d}z^3 \over \textrm{d}w/\textrm{d}z} - {3\over 2} \Biggl( {\textrm{d}^2 w/\textrm{d}z^2 \over \textrm{d}w/\textrm{d}z} \Biggr)^2
\end{equation}
with the Schwarzian derivative $S$ of $w(z)$, because the average $\langle T(w) \rangle_{\mathbb{H}_{\cal P}}$ in the half plane with arbitrary pattern ${\cal P}$ is known from Ref. \cite{EB_16}. For particles ${\rm P}$ with a {\it homogeneous} boundary $a$ the corresponding expression $\langle T(w) \rangle_{\mathbb{H}_{a}}$ vanishes \cite{Cardy_BCFT} so that in this case $\langle T(z) \rangle_{\textrm{P}} = S/24$.

The two eigenvalues $\lambda_{+}(x,y) > 0$ and $\lambda_{-}(x,y) \equiv - \lambda_{+}(x,y)$ of the $2 \times 2$ matrix $\langle T_{kl}(x,y) \rangle_{\textrm{P}}$  are given by 
\begin{equation}
\label{Teigenval}
\lambda_{\pm}(x,y) = \pm |\langle T(z) \rangle_{\textrm{P}}|/\pi
\end{equation}
because the product $\lambda_{+} \lambda_{-} = - \lambda_{+}^2 = - \lambda_{-}^2$ equals the determinant
\begin{equation}
- \langle T_{xx} \rangle_{\textrm{P}}^2 - \langle T_{xy} \rangle_{\textrm{P}}^2 \equiv - |\langle T(z) \rangle_{\textrm{P}}|^2 /\pi^2
\end{equation}
of the matrix.

We also consider the matrix with the elements $T_{rr}$, $T_{\theta \theta}$, $T_{r\theta}$, and $T_{\theta r}$ which arises from the Cartesian one by rotation to the local radial and tangential directions of the circular coordinates centered at the particle $\textrm{P}$ and which is also symmetric and traceless. One finds $T_{rr}=[(x^2 -y^2)T_{xx}+2xy T_{xy}]/r^2$ and $T_{r \theta}=[(x^2 -y^2)T_{xy}-2xy T_{xx}]/r^2$; their averages can be expressed by that of $T(z)$:
\begin{equation}
\label{radtang}
\langle T_{rr}(x,y) \rangle_{\rm P} = - {1 \over \pi |z|^2}{\rm Re}[z^2 \langle T(z) \rangle_{\rm P} ] \, , \quad \langle T_{r \theta} (x,y) \rangle_{\rm P} =  {1 \over \pi |z|^2}{\rm Im}[z^2 \langle T(z) \rangle_{\rm P} ] \, .
\end{equation}

\subsection{Symmetric circular Janus particle}
\label{symJanus}

\subsubsection{Profiles of the order parameter $\Phi$ and of the energy density $\varepsilon$}
\label{phiepsJ}
The surface universality classes + and $-$ on the upper and lower arc, $0<\theta<\pi$ and $\pi<\theta<2\pi$, respectively, of the boundary of the Janus particle ${\rm P=J}$ in Fig. \ref{fig02_01}(a) imply via Eq. (\ref{circby}) the simple half-plane boundary-pattern ${\cal P}$ with + and $-$ for $u>0$ and $u<0$, respectively (see Fig. \ref{fig03_01}). The corresponding one- and two-point correlation functions of $\Phi$ and $\varepsilon$ have been calculated in Ref. \cite{BX_91}. Denoting the present half-plane situation $\mathbb{H}_{\cal P}$ by $\mathbb{H}_{-+}$, the one-point correlation functions are
\begin{equation}
\label{03_05}
\langle\Phi(w,\bar{w})\rangle_{\mathbb{H}_{-+}} = {\cal A}_{\Phi}^{(+)} v^{-1/8} \cos\theta_{w}
\end{equation}
and
\begin{equation}
\label{03_06}
\langle\varepsilon(w,\bar{w})\rangle_{\mathbb{H}_{-+}} = {\cal A}_{\varepsilon}^{(+)} v^{-1} \left( 4 \cos^{2}\theta_{w} - 3 \right) \, ,
\end{equation}
with $\cos \theta_w \equiv \textrm{Re}\, w /|w|$. From Eqs. (\ref{03_08}) and (\ref{03_02}) one obtains the profiles around the Janus particle $\textrm{J}$:
\begin{equation}
\label{03_09}
\langle \Phi(z,\bar{z}) \rangle_{\textrm{J}} = \langle\Phi(z,\bar{z})\rangle_{{ \rm circle, + }} \cos\theta_{w} = 2^{1/4} \frac{2y}{(|z|^{2}-1)^{1/8} |z^{2}-1|}
\end{equation}
and
\begin{equation}
\label{03_10}
\langle \varepsilon(z,\bar{z}) \rangle_{\textrm{J}} = \langle\varepsilon(z,\bar{z})\rangle_{ \rm circle, + } \left( 4 \cos^{2}\theta_{w} - 3 \right) = -\frac{1}{(|z|^{2}-1)} \biggl[ \frac{16y^{2}}{|z^{2}-1|^{2}}-3\biggr] \, .
\end{equation}
Here $\langle {\cal O}(z,\bar{z}) \rangle_{ \rm circle, + } = {\cal A}_{\cal O}^{(+)} \,[2/(|z|^{2}-1)]^{x_{\cal O}}$ for ${\cal O}=\Phi , \, \varepsilon$ are the profiles around a unit circle with {\it homogeneous} boundary condition $+$, which via Eq. (\ref{03_08}) follow from Eqs. (\ref{03_03}), (\ref{03_04_01}), and (\ref{03_02}). 

\begin{figure*}[htbp]
\centering
        \begin{subfigure}[b]{0.328\textwidth}
            \centering
            \includegraphics[width=\textwidth]{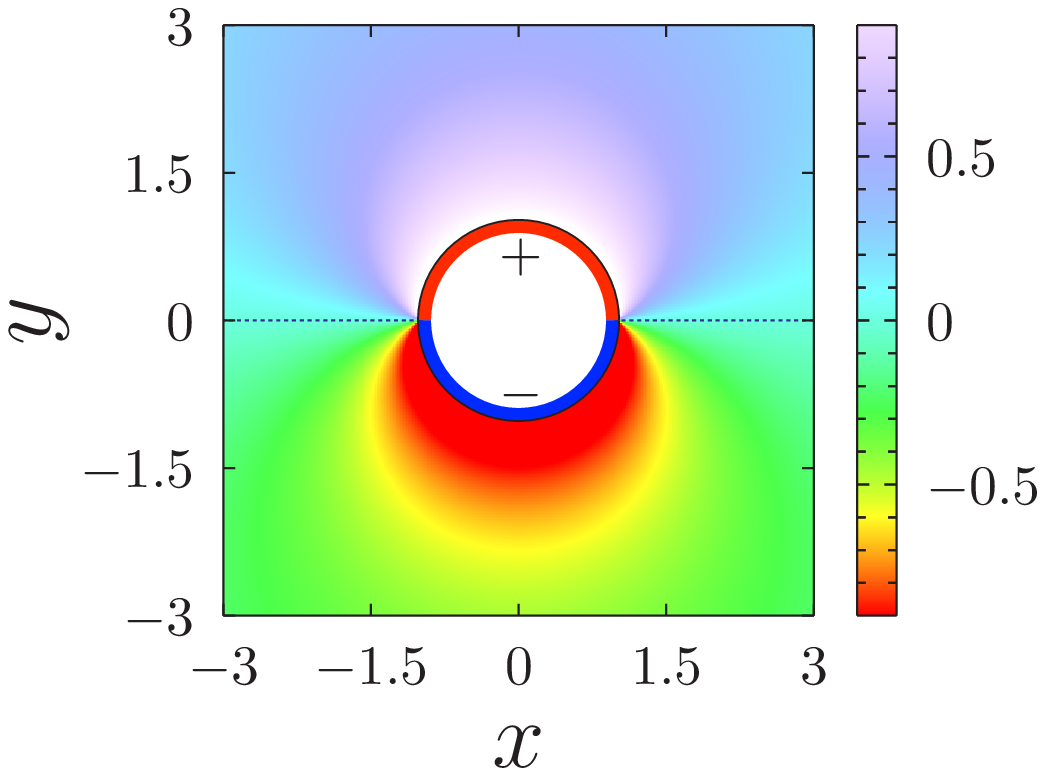}
            \caption[]
            {{\small }}    
            \label{plot02_01aa}
        \end{subfigure}
\hfill
        \begin{subfigure}[b]{0.328\textwidth}  
            \centering 
            \includegraphics[width=\textwidth]{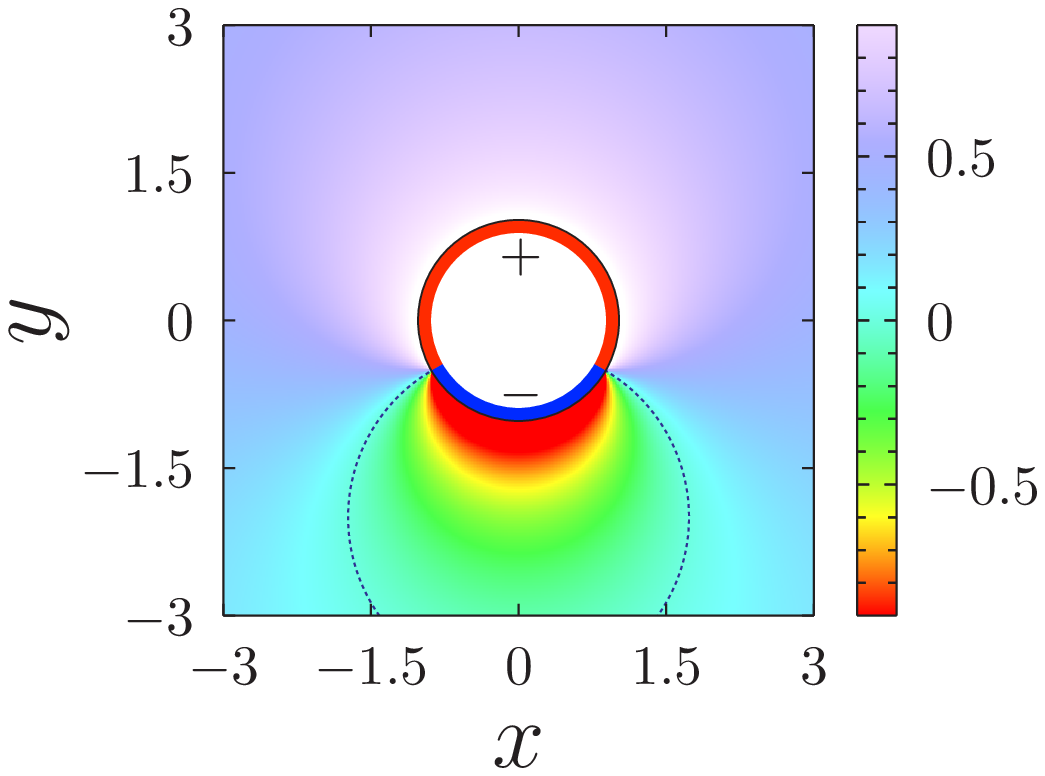}
            \caption[]
            {{\small }}    
            \label{plot02_01ba}
        \end{subfigure}
\hfill
        \begin{subfigure}[b]{0.328\textwidth}  
            \centering 
            \includegraphics[width=\textwidth]{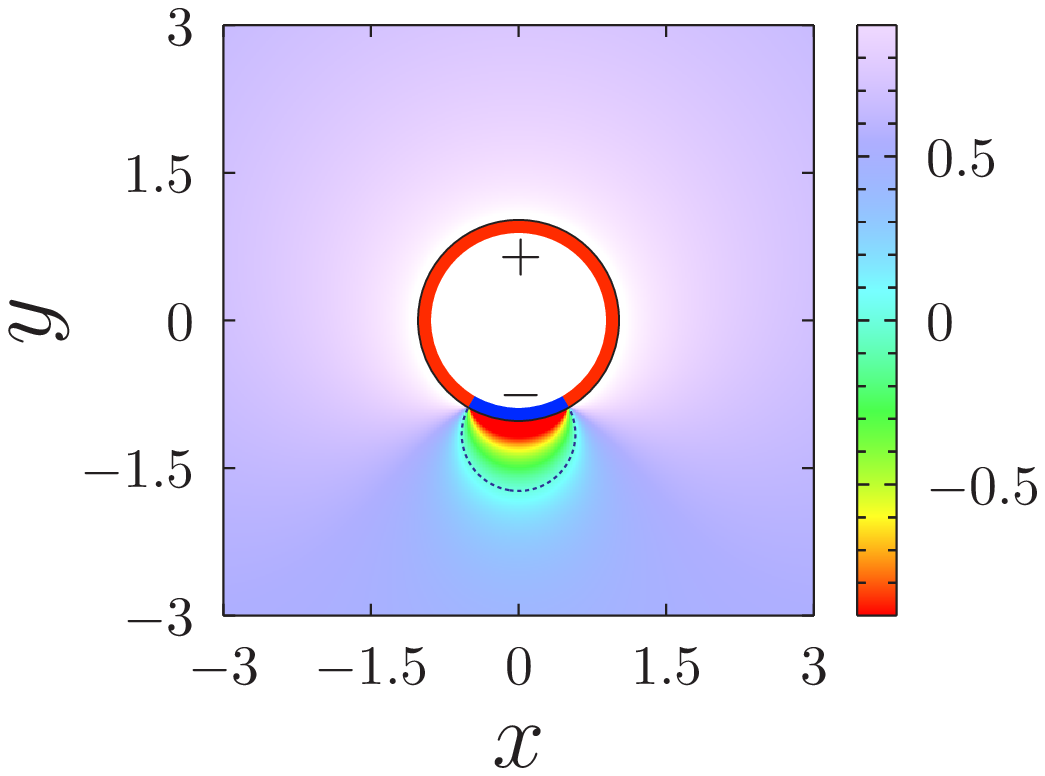}
            \caption[]
            {{\small }}    
            \label{plot02_01ca}
        \end{subfigure}
\vskip\baselineskip
\caption[] 
{\small Order parameter profiles for various realizations of the generalized Janus particle ${\rm J}_{\chi}$ in Fig.  \ref{fig02_01}(b) computed from Eq. (\ref{04_04}). The opening angle $\pi-2\chi$ for the range of the $-$ boundary condition (blue) takes the values $\pi$ (a), $2\pi/3$ (b), and $\pi/3$ (c). The dotted blue line represents the zero of the order parameter. The particle surface is indicated by the black circle $x^{2}+y^{2}=1$. The ranges of the boundary conditions $+$ and $-$ are marked as red and blue parts of the inner ring, respectively. We note that the order parameter profile diverges upon approaching the particle surface. There, for technical reasons, the color code is fixed to the maximum scale available in the legend.}
\label{plot02_01three}
\end{figure*}

\begin{figure*}[htbp]
\centering
        \begin{subfigure}[b]{0.328\textwidth}
            \centering
            \includegraphics[width=\textwidth]{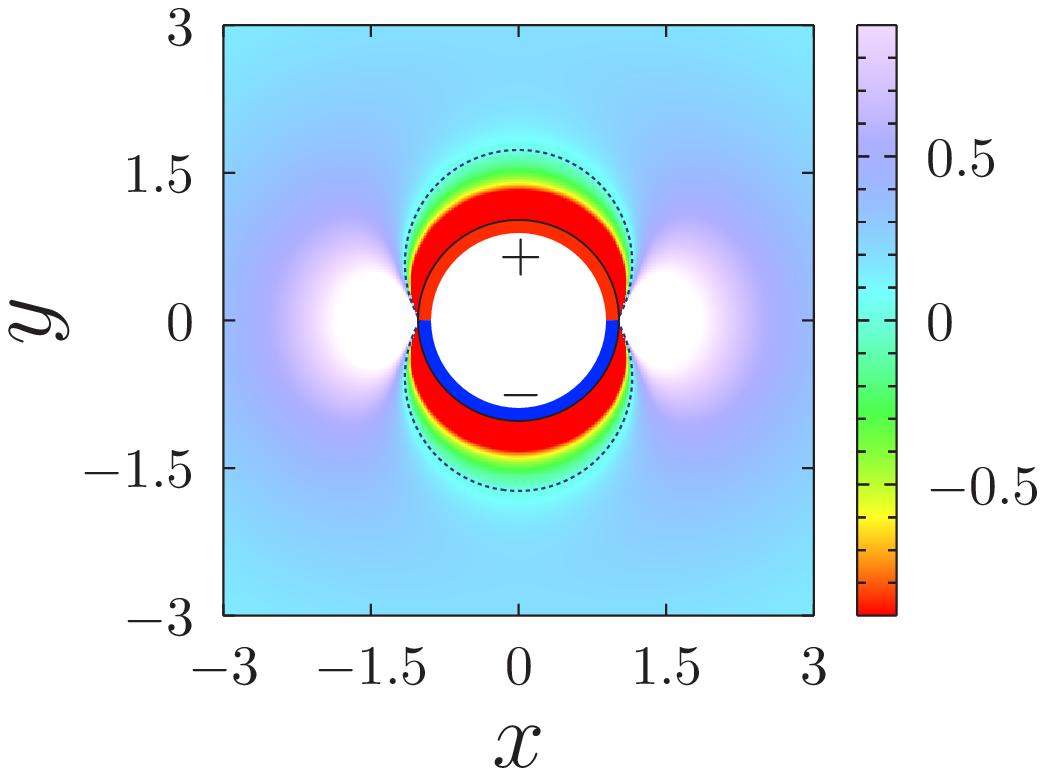}
            \caption[]%
            {{\small }}    
            \label{plot02_01ab}
        \end{subfigure}
\hfill
        \begin{subfigure}[b]{0.328\textwidth}  
            \centering 
            \includegraphics[width=\textwidth]{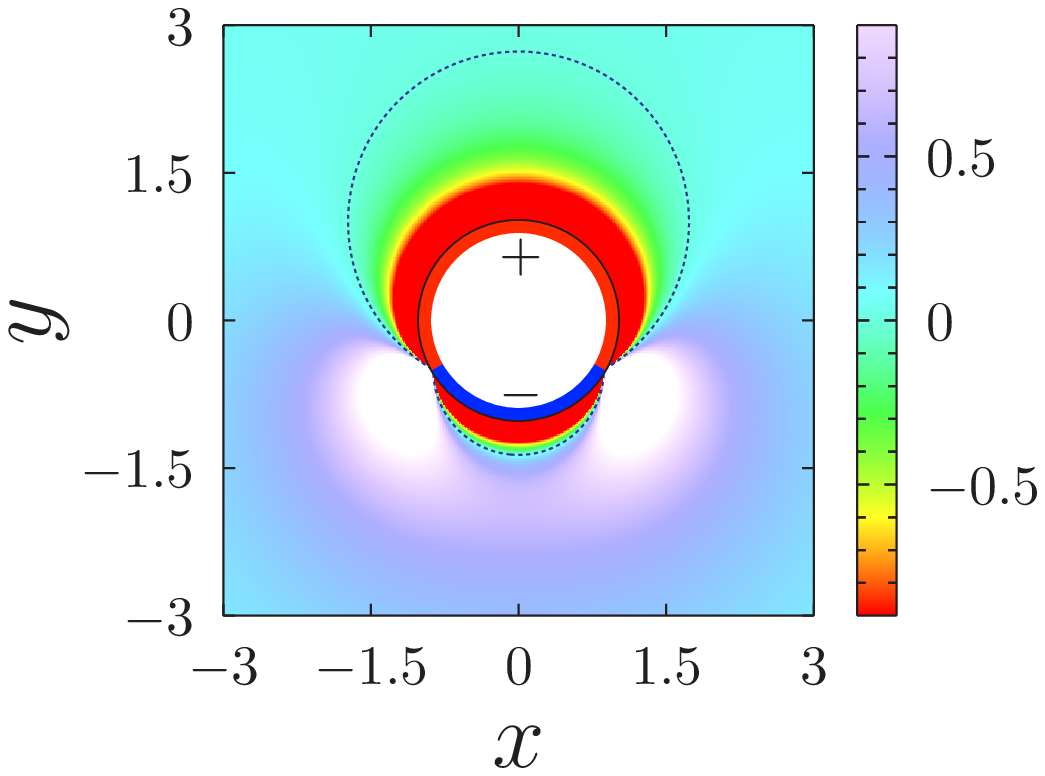}
            \caption[]%
            {{\small }}    
            \label{plot02_01bb}
        \end{subfigure}
\hfill
        \begin{subfigure}[b]{0.328\textwidth}  
            \centering 
            \includegraphics[width=\textwidth]{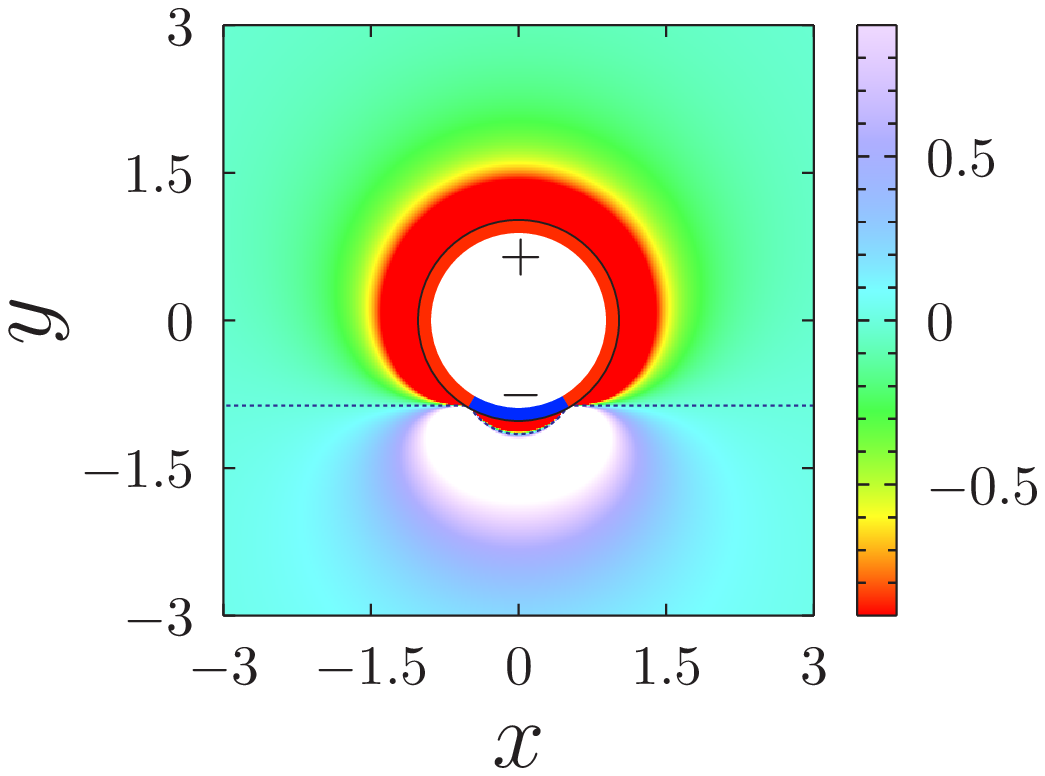}
            \caption[]%
            {{\small }}    
            \label{plot02_01cb}
        \end{subfigure}
\vskip\baselineskip
\caption[]
{\small Energy density profiles for various realizations of the generalized Janus particle ${\rm J}_{\chi}$ in Fig.  \ref{fig02_01}(b) computed from Eq. (\ref{04_05}). The opening angle $\pi-2\chi$ for the range of the $-$ boundary condition (blue) takes the values $\pi$ (a), $2\pi/3$ (b), and $\pi/3$ (c). Note the two regions of high energy density induced by the switching points; the unfolding divergence towards $+\infty$ of the energy density is indicated in white. The energy density diverges towards $-\infty$ upon approaching the particle surface; there this is signaled by the red regions. The dotted blue line represents the loci of vanishing energy density. The particle surface is indicated by the black circle $x^{2}+y^{2}=1$. The ranges of the boundary conditions $+$ and $-$ are marked as red and blue parts of the inner ring, respectively.}
\label{plot02_02}
\end{figure*}

The profiles in Eqs. (\ref{03_09}) and (\ref{03_10}) around the Janus particle in Fig. \ref{fig02_01}(a) are shown in Figs. \ref{plot02_01three}(a) and \ref{plot02_02}(a) below. They display the expected symmetries with $\langle \Phi(z,\bar{z}) \rangle_{\textrm{J}}$ being symmetric and antisymmetric by reflection about the $y$ and $x$ axis, respectively, and $\langle \varepsilon(z,\bar{z}) \rangle_{\textrm{J}}$ is symmetric about both axes.

For $|z| \searrow 1$, i.e., upon approaching the particle surface, the particle boundary attains the character of a straight line and the profiles show the behaviors corresponding to Eqs. (\ref{03_05}) and (\ref{03_06}). Their behaviors at large distances $|z| \gg 1$ will facilitate to find the SPOE for the Janus particle.

\subsubsection{Profile of the stress tensor}
\label{stressJ}
Taking the expression
\begin{equation}
\label{TH-+}
\langle T(w) \rangle_{\mathbb{H}_{-+}} = 1/(2w^{2}) 
\end{equation}
for the mean stress tensor for the half plane configuration ${\mathbb{H}_{-+}}$ from Ref. \cite{BX_91}, we obtain from Eq. (\ref{stresstrafo}) the corresponding average 
\begin{equation}
\label{03_12}
\langle T(z) \rangle_{\textrm{J}} = \frac{2}{\left( z^{2}-1 \right)^{2}} \, 
\end{equation}
in the region outside of the Janus particle in Fig. \ref{fig02_01}(a). The Schwarzian derivative $S$ vanishes for the present M$\ddot{\textrm{o}}$bius transformation in Eq. (\ref{03_02}). The right hand side of Eq. (\ref{03_12}) diverges at the switching points $z= \pm 1$.

In order to discuss eigenvectors and eigenvalues of the stress tensor, we start by considering certain special cases displaying features which are typical for all particles shown in Fig. \ref{fig02_01} and which follow directly from Eq. (\ref{03_12}) by using Eq. (\ref{radtang}). 

(i) For points $(x,y)=(\cos \theta, \sin \theta)$ right on the particle boundary the expression in the square brackets of Eq. (\ref{radtang}) is real so that $\langle T_{r\theta} \rangle$ vanishes, the radial-tangential matrix is diagonal, and the eigenvectors are parallel and perpendicular, respectively, to the boundary, with the eigenvalue $\langle T_{rr} \rangle_{\textrm{J}}$ of the vector perpendicular to the boundary being equal to\footnote{The vanishing of the off diagonal matrix elements, corresponding to the components parallel and perpendicular to the boundary, and the finiteness of the eigenvalues (in the diagonal) inside the segments with homogeneous boundary conditions, persist for correlation functions containing the stress tensor. This is in line with well known corresponding properties at the boundary of a half plane in arbitrary spatial dimensions.} $1/( 2 \pi \sin^2 \theta)>0$.

(ii) By the same argument one finds that along both the $x$ and the $y$ axis the eigenvectors point into radial and tangential directions, i.e., perpendicular and parallel to the boundary. For the two axes this can be inferred as well from the Cartesian matrix for which $\langle T_{xy} \rangle_{\textrm{J}}=0$, which can be expected from the symmetry of the Janus particle. For the two semi-infinite lines on the $y$ axis, which emanate from the centers $y= \pm 1$ of the + and $-$ segments, the eigenvalue $\langle T_{rr} \rangle_{\textrm{J}} \equiv \langle T_{yy} \rangle_{\textrm{J}}$ belonging to the vector perpendicular to the boundary, is given by $2/[\pi(y^2 +1)^2]>0$. For the two semi-infinite lines on the $x$ axis, emanating from the two switching points $x= \pm 1$, the eigenvalue $\langle T_{rr} \rangle_{\textrm{J}} \equiv \langle T_{xx} \rangle_{\textrm{J}}$ corresponding to the vector perpendicular to the boundary is given by $-2/[\pi(x^2 -1)^2]<0$.

\begin{figure*}[htbp]
\centering
        \begin{subfigure}[b]{0.285\textwidth}
            \centering
            \includegraphics[width=\textwidth]{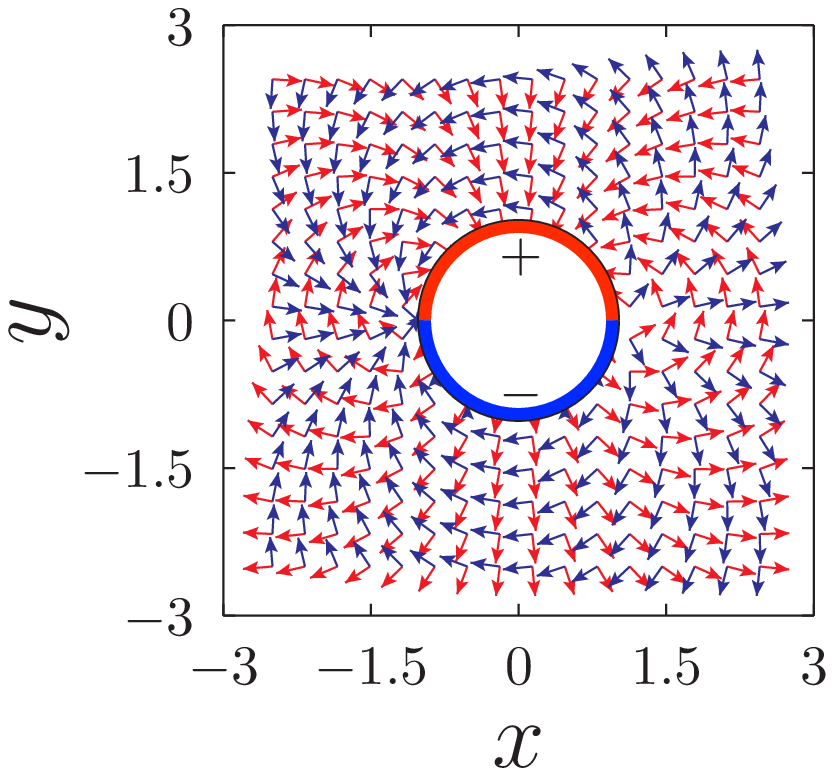}
            \caption[]%
            {{\small }}    
            \label{plot03_02a}
        \end{subfigure}
\hfill
        \begin{subfigure}[b]{0.345\textwidth}  
            \centering 
            \includegraphics[width=\textwidth]{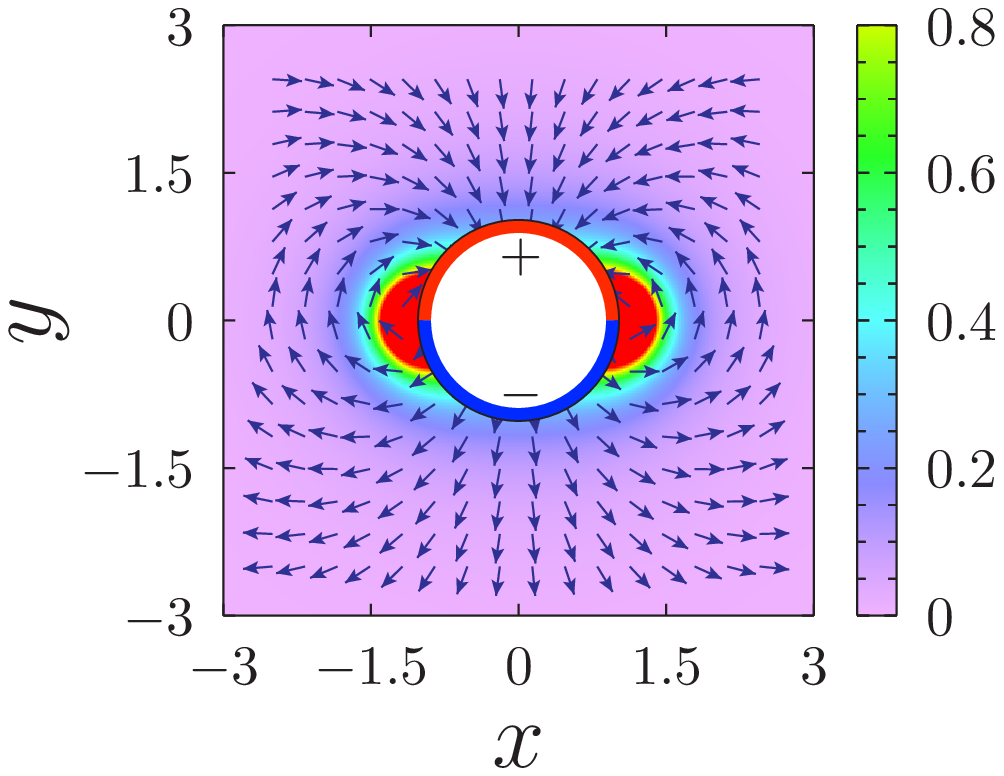}
            \caption[]%
            {{\small }}    
            \label{plot03_02b}
        \end{subfigure}
\hfill
        \begin{subfigure}[b]{0.345\textwidth}  
            \centering 
            \includegraphics[width=\textwidth]{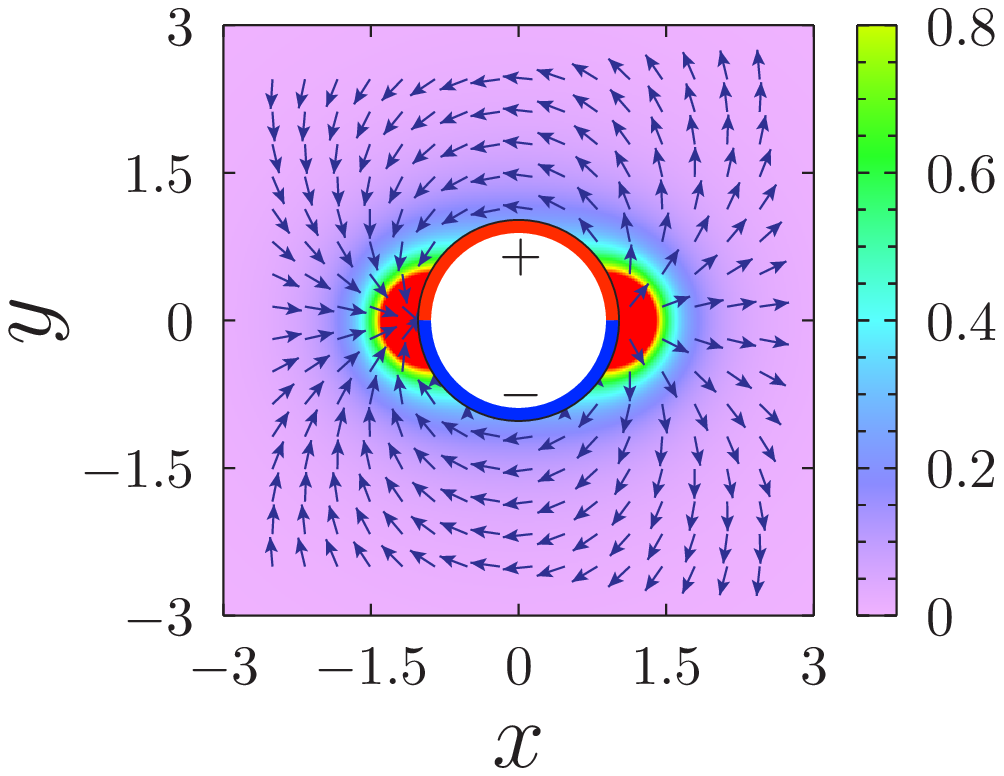}
            \caption[]%
            {{\small }}    
            \label{plot03_02c}
        \end{subfigure}
\caption[]
{\small Stress tensor profile around the circular, symmetric Janus particle. (a) The normalized eigenvectors $\widehat{\1{e}}_{+}(x,y)$ and $\widehat{\1{e}}_{-}(x,y)$ of the stress tensor corresponding to the positive and negative eigenvalue, in red and blue, respectively. (b) The normalized eigenvector $\widehat{\1{e}}_{+}(x,y)$ (blue arrows). (c) The normalized eigenvector $\widehat{\1{e}}_{-}(x,y)$ (blue arrows). The color plot in the background of panels (b) and (c) corresponds to the positive eigenvalue $\lambda_{\textrm{J}}(x,y)$. Like in the corresponding geometry of the upper half $w$-plane (Fig. \ref{fig03_01}(a)), along the particle surface and away from the two switching points the eigenvalue $\lambda_{\textrm{J}}$ is finite and $\widehat{\1{e}}_{+}$ is perpendicular to the particle surface, while upon approaching the switching points in radial direction, $\lambda_{\textrm{J}}$ diverges and $\widehat{\1{e}}_{+}$ is tangential to the particle surface. We mark the divergence of $\lambda_{\rm J}$, by fixing the corresponding color code in (b) and (c) to red.}
\label{plot03_02}
\end{figure*}

In order to obtain results for arbitrary points $(x,y)$ outside the particle $\textrm{J}$ we have calculated the Cartesian components
\begin{equation}
\label{03_15}
\langle T_{xx}(x,y) \rangle_{\textrm{J}} = - \frac{2}{\pi} \frac{1-2r^2 \cos2\theta+r^4 \cos4\theta}{\left(1-2 r^2\cos2\theta+r^4\right)^2}
\end{equation}
and
\begin{equation}
\label{03_16}
\langle T_{xy}(x,y) \rangle_{\textrm{J}} = - \frac{4}{\pi} \frac{\left(r^2 \cos2\theta-1\right)r^2\sin2\theta}{\left(1-2 r^2\cos2\theta+r^4\right)^2}
\end{equation}
by using Eqs. (\ref{cartcompstress}) and (\ref{03_12}) with $x=r\cos\theta$ and $y=r\sin\theta$. For the positive eigenvalues $\lambda_+ (x,y)$ and the corresponding normalized eigenvectors $\widehat{\1{e}}_{+}(x,y)$, this yields 
\begin{equation}
\label{03_14a}
\lambda_{+}(x,y) = \frac{2}{\pi(1-2r^{2}\cos2\theta+r^{4})} \, ,
\end{equation}
which is consistent with Eq. (\ref{Teigenval}),  and
\begin{equation}
\label{03_14c}
\widehat{\1{e}}_{+}(x,y) = \frac{\left( -r^{2}\sin2\theta, r^{2}\cos2\theta-1 \right)}{\sqrt{1-2r^{2}\cos2\theta+r^{4}}} \, ,
\end{equation}
in terms of the $x$ and $y$ components and in agreement with the previous notation $(x,y) = r (\cos \theta, \, \sin \theta)$. The eigenvectors $\widehat{\1{e}}_{-}(x,y)$ belonging to the negative eigenvalue $\lambda_- =-\lambda_+$ are perpendicular to $\widehat{\1{e}}_{+}(x,y)$. The eigenvectors and eigenvalues are visualized in Fig. \ref{plot03_02}.

\subsection{Janus needle}
\label{App_needle}
We now turn to the Janus needle ${\rm Jn}$ shown in Fig. \ref{fig02_01}(d) and take it to extend from $x=-1$ to $x=1$ on the real axis of the $z$ plane. Here the two boundary segments with boundary conditions + and $-$ are the upper and lower rims of the needle, respectively. Again, there is a conformal transformation $w(z)$ which maps ${\rm Jn}$ onto $\mathbb{H}_{-+}$, i.e., onto the upper half $w$-plane with boundary conditions + and $-$ along the positive and the negative real axis, respectively.

This function $w(z)$ is the analytic continuation of
\begin{equation}
\label{App_needle_map}
w(z) = i \sqrt{\frac{z+1}{z-1}}
\end{equation}
from $z = x > 1$ to the entire $z$-plane which is cut along the needle. For example, the regions
\begin{equation}
\label{App_needle_01}
z = \bigl\{ x>1 \, ; \, x< -1 \, ; \, iy \, ; \, -1<x<1 \, (\textrm{upper rim}) \, ; \, -1<x<1 \, (\textrm{lower rim}) \bigr\}
\end{equation}
in the $z$-plane are mapped to the regions
\begin{eqnarray} \nonumber
\label{App_needle_02}
w(z) & = & \bigl\{ i\sqrt{(x+1)/(x-1)} \, ; \, i\sqrt{(x+1)/(x-1)} \, ; \, (\textrm{sign}(y) + i|y|)/\sqrt{1+y^{2}} \, ; \\
& & \sqrt{(1+x)/(1-x)} \, ; \, - \sqrt{(1+x)/(1-x)} \bigr\}
\end{eqnarray}
in the upper half $w$-plane, respectively; all square roots are positive. Correspondingly, $\textrm{d}w/\textrm{d}z$ is the analytic continuation of $-w(z)/(z^2 -1)$ from $z=x>1$ to the entire $z$-plane which is cut along the needle.

\subsubsection{Profiles of $\Phi$ and $\varepsilon$}
\label{phiepsnJ}
Similar to Sec. \ref{symJanus} the profiles of ${\cal O}=\Phi$, $\varepsilon$ for the Janus needle are determined via the transformation formula in Eq. (\ref{03_08}) and the half plane expressions in Eqs. (\ref{03_05}) and (\ref{03_06}) by using the above relations to express $|\textrm{d}w/\textrm{d}z|$, $v$, and $\cos \theta_w$ in terms of $x$ and $y$.

For the needle particle with {\it homogeneous} boundary condition $+$, which we denote as ``needle,+'' we obtain, by using
\begin{eqnarray} \nonumber
\label{App_needle_03}
v & = & \textrm{Im}(w) = \bigl\{ \sqrt{(x+1)/(x-1)} \, ; \, \sqrt{(x+1)/(x-1)} \, ; \, |y|/\sqrt{1+y^{2}} \, ; \\
& & |y|/[\sqrt{1+x}(\sqrt{1-x})^{3}] \, ; \, |y|/[\sqrt{1+x}(\sqrt{1-x})^{3}] \bigr\} \, ,
\end{eqnarray}
the result
\begin{eqnarray} \nonumber
\label{App_needle_04}
\langle \mathcal{O}(z,\bar{z}) \rangle_{\textrm{needle},+} / \mathcal{A}_{\mathcal{O}}^{(+)} & = & \bigl\{ \left( x^{2}-1 \right)^{-x_{\mathcal{O}}} \, ; \, \left( x^{2}-1 \right)^{-x_{\mathcal{O}}} \, ; \, \left( |y|\sqrt{1+y^{2}} \right)^{-x_{\mathcal{O}}} \, ; \\
& & |y|^{-x_{\mathcal{O}}} \, ; \, |y|^{-x_{\mathcal{O}}} \bigr\} \, .
\end{eqnarray}
In the above regions of the $z$ plane we obtain for the Janus needle ${\rm Jn}$, by using the relations
\begin{equation}
\label{App_needle_05}
\cos\theta_{w} = \frac{\textrm{Re}(w)}{|w|} = \big\{ 0 \, ; 0 \, ; \, \textrm{sign}(y)/\sqrt{1+y^{2}} \, ; 1 \, ; -1 \bigr\} \, ,
\end{equation}
the profiles
\begin{equation}
\label{App_needle_06}
\langle \Phi(z,\bar{z}) \rangle_{\textrm{Jn}} / \mathcal{A}_{\Phi}^{(+)} = \big\{ 0 \, ; 0 \, ; \textrm{sign}(y)|y|^{-1/8}\left( \sqrt{1+y^{2}} \right)^{-9/8} \, ; |y|^{-1/8} \, ; -|y|^{-1/8} \bigr\}
\end{equation}
and
\begin{equation}
\label{App_needle_07}
\langle \varepsilon(z,\bar{z}) \rangle_{\textrm{Jn}} / \mathcal{A}_{\varepsilon}^{(+)} = \bigg\{ -\frac{3}{x^{2}-1} \, ; -\frac{3}{x^{2}-1} \, ; \frac{1-3y^{2}}{|y|\left( \sqrt{1+y^{2}} \right)^{3}} \, ; |y|^{-1} \, ; |y|^{-1} \bigg\} \, .
\end{equation}
These expressions for the five regions of the $z$ plane (see Eq. (\ref{App_needle_01})) are exact and instructive.

For our later construction of the SPOE we need the far-field behavior for $z$ pointing in an {\it arbitrary} direction. This is obtained by using the transformation law given in Eq. (\ref{03_08}) in which the conformal map given by Eq. (\ref{App_needle_map}), $w(z) =i \sqrt{1+[2/(z-1)]}$, is expanded for large $z$. This leads to
\begin{equation}
\label{App_needle_06bis}
\langle \Phi(z,\bar{z}) \rangle_{\textrm{Jn}} = -2^{1/8} |z|^{-1/4} \frac{z^{-1} - \bar{z}^{-1}}{2i} \biggl[ 1 + \frac{9}{64} \left( 2|z|^{-2} + 3 \left( z^{-2} + \bar{z}^{-2} \right) \right) \biggr]
\end{equation}
and
\begin{equation}
\label{App_needle_07bis}
\langle \varepsilon(z,\bar{z}) \rangle_{\textrm{Jn}} = \frac{3}{2} |z|^{-2} \biggl[ 1 + \frac{1}{24} \left( -10|z|^{-2} + 17 \left( z^{-2} + \bar{z}^{-2} \right) \right) \biggr] \, ,
\end{equation}
in leading and next-to-leading order. One may check that Eqs. (\ref{App_needle_06bis}) and (\ref{App_needle_07bis}) are consistent with Eqs. (\ref{App_needle_06}) and (\ref{App_needle_07}), respectively. The corresponding far-field expressions for the needle with a homogeneous boundary condition (``needle,+'') are given in Eq. (\ref{cf_009}).

\subsubsection{Stress tensor}
\label{stressnJ}
The profiles of the complex stress tensor follow from the transformation law in Eq. (\ref{stresstrafo}):
\begin{equation}
\label{App_needle_08}
\langle T(z) \rangle_{\textrm{needle},+} = \frac{1}{16} \frac{1}{\left( z^{2}-1 \right)^{2}} \, , \qquad \langle T(z) \rangle_{\textrm{Jn}} = \frac{9}{16} \frac{1}{\left( z^{2}-1 \right)^{2}} \, .
\end{equation}
For the needle with uniform $+$ boundary conditions on both rims the corresponding half plane boundary is homogeneous so that $\langle T(w) \rangle_{\mathbb{H}_+}$ vanishes - leaving only the Schwarzian contribution. For the Janus needle of Fig. \ref{fig02_01}(d) the stress profile of the corresponding $-+$ half plane boundary is nonvanishing and given by Eq. (\ref{TH-+}) so that both terms in Eq. (\ref{stresstrafo}) contribute to the leading behavior of the profiles in Eq. (\ref{App_needle_08}) which differ by a factor of $9$.

On the needle boundaries ($-1<x<1, \, y= \pm 0$), on the positive and negative $y$ semiaxis emanating from the centers ($x=0, y=\pm 0$) of the + and $-$ segments, and on the two semi-infinite lines on the $x$ axis emanating in radial directions from the two switching points $x=\pm1$, $\langle T(z) \rangle_{\textrm{Jn}}$ is real and thus $\langle T_{xy} \rangle_{\textrm{Jn}}$ vanishes and the eigenvectors point into the $x$ and $y$ directions. The eigenvalue $\langle T_{yy} \rangle_{\textrm{Jn}}$, associated with the eigenvector perpendicular to the boundaries of the $\textrm{Jn}$, is positive and equals $9/[16 \pi (1-x^2)^2]$. The eigenvalues $\langle T_{yy} \rangle_{\textrm{Jn}}$ and $\langle T_{xx} \rangle_{\textrm{Jn}}$ of the radial eigenvectors on the $y$ and $x$ axes are given by $9/[16 \pi (y^2 +1)^2]$ and $ - 9/[16 \pi (x^2 -1)^2]$, respectively. The qualitative features of these results for the Janus needle $\rm{Jn}$ resemble those of the circular Janus particle $\rm{J}$ discussed between Eqs. (\ref{03_12}) and (\ref{03_16}).

\subsection{Generalized Janus particle}
\label{genJanus}
The generalized Janus particle ${\rm P}={\rm J}_\chi$ of Fig. \ref{fig02_01}(b) has a circular shape and its two boundary segments + and $-$ extend along the angular intervals $-\chi < \theta < \pi+\chi$ and $\pi+\chi < \theta < 2\pi -\chi$, respectively, and we consider it for various values of $\chi$ within the interval $-\pi/2 < \chi < \pi/2$. For $\chi=0$ the particle degenerates to the Janus particle shown in Fig. \ref{fig02_01}(a) and for $\chi=\pi/2$ and $\chi=- \pi/2$ to the circular particles with homogeneous boundaries + and $-$, respectively. For all $\chi$ the particle is symmetric about the $y$ axis. Moreover, the particle configurations remain invariant under the combined changes
\begin{eqnarray} \label{combchange}
y \to -y \, , \quad \chi \to -\chi \, , \quad + \longleftrightarrow - \, \; \; {\rm on \, the \, boundary} \, ,
\end{eqnarray}
which are the counterpart of the antisymmetry about the $x$ axis of the Janus particle ${\rm P}={\rm J}$ of Fig. \ref{fig02_01}(a).

In order to calculate the profiles it is convenient to use instead of Eq. (\ref{03_02}) the conformal transformation
\begin{equation}
\label{composite}
w(z) = \tan \chi + i\frac{z\textrm{e}^{i \chi}+1}{z\textrm{e}^{i \chi}-1} \, , \qquad  \frac{\textrm{d}w(z)}{\textrm{d}z} = -\frac{2i}{(z\textrm{e}^{i \chi}-1)^{2}} \textrm{e}^{i\chi} \,
\end{equation}
in which the M$\ddot{\textrm{o}}$bius transformation (Eq. (\ref{03_02})) is preceded by a rotation of the particle and followed by a translation along the real axis of $\mathbb{H}$. This maps the two boundary segments of ${\rm J}_\chi$ with + and $-$ boundary conditions onto the positive and negative real axis of $\mathbb{H}$. Indeed, Eq. (\ref{composite}) implies that
\begin{equation}
\label{thetacomposite}
w(z=\textrm{e}^{ i\theta})=\tan \chi + \cot[(\chi+\theta)/2]
\end{equation}
(compare with Eq. (\ref{circby})) and in particular that $w\Bigl(z=\exp[i(-\chi \pm 0)] \Bigr)=\pm \infty$ and $w\Bigl(z=\exp[i(\pi+\chi)] \Bigr) =0$. The advantage of this approach is that the system $\mathbb{H}_{\cal P}$ on the right hand side of Eqs. (\ref{03_08}) and (\ref{stresstrafo}) is again $\mathbb{H}_{-+}$ as in Sec. \ref{symJanus} with the corresponding profiles given by Eqs. (\ref{03_05}), (\ref{03_06}), and (\ref{TH-+}).   

\subsubsection{Profiles of $\Phi$ and $\varepsilon$}
\label{phiepsgen} 
Since also the mapping according to Eq. (\ref{composite}) implies that ${\cal A}_{\cal O} (|\textrm{d}w/\textrm{d}z| /v)^{x_{\cal O}}$ equals the result $\langle {\cal O}(z,\bar{z}) \rangle_{+ \rm circle}$ for a {\it homogeneous} circle (given below Eq. (\ref{03_10})), the scale transformation given by Eqs. (\ref{03_08}), (\ref{03_05}), and (\ref{03_06}) yields
\begin{equation}
\label{04_04}
\langle \Phi(z,\bar{z}) \rangle_{\textrm{J}_{\chi}} = \langle \Phi(z,\bar{z}) \rangle_{ \rm circle,+ } \, \mathcal{C}
\end{equation}
and
\begin{equation}
\label{04_05}
\langle \varepsilon(z,\bar{z}) \rangle_{\textrm{J}_{\chi}} = \langle \varepsilon(z,\bar{z}) \rangle_{ \rm circle,+ } \, \left( 4\mathcal{C}^{2}-3 \right) \, ,
\end{equation}
where $\mathcal{C}=\mathcal{C}(\chi;x,y)$ is given by
\begin{equation}
\label{04_06}
\mathcal{C}(\chi;x,y) = \cos\theta_{w} \equiv {{\rm Re}\, w \over |w|} = \frac{2y + (x^{2}+y^{2}+1)\sin\chi}{\sqrt{[ 2y + (x^{2}+y^{2}+1)\sin\chi ]^{2} + (x^{2}+y^{2}-1)^{2}\cos^{2}\chi}}
\end{equation}
upon using Eq. (\ref{composite}).

As expected, the profiles in Eqs. (\ref{04_04}) and (\ref{04_05}) exhibit all symmetry properties discussed before Eq. (\ref{composite}); and their near surface behavior shows the same features as discussed for the Janus particle at the end of Sec. \ref{phiepsJ}. For various angles $\chi$ they are displayed in Fig. \ref{plot02_01three} and Fig. \ref{plot02_02}, respectively. As expected, the energy density strongly increases near the switches in the boundary condition.

\subsubsection{Stress tensor}
The complex stress tensor profile around a generalized Janus particle follows from the profile in $\mathbb{H}_{-+}$ (see Eq. (\ref{TH-+})) and from the transformation in Eq. (\ref{stresstrafo}). Using $\textrm{d}w/\textrm{d}z$ from Eq. (\ref{composite}) and realizing that the mapping in Eq. (\ref{composite}) can also be written as $w(z)=i\{[z+\exp (i\chi)]/[z-\exp (-i\chi)]\} / [\exp(i\chi) \cos \chi]$  yields
\begin{equation}
\label{04_07}
\langle T(z) \rangle_{\textrm{J}_{\chi}} = \frac{2\cos^{2}\chi}{\left( z-\textrm{e}^{-i\chi} \right)^{2} \left( z+\textrm{e}^{i\chi} \right)^{2}} \, ,
\end{equation}
which diverges at the switching points $z=\pm \exp{(\mp i \chi)}$ and reduces to Eq. (\ref{03_12}) for $\chi=0$. For a uniform $\pm$ boundary one has $\chi=\pm\pi/2$, respectively, and, as expected, the expression in Eq. (\ref{04_07}) vanishes.

\begin{figure*}[htbp]
\centering
        \begin{subfigure}[b]{0.285\textwidth}
            \centering
            \includegraphics[width=\textwidth]{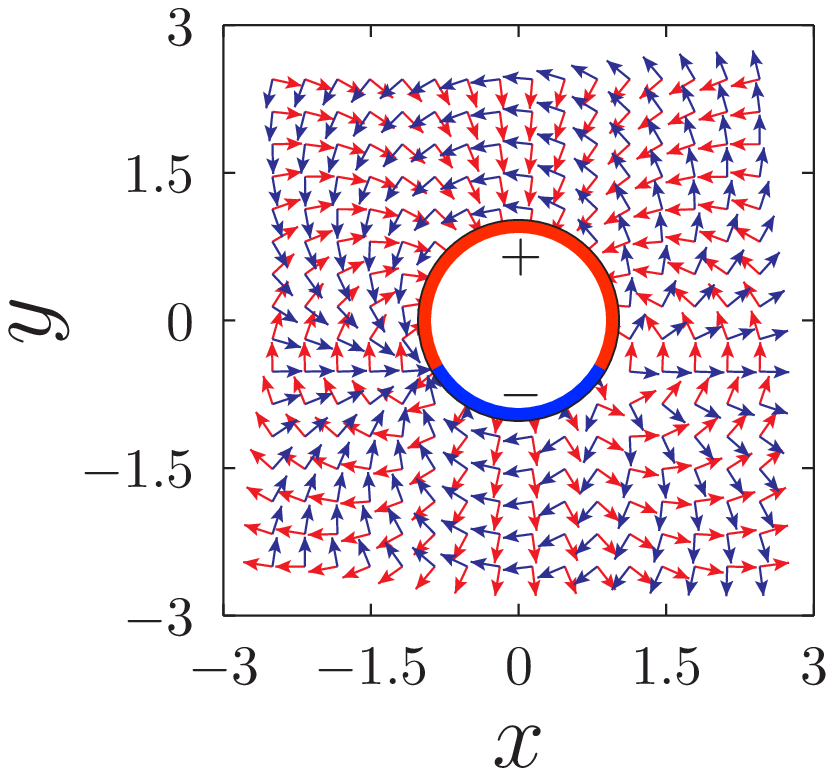}
            \caption[]%
            {{\small }}    
            \label{plot04_05a}
        \end{subfigure}
\hfill
        \begin{subfigure}[b]{0.345\textwidth}  
            \centering 
            \includegraphics[width=\textwidth]{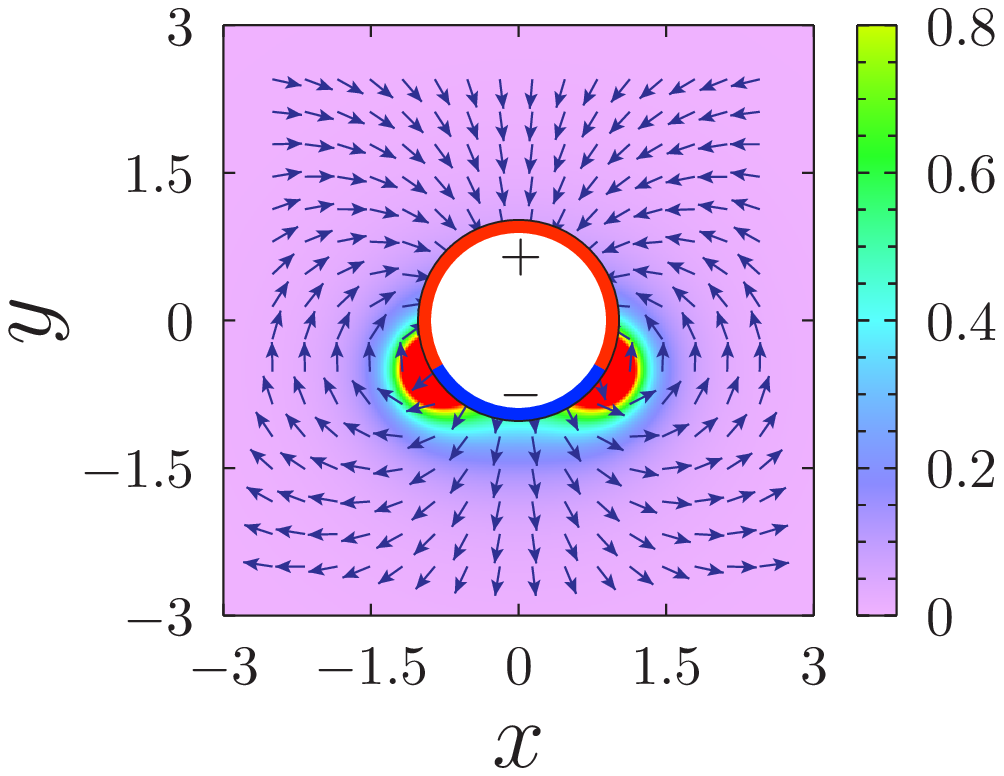}
            \caption[]%
            {{\small }}    
            \label{plot04_05b}
        \end{subfigure}
\hfill
        \begin{subfigure}[b]{0.345\textwidth}  
            \centering 
            \includegraphics[width=\textwidth]{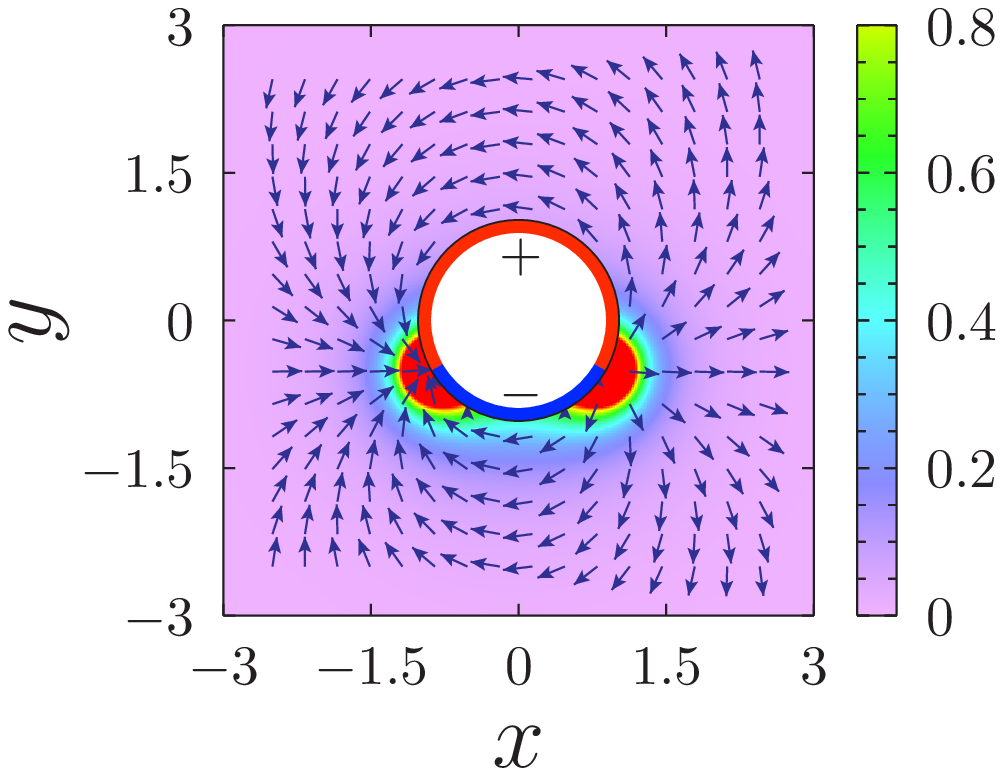}
            \caption[]%
            {{\small }}    
            \label{plot04_05c}
        \end{subfigure}
\vskip\baselineskip
\caption[]
{\small Stress tensor profile around a generalized circular Janus particle. The arrangement and the meaning of the panels are the same as in Fig. \ref{plot03_02} and the results shown correspond to an opening angle $2\pi/3$ of the blue segment.}
\label{plot04_05}
\end{figure*}

Proceeding for the Cartesian stress tensor as in Eqs. (\ref{03_15})-(\ref{03_14c}) we find
\begin{equation}
\label{gen_Janus_eigenvalue}
\lambda_{\textrm{J}_{\chi}}(\chi,x,y) = \frac{2 \cos^{2}\chi}{\pi  \Bigl[ \left(r^2+2 r \cos (\theta -\chi )+1\right) \left(r^2-2 r \cos (\theta +\chi )+1\right) \Bigr]}
\end{equation}
for its positive eigenvalue and
\begin{equation}
\widehat{\1{e}}_{+}^{(\chi)}(x,y) = \frac{1}{M}\left( -2r\cos\theta\left( r\sin\theta+\sin\chi \right) , r^{2}\cos2\theta-2r\sin\theta\sin\chi-1 \right)
\end{equation}
with
\begin{equation}
\label{ }
M = \sqrt{\left(r^2+2 r \cos (\theta -\chi )+1\right) \left(r^2-2 r \cos (\theta +\chi )+1\right)}
\end{equation}
for the corresponding normalized eigenvector. The field of the normalized eigenvectors is displayed in Fig. \ref{plot04_05}.

\subsection{Quadrupoles}
\label{quadrupoles}
Within the class of particles with a circular shape and inhomogeneous boundary conditions the Janus particle of Fig. \ref{fig02_01}(a) is the simplest representative. Here we go one step further and consider the spherical particle $\rm P=\rm Q$ of Fig. \ref{fig02_01}(c) with boundary conditions switching four times so that it exhibits certain features of a quadrupole.

\subsubsection{Profiles of $\Phi$ and $\varepsilon$}
These profiles are determined by following the procedure described in the introduction of Sec. \ref{single} by using the M$\ddot{\textrm{o}}$bius mapping (Eq. (\ref{03_02})). The switching points $\theta=0,\pi/2,\pi,$ and $3\pi/2$ of $\textrm{Q}$ (Fig. \ref{fig02_01} (c)) are transferred by means of Eq. (\ref{circby}) to those at the half plane boundary yielding the pattern $\cal P$ with $-,+,-,+$ along the intervals $-\infty<u< -1$, $-1<u<0$, $0<u<1$, and $1<u< +\infty$. This half plane system is denoted as ${\mathbb{H}_{-+-+}}$. The corresponding profiles of $\Phi$ and $\varepsilon$ are given by Eqs. (16a), (16b), and (17) in Ref. \cite{BG_93} in terms of Pfaffians of a $4 \times 4$ and a $6 \times 6$ matrix, with the results 
\begin{equation}
\label{07_01}
\langle \Phi(w,\bar{w}) \rangle_{\mathbb{H}_{-+-+}} = -\left( \frac{2}{v} \right)^{1/8} \frac{1-u^{2}-v^{2}}{|1-w^{2}|} \frac{u}{\sqrt{u^{2}+v^{2}}} \, 
\end{equation}
and
\begin{eqnarray} \nonumber
\label{07_02}
\langle \varepsilon(w,\bar{w}) \rangle_{\mathbb{H}_{-+-+}} & = & -\frac{1}{2v} + \frac{4v}{3} \biggl\{ \frac{1}{(u-1)^{2}+v^{2}} + \frac{1}{(u+1)^{2}+v^{2}} + \\
& + & \frac{1}{u^{2}+v^{2}} \biggl[ -\frac{1}{2} +  \frac{1}{(u-1)^{2}+v^{2}} + \frac{1}{(u+1)^{2}+v^{2}} \biggr] - \frac{2}{|1-w^{2}|^{2}} \biggr\} \,  
\end{eqnarray}
where 
\begin{equation}
\label{07_03}
|1-w^{2}| = \sqrt{(1+v^{2})^{2} - 2u^{2} (1-v^{2}) + u^{4}} \, .
\end{equation} 
Together with the relations
\begin{equation}
\label{07_04}
(u,v) = \frac{(2y,x^{2}+y^{2}-1)}{(x-1)^{2}+y^{2}} \, ,
\end{equation}
which follow from the M$\ddot{\textrm{o}}$bius mapping (Eq. (\ref{03_02})), the scale transformation in Eq. (\ref{03_08}) yields for the order parameter
\begin{equation}
\label{07_05}
\langle \Phi(z,\bar{z}) \rangle_{\textrm{Q}} = \langle \Phi(z,\bar{z}) \rangle_{\rm circle,+} \frac{4xy}{|z^{4}-1|} \, 
\end{equation}
with $\langle \Phi(z,\bar{z}) \rangle_{\rm circle,+}$ as defined below Eq. (\ref{03_10}). For the energy density one has
\begin{eqnarray}
\label{07_05bis}
\langle \varepsilon(z,\bar{z}) \rangle_{\textrm{Q}} & = & -\frac{1}{|z|^{2}-1} - \frac{4}{3}\left( |z|^{2}-1 \right) \left( \frac{1}{|z^{2}-1|^{2}} + \frac{1}{|z^{2}+1|^{2}} \right) + \frac{2\delta}{3} \, , \\ \nonumber
\delta & \equiv & (1+i)\biggl[ \left( \frac{1}{(z-1)(\bar{z}+i)}+\frac{1}{(z+i)(\bar{z}-1)} \right) + \left( (z,\bar{z}) \mapsto (-z,-\bar{z}) \right) \biggr] + \textrm{cc} \, .
\end{eqnarray} 

We note the expected symmetries under reflection (i) about the particle center, (ii) about the coordinate axes, and (iii) under a rotation by 90 degrees. While Eq. (\ref{07_05bis}) is invariant under all these symmetry operations, Eq. (\ref{07_05}) is invariant under (i) but changes its sign under (ii) and (iii).

Along the coordinate axes the form of Eq. (\ref{07_05bis}) simplifies and reads
\begin{equation}
\label{ }
\langle \varepsilon(z,\bar{z}) \rangle_{\textrm{Q}} \vert_{y=0} = \frac{3}{x^{2}-1} - \frac{4}{3} \frac{x^{2}-1}{\left( x^{2} + 1 \right)^{2}} \, .
\end{equation}
We point towards the leading, diverging behavior $\langle \varepsilon(z,\bar{z}) \rangle_{\textrm{Q}} \vert_{y=0} \rightarrow \frac{3}{2(x-1)}$, as $x \searrow 1$ approaches the switching point at $z=1$.

Expanding Eqs. (\ref{07_05}) and (\ref{07_05bis}) for large distances $r=\sqrt{x^{2}+y^{2}}\gg1$ to leading and next-to-leading order yields
\begin{equation}
\label{07_06}
\langle \Phi(z,\bar{z}) \rangle_{\textrm{Q}} = 2^{9/4} \frac{xy}{(x^{2}+y^{2})^{17/8}} \left( 1 + \frac{1}{8 \left( x^{2}+y^{2} \right) } \right) + O\left(r^{-25/4} \right)
\end{equation}
and
\begin{equation}
\label{07_07}
\langle \varepsilon(z,\bar{z}) \rangle_{\textrm{Q}} = \frac{5}{3} \frac{1}{x^{2}+y^{2}} + \frac{7}{ \left( x^{2}+y^{2} \right)^{2} } - \frac{11 x^4+150 x^2 y^2+11 y^4}{3 \left(x^2+y^2\right)^5} +  O\left( r^{-8} \right) \, .
\end{equation}

\begin{figure*}[htbp]
\centering
        \begin{subfigure}[htbp]{0.48\textwidth}
            \centering
            \includegraphics[width=\textwidth]{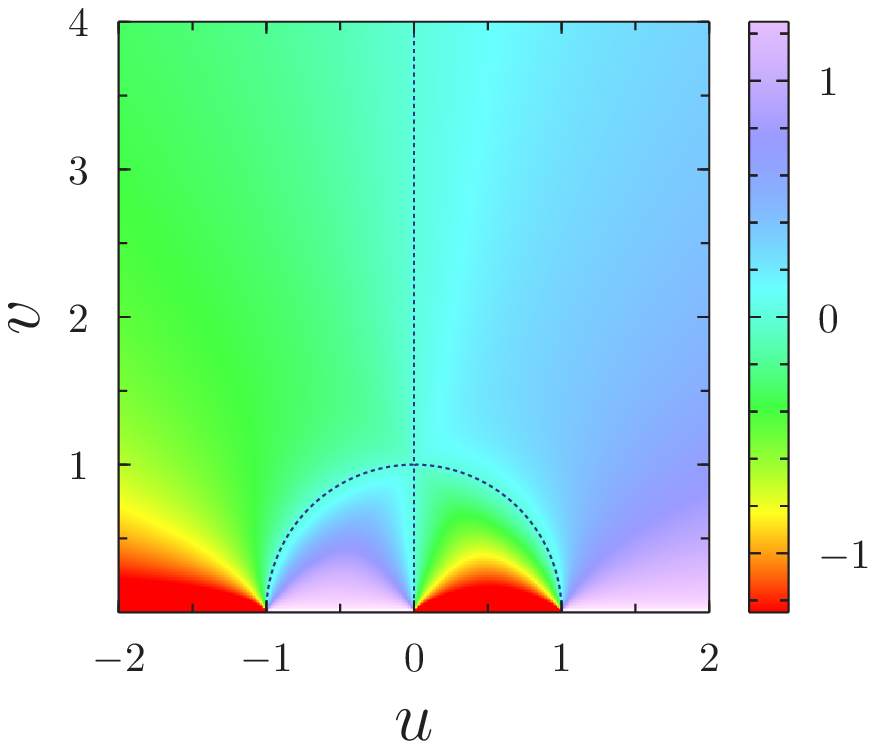}
            \caption[]%
            {{\small }}    
            \label{plot07_01a}
        \end{subfigure}
\hfill
        \begin{subfigure}[htbp]{0.51\textwidth}  
            \centering 
            \includegraphics[width=\textwidth]{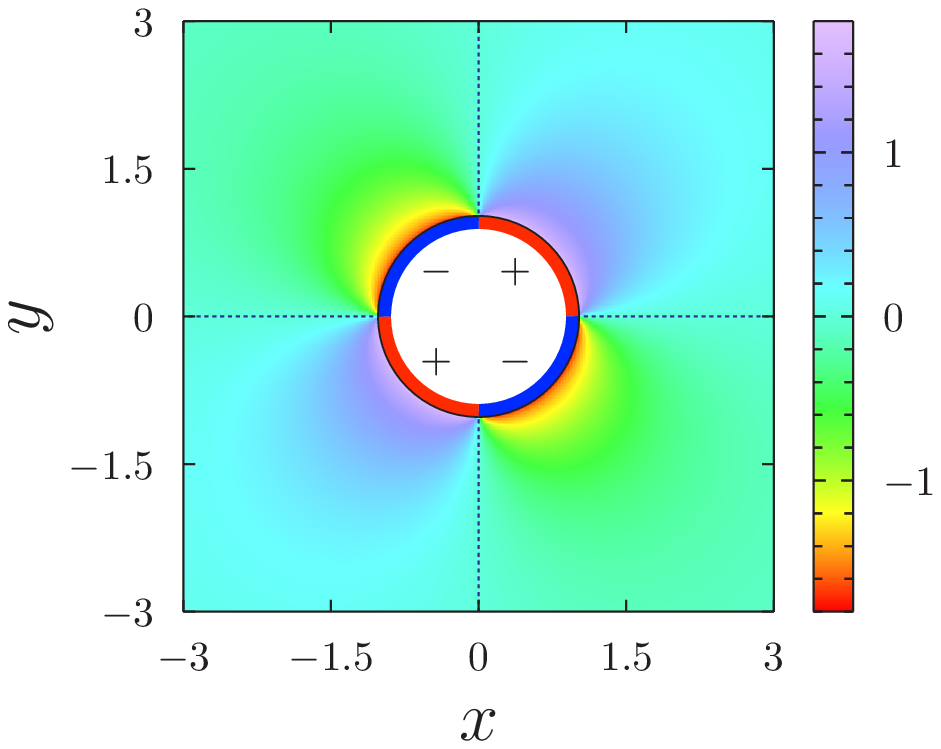}
            \caption[]%
            {{\small }}    
            \label{plot07_01b}
        \end{subfigure}
\vskip\baselineskip
\caption[]
{\small Order parameter profile $\langle\Phi(z,\bar{z})\rangle_{\textrm{Q}}$ around the quadrupole particle (b) and the corresponding profile $\langle\Phi(w,\bar{w})\rangle_{\mathbb{H}_{-+-+}}$ in the upper half $w$-plane (a). The order parameter vanishes along the dotted blue lines and diverges upon approaching the real axis in (a) and the particle surface in (b). Upon approaching divergences, the color code is fixed to the one corresponding to the maximum value available in the legend.}
\label{plot07_01}
\end{figure*}

\begin{figure*}[htbp]
\centering
        \begin{subfigure}[htbpb]{0.48\textwidth}
            \centering
            \includegraphics[width=\textwidth]{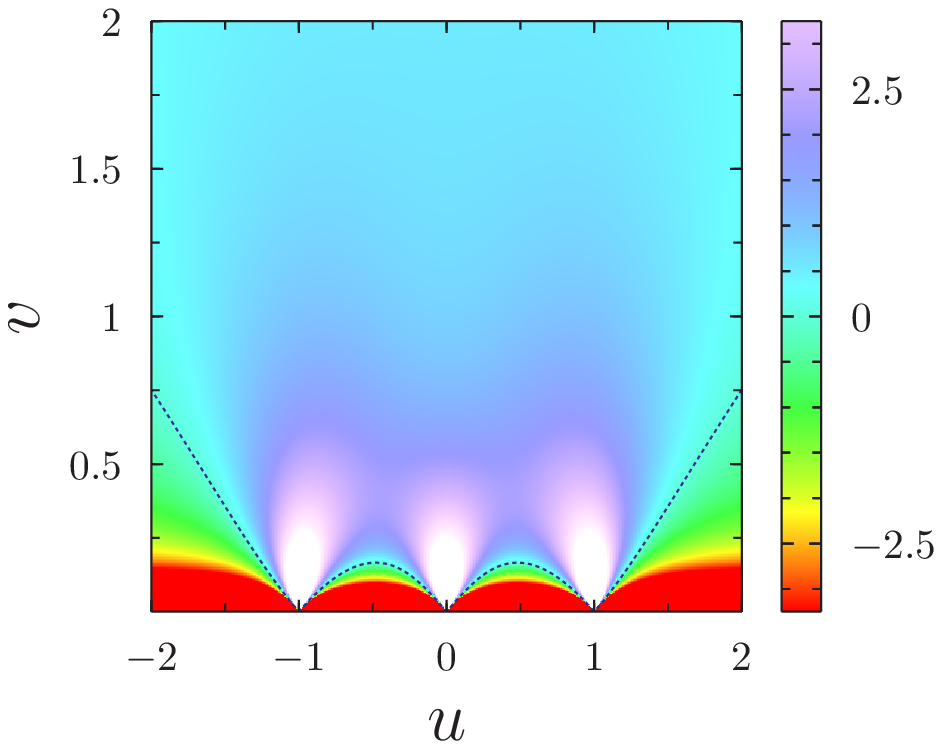}
            \caption[]%
            {{\small }}    
            \label{plot07_02a}
        \end{subfigure}
\hfill
        \begin{subfigure}[htbp]{0.51\textwidth}  
            \centering 
            \includegraphics[width=\textwidth]{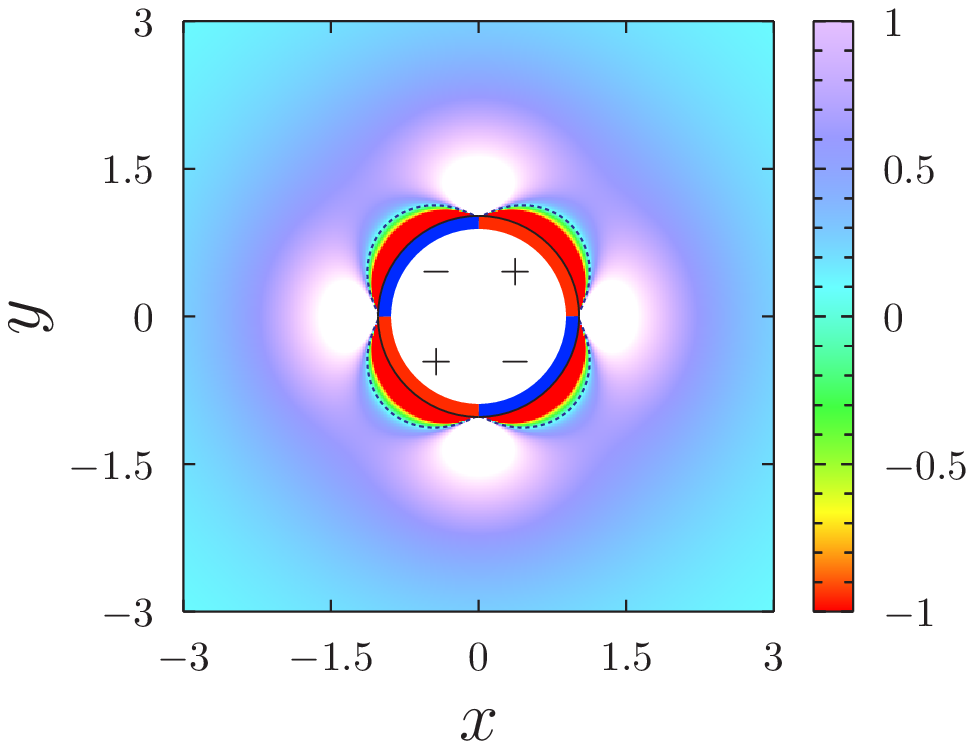}
            \caption[]%
            {{\small }}    
            \label{plot07_02b}
        \end{subfigure}
\vskip\baselineskip
\caption[]
{\small Energy density profile $\langle\varepsilon(z,\bar{z})\rangle_{\textrm{Q}}$ around a quadrupole particle (b) and the corresponding profile $\langle\varepsilon(w,\bar{w})\rangle_{\mathbb{H}_{-+-+}}$ in the upper half $w$-plane (a). The energy density vanishes along the dotted blue lines and diverges towards $+\infty$ at the switching points, as indicated by the white regions. The red regions close to the real axis in (a) and close to the particle surface in (b) correspond to the divergence towards $-\infty$ of the energy density. There, the color code is used as indicated in Fig. \ref{plot02_02}.}

\label{plot07_02}
\end{figure*}

The profiles $\langle \Phi \rangle_{\mathbb{H}_{-+-+}}$, $\langle \Phi \rangle_{\rm Q}$, $\langle \varepsilon \rangle_{\mathbb{H}_{-+-+}}$, and $ \langle \varepsilon \rangle_{\rm Q}$ in Eqs. (\ref{07_01}), (\ref{07_05}), (\ref{07_02}), and (\ref{07_05bis}) are visualized in Figs. \ref{plot07_01}(a) and \ref{plot07_01}(b) and in Figs. \ref{plot07_02}(a) and \ref{plot07_02}(b), respectively.

\newpage

\subsubsection{Stress tensor}
The result
\begin{equation}
\label{07_08}
\langle T(w) \rangle_{\mathbb{H}_{-+-+}} = \frac{1}{2w^{2}} \left( \frac{w^{2}+1}{w^{2}-1} \right)^{2} \, 
\end{equation}
for the stress tensor profile in the half plane follows from Eq. (D2) in Ref. \cite{EB_16} and yields, via Eqs. (\ref{stresstrafo}) and (\ref{03_02}), the profile
\begin{equation}
\label{07_09}
\langle T(z) \rangle_{\textrm{Q}} = \frac{8z^{2}}{(z^{4}-1)^{2}}
\end{equation}
outside the quadrupolar particle Q. This should be compared with its counterpart in Eq. (\ref{03_12}) outside the symmetric Janus particle $\textrm{J}$. The decay
\begin{equation}
\label{07_09bis}
\langle T(z) \rangle_{\textrm{Q}} \simeq \frac{8}{z^{6}} \, ,
\end{equation}
for large distances from the quadrupole $\textrm{Q}$ is faster than the corresponding decay $\propto z^{-4}$ implied by Eqs. (\ref{03_12}) and (\ref{04_07})  for the particles J and ${\rm J}_\chi$, respectively, which exhibit a dipolar character.

The discussion of the eigenvectors and eigenvalues for characteristic special cases of the quadrupole in Fig.~\ref{fig02_01}(c) proceeds like in Sec.~\ref{stressJ} for the circular Janus particle $\textrm{J}$. Due to Eq. (\ref{07_09}) the expression inside the square brackets in Eq. (\ref{radtang}) is real, (i) right on the boundary $(x,y)= (\cos \theta, \sin \theta)$, (ii) on the semi-infinite lines emanating from the segment centers $(x,y) =\pm (2^{-1/2}, 2^{-1/2})$ and $(x,y) =\pm (2^{-1/2}, -2^{-1/2})$ in radial directions, and (iii) on the semi-infinite lines on the $x$ and $y$ axis outside the particle, which emanate from the switching points $(x,y)=\pm (1,0)$ and $(x,y)=\pm (0,1)$ so that the eigenvectors point into the radial and tangential direction, respectively. The eigenvalues $\langle T_{rr} \rangle_{\textrm{Q}}$ belonging to the radial eigenvectors follow from Eq. (\ref{radtang}) as $2/[\pi \sin^2 (2 \theta)]$ for case (i), $8r^2 /[\pi (r^4 +1)^2]$ for case (ii), and $-8 r^2 /[\pi (r^4 -1)^2]$ for case (iii).

\begin{figure*}[htbp]
\centering
        \begin{subfigure}[b]{0.285\textwidth}
            \centering
            \includegraphics[width=\textwidth]{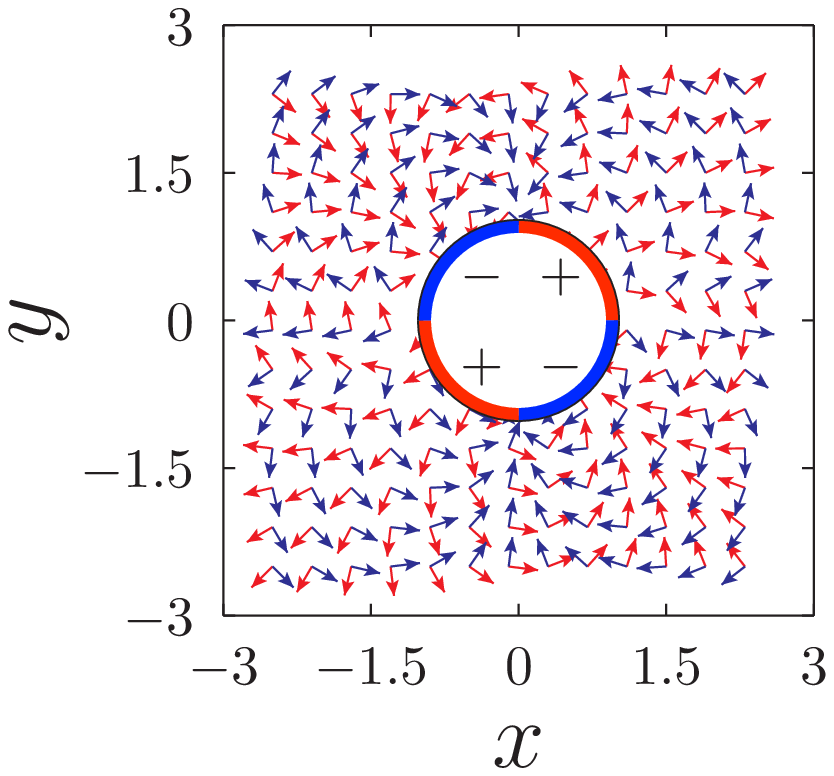}
            \caption[]%
            {{\small }}    
            \label{plot07_05a}
        \end{subfigure}
\hfill
        \begin{subfigure}[b]{0.345\textwidth}  
            \centering 
            \includegraphics[width=\textwidth]{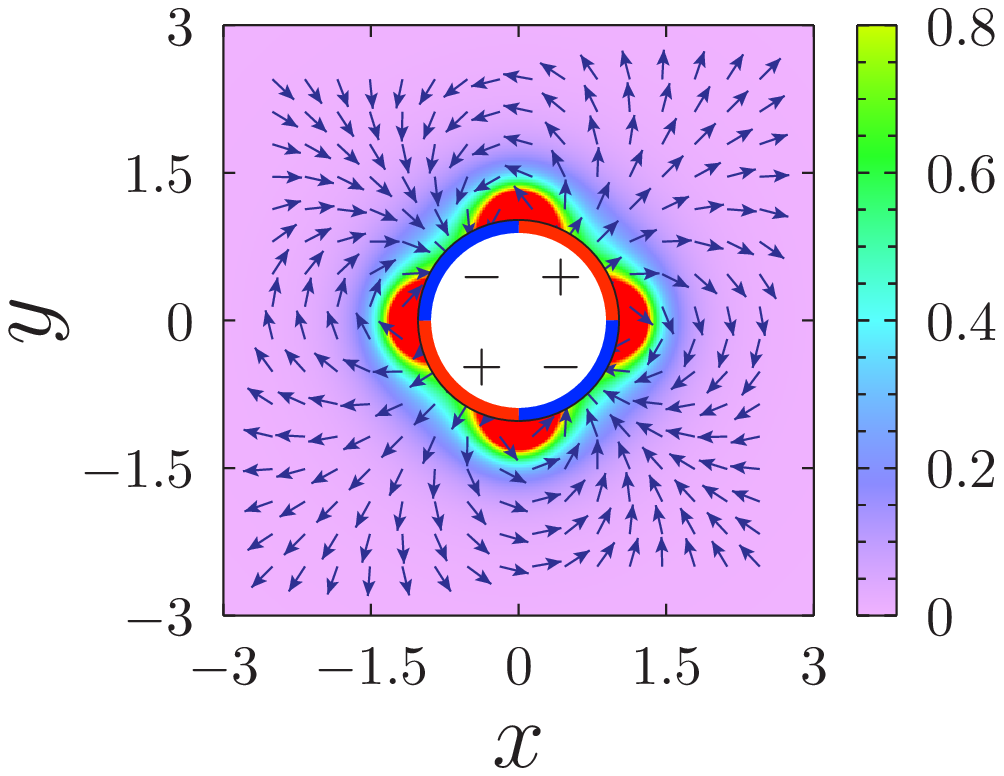}
            \caption[]%
            {{\small }}    
            \label{plot07_05b}
        \end{subfigure}
\hfill
        \begin{subfigure}[b]{0.345\textwidth}  
            \centering 
            \includegraphics[width=\textwidth]{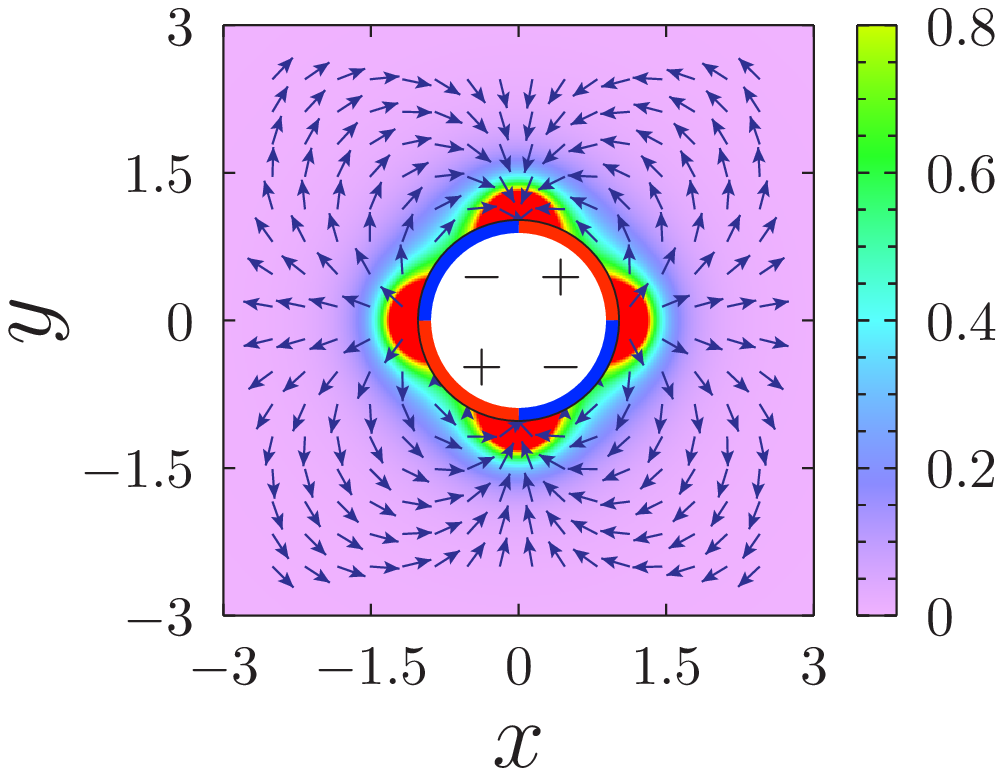}
            \caption[]%
            {{\small }}    
            \label{plot07_05c}
        \end{subfigure}
\vskip\baselineskip
\caption[]
{\small Stress tensor profile for the quadrupolar particle $\textrm{Q}$. The arrangement and meaning of the panels are like those in Figs. \ref{plot03_02} and \ref{plot04_05}.}
\label{plot07_05}
\end{figure*}

For arbitrary points $(x,y)$ we have calculated the Cartesian components
\begin{equation}
\label{TijQ}
\langle T_{xx} \rangle_{\textrm{Q}} = - \frac{8 r^2 \left(r^8 \cos6\theta +\left(1-2 r^4\right) \cos2\theta \right)}{\pi  \left(1-2 r^4 \cos4\theta +r^8\right)^2}
\end{equation}
and
\begin{equation}
\langle T_{xy} \rangle_{\textrm{Q}} = \frac{ 8r^{2} \left( \left(2 r^4+1\right) \sin2\theta -r^{8} \sin6\theta \right) }{\pi  \left(1-2 r^4 \cos4\theta+r^8\right)^2}
\end{equation}
from Eqs. (\ref{cartcompstress}) and (\ref{07_09}), and we find for the positive eigenvalue and its corresponding eigenvector
\begin{equation}
\label{Q_eigenvalue}
\lambda_{\textrm{Q}}(x,y) = \frac{ 8r^{2} }{ \pi \left( 1 - 2r^{4} \cos4\theta + r^{8} \right) }
\end{equation}
and
\begin{eqnarray}
\label{e+Q}
\widehat{\1{e}}_{+}(x,y) & = & \frac{\left( \sin\theta + r^{4}\sin3\theta , \cos\theta - r^{4}\cos3\theta \right)}{\sqrt{1 - 2r^{4} \cos4\theta + r^{8}}} \, ,
\end{eqnarray}
respectively. The vector fields $\widehat{\1{e}}_{\pm}(x,y)$ are visualized in Fig. \ref{plot07_05}. Equations (\ref{TijQ})-(\ref{e+Q}) should be compared with the corresponding expressions in Eqs. (\ref{03_15})-(\ref{03_14c}) for the Janus particle $\textrm{J}$.

\newpage
\section{Small particle operator expansion (SPOE)}
\label{SPOE}
Here we present the SPOEs for circular Janus particles, Janus needles, generalized circular Janus particles, and circular quadrupoles (see Fig. \ref{fig02_01}). They are denoted as ${\rm P}=[{\rm J},  \, {\rm Jn}, \, {\rm J}_\chi, \, {\rm Q}]$, respectively. These expansions make it possible to determine the free energies of interaction between these particles and distant objects. The presence of a small particle $\mathrm{P}$ inserted at the position $(z_{0},\bar{z}_{0})$ and suspended in a critical medium can be regarded, within a coarse-grained picture, as a local modification of the Boltzmann weight:
\begin{equation}
\label{def_SPOE}
\frac{ \textrm{e}^{-\delta\mathcal{H}_{\rm P}} }{ \langle \textrm{e}^{-\delta\mathcal{H}_{\rm P}} \rangle} = 1 + s_{\textrm{P}} \equiv 1 + s_{\rm P}^{(\varepsilon)} + s_{\rm P}^{(\Phi)} + s_{\rm P}^{(\mathbb{I})} \, ,
\end{equation}
where $\delta\mathcal{H}_{\rm P}(z_{0},\bar{z}_{0})$ stands for the effective Hamiltonian, measured in units of $k_{\rm B}T_{\rm c}$, of the particle $\mathrm{P}$. In Eq. (\ref{def_SPOE}) the \emph{s}mall deviation from unity is quantified by $s_{\mathrm{P}}$, with the latter being expanded in terms of a complete basis of operators of the corresponding bulk CFT. For the particles shown in Fig. \ref{fig02_01}, we have the expressions
\begin{eqnarray}
\label{J_spoe_en}
s_{\textrm{J}}^{(\varepsilon)} & = & R \biggl\{ 3 - R^{2} \biggl[ 5 \partial_{z_{0}}\partial_{\bar{z}_{0}} - 2 \left( \textrm{e}^{2i\alpha}\partial_{z_{0}}^{2} + \textrm{e}^{-2i\alpha}\partial_{\bar{z}_{0}}^{2} \right) \biggr] \biggr\} \varepsilon(z_{0},\bar{z}_{0}) + O(R^{5}) \, , \\ \nonumber
\label{J_spoe_phi}
s_{\textrm{J}}^{(\Phi)} & = & 2^{13/4} R^{9/8} i \biggl\{ \left( \textrm{e}^{i\alpha}\partial_{z_{0}} - \textrm{e}^{-i\alpha}\partial_{\bar{z}_{0}} \right) + R^{2} \biggl[ - \frac{11}{7} \left( \textrm{e}^{3i\alpha}L_{-3}-\textrm{e}^{-3i\alpha}\bar{L}_{-3} \right) + \\
& + & \frac{32}{21} \left( \textrm{e}^{3i\alpha} \partial_{z_{0}}^{3} - \textrm{e}^{-3i\alpha} \partial_{\bar{z}_{0}}^{3} \right) - \frac{8}{3} \left( \textrm{e}^{i\alpha}\partial_{z_{0}}^{2}\partial_{\bar{z}_{0}} - \textrm{e}^{-i\alpha}\partial_{\bar{z}_{0}}^{2}\partial_{z_{0}} \right) \biggr] \Phi(z_{0},\bar{z}_{0}) +O\left( R^{5+1/8} \right) , \qquad
\end{eqnarray}
and
\begin{equation}
\label{J_spoe_id}
s_{\textrm{J}}^{(\mathbb{I})} = 8R^{2} \biggl[ \textrm{e}^{2i\alpha}T(z_{0})+\textrm{e}^{-2i\alpha}\bar{T}(\bar{z}_{0}) \biggr] + O\left( R^{4} \right)
\end{equation}
which hold for symmetric \emph{circular Janus} particles;
\begin{eqnarray}
\label{nJ_spoe_en}
s_{\textrm{Jn}}^{(\varepsilon)} & = & \frac{D}{2} \biggl\{ \frac{3}{2} + \left( \frac{D}{2} \right)^{2} \biggl[ -\frac{5}{8} \partial_{z_{0}}\partial_{\bar{z}_{0}} + \frac{17}{32} \left( \textrm{e}^{2i\alpha}\partial_{z_{0}}^{2} + \textrm{e}^{-2i\alpha}\partial_{\bar{z}_{0}}^{2} \right) \biggr] \biggr\} \varepsilon(z_{0},\bar{z}_{0}) + O\left( D^{5} \right) \, , \\ \nonumber
\label{nJ_spoe_phi}
s_{\textrm{Jn}}^{(\Phi)} & = & 2^{17/8} \left( \frac{D}{2} \right)^{9/8} i \biggl\{ \left( \textrm{e}^{i\alpha}\partial_{z_{0}} - \textrm{e}^{-i\alpha}\partial_{\bar{z}_{0}} \right) + \left( \frac{D}{2} \right)^{2} \biggl[ - \frac{9}{14} \left( \textrm{e}^{3i\alpha}L_{-3}-\textrm{e}^{-3i\alpha}\bar{L}_{-3} \right) + \\
& + & \frac{5}{7} \left( \textrm{e}^{3i\alpha} \partial_{z_{0}}^{3} - \textrm{e}^{-3i\alpha} \partial_{\bar{z}_{0}}^{3} \right) - \left( \textrm{e}^{i\alpha}\partial_{z_{0}}^{2}\partial_{\bar{z}_{0}} - \textrm{e}^{-i\alpha}\partial_{\bar{z}_{0}}^{2}\partial_{z_{0}} \right) \biggr] \Phi(z_{0},\bar{z}_{0}) +O\left( D^{5+1/8} \right) \, ,
\end{eqnarray}
and
\begin{equation}
\label{nJ_spoe_id}
s_{\textrm{Jn}}^{(\mathbb{I})} = \frac{9}{4} \left( \frac{D}{2} \right)^{2} \biggl[ \textrm{e}^{2i\alpha}T(z_{0})+\textrm{e}^{-2i\alpha}\bar{T}(\bar{z}_{0}) \biggr] + O\left( D^{4} \right) \, ,
\end{equation}
which hold for Janus \emph{needles};
\begin{eqnarray}
\label{genJ_spoe_en} \nonumber
s_{\textrm{J}_{\chi}}^{(\varepsilon)} & = & R \biggl[ 3-4\sigma^{2} - 8 i \sigma \gamma^{2} R \left( \textrm{e}^{i\alpha} \partial_{z_{0}} - \textrm{e}^{-i\alpha} \partial_{\bar{z}_{0}} \right) + R^{2} \left( -1 + 12\gamma^{2} - 16 \gamma^{4} \right) \partial_{z_{0}}\partial_{\bar{z}_{0}} + \\
& + & R^{2} (8\gamma^{4}-6\gamma^{2}) \left( \textrm{e}^{2i\alpha}\partial_{z_{0}}^{2} + \textrm{e}^{-2i\alpha}\partial_{\bar{z}_{0}}^{2} \right) \biggr] \varepsilon(z_{0},\bar{z}_{0}) + O(R^{4}) \, , \\ \nonumber
\label{genJ_spoe_phi}
s_{\textrm{J}_{\chi}}^{(\Phi)} & = & 2^{1/4} R^{1/8} \biggl[ \sigma  + 8 i \gamma^{2} R \left( \textrm{e}^{i\alpha}\partial_{z_{0}} - \textrm{e}^{-i\alpha}\partial_{\bar{z}_{0}} \right) + 8\sigma R^{2} \left( 1-8\gamma^{2} \right) \partial_{z_{0}}\partial_{\bar{z}_{0}} + \\
& + & (32/3)\sigma \gamma^{2} R^{2} \left( \textrm{e}^{2i\alpha}\partial_{z_{0}}^{2} + \textrm{e}^{-2i\alpha}\partial_{\bar{z}_{0}}^{2} \right) \biggr] \Phi(z_{0},\bar{z}_{0}) + O\left( R^{3+1/8} \right) \, ,
\end{eqnarray}
and
\begin{equation}
\label{genJ_spoe_id}
s_{\textrm{J}_{\chi}}^{(\mathbb{I})} = 8 \gamma^{2} R^{2} \biggl[ \textrm{e}^{2i\alpha}T(z_{0})+\textrm{e}^{-2i\alpha}\bar{T}(\bar{z}_{0}) - i \sigma R \left( \textrm{e}^{3i\alpha}\partial_{z_{0}}^{3} - \textrm{e}^{-3i\alpha}\partial_{\bar{z}_{0}}^{3} \right) \biggr] + O\left( R^{4} \right)
\end{equation}
with $\gamma \equiv \cos\chi$, $\sigma \equiv \sin\chi$, for the \emph{generalized circular Janus} particles; and for \emph{circular quadrupole} particles one has
\begin{eqnarray}
\label{Q_spoe_en}
s_{\textrm{Q}}^{(\varepsilon)} & = & R \biggl[ \frac{5}{3} + 7 R^{2} \partial_{z_{0}}\partial_{\bar{z}_{0}} \biggr] \varepsilon(z_{0},\bar{z}_{0}) + O(R^{5}) \, , \\
\label{Q_spoe_phi}
s_{\textrm{Q}}^{(\Phi)} & = & \frac{2^{25/4}}{9} R^{17/8} i \biggl[ \textrm{e}^{2i\alpha}\partial_{z_{0}}^{2} - \textrm{e}^{-2i\alpha}\partial_{\bar{z}_{0}}^{2} \biggr]  \Phi(z_{0},\bar{z}_{0}) + O\left( R^{4+1/8} \right) \, ,
\end{eqnarray}
and
\begin{eqnarray} \nonumber
\label{Q_spoe_id}
s_{\textrm{Q}}^{(\mathbb{I})} & = & R^{4} \biggl[ \frac{140}{3} L_{-2}\bar{L}_{-2} - \frac{128}{21} \left( \textrm{e}^{4i\alpha} L_{-2}^{2} + \textrm{e}^{-4i\alpha} \bar{L}_{-2}^{2} \right) + \frac{48}{7} \left( \textrm{e}^{4i\alpha} L_{-4} + \textrm{e}^{-4i\alpha} \bar{L}_{-4} \right) \biggr] \mathbb{I}(z_{0},\bar{z}_{0}) + \\
& + & O\left( R^{6} \right) \, .
\end{eqnarray}

Concerning the derivation of Eqs. (\ref{J_spoe_en})-(\ref{nJ_spoe_id}) we refer to Appendix \ref{correlations}, of Eqs. (\ref{genJ_spoe_en})-(\ref{genJ_spoe_id}) to Appendix \ref{appendix_C}, and of Eqs. (\ref{Q_spoe_en})-(\ref{Q_spoe_id}) to Appendix \ref{appendix_D}. For the complex coordinates defined below Eq. (\ref{03_08}), the complex derivatives $\partial_{z_{0}}$ and $\partial_{\bar{z}_{0}}$ are related to the Cartesian ones according to
\begin{equation}
\label{ }
\partial_{x_{0}} = \partial_{z_{0}} + \partial_{\bar{z}_{0}} \, , \qquad \partial_{y_{0}} = i \left( \partial_{z_{0}} - \partial_{\bar{z}_{0}} \right) \, .
\end{equation}
Here $z_{0}$ is the position of the particle center in the $z$ plane and $R$ is the radius of a circular Janus particle, of a generalized Janus particle, or of a quadrupolar particle. $D$ is the length of a Janus needle. The particle orientation is characterized by the angle $\alpha$ of the counterclockwise rotation necessary in order to obtain the actual orientation from the standard one illustrated in Fig. \ref{fig02_01}.
\begin{figure}[htbp]
\centering
\includegraphics[width=16.5cm]{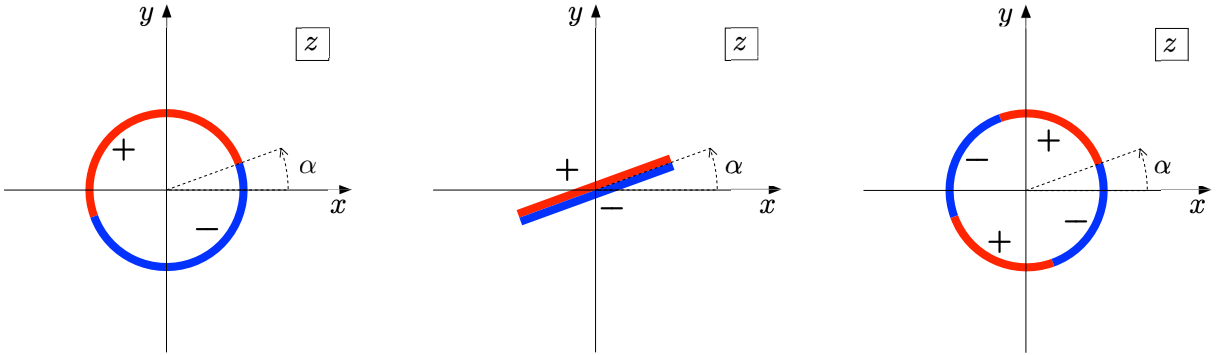}
\caption{Illustration of the rotation angle $\alpha$.}
\label{fig_angles}
\end{figure}

The meaning of the operators $L_{-3} \Phi$ and $\bar{L}_{-3} \Phi$ in Eqs. (\ref{J_spoe_phi}) and (\ref{nJ_spoe_phi}), respectively, and of the operators in (\ref{Q_spoe_id}) follows from the definitions \cite{BPZ_1,BPZ_2,Cardy_review,Cardy_1986}
\begin{eqnarray}
\label{def_descendant}
L_{p}\Psi(z_{0},\bar{z}_{0}) \equiv \int_{\mathcal{C}_{z_{0}}} \frac{\textrm{d}z}{2\pi i} (z-z_{0})^{p+1} T(z) \Psi(z_{0},\bar{z}_{0}), \nonumber \\
\bar{L}_{p}\Psi(z_{0},\bar{z}_{0}) \equiv \int_{\mathcal{C}_{\bar{z}_{0}}} \frac{\textrm{d}\bar{z}}{2\pi i} (\bar{z}-\bar{z}_{0})^{p+1} \bar{T}(\bar{z}) \Psi(z_{0},\bar{z}_{0})
\end{eqnarray}
where $p = 0,\pm 1,\pm 2, \dots$, $\Psi$ is any local operator (primary or not) of the theory, and where the closed integration paths $\mathcal{C}_{z_{0}}$ and ${\mathcal{C}_{\bar{z}_{0}}}$ enclose counterclockwise the points $z_{0}$ and $\bar{z}_{0}$, respectively. Obviously one has $L_{-2}\mathbb{I}(z_{0},\bar{z}_{0})=T(z_{0})$, $L_{-3}\mathbb{I}(z_{0},\bar{z}_{0})= \partial_{z_0}T(z_{0})$, etc. The meaning of $L_{-2}^2 \mathbb{I}(z_{0},\bar{z}_{0}) \equiv L_{-2} T(z_0)$ is also provided by Eq. (\ref{def_descendant}) as well as of $L_{-1}\Psi(z_{0},\bar{z}_{0})=\partial_{z_{0}}\Psi(z_{0})$ due to the shift property of the stress tensor integral (compare, e.g., the conformal Ward identity given in Eq. (\ref{cf_011})). We note that $L_{-1}L_{-2}\mathbb{I}(z_{0},\bar{z}_{0})= \partial_{z_0} T(z_{0})$ while $L_{-2}L_{-1}\mathbb{I}(z_{0},\bar{z}_{0})$ vanishes because $\partial_{z_{0}} \mathbb{I}(z_{0},\bar{z}_{0}) =0$. This is a simple example of the non-commutativity of the operations $L$ and is in line with their Virasoro algebra \cite{BPZ_1,BPZ_2}:
\begin{equation}
\begin{aligned}
\label{VAextended}
[L_{p}, L_{q}] & = (p-q) L_{p+q} + \frac{c}{12} q(q^{2}-1)\delta_{p+q,0} \\
[\bar{L}_{p}, \bar{L}_{q}] & = (p-q) \bar{L}_{p+q} + \frac{c}{12} q(q^{2}-1)\delta_{p+q,0} \\
[L_{p}, \bar{L}_{q}] & = 0 \, ,
\end{aligned}
\end{equation}
with $c=1/2$ for the Ising model. Here the square brackets denote a commutator\footnote{When acting with  $[L_p, L_q] \equiv L_{p}L_{q}-L_{q}L_{p}$ on an arbitrary operator $\Psi(z_{0},\bar{z}_{0})$ and taking into account Eq.  (\ref{def_descendant}), the integration path $\int {\rm d}z_1$ belonging to $L_{p}$ encircles $z_0$ outside of and inside of the corresponding path $\int {\rm d}z_2$ belonging to $L_q$ in the first and second term of the commutator, respectively. The difference can be determined, e.g., by evaluating for fixed $z_1$ the change in $\int {\rm d}z_2$ when it encircles $z_0$ outside rather than inside of $z_1$. This change is determined by the two singular terms $T(z_1) T(z_2) \to (c/2) (z_1 -z_2)^{-4} + 2 T((z_1 +z_2)/2) (z_1 - z_2)^{-2}$ in the OPE of $T(z_1)$ and $T(z_2)$ which finally lead to the two contributions arising when the right hand side of the first Eq. (\ref{VAextended}) acts on $\Psi (z_0 , \bar{z}_0)$.}

In our operator terms in Eqs. (\ref{J_spoe_en})-(\ref{Q_spoe_id}) the dependence on the particle sizes $R$ or $D$ enters via a prefactor $R^x$ or $D^x$ with the scaling dimension $x$ of the operator and the dependence on the particle orientation $\alpha$ via another prefactor $\exp(i \sigma \alpha)$ with the ``spin'' $\sigma$ of the operator. While the spin vanishes for the primary, scalar operators ${\cal O}= \varepsilon, \Phi$, or $\mathbb{I}$, for a descendant operator $\Psi$ of the form $\Psi \equiv L_{-n_1} L_{-n_2} \cdots \bar{L}_{-m_1} \bar{L}_{-m_2} \cdots {\cal O}$ with positive integers $n_1, n_2, \dots, m_1, m_2, \dots$ the scaling dimension equals $x_{\Psi}= n_1 +n_2 + \dots +m_1+m_2 + \dots + x_{\cal O}$ and the spin equals $\sigma_{\Psi}= n_1 +n_2 + \dots -m_1 -m_2 - \dots$. Like $L_{-1}$ and $\bar{L}_{-1}$, which refer to translations (shifts), the operations $L_0$, $\bar{L}_0$ refer to dilatations and rotations and the scaling dimension $x_{\Psi}$ and the spin $\sigma_{\Psi}$ of an operator $\Psi$ appear in the relations $L_0 \Psi = \Delta_{\Psi} \Psi$ and $\bar{L}_0 \Psi = \bar{\Delta}_{\Psi} \Psi$  where $\Delta_{\Psi}=((x_{\Psi}+\sigma_{\Psi})/2)$ and $\bar{\Delta}_{\Psi} = ((x_{\Psi}-\sigma_{\Psi})/2)$ \cite{BPZ_1,BPZ_2}.\footnote{For the primary operators ${\cal O}=\varepsilon, \Phi$ these relations are consistent with the conformal Ward identity (see Eq. (\ref{cf_011})) and the corresponding relation $L_0 \mathbb{I} =0$ is obvious from Eq. (\ref{def_descendant}). For descendant operators the relations follow by commuting $L_0$ or $\bar{L}_0$ --- by means of the Virasoro algebra --- through all the $L_{-n}$ or all the $\bar{L}_{-m}$ to the right until they arrive in front of ${\cal O}$.  A simple example is the descendant operator $\Psi \equiv L_{-n} {\cal O}$ for which  $x_{\Psi}=n+x_{\cal O}$ and $\sigma_{\Psi}=n$ is corroborated by the Virasoro algebra, which yields $L_0 \Psi = L_0 L_{-n} {\cal O} = (L_{-n}L_0 +n L_{-n}){\cal O} = ((x_{\cal O}/2)+n) L_{-n} {\cal O} = ((x_{\Psi}+\sigma_{\Psi})/2) \Psi$.}

Apart from these general properties of a $d=2$ conformal field theory, in the Ising model the two primary operators $\varepsilon$ and $\Phi$ are ``degenerate on level 2'' \cite{BPZ_1,BPZ_2}. In Appendix \ref{discussion_degeneracy} we comment upon this degeneracy. In the Ising model the operations $L_{-2}$ and $L_{-1}^{2}$ are not completely independent because their action on on the primary operators $\varepsilon$ and $\Phi$ gives the same result up to a multiplicative factor:
\begin{equation}
\label{degeneracy}
L_{-2} {\cal O}(z_{0},\bar{z}_{0}) = {3 \over 2(1+x_{\cal O})} L_{-1}^2 {\cal O}(z_{0},\bar{z}_{0}) \, , \qquad {\cal O}=\varepsilon, \Phi \, ,
\end{equation}
which implies the proportionalities  $L_{-2}\varepsilon(z_{0},\bar{z}_{0})=(3/4)\partial_{z_{0}}^{2}\varepsilon(z_{0},\bar{z}_{0})$ and, similarly, $L_{-2}\Phi(z_{0},\bar{z}_{0})=(4/3)\partial_{z_{0}}^{2}\Phi(z_{0},\bar{z}_{0})$ of their descendants of level two. In our presentation of Eqs. (\ref{J_spoe_en})-(\ref{Q_spoe_id}) we have opted in favor of $\partial_{z_{0}}^{2}\varepsilon$ and $\partial_{z_{0}}^{2}\Phi$ whereas $L_{-2}\varepsilon$ and $L_{-2}\Phi$ do not appear.

The above quantities $s_{\rm P}$ are series of operators with increasing scaling dimensions and being consistent with the particle symmetries. The prefactors of the operators are fixed so that all $n$-point correlation functions $\langle \Psi_{1}\Psi_{2} \cdots \Psi_{n} \rangle_{\rm P}$ in the presence of the particle $\textrm{P}$ are represented, at large distances from it, via
\begin{equation}
\label{Psidotdot}
\langle \Psi_{1}\Psi_{2} \cdots \Psi_{n} \rangle_{\rm P} \longrightarrow \langle (1+s_{\rm P})\Psi_{1}\Psi_{2} \cdots \Psi_{n} \rangle \, ,
\end{equation}
in terms of the series $\langle (1+s_{\rm P})\Psi_{1}\Psi_{2} \cdots \Psi_{n} \rangle$ of $(n + 1)$-point bulk correlation functions. Since for a given particle the same operator series $s_{\rm P}$ applies to all the various multipoint correlation functions, the SPOE is a nontrivial property which is quite similar to the well known operator product expansion in the bulk where the role of the ``small'' particle is taken by the product of two ``nearby'' operators. We recall that $\langle \cdots \rangle$ stands for a thermal average in \emph{bulk}.

We have determined the prefactors from the comparison of the effect of the particle onto the profiles (i.e., one-point correlation functions) obtained in Secs. \ref{symJanus}, \ref{App_needle}, and \ref{quadrupoles}. There the product $\Psi_{1}\Psi_{2}\cdots \Psi_{n}$ in Eq. (\ref{Psidotdot}) is replaced by $\varepsilon$, $\Phi$, or $T$ as well as in two-point correlation functions, obtained in Appendices \ref{correlations} and \ref{appendix_D}, where $\Psi_{1}\Psi_{2}\cdots \Psi_{n}$ is replaced by $\Phi\varepsilon$, $\Phi\Phi$, or $\varepsilon\varepsilon$, order by order in their large distance expansions. While these prefactors refer to particles in the standard orientation $\alpha=0$ and for the standard size $R=D/2=1$, the prefactors for arbitrary orientations $\alpha$ and sizes $R$ and $D$, as given in Eqs. (\ref{J_spoe_en})-(\ref{Q_spoe_id}), follow from the dilatation and rotation properties of the operators appearing in $s_{\rm P}$ which are determined by their scaling dimensions and spins\footnote{The corresponding transformation for, e.g. a Janus circle J is $z'(z) = (z-z_0) /[R \exp (i \alpha)]$, which maps the Janus J($z_0, R, \alpha$) in the $z$ plane to the standard one J($0,1,0$) in the $z'$ plane. One may check that our SPOE given in Eqs. (\ref{J_spoe_en})-(\ref{J_spoe_id}) is consistent with the corresponding local scale transformation which, e.g., for the profile of a nonscalar, descendant operator $\Psi$  reads \cite{BPZ_1,BPZ_2} $\langle \Psi (z, \bar{z}) \rangle_{{\rm J}(R,\alpha)}= (\textrm{d}z'/\textrm{d}z)^{\Delta_{\Psi}} (\textrm{d} \bar{z}'/\textrm{d} \bar{z})^{\bar{\Delta}_{\Psi}} \langle \Psi (z', \bar{z}') \rangle_{{\rm J}(1,0)}$. This form is consistent with the scale transformation given in Eq. (\ref{stresstrafo}) for $\Psi = T$ because $x_{T} = \sigma_T =2$ and the Schwarzian vanishes for the present transformation. It reduces to the corresponding scale transformation of the form given in Eq. (\ref{03_08}) if $\Psi$ is a scalar operator with $\sigma_{\Psi}=0$.}.

We close this section with a few remarks.
\begin{enumerate}
\item As expected, the operators in the SPOEs of the two Janus particles $\rm{J}$ and $\rm{Jn}$ are the same, only their prefactors are different.
\item Both for Janus circles $\textrm{J}$ and for Janus needles $\rm Jn$ as well as for the quadrupolar particle $\textrm{Q}$ there is no order parameter ``monopole'' contribution $\propto \Phi$ in their operator expansions. But --- in agreement with their profiles $\langle \Phi \rangle$ in Eqs. (\ref{03_09}), (\ref{App_needle_06bis}), and (\ref{07_05}) --- the expansions in Eqs. (\ref{J_spoe_phi}) and (\ref{nJ_spoe_phi}) start with a dipole and in Eq. (\ref{Q_spoe_phi}) with a quadrupole, which for $\alpha=0$ is proportional to $i \left( \partial_{z_{0}} - \partial_{\bar{z}_{0}} \right) \Phi = \partial_{y}\Phi$ and to $i \left( \partial_{z_{0}}^{2} - \partial_{\bar{z}_{0}}^{2} \right)\Phi = \partial_{x}\partial_{y}\Phi$, respectively. This should be compared with the presence of a $\Phi$ monopole in the SPOE of a generalized Janus particle in Eq. (\ref{genJ_spoe_phi}).
\item The even and odd rotational invariances $\alpha \mapsto \alpha + \pi$ for $s_{\rm P}^{(\varepsilon)}$, $s_{\rm P}^{(\mathbb{I})}$, and $s_{\rm P}^{(\Phi)}$ in the case of each of the two Janus particles $\textrm{J}$ and $\textrm{Jn}$, and $\alpha \mapsto \alpha+\pi/2$ in the case of the quadrupolar particle show up clearly in the SPOEs.
\item It is interesting to compare the prefactors of the energy-density operator $\varepsilon$ in the SPOEs for the circular unit disk with (i) homogeneous boundary condition $+$, (ii) Janus boundary conditions, (iii) quadrupolar boundary conditions, and (iv) homogeneous ``ordinary'' (or free) boundary conditions. These prefactors are $-1$, $3$, $5/3$, and $1$, respectively. Their signs are as expected because the homogeneous ordering in (i) reduces the energy with respect to the energy in the bulk. In the other cases, the energy is increased due to the forced change in the order in the cases (ii) and (iii), and due to the disorder at a free boundary in the case (iv). Moreover, one expects that for a circular boundary consisting of more and more alternating $+$ and $-$ sections of equal length (reminiscent of higher and higher ``multipoles'') finally behaves effectively like a system with a homogeneous ordinary boundary. This is in line with the prefactors decreasing monotonically as one moves from (ii) via (iii) to (iv).
\end{enumerate}

Finally we note for later use the SPOE for a needle $\textrm{N}$ with homogeneous ordinary boundary conditions. This expansion is well known (see, e.g., Ref. \cite{VED}) and can be written in the form of Eq. (\ref{def_SPOE}) with $s_{\textrm{N}}^{(\Phi)}=0$:
\begin{equation}
\label{N_spoe_en}
s_{\textrm{N}}^{(\varepsilon)} = \frac{D/2}{2} \biggl\{ 1 + \left( \frac{D}{2} \right)^{2} \biggl[ \frac{1}{4} \partial_{z_{0}}\partial_{\bar{z}_{0}} + \frac{3}{16} \left( \textrm{e}^{2i\alpha}\partial_{z_{0}}^{2} + \textrm{e}^{-2i\alpha}\partial_{\bar{z}_{0}}^{2} \right) \biggr] \biggr\} \varepsilon(z_{0},\bar{z}_{0}) + O\left( D^{5} \right) \, ,
\end{equation}
and
\begin{equation}
\label{N_spoe_id}
s_{\textrm{N}}^{(\mathbb{I})} = \frac{(D/2)^{2}}{4} \biggl[ \textrm{e}^{2i\alpha}T(z_{0}) + \textrm{e}^{-2i\alpha}\bar{T}(\bar{z}_{0}) \biggr] + O\left( D^{4} \right) \, .
\end{equation}

\section{Interaction between particles}
\label{wall}
The free energy required to transfer the Janus or the quadrupolar particle $\textrm{P}$ from the bulk phase to a phase, which exhibits boundaries at large distances from the particle, is given by
\begin{eqnarray} \label{interact}
\textrm{e}^{-\delta {\cal F}} = 1+\langle s_{\rm P} \rangle_{\rm plane \, with \, distant \, boundaries} \, .
\end{eqnarray}
The quantity $\delta {\cal F}$ denotes the free energy in units of $k_{\rm B} T_{\rm c}$ and it vanishes in the bulk because the bulk average $\langle s_{\rm P} \rangle =0$. In Secs. \ref{sec_J_half} and \ref{wall_section2} we shall use this relation in order to determine the interaction of the particles shown in Fig. \ref{fig02_01} with boundaries of infinite extent belonging to a confined geometry ${\mathscr{G}}$ such as a half plane, strips or wedges in which the particle is embedded. In this case Eq. (\ref{interact}) turns into
\begin{equation}
\label{int_01}
\textrm{e}^{-\delta\mathcal{F}} = 1 + \langle s_{\rm P} \rangle_{\mathscr{G}} \, .
\end{equation}
In Sec. \ref{interaction_JJ} we shall study the opposite case in which the boundaries are those of a small distant object such as another particle $\textrm{P}^{\prime}$, which itself can be represented by a small particle expansion $\exp(-\delta\mathcal{H}_{\rm P'}) \propto 1 + s_{\rm P'}$ and for which Eq. (\ref{interact}) implies
\begin{equation}
\label{int_02}
\textrm{e}^{-\delta\mathcal{F}} = 1 + \langle s_{\rm P} s_{\rm P'} \rangle \, .
\end{equation}
As before, $\langle \cdots \rangle$ without subscript stands for \emph{bulk} statistical averages.

Concerning the interaction of the Janus and the quadrupolar particles with a distant object, the dependence on distance is dominated by the operator $\varepsilon$ in the SPOEs because $\langle\varepsilon\rangle_{\mathscr{G}}$ and $\langle s_{\textrm{object}}\varepsilon \rangle$ are in general nonzero. The dependence on orientation is dominated by $s_{\rm P}^{(\Phi)}$ if the object breaks the $\pm$ symmetry. If it does not, the issue of which is the dominating operator depends on further details: for the present particles in a strip or wedge with ordinary boundaries, the orientation dependence is dominated by $s_{\rm P}^{(\mathbb{I})}$ so that the orientation-dependent free energy is proportional to $R^2$ or $D^2$ and $R^4$ for Janus and quadrupolar particles, respectively. However, in the half plane the averages of the orientation dependent terms in $s_{\rm P}^{(\mathbb{I})}$ vanish and for Janus particles it is $\left( \exp(2i\alpha)\partial_{z_{0}}^{2}+\exp(-2i\alpha)\partial_{\bar{z}_{0}}^{2} \right)\varepsilon$ so that the orientation-dependent free energy is proportional to $R^3$ or $D^3$. In the following sections we study these effective interactions in more detail.

\section{Janus particle in a half plane, in a strip, and in a wedge}
\label{sec_J_half}
The cost of free energy for transferring one of the present particles from the bulk into a bounded system (Eq. (\ref{int_01})) depends, apart from the location $z_{0}$ and the orientation $\alpha$ of the particle, on the boundary conditions and on the shape of the embedding system. As paradigmatic shapes we consider the right half $z$ plane as well as a strip and a wedge - all three of them with the real $z$ axis as their midlines. The averages of  $s_{\rm P}$, appearing in Eq. (\ref{int_01}) for these geometries in the $z$ plane, follow from the corresponding averages in the upper half $w$ plane $\mathbb{H}$ via suitable conformal mappings $w=w(z)$.

For the right half $z$ plane $\widetilde{\mathbb{H}}$ the mapping is simply the rotation
\begin{equation}
\label{rotation}
w(z) = iz \, ,
\end{equation}
while for the strip ${\mathbb{S}}$ of width $\mathcal{W}$ the function
\begin{equation}
\label{map_strip}
w(z) = i \textrm{e}^{\pi z/\mathcal{W}}
\end{equation}
maps the lower and upper boundary $y = -\mathcal{W}/2$ and $y = \mathcal{W}/2$ to the positive and negative real $w$ axis, respectively,  and the strip-region $-\mathcal{W}/2 \leqslant y \leqslant \mathcal{W}/2$ onto the upper half $w$ plane. For the wedge ${\mathbb{W}}$ with its apex at $z = 0$ and with opening angle $\pi/\Xi$ the function
\begin{equation}
\label{map_wedge}
w(z) = iz^{\Xi}
\end{equation}
maps the half lines $z = |z| \exp[-i\pi/(2\Xi)]$ and $z = |z| \exp[i\pi/(2\Xi)]$ onto the positive and negative real $w$ axis, respectively, and the region $z = |z| \exp(i\theta)$ with $-\pi/(2\Xi) \leqslant \theta \leqslant \pi/(2\Xi)$ onto the upper half $w$ plane.

The wedge geometry is more general than the half plane and the strip geometry in that it contains both of them as special cases. The opening angle $\pi/\Xi$ can vary between $0$ and $2\pi$ and the embedding region degenerates for $\Xi = 1$ and $\Xi = 1/2$ to the right half $z$ plane and the region outside a semi-infinite needle extending along the negative real $z$ axis, respectively. For $\pi/\Xi=0$ there is no point $z=z_{0}$ left where to embed the particle. However, by combining the decrease $\pi/\Xi \searrow 0$ in opening angle with an increase $\textrm{Re}(z_{0})=\mathcal{W}\Xi/\pi \nearrow \infty$ of the real part of the distance vector $z_{0}$ from the apex to the insertion point (keeping its imaginary part $\textrm{Im}(z_{0})$ and $\mathcal{W}>2|\textrm{Im}(z_{0})|$ fixed) one reaches the situation in which the particle is embedded in the horizontal strip of width $\mathcal{W}$ at a distance $\textrm{Im}(z_{0})$ from the midline. Clearly, in these limits the polar angle $\theta_{0} \equiv \tan^{-1}\left( \textrm{Im}(z_{0})/\textrm{Re}(z_{0}) \right)$ of $z_{0}=|z_{0}|\exp(i\theta_{0})$ approaches $\pi \textrm{Im}(z_{0})/(\mathcal{W}\Xi)$ and vanishes.

\subsection{Symmetric Janus circle and Janus needle in a half plane}
\label{JnJ_HP}
The free energy $\delta\mathcal{F}$ required to transfer the Janus particle from the bulk to the right half plane with a boundary of universality class $a$ follows from Eq. (\ref{int_01}) and the SPOEs (\ref{def_SPOE})-(\ref{nJ_spoe_id}) as
\begin{equation}
\label{energy_hp}
\textrm{e}^{-\delta\mathcal{F}} = 1 + \langle s_{\rm P}^{(\varepsilon)} + s_{\rm P}^{(\Phi)} \rangle_{\widetilde{\mathbb{H}}_{a}} \, ;
\end{equation}
the symbol $\widetilde{\mathbb{H}}_{a}$ stands for the half-plane geometry $\textrm{Re}(z)>0$, with uniform boundary condition $a\in\{ O, +\}$ along the boundary $\textrm{Re}(z)=0$.

Since the average of the stress tensor vanishes in the half plane, $s_{\rm P}^{(\mathbb{I})}$ does not contribute within the orders considered in Eqs. (\ref{J_spoe_en})-(\ref{J_spoe_id}). Here one finds\footnote{Equations (\ref{wall_J_eps})-(\ref{wall_nJ_phi}) are obtained most easily within the Cartesian language in which $\partial_{y_{0}}$ is replaced by $ - (\sin\alpha)\partial_{x_{0}} + (\cos\alpha)\partial_{y_{0}} $ if $\alpha\neq0$.}
\begin{equation}
\label{wall_J_eps}
\langle s_{\textrm{J}}^{(\varepsilon)} \rangle_{\widetilde{\mathbb{H}}_{a}} = \mathcal{A}_{\varepsilon}^{(a)} \frac{R}{x_{0}} \biggl\{ 3 - \left( \frac{R}{x_{0}} \right)^{2} \biggl[ \frac{1}{2} + 4 \sin^{2}\alpha \biggr] \biggr\}
\end{equation}
and
\begin{equation}
\label{wall_J_phi}
\langle s_{\textrm{J}}^{(\Phi)} \rangle_{\widetilde{\mathbb{H}}_{a}} = \mathcal{A}_{\Phi}^{(a)} 2^{1/4} \left( \frac{R}{x_{0}} \right)^{9/8} \left( \sin\alpha \right) \biggl\{ 1 - \frac{1}{32} \left( \frac{R}{x_{0}} \right)^{2} \biggl[ 39 + 16 \sin^{2}\alpha \biggr] \biggr\}
\end{equation}
for the symmetric Janus particle and
\begin{equation}
\label{wall_nJ_eps}
\langle s_{\textrm{Jn}}^{(\varepsilon)} \rangle_{\widetilde{\mathbb{H}}_{a}} = \mathcal{A}_{\varepsilon}^{(a)} \frac{\mathcal{D}}{2} \biggl\{ 3 + \frac{\mathcal{D}^{2}}{16} \biggl[ 7 - 34 \sin^{2}\alpha \biggr] \biggr\}
\end{equation}
and
\begin{equation}
\label{wall_nJ_phi}
\langle s_{\textrm{Jn}}^{(\Phi)} \rangle_{\widetilde{\mathbb{H}}_{a}} = \mathcal{A}_{\Phi}^{(a)} 2^{-7/8} \mathcal{D}^{9/8} \left( \sin\alpha \right) \biggl\{ 1 - \frac{9\mathcal{D}^{2}}{64} \biggl[ 2 + 3 \sin^{2}\alpha \biggr] \biggr\} \, , \qquad \mathcal{D} \equiv \frac{D}{2x_{0}}
\end{equation}
for the Janus needle. $\mathcal{A}_{\mathcal{O}}^{(a)}$ is the universal amplitude introduced in Eq. (\ref{03_04_01}) and $x_{0}=\textrm{Re}(z_{0})$ is the distance of the center of the Janus particle from the boundary of the right half plane. Equation (\ref{cf_014}) is useful for deriving the averages of $s_{\rm P}^{(\Phi)}$ in Eqs. (\ref{wall_J_eps})-(\ref{wall_nJ_phi}).

\subsubsection{Janus particle interacting with an ordinary boundary}
\label{sec_J_half_ord}
Since $s_{\rm P}^{(\Phi)}$ does not contribute in a half plane with ordinary boundary, i.e., $a=O$, Eqs. (\ref{energy_hp})-(\ref{wall_nJ_phi}) yield
\begin{equation}
\label{Januswall1}
\delta\mathcal{F}_{\textrm{wall,O};\textrm{J}}(x_{0}/R,\alpha) = - \ln \biggl\{ 1 + \frac{R}{2x_{0}} \biggl[ 3 - \left( \frac{R}{x_{0}} \right)^{2} \left( \frac{1}{2} + 4\sin^{2}\alpha \right) \biggr] \biggr\}
\end{equation}
for the Janus circle and
\begin{equation}
\label{nJanuswall1}
\delta\mathcal{F}_{\textrm{wall,O};\textrm{Jn}}(\mathcal{D},\alpha) = - \ln \biggl\{ 1 + \frac{\mathcal{D}}{4} \biggl[ 3 + \frac{\mathcal{D}^{2}}{16} \left( 7 - 34\sin^{2}\alpha \right) \biggr] \biggr\}
\end{equation}
for the Janus needle. The orientation $\alpha$ corresponding to the lowest free energy for fixed $x_{0}$ minimizes the right hand sides of Eqs. (\ref{Januswall1}) and (\ref{nJanuswall1}) at $\alpha=0$ or $\alpha=\pi$. This agrees with the intuition that the switching points at the surface of a Janus particle, where $+$ and $-$ boundary conditions meet and where the local energy $\langle \varepsilon \rangle_{\textrm{J}}$ attains its maximal surface value, tend to be as close as possible to the half plane boundary where the local energy $\langle \varepsilon \rangle_{\widetilde{\mathbb{H}}_{O}}$ is enhanced over the bulk value.

\begin{figure*}[htbp]
\centering
        \begin{subfigure}[htbp]{0.48\textwidth}
            \centering
            \includegraphics[width=\textwidth]{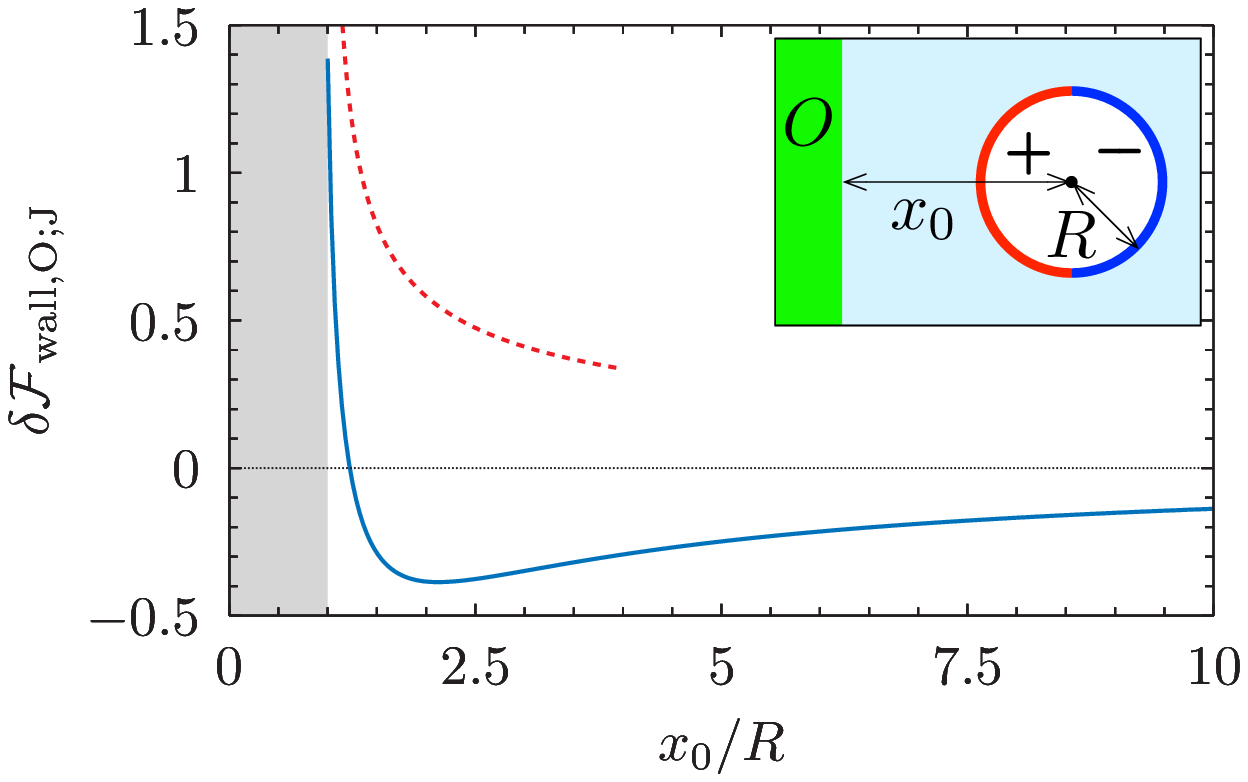}
            \caption[]%
            {{\small }}    
            \label{fig07_01a}
        \end{subfigure}
\hfill
        \begin{subfigure}[htbp]{0.48\textwidth}  
            \centering 
            \includegraphics[width=\textwidth]{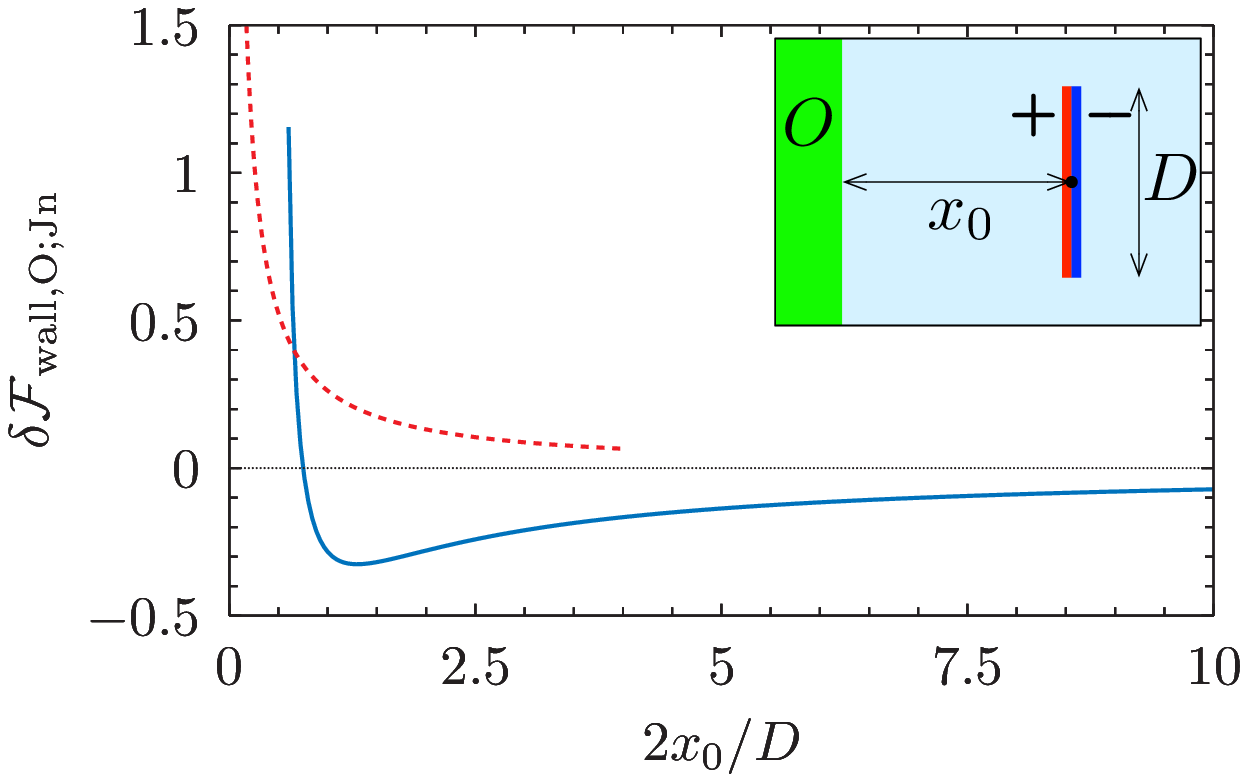}
            \caption[]%
            {{\small }}    
            \label{fig07_01b}
        \end{subfigure}
\caption[]
{\small Distance dependence of the insertion free energy $\delta\mathcal{F}$ of a Janus circle (a) and a Janus needle (b) with fixed orientation $\alpha=\pi/2$ so that it faces with its $+$ segment an ordinary boundary wall of the embedding right half plane. Combining the (extrapolated) large distance behaviors from Eq. (\ref{Januswall1}) and Eq. (\ref{nJanuswall1}) from SPOE (solid blue line) with the short distance Derjaguin behaviors (Eqs. (\ref{Januswall2}) and (\ref{nJanuswall2})) (dashed red line) strongly suggests a non-monotonic behavior of $\delta\mathcal{F}$ with a minimum at a distance $x_{0}$ of the particle center from the wall of the order of the particle size. The shaded  area in (a) indicates hard core repulsion for $x_{0} \leqslant R$. Note that $\delta\mathcal{F}$ in (a) and (b) depends on $R$ and $D$ only via the dimensionless ratios $x_{0}/R$ and $x_{0}/D$, respectively.}
\label{fig07_01}
\end{figure*}

It is instructive to compare the behavior of the free energy in Eqs. (\ref{Januswall1}) and (\ref{nJanuswall1}) at large distances, i.e, $x_{0}\gg R$ and $x_{0}\gg D$, respectively, with the behavior when the Janus particle nearly touches the boundary of the half plane. This can be done easily for the orientation $\alpha=\pm\pi/2$ (which is the orientation of maximal free energy), because in this case the region with $+$ (or $-$) boundary conditions midway between the switching points faces the half plane boundary; at close distance it is only this homogeneous region which matters quantitatively. Thus for the Janus circle and the Janus needle the free energy follows from a Derjaguin formula for a circle with a \emph{homogeneous} boundary and from an expression for a strip of finite length $D$ with homogeneous boundaries, respectively. In the present case this reads
\begin{equation}
\label{Januswall2}
\delta\mathcal{F}_{\textrm{wall,O};\textrm{J}}(x_{0}/R,\alpha=\pm\pi/2) = \pi \Delta_{O,\pm} \sqrt{\frac{2R}{x_{0}-R}} \, ,
\end{equation}
for the Janus circle with $x_{0}/R \rightarrow 1^{+}$ and
\begin{equation}
\label{nJanuswall2}
\delta\mathcal{F}_{\textrm{wall,O};\textrm{Jn}}(\mathcal{D},\alpha=\pm\pi/2) = \Delta_{O,\pm} \frac{D}{x_{0}}
\end{equation}
for the Janus needle with $x_{0}/D \rightarrow 0^{+}$. Here $\Delta_{O,\pm} = \pi/24$ is the corresponding Casimir amplitude \cite{Cardy_review}; see also Eq. (\ref{Casimir_plates'}) and the last relation in Eq. (\ref{f_strip_average}) and the remark below it. The comparison of the large-distance results in Eqs. (\ref{Januswall1}) and (\ref{nJanuswall1}) with the short-distance ones in Eqs. (\ref{Januswall2}) and (\ref{nJanuswall2}) is displayed in the two panels of Fig. \ref{fig07_01}. For both the Janus circle and the Janus needle this implies a non-monotonic behavior of the free energy exhibiting a minimum. Although our results do not allow us to determine the position of these minima quantitatively, they suggest that at the minima the closest distances between the half plane boundary and the Janus particles are of the order of their sizes.

We now turn to the stable orientations $\alpha=0$ of Janus circles and Janus needles. While in the \emph{circular} case $\delta\mathcal{F}$ exhibits the same qualitative behavior as for $\alpha=\pi/2$, i.e., attraction at large and repulsion at small distances, the perpendicular Janus \emph{needle} is attracted by the ordinary wall not only at large but also at small distances. Here the interaction is dominated by the high energy region around the closer tip of the needle, fitting well to the enhanced disorder at the ordinary wall while the ordering effects at the two sides of the needle are of minor importance (see the discussion at the end of Appendix \ref{appendix_F}).

\subsubsection{Janus particle interacting with a $+$ boundary}
\label{sec_J_half_plus}
For $a=+$ both $s_{\rm P}^{(\varepsilon)}$ and $s_{\rm P}^{(\Phi)}$ contribute to Eq. (\ref{int_01}). The explicit expression for the interaction free energy following from Eqs. (\ref{wall_J_eps})-(\ref{wall_nJ_phi}) shows that the free energy attains its minimum at the orientation $\alpha=\pi/2$, where the $+$ side of the Janus circle faces the $+$ wall, as expected intuitively. Now we consider the large-distance dependence for this orientation, which is given by
\begin{equation}
\label{Januswall3}
\delta\mathcal{F}_{\textrm{wall,+};\textrm{J}}(x_{0}/R,\alpha=\pi/2) = - \ln \biggl\{ 1 - \frac{3R}{2x_{0}} \biggl[ 1 - \frac{3}{2}\left( \frac{R}{x_{0}} \right)^{2} \biggr] + 2^{3/8} \left(\frac{R}{x_{0}}\right)^{9/8} \biggl[ 1 - \frac{55}{32}\left( \frac{R}{x_{0}} \right)^{2} \biggr] \biggr\}
\end{equation}
for the Janus circle and
\begin{equation}
\label{nJanuswall3}
\delta\mathcal{F}_{\textrm{wall,+};\textrm{Jn}}(\mathcal{D},\alpha=\pi/2) = - \ln \biggl\{ 1 - \frac{3\mathcal{D}}{4} \biggl[ 1 - \frac{9\mathcal{D}^{2}}{16} \biggr] + 2^{-3/4} \mathcal{D}^{9/8}  \biggl[ 1 - \frac{45\mathcal{D}^{2}}{64} \biggr] \biggr\}
\end{equation}
for the Janus needle\footnote{In Eqs. (\ref{Januswall3}) and (\ref{nJanuswall3}) the contributions from the energy density and from the order parameter are antagonistic, but the contribution from the energy density dominates. Concerning the order parameter, the half plane boundary $+$ and the opposing $+$ side of the Janus circle fit together and favor attraction while the increase of the energy density induced by the Janus circle does not fit to the energy decrease of the half plane boundary $+$ and thus favors repulsion.}. This should be compared with the short-distance dependences
\begin{equation}
\label{Januswall4}
\delta\mathcal{F}_{\textrm{wall,+};\textrm{J}}(x_{0}/R,\alpha=\pi/2) = \pi \Delta_{++} \sqrt{\frac{2R}{x_{0}-R}}
\end{equation}
with $x_{0}\rightarrow R^{+}$ for the Janus circle and
\begin{equation}
\label{nJanuswall4}
\delta\mathcal{F}_{\textrm{wall,+};\textrm{Jn}}(\mathcal{D},\alpha=\pi/2) =  \Delta_{++} \frac{D}{x_{0}}
\end{equation}
with $x_{0}\rightarrow0^{+}$ for the Janus needle and with the corresponding critical Casimir amplitude $\Delta_{++}=-\pi/48$ \cite{Cardy_review}. 

The interaction of both types of Janus particles with the half plane boundary displays the same remarkable qualitative features. While being attractive at short distances, the interaction is repulsive at large distances so that at intermediate distances (not covered by Eqs. (\ref{Januswall3})-(\ref{nJanuswall4})) the interaction free energy must have a maximum. It would be interesting to determine the whole distance dependence by means of simulations which on a square Ising lattice should be easier to carry out for the Janus needle than for the Janus circle.

\begin{figure*}[htbp]
\centering
        \begin{subfigure}[htbp]{0.48\textwidth}
            \centering
            \includegraphics[width=\textwidth]{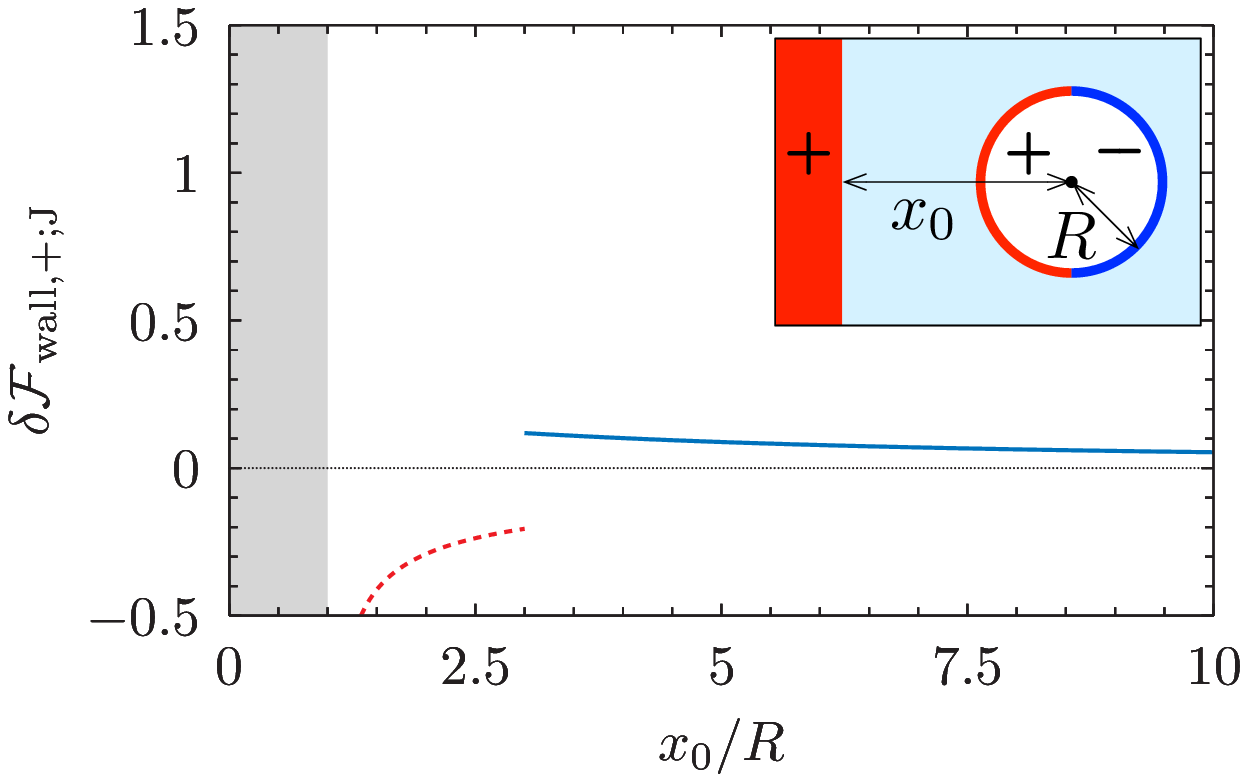}
            \caption[]%
            {{\small }}    
            \label{fig07_02a}
        \end{subfigure}
\hfill
        \begin{subfigure}[htbp]{0.48\textwidth}  
            \centering 
            \includegraphics[width=\textwidth]{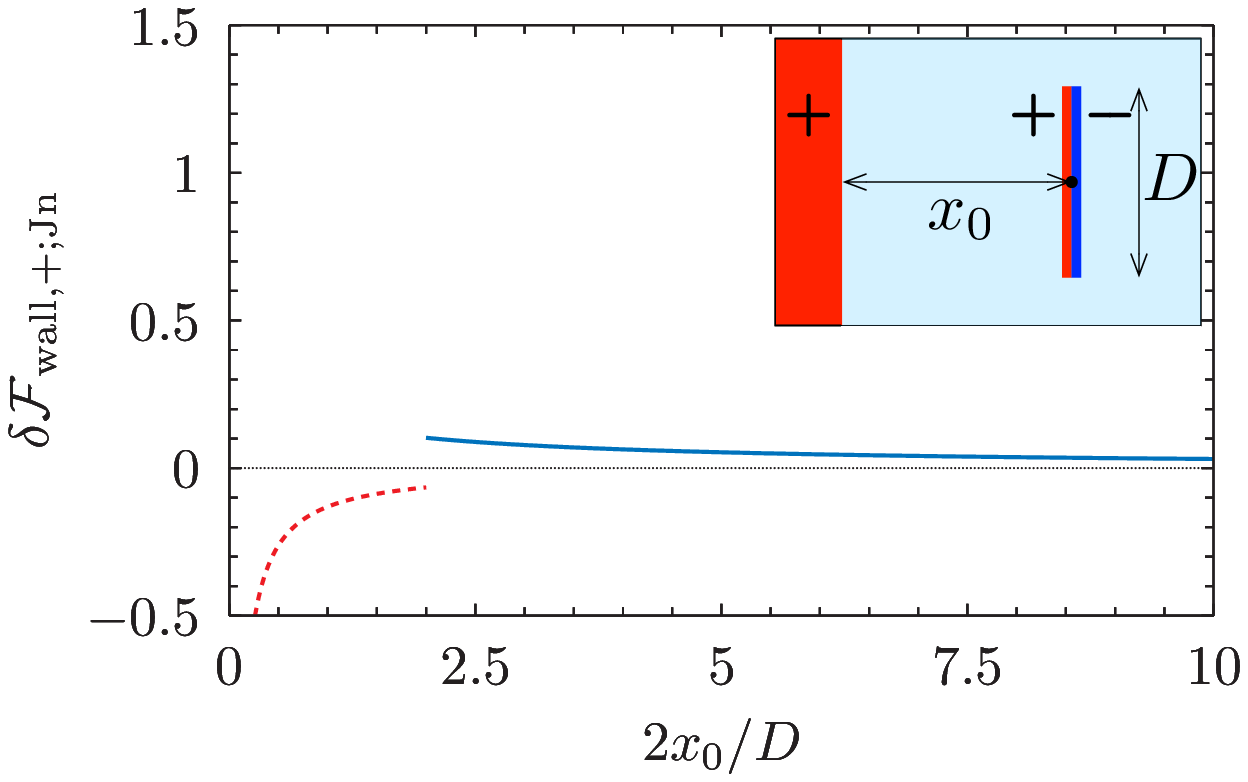}
            \caption[]%
            {{\small }}    
            \label{fig07_02b}
        \end{subfigure}
\caption[]
{\small Same as Fig. \ref{fig07_01} except that the universality class of the wall is now $+$. Thus the Janus orientation again being fixed at $\alpha=\pi/2$ is now the equilibrium orientation (i.e., with the lowest $\delta\mathcal{F}$ for any given $x_{0}$). Combining the large (Eqs. (\ref{Januswall3}) and (\ref{nJanuswall3})) and short (Eqs. (\ref{Januswall4}) and (\ref{nJanuswall4})) distance behaviors suggests a non-monotonic distance dependence of $\delta\mathcal{F}$ with a maximum.}
\label{fig07_02}
\end{figure*}

The asymptotic results for large and short distances (see Eqs. (\ref{Januswall3}) and (\ref{nJanuswall3}) as well as (\ref{Januswall4}) and (\ref{nJanuswall4}), respectively) are displayed in the two panels of Fig. \ref{fig07_02}.

\begin{figure*}[htbp]
\centering
        \begin{subfigure}[htbp]{0.48\textwidth}
            \centering
            \includegraphics[width=\textwidth]{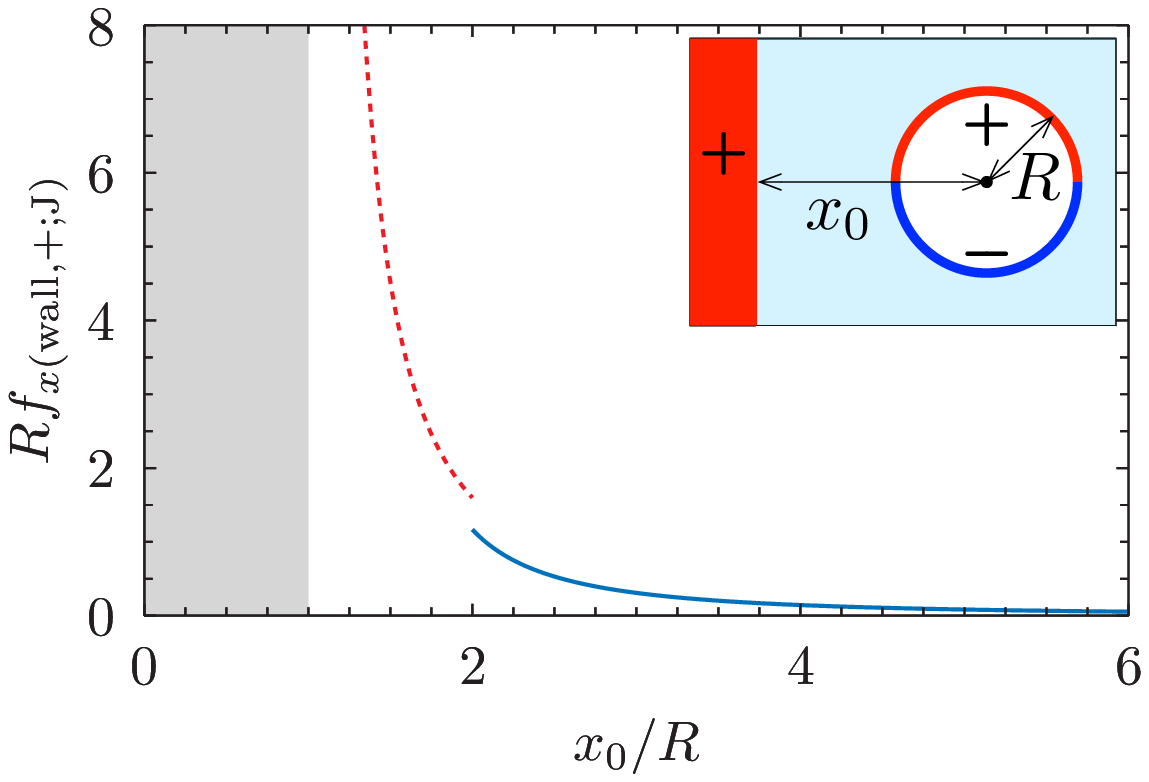}
            \caption[]%
            {{\small }}    
            \label{fig07_03a}
        \end{subfigure}
\hfill
        \begin{subfigure}[htbp]{0.48\textwidth}  
            \centering 
            \includegraphics[width=\textwidth]{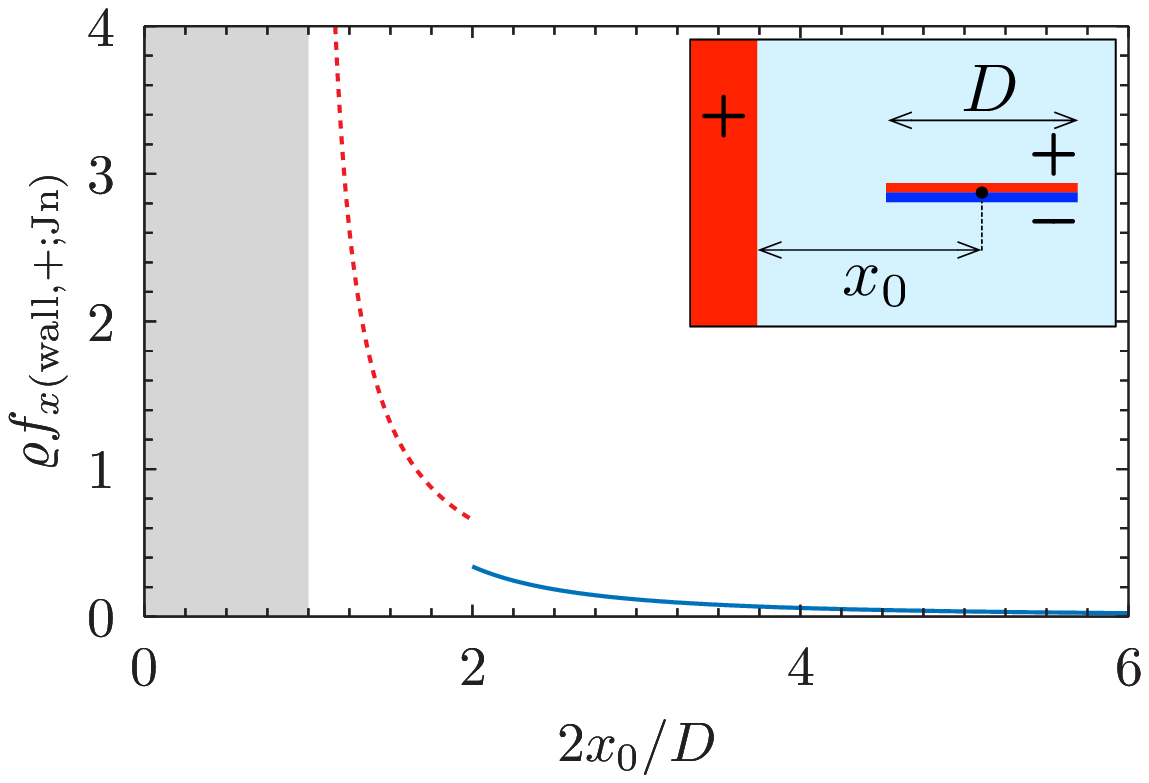}
            \caption[]%
            {{\small }}    
            \label{fig07_03b}
        \end{subfigure}
\caption[]
{\small Same as Fig. \ref{fig07_02} but with the Janus particles forced to face the wall with their switching points (i.e., with orientation $\alpha=0$). Comparing large (Eqs. (\ref{Januswall5}) and (\ref{nJanuswall5})) and small (Eqs. (\ref{Januswall6}) and (\ref{nJanuswall6})) distance results shows that $\delta\mathcal{F}$ is repulsive at all distances. The shaded areas indicate the hard core repulsion for $x_{0}\leqslant R$ in (a) and for $x_{0}\leqslant D/2$ in (b).}
\label{fig07_03}
\end{figure*}

Finally we consider the distance dependences of a Janus circle and a Janus needle which are forced to a fixed orientation with $\alpha=0$ or $\alpha=\pi$ so that the Janus particle faces the half plane boundary with its switching point. Here $s_{\rm J}^{(\Phi)}$ and $s_{\rm Jn}^{(\Phi)}$ do not contribute, due to the factor $\sin\alpha$ in Eqs. (\ref{wall_J_phi}) and (\ref{wall_nJ_phi}). At large distances, the force $f_{x}$ in $x$ direction on the Janus circle and the Janus needle is given by
\begin{equation}
\label{Januswall5}
R f_{x (\rm{wall,+;J})} \equiv - R \partial_{x_{0}}\delta\mathcal{F}_{\textrm{wall,+};\textrm{J}}(x_{0}/R,\alpha=0) = R \partial_{x_{0}} \ln\biggl[ 1-\frac{R}{2x_{0}} \left( 3 - \frac{R^{2}}{2x_{0}^{2}}  \right) \biggr]
\end{equation}
and
\begin{equation}
\label{nJanuswall5}
(D/2) f_{x (\rm{wall,+;Jn})} \equiv - (D/2) \partial_{x_{0}}\delta\mathcal{F}_{\textrm{wall,+};\textrm{Jn}}(\mathcal{D},\alpha=0) = \partial_{\mathcal{D}^{-1}} \ln\biggl[ 1-\frac{\mathcal{D}}{4} \left( 3 + \frac{7\mathcal{D}^{2}}{16}  \right) \biggr] \, ,
\end{equation}
respectively, while for small distances it is given by
\begin{equation}
\label{Januswall6}
R f_{x (\rm{wall,+;J})} = \frac{11}{24} \pi^{2} \biggl[ 2 \left( \frac{x_{0}}{R} - 1 \right) \biggr]^{-3/2}
\end{equation}
and
\begin{equation}
\label{nJanuswall6}
(D/2) f_{x (\rm{wall,+;Jn})} = \frac{21/32}{\mathcal{D}^{-1}-1} \, ,
\end{equation}
respectively. The result in Eq. (\ref{Januswall6}) follows upon replacing the critical Casimir amplitude $\Delta_{+a}$ in the corresponding Derjaguin expression for a circle with \emph{homogeneous} boundary condition $a$ by $(\Delta_{++}+\Delta_{+-})/2=11\pi/48$. This is plausible and is demonstrated in Appendix \ref{appendix_F}. In order to derive Eq. (\ref{nJanuswall6}) we have used Eqs. (D10) and (D12) in Ref. \cite{EB_16} for a semi-infinite Janus needle. The expressions for the force at large and at short distances in the case of the Janus circle (Eqs. (\ref{Januswall5}) and (\ref{Januswall6}), respectively) and in the case of the Janus needle (Eqs. (\ref{nJanuswall5}) and (\ref{nJanuswall6}), respectively) are displayed in the two panels of Fig. \ref{fig07_03}.
They imply a force on the two Janus particles which is repulsive at all distances, in agreement with intuition: the increased energy density near the switching points is incompatible with the reduced energy density near the $+$ boundary of the half plane. Moreover, for the Janus circle close to the wall, the repulsive contribution coming from its $-$ segment dominates the attractive one from its + segment.

\subsection{Janus particle in a strip}
\label{sec_J_strip}
Here we consider a horizontal strip $\mathbb{S}$ of macroscopic length and width $\mathcal{W}$ centered at the real axis in the $z$-plane as introduced below Eq. (\ref{map_strip}). Its lower and upper boundary lines are endowed with boundary universality classes $a$ and $b$, respectively, and the system is denoted as $\mathbb{S}_{ab}$. Applying the conformal mapping given by Eq. (\ref{map_strip}) to the upper half $w$-plane  together with using the averages of $\varepsilon$, $\Phi$, and $T$ in the upper half $w$-plane $\mathbb{H}_{ab}$ (with corresponding boundary conditions $(a,b)$ for $(u=\textrm{Re} \, w > 0, u<0)$) provided by Ref. \cite{BX_91} implies, via the transformation formulas in Eqs. (\ref{03_08}) and (\ref{stresstrafo}), the averages
\begin{equation}
\begin{aligned}
\label{PsiS}
\langle \varepsilon(z,\bar{z}) \rangle_{\mathbb{S}_{ab}} & = (\pi/\mathcal{W}) f_{ab}^{(\varepsilon)}(Y) \, , \\
\langle \Phi(z,\bar{z}) \rangle_{\mathbb{S}_{ab}} & = (\pi/\mathcal{W})^{1/8} f_{ab}^{(\Phi)}(Y) \, , \\
\langle T(z) \rangle_{\mathbb{S}_{ab}} & = (\pi/\mathcal{W})^{2} f_{ab}^{(T)} \, ,
\end{aligned}
\end{equation}
in the strip. Here $Y\equiv \pi y/\mathcal{W}$ and
\begin{align}\nonumber 
\label{f_strip_average}
f_{OO}^{(\varepsilon)} & = (1/2)(\cos Y)^{-1} \, , && f_{OO}^{(\Phi)}=0 \, , && f_{++}^{(\varepsilon)}=-f_{OO}^{(\varepsilon)} \, , && \\ \nonumber
f_{++}^{(\Phi)} & = 2^{1/8}(\cos Y)^{-1/8} \, , && f_{+-}^{(\varepsilon)}=(4\cos Y-(\cos Y)^{-1})/2 \, , && f_{+-}^{(\Phi)}=-2^{1/8}(\cos Y)^{-1/8} \sin Y \, , &&\\
f_{+O}^{(\varepsilon)} & = (\tan Y)/2 \, , && f_{+O}^{(\Phi)}=(2\cos Y)^{-1/8}(1-\sin Y)^{1/4} \, , && f_{ab}^{(T)}= \Delta_{ab} / \pi \, , &&
\end{align}
with $\Delta_{ab}$ from Eq. (\ref{Casimir_plates'}).

The free energy needed to transfer a Janus particle from the bulk into the strip with the particle center located at $(x_{0},y_{0})$ follows from Eq. (\ref{int_01}). In the case of a symmetric \emph{circular} Janus particle for $\langle s_{\textrm{J}}\rangle$ in Eq. (\ref{int_01}) one inserts the sum of the following strip averages\footnote{For the derivation of Eq. (\ref{J_strip_average}) we have used the relations
 $\partial_{z}=(-i\pi/(2\mathcal{W}))\partial_{Y}$ and $\partial_{\bar{z}}=(i\pi/(2\mathcal{W}))\partial_{Y}$. Similar relations have been used for Eq. (\ref{nJ_strip_average}).}:
\begin{equation}
\begin{aligned}
\label{J_strip_average}
\langle s_{\textrm{J}}^{(\varepsilon)} \rangle_{\mathbb{S}_{ab}} & = \mathcal{J} \bigl[ 3 - \mathcal{J}^{2}\left( 1/4+2c^{2} \right) \partial_{Y_{0}}^{2} \bigr] f_{ab}^{(\varepsilon)}(Y_{0}) + O\left( \mathcal{J}^{5} \right) \, , \\
\langle s_{\textrm{J}}^{(\Phi)} \rangle_{\mathbb{S}_{ab}} & = 2^{13/4} \mathcal{J}^{9/8} c\partial_{Y_{0}} f_{ab}^{(\Phi)}(Y_{0}) + \langle s_{\textrm{J}}^{(\Phi)} \rangle_{\mathbb{S}_{ab}}^{(\textrm{nl})} + O\left( \mathcal{J}^{5+1/8} \right) \, , \\
\langle s_{\textrm{J}}^{(\mathbb{I})} \rangle_{\mathbb{S}_{ab}} & = 16\mathcal{J}^{2} \left( 2c^{2}-1 \right) f_{ab}^{(T)} + O\left( \mathcal{J}^{4} \right) \, ,
\end{aligned}
\end{equation}
with $\mathcal{J} \equiv \pi R/\mathcal{W}$, $c \equiv \cos\alpha$, and $Y_{0} \equiv \pi y_{0}/\mathcal{W}$, which follow from Eqs. (\ref{J_spoe_en})-(\ref{J_spoe_id}). In the case of the Janus \emph{needle} one inserts the sum of
\begin{equation}
\begin{aligned}
\label{nJ_strip_average}
\langle s_{\textrm{Jn}}^{(\varepsilon)} \rangle_{\mathbb{S}_{ab}} & = \mathcal{N} \biggl[ \frac{3}{2} + \mathcal{N}^{2}\left( \frac{7}{64} - \frac{17}{32} c^{2} \right) \partial_{Y_{0}}^{2} \biggr] f_{ab}^{(\varepsilon)}(Y_{0}) + O\left( \mathcal{N}^{5} \right) \, , \\
\langle s_{\textrm{Jn}}^{(\Phi)} \rangle_{\mathbb{S}_{ab}} & = 2^{17/8}\mathcal{N}^{9/8} c\partial_{Y_{0}} f_{ab}^{(\Phi)}(Y_{0}) +  \langle s_{\textrm{Jn}}^{(\Phi)} \rangle_{\mathbb{S}_{ab}}^{(\textrm{nl})} + O\left( \mathcal{N}^{5+1/8} \right) \, , \\
\langle s_{\textrm{Jn}}^{(\mathbb{I})} \rangle_{\mathbb{S}_{ab}} & = \frac{9}{2} \mathcal{N}^{2} \left( 2c^{2}-1 \right) f_{ab}^{(T)} + O\left( \mathcal{N}^{4} \right) \, ,
\end{aligned}
\end{equation}
with $\mathcal{N} \equiv \pi D/(2\mathcal{W})$, which follow from Eqs. (\ref{nJ_spoe_en})-(\ref{nJ_spoe_id}). Here $\langle s_{\rm P}^{(\Phi)} \rangle_{\mathbb{S}_{ab}}^{(\textrm{nl})}$ are the next-to-leading contributions to $\langle s_{\rm P}^{(\Phi)} \rangle_{\mathbb{S}_{ab}}$ which are of the order of $\mathcal{J}^{3+1/8}$ and $\mathcal{N}^{3+1/8}$, respectively, for $\textrm P=\textrm{J}$ and $\textrm P=\textrm{Jn}$.

For $ab=++$ we have obtained the explicit expressions
\begin{equation}
\begin{aligned}
\label{nl_averages}
\langle s_{\textrm J}^{(\Phi)} \rangle_{\mathbb{S}_{++}}^{(\textrm{nl})} & = - \frac{2^{1/4}}{672} \bigl[ 16(53 t_{0} + 21 t_{0}^{3})c^{3} + (428t_{0}+819t_{0}^{3})c \bigr] \mathcal{J}^{3+1/8} f_{++}^{(\Phi)}(Y_{0}) \, , \\
\langle s_{\textrm{Jn}}^{(\Phi)} \rangle_{\mathbb{S}_{++}}^{(\textrm{nl})} & = - \frac{2^{9/8}}{1792} \bigl[ (328 t_{0} +189 t_{0}^{3})c^{3} + (20 t_{0} + 126 t_{0}^{3})c \bigr] \mathcal{N}^{3+1/8} f_{++}^{(\Phi)}(Y_{0}) \, ,
\end{aligned}
\end{equation}
for a Janus circle and a Janus needle, respectively. Here $t_{0}=\tan Y_{0}$, and we have used Eqs. (\ref{cf_010}) and (\ref{cf_014_nl}).

In the subsequent subsections we conclude with observations for the various pairs $ab$ of strip boundary conditions, mainly concerning the orientation of the Janus particles.

\subsubsection{Janus particle in a strip with two ordinary boundaries}
\label{sec_J_strip_ord}
If both boundaries belong to the ordinary surface universality class, the strip average $\langle \Phi \rangle_{\mathbb{S}_{OO}}$ vanishes and the orientation-dependence of a small Janus circle and of a Janus needle stems in leading order from
\begin{equation}
\label{Janus_ordinary}
\langle s_{\textrm{P}}^{(\mathbb{I})} \rangle_{\mathbb{S}_{OO}} = -[16R^{2},(9/8)D^{2}] \frac{\pi^{2}}{48\mathcal{W}^{2}} \cos2\alpha \, ,
\end{equation}
for $\textrm{P}=[\textrm{J},\textrm{Jn}]$. This implies that the orientation with the lowest free energy, i.e., of maximal $\langle s_{\textrm{J}, \textrm{Jn} }^{(\mathbb{I})} \rangle_{\mathbb{S}_{OO}}$, is $\alpha=\pm\pi/2$ for which the dipole direction of the Janus particle is parallel to the strip axis. This is easy to understand intuitively in terms of the best fit between the stress tensor in the empty strip and the stress tensor induced at large distances around an isolated Janus particle in the entire plane without the strip.

As discussed between Eqs. (\ref{03_12}) and (\ref{03_15}) and at the end of Sec. \ref{stressnJ}, for Janus circles and Janus needles with $\alpha =0$ the eigenvector $\widehat{\bm{e}}_{+}(0,y)$ has the orientation of the $y$ axis and $\widehat{\bm{e}}_{-}(x,0)$ has the orientation of the $x$ axis (see Fig. \ref{plot03_02}). The Cartesian stress tensor of the horizontal empty strip with equal boundaries has eigenvectors $\widehat{\bm{e}}_{+}$ (with positive eigenvalue) and $\widehat{\bm{e}}_{-}$ (with negative eigenvalue) pointing along the $x$ and $y$ axes, respectively (see Eqs. (\ref{PsiS}), (\ref{f_strip_average}), and (\ref{cartcompstress})). It follows that for $\alpha=0$ the eigenvector fields associated with the Janus particle and with the empty strip do not agree. However, rotating the Janus particle to $\alpha=\pm\pi/2$, the signs of the corresponding eigenvalues do agree.

This orientational preference is supported by the next-higher order via the energy term $\langle s_{\textrm{J}}^{(\varepsilon)} \rangle_{\mathbb{S}_{OO}}$ which also favors $\alpha=\pm\pi/2$. Again this can be understood intuitively: the region with enhanced energy near the switching points prefers the neighborhood of the ordinary strip edges with their enhanced energy. 

\subsubsection{Janus circle in a strip with $++$ boundaries}
\label{sec_J_strip_plus}
For $Y_{0} \neq 0$ the orientation of the small Janus particle with the lowest free energy is determined by the contribution $\langle s_{\textrm{J}, \textrm{Jn}}^{(\Phi)} \rangle_{\mathbb{S}_{++}}$, predicting $\alpha=0$ for $Y_{0} <0$ and $\alpha=\pi$ for $Y_{0} <0$. As expected, with these orientations the $+$ half circle of the Janus circle faces that one of the two $+$ strip boundaries, which is closest. For the center of the Janus circle on the midline of the strip $Y_{0} =0$, $\langle s_{\textrm{J}, \textrm{Jn}}^{(\Phi)} \rangle_{\mathbb{S}_{++}}$ vanishes for all $\alpha$ and it is $\langle s_{\textrm{J}, \textrm{Jn}}^{(\mathbb{I})} \rangle_{\mathbb{S}_{++}}$ which determines the equilibrium orientation. Due to $f_{++}^{(T)} = f_{OO}^{(T)}$ this expression is the same as for the $OO$ strip and the orientation is $\alpha = \pm \pi/2$, as discussed in the previous subsection. However, $f_{++}^{(\varepsilon)} = - f_{OO}^{(\varepsilon)}$ so that this orientational preference is reduced rather than supported by the next-higher order term which is the energy term.

\subsubsection{Janus particle in a strip with $+-$ or $+O$ boundaries}
Here the orientation with the lowest free energy is $\alpha=\pi$ for all $ Y \in (-\pi/2,\pi/2)$, because the leading contribution of $\langle s_{\textrm{J}, \textrm{Jn} }^{(\Phi)} \rangle_{\mathbb{S}_{ab}}$ $\propto R^{9/8}$ or $\propto D^{9/8}$ with $ab \in \{ +-, +O\}$ is negative for all $Y_{0} $. Thus the $+$ half circle of the small Janus particle always faces the lower $+$ boundary of the strip.

\subsection{Janus particle in a wedge}
\label{sec_J_wedge}
\begin{figure*}[htbp]
\centering
        \begin{subfigure}[b]{0.42\textwidth}
            \centering
            \includegraphics[width=\textwidth]{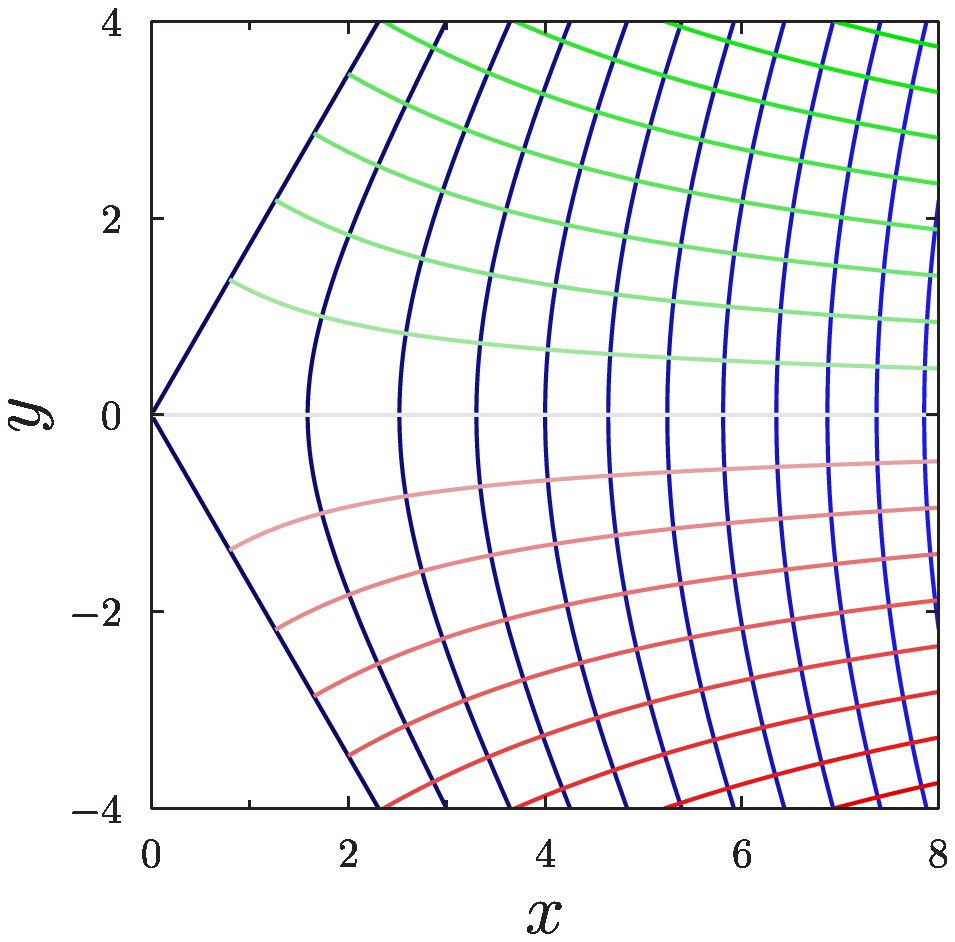}
            \caption[]%
            {{\small }}    
            \label{fig08_01a}
        \end{subfigure}
\hfill
        \begin{subfigure}[b]{0.42\textwidth}  
            \centering 
            \includegraphics[width=\textwidth]{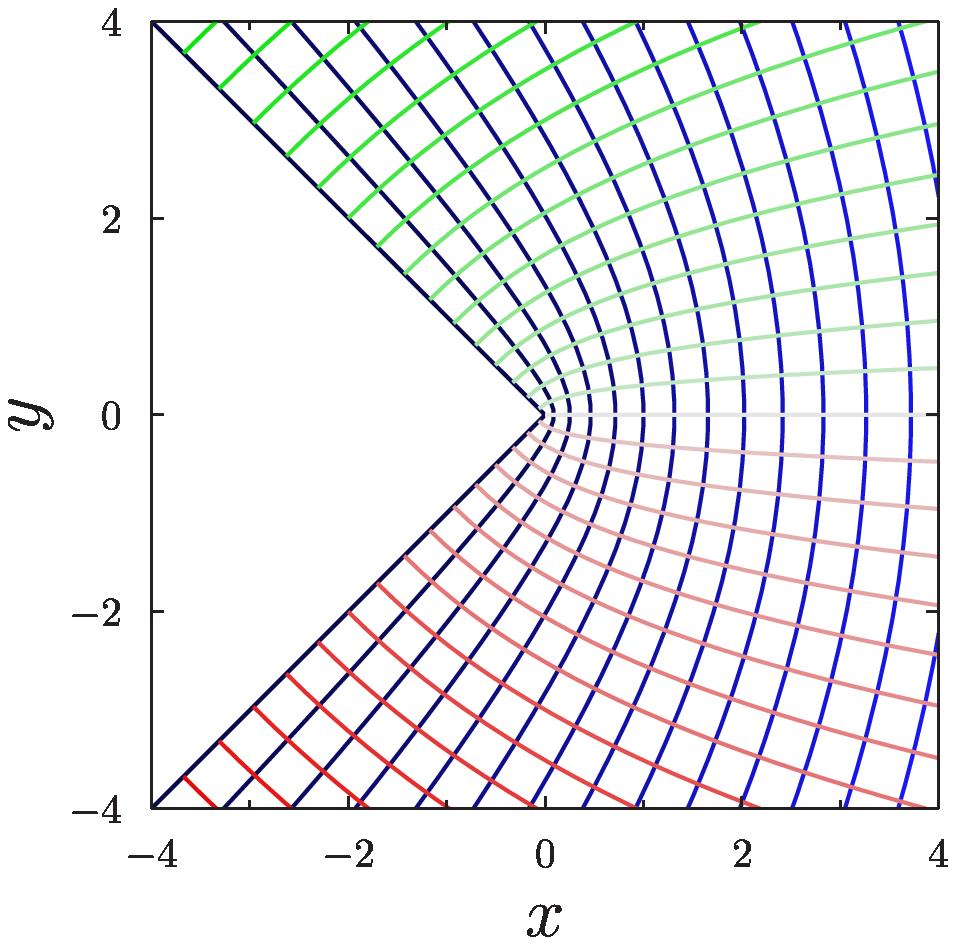}
            \caption[]%
            {{\small }}    
            \label{fig08_01c}
        \end{subfigure}
\caption[]
{\small A wedge with $\pi/\Xi=2\pi/3$ (a) and a wedge with $\pi/\Xi=3\pi/2$ (b) in the $z$ plane. The latter corresponds to an acute opening angle of $\pi/2$ at the apex in the inaccessible region (white). We show the curved rectangular grid which, via Eq. (\ref{map_wedge}), corresponds to the grid of  straight lines with constant $u$ and $v$ in the upper half $w=u+iv$ plane. The boundary of the wedge (dark blue) corresponds to the boundary $v=0$ and the light blue lines in the interior of the wedge correspond to $v>0$. The midline (depicted in grey) and the green and red lines in the upper and lower half of the wedge correspond to $u=0$ and to $u<0$ and $u>0$, respectively.}
\label{fig08_01}
\end{figure*}

In order to calculate the necessary operator averages in the wedge geometry $\mathbb{W}$ as introduced in Sec. \ref{sec_J_half}, we use the conformal mapping in Eq. (\ref{map_wedge}) which relates the accessible region to the upper half $w$ plane. The wedge is depicted in Fig. \ref{fig08_01}.

We start by discussing the stress tensor in the case of a wedge $\mathbb{W}_{ab}$ with one edge carrying the boundary condition $a$ and the other edge exhibiting the boundary condition $b$. In this case the transformation formula in Eq. (\ref{stresstrafo}) and the corresponding half plane expression $\langle T(w) \rangle_{\mathbb{H}_{ab}}= t_{ab} /w^2$ yields
\begin{equation}
\label{wall_04}
\langle T(z) \rangle_{\mathbb{W}_{ab}} = \frac{ t_{ab} \Xi^{2}+ \left( 1-\Xi^{2} \right)/48 }{z^{2}} \, .
\end{equation}
Here ${\mathbb{H}_{ab}}$ is a generalization of ${\mathbb{H}_{-+}}$ with boundary conditions $a$ and $b$ for $u>0$ and $u<0$, respectively; $t_{aa}=0$ and $t_{-+}=1/2$ (compare Eq. (\ref{TH-+})). The corresponding Cartesian stress tensor $\langle T_{kl} \rangle_{\mathbb{W}_{ab}}$ at a point $z = |z| \exp(i\theta)$ has an eigenvector $(\cos\theta,\sin\theta)$ with an eigenvalue given by the right hand side of Eq. (\ref{wall_04}) times $-z^2/(\pi |z|^2)$. The stress tensor averages reproduce the special cases of the half plane and the strip.

Next we turn to the profiles of the primary operators $\mathcal{O}=\Phi$ or $\mathcal{O}=\varepsilon$. For simplicity we consider equal boundary conditions $a$ on both edges of the wedge. Combining the corresponding expression in Eq. (\ref{03_03}) for $ \langle \mathcal{O}(w,\bar{w}) \rangle_{\mathbb{H}_{a}}$ in the half plane with the conformal mapping in Eq. (\ref{map_wedge}), the transformation formula in Eq. (\ref{03_08}) yields the result
\begin{equation}
\label{wedge_1pf}
\langle \mathcal{O}(z,\bar{z}) \rangle_{\mathbb{W}_{aa}} =  \mathcal{A}_{\mathcal{O}}^{(a)} \left( \frac{\Xi}{|z|\cos(\Xi\theta)} \right)^{x_{\mathcal{O}}} \, .
\end{equation}
It can be verified that for $\Xi=1$ the average in the wedge as given in Eq. (\ref{wedge_1pf}) reduces to the average in the half plane with uniform boundary conditions of type $a$. Analogously, the average in a strip with equal boundary conditions $a$,
\begin{equation}
\label{ }
\langle \mathcal{O}(z,\bar{z}) \rangle_{\mathbb{S}_{aa}} =  \mathcal{A}_{\mathcal{O}}^{(a)} \left( \frac{\mathcal{W}}{\pi} \cos\frac{\pi\textrm{Im}(z)}{\mathcal{W}} \right)^{-x_{\mathcal{O}}} \, ,
\end{equation}
is retrieved by taking the limit of acute wedges, i.e., $\Xi \nearrow \infty$ in Eq. (\ref{wedge_1pf}) together with $\textrm{Re}(z) \rightarrow \infty$, in which the strip width is identified with $\mathcal{W} \rightarrow \pi \textrm{Re}(z)/\Xi$, while the polar angle is identified with $\theta \rightarrow \pi\textrm{Im}(z)/(\mathcal{W}\Xi)$.

\subsubsection{Janus particle in an ordinary wedge}
\label{sec_J_wedge_ord}
If the wedge has ordinary boundaries, which do not break the $\pm$ Ising symmetry, one has $\langle \Phi \rangle_{\mathbb{W}_{OO}} = 0$, so that $s_{\textrm{J}}^{(\Phi)}$ in Eq. (\ref{J_spoe_phi}) and $s_{\textrm{Jn}}^{(\Phi)}$ in Eq. (\ref{nJ_spoe_phi}) do not contribute to the embedding free energy in Eq. (\ref{int_01}) so that the dependence of the interaction on the orientation is dominated by the operators $s_{\rm J}^{(\mathbb{I})}$ and $s_{\rm Jn}^{(\mathbb{I})}$ yielding, with Eq. (\ref{wall_04}),
\begin{equation}
\label{JanWOO}
\langle s_{\textrm{P}}^{(\mathbb{I})} \rangle_{\mathbb{W}_{OO}} = [8R^{2}, (9/16)D^{2} ] \frac{1-\Xi^{2}}{24|z_{0}|^{2}} \cos(2\alpha-2\theta_{0})
\end{equation}
for $\textrm{P}=[\textrm{J},\textrm{Jn}]=[\textrm{Janus particle}, \, \textrm{Janus needle}]$ where we have introduced the polar angle $\theta_{0}$ of the point $z_{0} \equiv |z_{0}|\exp(i\theta_{0})$ where the center of the Janus particle is located. The orientation associated with the minimum of the free energy, i.e., the maximum of $\langle s_{\rm P} \rangle_{\mathbb{W}_{OO}}$, depends on whether the opening angle $\pi/\Xi$ of the wedge is larger or smaller than $\pi$. If the center $z_{0}$ of the Janus particle is located on the positive real axis, i.e., $\theta_{0}=0$, this optimal orientation is given by $\alpha=0$ or $\alpha=\pi$ (i.e., the dipole vector is perpendicular to the real axis) in the case $\Xi<1$ (as in Fig. \ref{fig08_01}(b)), while it is $\alpha=\pi/2$ or $\alpha=-\pi/2$ (i.e., the dipole vector is antiparallel or parallel to the direction of the real axis) in the case $\Xi>1$ (as in Fig. \ref{fig08_01}(a)). As expected, it is for these orientations that the stress tensor field of an \emph{isolated} Janus particle ``fits best'' the corresponding field of the isolated wedge; see the discussion below Eq. (\ref{Janus_ordinary}).

\subsubsection{Janus particle in a $+$ wedge}
\label{sec_J_wedge_plus}
For a wedge with $+$ boundaries the dependence of the interaction on the orientation is dominated by the operator $s_{\textrm{P}}^{(\Phi)}$ in Eqs. (\ref{J_spoe_phi}) and (\ref{nJ_spoe_phi}) yielding
\begin{equation}
\label{Janus_100}
\langle s_\textrm{P}^{(\Phi)} \rangle_{\mathbb{W}_{++}} = 2^{3/8} \Xi^{1/8} \left( \frac{ \bigl[ R, D/4 \bigr] }{ |z_{0}|\cos(\Xi\theta_{0}) }  \right)^{9/8} \Bigl( \cos(\Xi\theta_0)\sin(\alpha-\theta_0) + \Xi\cos(\alpha-\theta_0)\sin(\Xi\theta_0) \Bigr) \, ,
\end{equation}
for $\textrm{P}=[\textrm{J},\textrm{Jn}]$, where we have focused the discussion on the leading term and have used Eq. (\ref{wedge_1pf}) with $\mathcal{O}=\Phi$. For $\Xi = 1$, Eq. (\ref{Janus_100}) reduces to the corresponding half plane expressions in Eqs. (\ref{wall_J_phi}) and (\ref{wall_nJ_phi}) and for $\Xi \nearrow \infty$, as described in Sec. \ref{sec_J_half}, it reduces to the corresponding expressions in Eqs. (\ref{J_strip_average}) and (\ref{nJ_strip_average}) for the strip. Considering for the wedge again the special case in which the Janus particle is centered on the positive real axis, the right hand side of Eq. (\ref{Janus_100}) becomes proportional to $\sin\alpha$ (with a positive proportionality factor), which is maximal for $\alpha=\pi/2$, i.e., for the dipole vector pointing towards the apex of the wedge at the origin, as expected intuitively.

\section{Quadrupolar particle in a half plane, in a strip, or in a wedge}
\label{wall_section2}
Here we use Eq. (\ref{int_01}) together with the SPOE (Eq. (\ref{def_SPOE})) and Eqs. (\ref{Q_spoe_en})-(\ref{Q_spoe_id}) in order to calculate the free energy expense to embed the quadrupolar particle $\textrm{Q}$ shown in Fig. \ref{fig02_01}(c) into the three confined geometries ${\mathscr{G}}$ forming a half plane, a strip, or a wedge. While the corresponding averages $\langle s_{\textrm{Q}}^{(\varepsilon)} \rangle_{\mathscr{G}}$ and $\langle s_{\textrm{Q}}^{(\Phi)} \rangle_{\mathscr{G}}$ follow from $\langle \varepsilon \rangle_{\mathscr{G}}$ and $\langle \Phi \rangle_{\mathscr{G}}$, as shown in the above sections on Janus particles, the corresponding averages  $\langle  \cdot \rangle_{\mathscr{G}}$ of the three terms in the expression for $s_{\textrm{Q}}^{(\mathbb{I})}$ (Eq. (\ref{Q_spoe_id})) can be taken from Appendix \ref{appendix_E} in which they are calculated for each of the three geometries.

\subsection{Quadrupole in a half plane}
\label{sec_Q_half}
The free energy, which must be invested in order to transfer a quadrupolar particle from the bulk into the right half $z$ plane $\widetilde{\mathbb{H}}_{a}$ with boundary condition $a=O$ or $a=+$ on its boundary line $x = 0$, follows from Eq. (\ref{int_01}) with its right hand side given by the sum of
\begin{eqnarray}
\label{Q_eps_rhp}
\langle s_{\textrm{Q}}^{(\varepsilon)} \rangle_{\widetilde{\mathbb{H}}_{a}} & = & \left( \frac{5}{3} + \frac{7}{4}\partial_{\mathcal{X}_{0}}^{2} \right) \frac{\mathcal{A}_{\varepsilon}^{(a)}}{\mathcal{X}_{0}} + O\left( \mathcal{X}_{0}^{-5} \right) \, , \\
\label{Q_phi_rhp}
\langle s_{\textrm{Q}}^{(\Phi)} \rangle_{\widetilde{\mathbb{H}}_{a}} & = & -2^{1/4} \frac{32}{9} \left( 1 + \frac{2}{17}\partial_{\mathcal{X}_{0}}^{2} \right) (\sin2\alpha) \partial_{\mathcal{X}_{0}}^{2} \frac{\mathcal{A}_{\Phi}^{(a)}}{\mathcal{X}_{0}^{1/8}} + O\left( \mathcal{X}_{0}^{-(6+1/8)} \right) \, , \\
\label{Q_id_rhp}
\langle s_{\textrm{Q}}^{(\mathbb{I})} \rangle_{\widetilde{\mathbb{H}}_{a}} & = & \frac{35}{48} \frac{1}{\mathcal{X}_{0}^{4}} + O\left( \mathcal{X}_{0}^{-6} \right) \, ,
\end{eqnarray} 
with $\mathcal{X}_{0} \equiv x_{0}/R$ where $x_{0}$ is the distance of the center of the quadrupole from the boundary. Here we have used the SPOE in Eqs. (\ref{Q_spoe_en})-(\ref{Q_spoe_id}), and for Eq. (\ref{Q_id_rhp}) the results in Appendix \ref{appendix_E}. For a boundary with $a=+$, the amplitude $\mathcal{A}_{\Phi}^{(a)}$ is positive and the orientation $\alpha$ of the quadrupole, which minimizes the free energy, is given by the minimum of $\sin2\alpha$ for which $\langle s_{\textrm{Q}}^{(\Phi)} \rangle_{\widetilde{\mathbb{H}}_{+}}$ is maximal. As expected, this yields $\alpha=-\pi/4$ for which the Janus particle faces the $+$ boundary with one of its $+$ surface segments. For an ordinary boundary $a=O$ of the half plane, for which $\mathcal{A}_{\Phi}^{(a)}$ vanishes, Eq. (\ref{Q_phi_rhp}) yields no orientation dependence up to the orders studied.

\subsection{Quadrupole in a strip}
\label{sec_Q_strip}
Here we focus on strips with $OO$ or $++$ boundaries. In addition to the corresponding averages of $\varepsilon$ and $\Phi$ given in Eq. (\ref{f_strip_average}) one also needs the strip averages $\langle T\bar{T} \rangle_{\mathbb{S}_{aa}}$ and $\langle L_{-2}^{2}\mathbb{I} \rangle_{\mathbb{S}_{aa}}$ which are derived in Appendix \ref{appendix_E} and are given in Eqs. (\ref{strip_01}) and (\ref{des_strip}); they are the same for $OO$ and $++$ boundary conditions. The insertion free energy follows from Eq. (\ref{int_01}) with its right hand side being the sum of
\begin{eqnarray}
\label{SQE}
\langle s_{\textrm{Q}}^{(\varepsilon)} \rangle_{\mathbb{S}_{aa}} & = & \mathcal{J} \biggl[ \frac{5}{3} + \frac{7}{4} \mathcal{J}^{2}\partial_{Y_{0}}^{2} \biggr] f_{aa}^{(\varepsilon)}(Y_{0}) + O\left( \mathcal{J}^{5} \right) \, , \\
\label{SQP}
\langle s_{\textrm{Q}}^{(\Phi)} \rangle_{\mathbb{S}_{aa}} & = & 2^{1/4} \frac{32}{9} \mathcal{J}^{2+1/8} \biggl[ 1 + \frac{2}{17} \mathcal{J}^{2}\partial_{Y_{0}}^{2} \biggr] (\sin2\alpha) \partial_{Y_{0}}^{2} f_{aa}^{(\Phi)}(Y_{0}) + O\left( \mathcal{J}^{6+1/8} \right) \, ,
\end{eqnarray}
and
\begin{equation}
\label{SQI}
\langle s_{\textrm{Q}}^{(\mathbb{I})} \rangle_{\mathbb{S}_{aa}} = \mathcal{J}^{4} \biggl\{ \frac{35}{48}\biggl[ \frac{1}{\cos^{4}Y_{0}} + \frac{1}{36} \biggr] - \frac{7}{135}\cos4\alpha \biggr\} + O\left( \mathcal{J}^{6} \right) \, ,
\end{equation}
with $\cal J$ and $Y_{0}$ defined below Eq. (\ref{J_strip_average}).

\subsubsection{Ordinary boundaries}
In the case $aa=OO$, for which the scaling function $f_{aa}^{(\Phi)}$ vanishes (see Eqs. (\ref{PsiS}) and (\ref{f_strip_average})), the leading orientation dependence stems from $\langle s_{\textrm{Q}}^{(\mathbb{I})} \rangle_{\mathbb{S}_{aa}}$. This contribution predicts that the free energy is minimal for the orientations $\alpha = \pm \pi/4$ for which the square, formed by the four switching points of the quadrupole, has its edges oriented parallel and perpendicular to the boundary lines of the strip. An intuitive reasoning, similar to the one given for the Janus particle in a strip with ordinary boundaries, can be provided. In the empty strip with $OO$ boundaries the eigenvector parallel to the strip axis has a positive eigenvalue. This should be compared with the eigenvector field around an isolated quadrupolar particle which has been discussed in Fig. \ref{plot07_05} and between Eqs. (\ref{07_09bis}) and (\ref{TijQ}). What matters in the comparison are the eigenvectors of the quadrupole on a line passing through its center and being parallel to the strip axis, because it is only in this direction where long-distance correlations between the quadrupole and the strip can build up. For the orientations $\alpha=\pm \pi/4$ of the quadrupole this line passes through the midpoints of the $+$ and $-$ sections of the quadrupole (compare the rightmost panel of Fig. \ref{fig_angles}), and the eigenvector $\widehat{\bm{e}}_{+}$ (with positive eigenvalue) on this line is oriented parallel to it, i.e., parallel to the strip axis, which fits the eigenvector of the empty strip. Conversely, for the orientation $\alpha=0$ the line passes through the switching points and the eigenvector on this line, being oriented parallel to it (and to the strip), is $\widehat{\bm{e}}_{-}$ with a negative eigenvalue (see Fig. \ref{plot07_05}(c)), which does not fit the eigenvectors of the empty strip.

\subsubsection{Strip with $++$ boundaries}
\label{sec_Q_strip_plus}
For identical boundaries with symmetry breaking boundary conditions $+$, the leading orientational dependence of the free energy is dominated by the order parameter contribution given in Eq. (\ref{SQP}):
\begin{equation}
\langle s_{\textrm{Q} }^{(\Phi)} \rangle_{\mathbb{S}_{++}} = 2^{1/4+1/8} \frac{32}{9} \mathcal{J}^{17/8} \frac{17-\cos2Y_{0}}{ 128 (\cos Y_{0})^{17/8} } \sin2\alpha \, .
\end{equation}
For any position in the strip the factor multiplying $\sin2\alpha$ is positive. Therefore, as expected, the orientation $\alpha=\pi/4$ (for which the $+$ segments of the quadrupole face the strip boundaries) minimizes the free energy.

\subsection{Orientations of a quadrupole in a wedge}

\subsubsection{Ordinary boundaries}
\label{sec_Q_wedge_ord}
For a wedge ${\mathbb{W}}_{OO}$ with two ordinary edges, $\langle \Phi \rangle_{\mathbb{W}_{OO}}=0$ and the orientational dependence of the interaction of the quadrupole with the wedge is dominated by $\langle s_{\textrm{Q}}^{(\mathbb{I})} \rangle_{\mathbb{W}}$, which has the form
\begin{equation}
\label{Q_Owedge}
\langle s_{\textrm{Q}}^{(\mathbb{I})} \rangle_{\mathbb{W}_{OO}} = \left( \frac{R}{|z_{0}|} \right)^{4} \biggl\{ \frac{35}{48} \biggl[ \frac{\Xi^{4}}{\cos^{4}\Xi\theta_{0}} + \frac{ \left( \Xi^{2}-1 \right)^{2} }{36} \biggr] - \left( \Xi^{2}-1\right)\frac{49\Xi^{2}+329}{945}\cos(4\alpha-4\theta_{0}) \biggr\} \, .
\end{equation}
This follows from averaging $s^{(\mathbb{I})}$ in Eqs. (\ref{Q_spoe_en})-(\ref{Q_spoe_id}) with the aid of Eqs. (\ref{wedge_desc_01}) and (\ref{wedge_desc_02}). Here $\theta_0$ is the angle introduced below Eq. (\ref{JanWOO}). Thus, following from Eq. (\ref{int_01}), the orientations $\alpha$ of the quadrupole, associated with the lowest free energy and corresponding to the maximum of $\langle s^{(\mathbb{I})} \rangle_{\mathbb{W}_{OO}}$, are $\alpha-\theta_{0} \in \{\pm\pi/4, \pm 3\pi/4 \}$ for a wedge with opening angle $\pi/\Xi$ smaller than $\pi$ and $\alpha-\theta_{0} \in \{0, \pm \pi/2 , \pi\}$ for an opening angle larger than $\pi$. These conclusions again agree with the ``best fit'' scenario between the stress tensor fields of the isolated quadrupole and the empty wedge.

\subsubsection{$+$ boundaries}
\label{sec_Q_wedge_plus}
Here the orientational dependence of the interaction of the quadrupole with a wedge is dominated by $\langle s^{(\Phi)} \rangle_{\mathbb{W}}$ which in leading order is given by
\begin{equation}
\label{wedge_019}
\langle s_{\textrm{Q}}^{(\Phi)} \rangle_{\mathbb{W}_{++}} = 2^{9/4} \frac{16}{9} R^{17/8} \biggl[ \textrm{e}^{2i(\alpha+\pi/4)}\partial_{z_{0}}^{2} \langle \Phi(z_{0},\bar{z}_{0}) \rangle_{\mathbb{W}_{++}} + \textrm{cc} \biggr] \, .
\end{equation}
In the special case that the quadrupole is centered on the positive real axis, i.e., $z_{0}=\bar{z}_{0}\equiv x_{0}$, the mean value in Eq. (\ref{wedge_019}) reduces to
\begin{equation}
\label{wedge_020}
\langle s_{\textrm{Q}}^{(\Phi)} \rangle_{\mathbb{W}_{++}} = \frac{\Xi^{1/8}}{9 \times 2^{5/8}} \left( \frac{R}{x_{0}} \right)^{17/8} \left( 8\Xi^{2}-17 \right) \sin2\alpha \, .
\end{equation}
The quadrupole orientations $\alpha$ with the lowest free energy, which are determined by the maximum of $\langle s_{\textrm{Q}}^{(\Phi)} \rangle_{\mathbb{W}_{++}}$, are given by $\alpha \in \{ -\pi/4, 3\pi/4\}$ (i.e., the $+$ segments of the quadrupole are perpendicular to the real axis) for large opening angles of the wedge with $\Xi^{-1} > \sqrt{8/17}$ so that Eq. (\ref{wedge_020}) is negative. For small angles $\Xi^{-1} < \sqrt{8/17}$, which render Eq. (\ref{wedge_020}) positive, they are given by $\alpha \in \{ \pi/4, -3\pi/4\}$ (i.e., the $+$ segments are parallel to the real axis). The marginal case corresponds to the critical angle $\pi\sqrt{8/17} \approx 123.5^{\circ}$. This agrees with the expectation that the $+$ segments are as close as possible to the $+$ boundaries of the wedge.

\section{Interaction between two particles}
\label{interaction_JJ}
The interaction between two small particles follows from Eq. (\ref{int_02}). More generally, the free energy cost $\delta {\mathcal F}_{{\rm P}_1 \cdots {\rm P}_n}$ to transfer $n$ small colloids  ${\rm P}_{1}, \dots, {\rm P}_{n}$, each of them codified by a SPOE of the type given in Eq. (\ref{def_SPOE}), from infinite mutual distances to their actual positions $\1{x}_{1}, \dots,  \1{x}_{n}$ of their centers in the bulk, is given by \cite{BE_95}
\begin{equation}
\label{OPE_10_new}
\delta \mathcal{F}_{{\rm P}_{1} , \dots , {\rm P}_{n}} = -\ln \Bigg\langle \prod_{j=1}^{n} \Bigl( 1+s_{{\rm P}_j}(\1{x}_{j}, \alpha_j) \Bigr) \Bigg\rangle \, ,
\end{equation}
where $\alpha_j$ denotes the orientation of particle $\textrm{P}_{j}$ and the angular brackets without subscript denote as usual an average in the unperturbed bulk. Due to the multi-point correlation functions appearing on the right hand side of Eq. (\ref{OPE_10_new}), the critical system induces many-body interactions between the particles. Discussing them is interesting but beyond the scope of the present study. Upon increasing all mutual distances $|\1{x}_{j} - \1{x}_{k}|$ between the particles, all correlations but the two-point ones of the leading operators can be neglected and the interaction reduces to a sum of two-body terms\footnote{A simple interesting example is the three-body interaction arising between two circular Janus particles $\textrm{J}_{1}$ and $\textrm{J}_{2}$, and a circular particle C with a homogeneous ordinary boundary, all three of them with the same radius  $R$. Since $\langle \varepsilon_1 \varepsilon_2 \varepsilon_3 \rangle$ vanishes, the leading three-body term in $\delta {\cal F}_{\rm \textrm{J}_{1}, \textrm{J}_{2}, C}$ for $R \to 0$ is of the order $R^{3+1/4}$ and arises from $-\langle s_{\textrm{J}_{1}}^{(\Phi)} s_{\textrm{J}_{2}}^{(\Phi)} s_{\rm C}^{} \rangle$ where $s_{\rm C} \to R \varepsilon(\1{x}_{3}) $ is the leading operator contribution for the ordinary circle. This term with its dependence on $\1{x}_{12}, \1{x}_{13}, \1{x}_{23}$ and $\alpha_1, \alpha_2$ can be determined by appropriate differentiations of the bulk three-point correlation function $\langle \Phi_1 \Phi_2 \varepsilon_3 \rangle = -(1/2)   |\1{x}_{12}|^{3/4} |\1{x}_{13}|^{-1} |\1{x}_{23}|^{-1}$. This should be compared with the two-body terms $\propto \langle \varepsilon_{1} \varepsilon_{2} \rangle, \langle \varepsilon_{1} \varepsilon_{3} \rangle, \langle \varepsilon_{2} \varepsilon_{3} \rangle$, and $\langle \partial \Phi_{1} \partial \Phi_{2} \rangle$ stemming from the energy monopole-monopole and spin dipole-dipole channels, which are of the order $R^{2}$ and $R^{2+1/4}$, respectively, and dominate for $R \to 0$. Three-body interactions between two small Janus particles and a large particle, such as the boundary of a half plane, can also be determined by the SPOE approach. Here one has to suitably differentiate the two-point order parameter correlation function in the half plane.}.

Here we focus on the interaction between two particles $\textrm{P}_{1}$ and $\textrm{P}_{2}$ for which their SPOEs in Eq. (\ref{def_SPOE}) yield the expression
\begin{equation}
\label{int_04}
\textrm{e}^{-\delta\mathcal{F}_{{\rm P}_1; {\rm P}_2}} = 1 + \langle s_{{\rm P}_1} s_{{\rm P}_2} \rangle = 1 + \langle s_{{\rm P}_1}^{(\varepsilon)} s_{{\rm P}_2}^{(\varepsilon)} \rangle + \langle s_{{\rm P}_1}^{(\Phi)} s_{{\rm P}_2}^{(\Phi)} \rangle + \langle s_{{\rm P}_1}^{(\mathbb{I})} s_{{\rm P}_2}^{(\mathbb{I})} \rangle \, ,
\end{equation}
because the descendants of $\varepsilon$, $\Phi$, and $\mathbb{I}$ are ``orthogonal in the bulk''. We recall that a bulk two-point correlation function of primary operators vanishes unless they have the same scaling dimension. Accordingly, the correlation function of descendant operators belonging to different conformal families also vanishes \cite{CFT_books_1}. In the following we shall address three representative cases: two circular Janus particles; a Janus circle or a Janus needle and an ordinary needle; and two quadrupoles. Besides the interaction at \emph{large} interparticle distances given by Eq. (\ref{int_04}) with the three averages on the right hand side to be calculated from the explicit expressions for the quantities $s_{\rm P}$ in Sec. \ref{SPOE}, we shall, for certain cases, also consider the interaction if the two particles \emph{nearly touch} each other.
	
Without losing information we position the two particles $\textrm{P}_{1}$ and $\textrm{P}_{2}$ with their centers at $z_{1}=-r/2$ and $z_{2}=r/2$ on the negative and positive real axis, respectively.

\subsection{Two Janus circles}
\label{2J}
\begin{figure}[htbp]
\centering
\includegraphics[width=10.5cm]{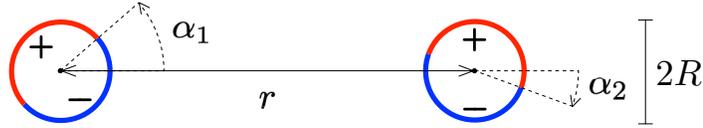}
\caption{Two identical Janus particles with radii $R$ and center-to-center distance $r$. As explained in Fig. \ref{fig_angles}, the orientations $\alpha_{j}$ with $j=1,2$ are measured relative to the reference configuration $\alpha=0$ shown in Fig. \ref{fig02_01} (a). Here, the angle $\alpha_{1}$ is positive while the angle $\alpha_{2}$ is negative.}
\label{fig05_01}
\end{figure}

Consider two circular Janus particles ${\rm P}_1 = {\rm J}_1$ and ${\rm P}_2 = {\rm J}_2$ with equal radii $R$ and with their orientations rotated away from the standard orientation (see  Fig. \ref{fig02_01}) by the angles $\alpha_{1}$ and $\alpha_{2}$, as illustrated in Fig. \ref{fig05_01}.

For large separations, i.e., a small size over distance ratio $R/r$, the averages on the right hand side of the expression in Eq. (\ref{int_04}) for the interaction free energy $\delta {\cal F}_{\rm J_1 ; J_2}$ are determined by Eqs. (\ref{J_spoe_en})-(\ref{J_spoe_id}) and give rise to the following expansions, which include powers up to $(R/r)^{4+1/4}$:
\begin{equation}
\begin{aligned}
\label{JJ_result_interaction}
\langle s_{\textrm{J}_{1}}^{(\varepsilon)} s_{\textrm{J}_{2}}^{(\varepsilon)} \rangle & = 3 \left( \frac{R}{r} \right)^{2} \biggl\{ 3 + 2 \left( \frac{R}{r}\right)^{2} \Bigl[ -5 + 4(\cos2\alpha_{1}+\cos2\alpha_{2}) \Bigr] \biggr\} \, , \\
\langle s_{\textrm{J}_{1}}^{(\Phi)} s_{\textrm{J}_{2}}^{(\Phi)} \rangle & = 2^{3/2} \left( \frac{R}{r} \right)^{9/4} \biggl\{ \mathscr{A}(\alpha_{1},\alpha_{2}) + \frac{1}{4} \left( \frac{R}{r} \right)^{2} \biggl[ 4\mathscr{B}(\alpha_{1},\alpha_{2}) + 3\mathscr{C}(\alpha_{1},\alpha_{2}) \biggr] \biggr\} \, , \\
\langle s_{\textrm{J}_{1}}^{(\mathbb{I})} s_{\textrm{J}_{2}}^{(\mathbb{I})} \rangle & = 32 \left( \frac{R}{r} \right)^{4} \cos(2\alpha_{1}+2\alpha_{2}) \, ,
\end{aligned}
\end{equation}
where
\begin{equation}
\begin{aligned}
\label{JJ_result_interaction'}
\mathscr{A}(\alpha_{1},\alpha_{2}) & = 9 \cos(\alpha_{1}+\alpha_{2}) - \cos(\alpha_{1}-\alpha_{2}) \, , \\
\mathscr{B}(\alpha_{1},\alpha_{2}) & = 25 \cos\bigl[ 2(\alpha_{1} + \alpha_{2}) \bigr] \cos\bigl[ \alpha_{1} - \alpha_{2} \bigr] -  \cos\bigl[ \alpha_{1} + \alpha_{2} \bigr] \cos\bigl[ 2 (\alpha_{1} - \alpha_{2}) \bigr] \, , \\
\mathscr{C}(\alpha_{1},\alpha_{2}) & = -17 \cos(\alpha_{1}+\alpha_{2}) + 9 \cos(\alpha_{1}-\alpha_{2}) \, .
\end{aligned}
\end{equation}

$\delta\mathcal{F}_{\rm J_1 ; J_2} (r/R ;\alpha_{1},\alpha_{2})$ is invariant under the transformations $(\alpha_{1},\alpha_{2}) \mapsto (\alpha_{1}+\pi,\alpha_{2}+\pi)$ and $(\alpha_{1},\alpha_{2}) \mapsto (\alpha_{2},\alpha_{1})$. These invariances hold to any order in $r/R$, because the first of these transformations, i.e., rotating each of the two circular Janus particles about their center by $180$ degrees, amounts to exchanging the $+$ and the $-$ boundary conditions in all boundary pieces of the Janus circles. Due to the $+ \leftrightarrow -$ symmetry of the Ising model this transformation does not change the partition sum and thus the free energy. Combining the first transformation with an overall rotation of the system by $180$ degrees yields the second transformation, obviously without changing the free energy.

\begin{figure}[htbp]
\centering
\includegraphics[width=15.5cm]{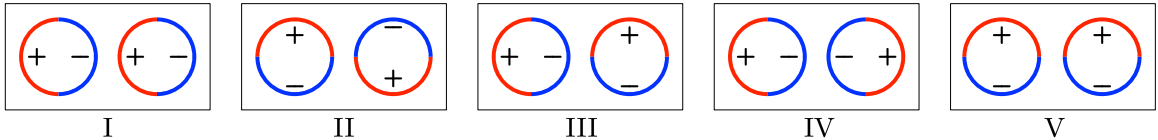}
\caption{Five special angular configurations of two Janus particles.}
\label{angles}
\end{figure}

In Ref. \cite{BPK} the effective interaction of two Janus particles has been obtained by simulations on a square lattice. The Janus particles have been treated as pieces of material, each composed of neighboring elementary squares, thereby approximating a circular particle shape with piecewise straight Janus boundaries. We provide analytical results for the five special configurations considered in Ref. \cite{BPK} which in our notation correspond to $(\alpha_1 , \alpha_2)$ equal to
\begin{align}
\label{}
(\pi/2,\pi/2) \,\,  [\textrm{I}] \, ,   && (0,\pi) \,\, [\textrm{II}] \, ,   &&	(\pi/2,0) \,\, [\textrm{III}] \, ,   && (\pi/2,-\pi/2) \,\, [\textrm{IV}] \, ,   && (0,0) \,\, [\textrm{V}] \, 
\end{align}
(see Fig. \ref{angles}). 

\begin{figure}[htbp]
\centering
\includegraphics[width=12.5cm]{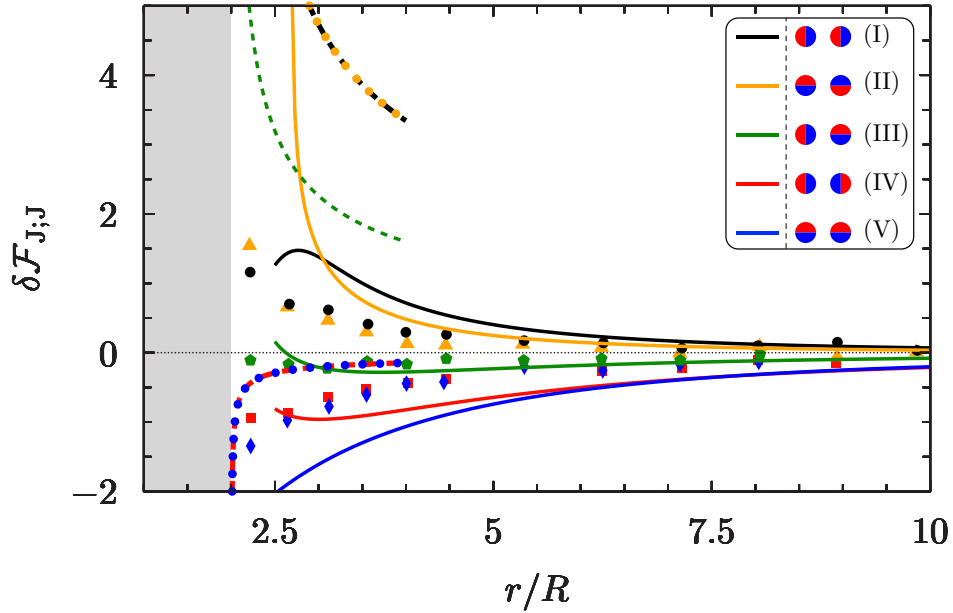}
\caption{The free energy of interaction (per $k_{\rm B}T_{\rm c}$) for two circular Janus colloids for the five orientations shown in Fig. \ref{angles}. The analytical results from Eq. (\ref{JJ_result_interaction}) for the interaction energy at large distances, $\delta\mathcal{F}_{\textrm{J;J}}(r/R;\alpha_{1},\alpha_{2})$, as a function of the rescaled separation $r/R$ for various relative alignments, are illustrated by solid lines. The leading short-distance asymptotic results in Eq. (\ref{short04}) are shown as dotted and dashed lines with the same color used for large separations. The dashed and dotted lines for I and II as well as those for IV and V de facto coincide (see Eq. (\ref{short05})). The symbols correspond to simulation data extracted from Fig. 7(b) in Ref. \cite{BPK}. The color code of the symbols corresponds to that of the lines. The shaded region corresponds to hard core repulsion.}
\label{fig05_02}
\end{figure}

As in Secs. \ref{sec_J_half} and \ref{wall_section2} we augment, for these configurations, our large distance results in Eqs. (\ref{JJ_result_interaction}) and (\ref{JJ_result_interaction'}), which are inserted into Eq. (\ref{int_04}), by the short distance Derjaguin-like results:
\begin{equation}
\label{short04}
\delta\mathcal{F}_{\rm J_1 , J_2} \bigl( r/R \, ; \, \alpha_{1}(j),\alpha_{2}(j) \bigr) = \frac{\pi \mathscr{D}_{j}}{\sqrt{r/R-2}} \, ,
\end{equation}
with $j \in \{ \textrm{I}, \dots , \textrm{V}\}$ and
\begin{align}
\label{short05}
&\mathscr{D}_{\textrm{I}} = \mathscr{D}_{\textrm{II}} = \Delta_{+-} \, , &		& \mathscr{D}_{\textrm{III}} = \frac{\Delta_{+-}+\Delta_{--}}{2} \, , &		& \mathscr{D}_{\textrm{IV}} = \mathscr{D}_{\textrm{V}} = \Delta_{++} \, , &
\end{align}
which become asymptotically exact for nearly touching Janus particles, i.e., for $r/R \rightarrow 2^{+}$ as discussed in Appendix \ref{appendix_F}. Concerning the values of the well-known \cite{Cardy_review} critical Casimir amplitudes $\Delta_{ab}$ see Eq. (\ref{Casimir_plates'}). In Fig. \ref{fig05_02} we compare our analytical predictions at large and small distances with the corresponding simulation results from Ref. \cite{BPK}.

The theoretical predictions and the numerical results from Ref. \cite{BPK} agree on a qualitative level for large separations. Concerning the proximal regime, our findings of repulsion in the cases I and II and of attraction in the cases IV and V are consistent with the numerical data. In case III the present large distance behavior agrees rather well with the simulations. However, the repulsion at short distances is not visible in the numerical data, presumably because this small scale of separations is not sufficiently resolved. We cannot expect \emph{quantitative} agreement, because our predictions are valid for the universal scaling region where the size of the particles and the distances between them are large on the scale of the lattice constant. Within the simulation model, these requirements are not met by the particle size of 4 lattice constants.

\subsection{Two Janus needles}
\begin{figure}[htbp]
\centering
\includegraphics[width=10.5cm]{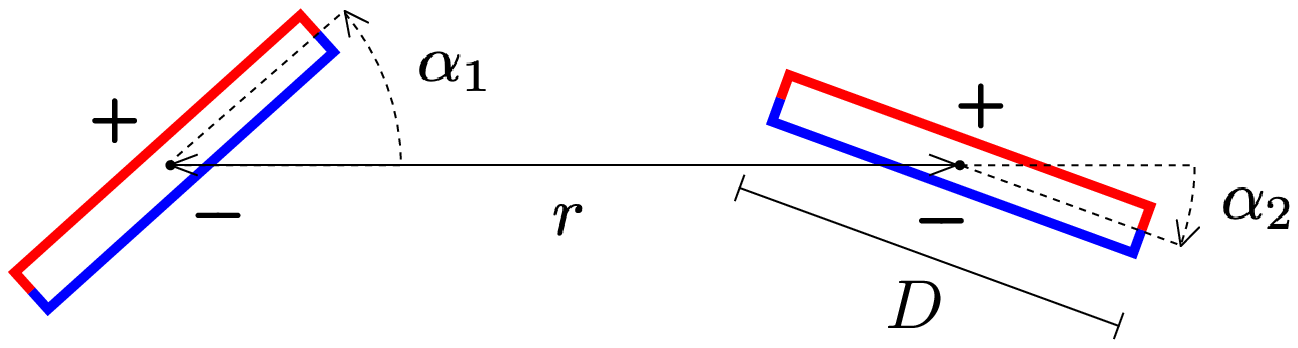}
\caption{Two identical Janus needles with length $D$ and center-to-center distance $r$. A nonzero needle thickness has been introduced for reasons of clarity. The orientations $\alpha_{j}$ with $j=1,2$ are measured relative to the reference directions, as described in Fig. \ref{fig05_01}}.
\label{fig05_01_bis}
\end{figure}

The interaction potential at large distances between two Janus needles ${\rm P}_1 = {\rm Jn}_1$ and ${\rm P}_2 = {\rm Jn}_2$ of equal length $D$ (see Fig. \ref{fig05_01_bis})
follows from the interaction between the Janus circles studied in Subsection \ref{2J} upon replacing Eq. (\ref{JJ_result_interaction}) by   
\begin{equation}
\begin{aligned}
\label{}
\langle s_{\textrm{Jn}_{1}}^{(\varepsilon)} s_{\textrm{Jn}_{2}}^{(\varepsilon)} \rangle & = \left( \frac{D}{2r} \right)^{2} \biggl\{ \frac{9}{4} + \frac{3}{16} \left( \frac{D}{2r}\right)^{2} \Bigl[ - 10 + 17 (\cos2\alpha_{1}+\cos2\alpha_{2}) \Bigr] \biggr\} \, , \\
\langle s_{\textrm{Jn}_{1}}^{(\Phi)} s_{\textrm{Jn}_{2}}^{(\Phi)} \rangle & = 2^{-3/4} \left( \frac{D}{2r} \right)^{9/4} \biggl\{ \mathscr{A}(\alpha_{1},\alpha_{2}) + \frac{9}{64}  \left( \frac{D}{2r} \right)^{2} \biggl[ 3\mathscr{B}(\alpha_{1},\alpha_{2}) + \mathscr{C}(\alpha_{1},\alpha_{2}) \biggr] \biggr\} \, , \\
\langle s_{\textrm{Jn}_{1}}^{(\mathbb{I})} s_{\textrm{Jn}_{2}}^{(\mathbb{I})} \rangle & = \frac{81}{32} \left( \frac{D}{2r} \right)^{4} \cos(2\alpha_{1}+2\alpha_{2}) \, .
\end{aligned}
\end{equation}

\begin{figure}[htbp]
\centering
\includegraphics[width=12.5cm]{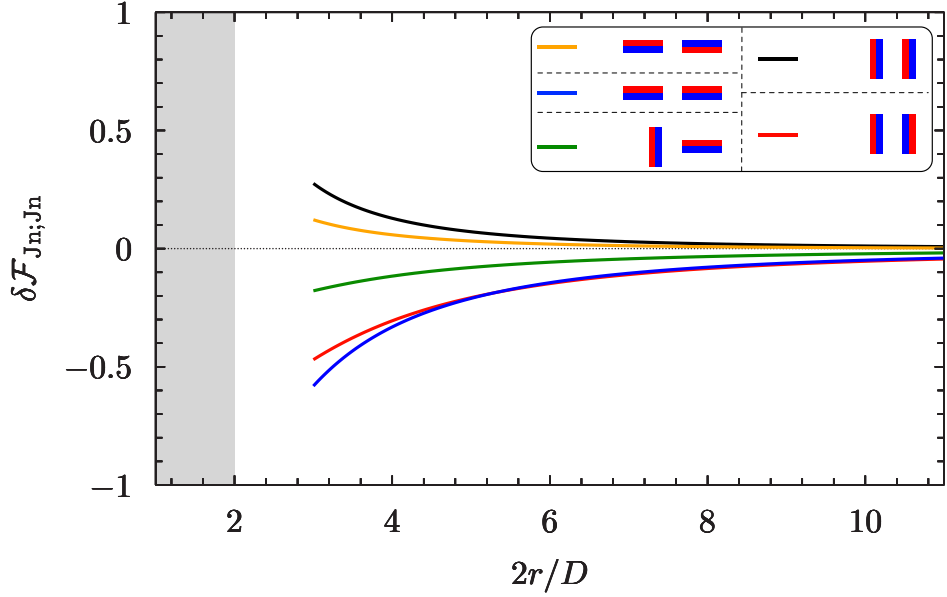}
\caption{Large distance behavior of the interaction energy $\delta\mathcal{F}_{\textrm{Jn;Jn}}$ between two Janus needles (per $k_{\rm B}T_{\rm c}$) as function of the rescaled separation $2r/D$. The shaded region corresponds to hard core repulsion of the needles in the tip-to-tip configuration.}
\label{interaction_nJnJ}
\end{figure}

In Fig. \ref{interaction_nJnJ} we plot the interaction potential between two Janus needles for various orientations.

\newpage

\subsection{Homogeneous needle interacting with a Janus circle or with a Janus needle}
\begin{figure}[htbp]
\centering
\includegraphics[width=10.5cm]{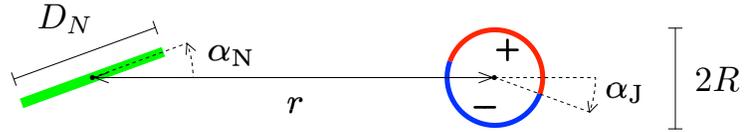}
\caption{Configuration of an ordinary needle $\textrm{N}$ of length $D_{N}$ and a symmetric Janus circle of radius $R$ with center-to-center distance $r$. The orientations $\alpha_{\textrm{N}}$ and $\alpha_{\textrm{J}}$ are measured relative to the reference directions, as described in Fig. \ref{fig05_01}.}
\label{fig09_01}
\end{figure}

The interaction potential at large distances between the needle $\textrm{P}_{1}=\textrm{N}$ with a homogeneous ordinary boundary and the Janus circle ${\rm P}_2 = {\rm J}$ as shown in Fig. \ref{fig09_01} follows from

\begin{eqnarray} \label{NJ_int}
\langle s_{\textrm{N}}^{(\varepsilon)} s_{\textrm{J}}^{(\varepsilon)} \rangle & = & \frac{R D_{\textrm{N}}}{4r^{2}} \biggl\{ 3 + \frac{3}{4} \left( \frac{D_{\textrm{N}}/2}{r} \right)^{2} \bigl[ 1+3\cos(2\alpha_{\textrm{N}})\bigr] + \left( \frac{R}{r} \right)^{2} \bigl[ -5+8\cos(2\alpha_{\textrm{J}})\bigr] \biggr\} \, , \nonumber \\
\langle s_{\textrm{N}}^{(\mathbb{I})} s_{\textrm{J}}^{(\mathbb{I})} \rangle & = & \left( \frac{R D_{\textrm{N}}}{2r^{2}} \right)^{2} \cos[2(\alpha_{\textrm{N}}+\alpha_{\textrm{J}})] \, .
\end{eqnarray}
The corresponding interaction between the above homogeneous ordinary needle ${\rm P}_1 = {\rm N}$ of length $D_{N}$ and a Janus needle ${\rm P}_2 = {\rm Jn}$ of length $D$ is obtained from
\begin{equation}
\begin{aligned}
\label{N,Jn}
\langle s_{\textrm{N}}^{(\varepsilon)} s_{\textrm{Jn}}^{(\varepsilon)} \rangle & = \frac{D_{\textrm{N}}D}{16r^{2}} \biggl\{ 3 + \frac{1}{4} \left( \frac{D}{2r} \right)^{2} \biggr\} \Bigl[ -5+17\cos(2\alpha_{\textrm{Jn}}) \Bigr] + \frac{3}{4} \left( \frac{D_{\textrm{N}}}{2r} \right)^{2} \biggr\} \Bigl[ 1+3\cos(2\alpha_{\textrm{N}}) \Bigr] \biggr\}   \\
\langle s_{\textrm{N}}^{(\mathbb{I})} s_{\textrm{Jn}}^{(\mathbb{I})} \rangle & = \frac{9}{32} \left( \frac{D_{\textrm{N}}D}{4r^{2}} \right)^{2} \cos\left( 2\alpha_{\textrm{N}} + 2\alpha_{\textrm{Jn}} \right) \, .
\end{aligned}
\end{equation}
Equations (\ref{NJ_int}) and (\ref{N,Jn}) apply to the first and third average on the right hand side of Eq. (\ref{int_04}) while therein the second average vanishes. Here the expressions in Eqs. (\ref{N_spoe_en}) and (\ref{N_spoe_id}) for $s_{\textrm{Jn}}^{(\varepsilon)}$ and $s_{\textrm{Jn}}^{(\mathbb{I})}$, respectively, as well as $s_{\rm N}^{(\Phi)}=0$ have been used. 

\begin{figure}[htbp]
\centering
\includegraphics[width=12.5cm]{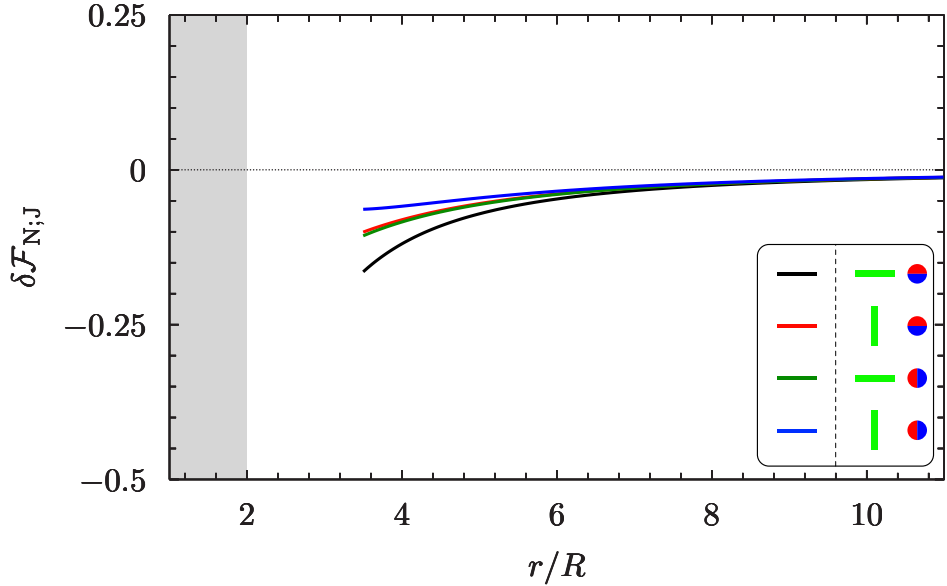}
\caption{Large distance behavior of the free energy of interaction (per $k_{\rm B}T_{\rm c}$) between a homogeneous ordinary needle of length $D_{N}$ and a symmetric Janus circle of radius $R$. The curves correspond to $D_{N}/R=2$ (the symbols in the inset legend are not drawn to scale) and provide the large-separation result in Eq. (\ref{NJ_int}). The shaded region corresponds to the hard core repulsion between the needle and the Janus circle if $\alpha_{N}=0$, in this case corresponding to the center-to-center distance equal to $2R$.}
\label{fig_NJ_int}
\end{figure}

The interaction free energy for the pair $(\textrm{N},\textrm{J})=(\textrm{homogeneous ordinary needle}, \, \textrm{Janus circle})$ is shown in Fig. \ref{fig_NJ_int}.  

\newpage
\subsection{Two quadrupoles}
\begin{figure}[htbp]
\centering
\includegraphics[width=10.5cm]{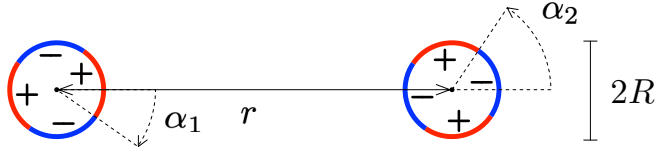}
\caption{Two identical quadrupolar particles with radii $R$ and center-to-center distance $r$. The orientations $\alpha_{j}$ with $j=1,2$ are measured relative to the reference directions, as described in Fig. \ref{fig05_01}.}
\label{configuration_QQ}
\end{figure}

Finally, we consider the interaction $\delta {\cal F}_{\textrm{Q}_{1};\textrm{Q}_{2}}$ between the two quadrupolar particles $\textrm{P}_{1}=\textrm{Q}_{1}$ and $\textrm{P}_{2}=\textrm{Q}_{2}$ of equal size, as shown in Fig. \ref{configuration_QQ}.

For large separations, $\exp (- \delta {\cal F}_{\textrm{Q}_{1};\textrm{Q}_{2}})$ is also given by the right hand side of Eq. (\ref{int_04}) with
\begin{eqnarray} \label{interactQQ} \nonumber
\langle s_{\rm Q_1}^{(\varepsilon)} \, s_{\rm Q_2}^{(\varepsilon)} \rangle &=&  {25 \over 9} \Bigl( {R \over r} \Bigr)^2  +{70 \over 3} \Bigl( {R \over r} \Bigr)^4 +O \Bigl( (R/r)^6 \Bigr) \, , \\ \nonumber
\langle s_{\rm Q_1}^{(\Phi)} \, s_{\rm Q_2}^{(\Phi)} \rangle &=& -2^{3/2}  \biggl[{425 \over 9} \cos( 2\alpha_1 +2 \alpha_2)  - \cos ( 2\alpha_1 - 2\alpha_2) \biggr] \Bigl( {R \over r} \Bigr)^{17/4} + O \Bigl( (R/r)^{6+1/4} \Bigr) \, , \\
\langle s_{\rm Q_1}^{(\mathbb{I})} \, s_{\rm Q_2}^{(\mathbb{I})} \rangle &=& O \bigl( (R/r)^8 \bigr) \, ,
\end{eqnarray}
which follows from Eqs. (\ref{Q_spoe_en})-(\ref{Q_spoe_id}). Orientations, which correspond to free energy minima, are those with like segments facing each other.

Due to the invariance of the Ising model under the exchange of $+$ and $-$ configurations, the interaction free energy $\delta \mathcal{F}_{\textrm{Q}_{1};\textrm{Q}_{2}}(r/ R ;\alpha_{1},\alpha_{2})$ is invariant with respect to the transformation $(\alpha_1 , \alpha_2) \mapsto (\alpha_1 + \pi /4, \alpha_2 + \pi /4)$. Moreover it is periodic with period $\pi$ in $\alpha_1$ and in $\alpha_2$ separately, because a quadrupole remains invariant upon changing its orientation $\alpha \mapsto \alpha \pm \pi$.

\begin{figure}[htbp]
\centering
\includegraphics[width=14.2cm]{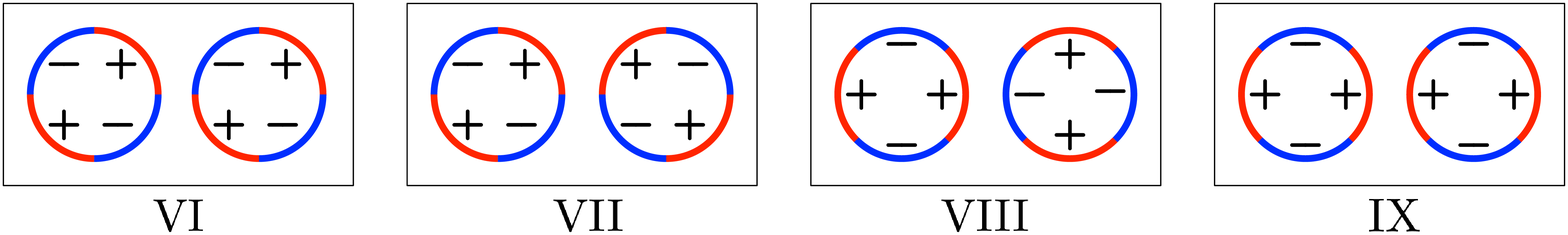}
\caption{Four special angular configurations of quadrupolar particles.}
\label{angles2}
\end{figure}

In order to present the interaction, we focus on those special orientations for which $(\alpha_1, \alpha_2)$ is equal to
\begin{align}
\label{}
(0,0) \, \, [\textrm{VI}] \, ,   &&(0,\pi/2) \, \, [\textrm{VII}] \, , && (-\pi/4,\pi/4) \, \, [\textrm{VIII}] \, , && (-\pi/4,-\pi/4) \, \, [\textrm{IX}] \, ,
\end{align}
which are shown in Fig. \ref{angles2}.
For these the asymptotic behavior of $\delta \mathcal{F}_{\textrm Q_1 , Q_2}$ at short distances is given by the right hand side of Eq. (\ref{short04}) with $j \in \{ \textrm{VI}, \dots , \textrm{IX}\}$, $\mathscr{D}_{\textrm{VI}} = \mathscr{D}_{\textrm{VIII}}= \Delta_{+-}$, and $\mathscr{D}_{\textrm{VII}} =\mathscr{D}_{\textrm{IX}}= \Delta_{++}$.

\begin{figure}[htbp]
\centering
\includegraphics[width=12.5cm]{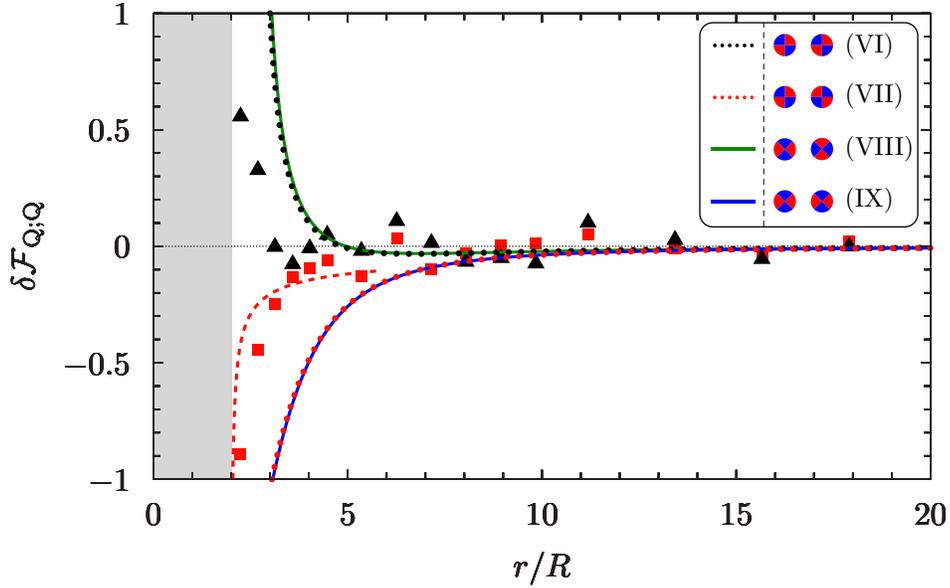}
\caption{The free energy of interaction (per $k_{\rm B}T_{\rm c}$) for two quadrupoles with the special orientations shown in Fig. \ref{angles2}. The analytical result in Eq. (\ref{int_04}) with Eq. (\ref{interactQQ}) for the interaction energy $\delta\mathcal{F}_{\textrm{Q};\textrm{Q}}(r/R;\alpha_{1},\alpha_{2})$ at large distances is shown as full curves (green and blue, VIII and IX, respectively) and as dotted curves (black and red, VI and VII, respectively). The black triangles and the red squares indicate the numerical data extracted from Fig. 8(b) of Ref. \cite{BPK}, corresponding, respectively, to the configurations VI and VII in Fig. \ref{angles2}. The leading asymptotic short-distance result (Eq. (\ref{short04})) for the configuration VII is indicated by a dashed red line. The short-distance result in Eq. (\ref{short04}) for the configuration VI is not visible, because already for $r/R=5$ the Derjaguin formula in Eq. (\ref{short04}) gives $\delta\mathcal{F}_{\textrm{Q};\textrm{Q}} \approx 2.7$. The results at large distances corresponding to the configurations VIII (full green) and IX (full blue) fall almost on top of the results for the configurations VI (dotted black) and VII (dotted red), respectively. The shaded region corresponds to the hard core repulsion between the two circular particles.}
\label{plot07_03}
\end{figure}

In Fig. \ref{plot07_03} we present the theoretical predictions given above for both the short and the large separation regime. As one can infer from Fig. \ref{plot07_03}, the theoretical results are in qualitative agreement with the numerical data from Ref. \cite{BPK}.

Evidently, a quantitative agreement between the presently available numerical data and our theoretical results cannot be expected (see the remarks at the end of Subsec. \ref{2J}). In any case, accurate analyses of the tails ($r\gg R$) are needed to facilitate a meaningful comparison with our predictions concerning the behaviors at large distances.

\section{Summary and concluding remarks}
\label{conclusions}
Critical fluids, such as binary liquid mixtures near their critical point of demixing, induce effective interactions between suspended mesoscopic particles which are of long range and of universal character. Here, we have considered particles with chemically inhomogeneous surfaces suspended in a two-dimensional critical fluid belonging to the Ising universality class. Motivated by the ability to fabricate ``patchy colloids'' \cite{fabrication_1,fabrication_2,fabrication_3,LLD_16}, we have studied planar particles with one-dimensional surfaces which consist of consecutive segments with alternating preferences for one or the other component of the binary liquid mixture. In the language of magnetism such surfaces carry infinitely strong surface magnetic fields the direction of which alternate between the spin up (+) and the spin down ($-$) direction, known as ``normal'' boundary conditions $+$ and $-$ (see, e.g., Ref. \cite{HWD_review}). We have studied in detail the simple representatives of such kind of particles as shown in Figs. \ref{fig02_01}(a) - \ref{fig02_01}(d). We denote them as a (symmetric) Janus circle, a generalized Janus circle, a circular quadrupole, and a Janus needle. If perturbations (such as confining walls or other particles) are absent, the dipolar and quadrupolar symmetries of the boundary conditions introduced in Figs. \ref{fig02_01}(a), \ref{fig02_01}(d), and \ref{fig02_01}(c), respectively, are reflected in corresponding symmetries of the profiles of the order parameter (i.e., the magnetization in the Ising model), see Fig. \ref{plot02_01three}(a) and Fig. \ref{plot07_01}(b). 

We consider the above systems right at their bulk critical point where they are invariant not only under scale transformations, but also under conformal or angle-preserving coordinate transformations. This enables us to obtain the averages and multi-point correlation functions of the densities $\varepsilon$, $\Phi$, and $T$ of the energy, the order parameter, and the stress tensor, respectively, in the presence of a {\it single} particle with an inhomogeneous surface in the entire bulk plane. This is accomplished via conformal transformations from the corresponding quantities in the upper half plane with inhomogeneous boundary conditions on the real axis, which Burkhardt, Guim, and Xue have derived in Refs. \cite{BX, BX_91, BG_93}. These are exact results valid in the whole upper half plane and, correspondingly, in the entire plane outside the particle,  which is a simply connected region.  

As explained in the Introduction, at present there are no similar results available for a double- or multiply- 
connected region with an inhomogeneous boundary. This precludes access to exact results valid at \emph{all} distances of the interaction of the particles shown in Fig. \ref{fig02_01} with each other or with a confinement. However, by using the small particle operator expansion (SPOE) \cite{BE_95, Eisenriegler_04}, asymptotically exact results can be obtained for the interaction of the aforementioned particles with distant objects, while for a close object --- depending on the geometry of the region of close approach --- other asymptotic methods (such as the Derjaguin approximation) are available.

Using these approaches we have investigated the dependence on particle positions and orientations of the cost of free energy to transfer Janus circles, needles, as well as quadrupoles from the bulk into a half plane, a strip, or a wedge. We have also determined the free energy of interaction between two of these particles in bulk. Moreover, we have considered the case that these particles interact with a needle, the two sides of which belong to the ordinary surface universality class.

In the following we provide a summary of our main results.

\begin{itemize}
\item[(i)] \emph{One-point correlation functions}. In Secs. \ref{symJanus}-\ref{quadrupoles} we have determined the profiles (i.e., the one-point correlation functions) of the densities $\varepsilon$, $\Phi$, and $T$ of the energy, the order parameter, and the stress tensor, respectively, around each of the particles shown in Fig. \ref{fig02_01} for the entire bulk plane. For the circular particles of Figs. \ref{fig02_01}(a), \ref{fig02_01}(b), and Fig. \ref{fig02_01}(c) these are shown in Figs. \ref{plot02_01three}-\ref{plot07_05}. 
The $+$ and $-$ segments on the particle surfaces induce local {\it order}, leading in their surrounding to positive and negative values of $\langle \Phi \rangle_{\rm particle}$ (compare Fig. \ref{plot02_01three} and Fig. \ref{plot07_01}(b)) and to negative values of $\langle \varepsilon \rangle_{\rm particle}$. However, the switching points at the particle surface induce {\it disorder} in their surrounding, clearly visible via the large positive values of $\langle \varepsilon \rangle_{\rm particle}$ there (Fig. \ref{plot02_02} and Fig. \ref{plot07_02}). At large distances from the particle, $\langle \varepsilon \rangle_{\rm particle}$ is also positive for all particles in Fig. \ref{fig02_01}, except for the generalized Janus particle with $|\chi|> \pi /3$, i.e., if one of the two segments $+$ or $-$ extends over less than 60 degrees. For most of the particles studied, the behavior of the eigenvalues and eigenvectors of the averages $\langle T_{kl} \rangle_{\rm particle}$ of the Cartesian stress tensor as introduced in Eq. (\ref{cartcompstress}), are shown in Figs. \ref{plot03_02}, \ref{plot04_05}, and \ref{plot07_05}. The eigenvalues of the stress tensor turn out to be finite on the homogeneous patches of the particle boundary, while they diverge at the switching points. Still on the particle boundary, but away from the switching points, the eigenvector belonging to the positive eigenvalue turns out to be orthogonal to the boundary, while the eigenvector belonging to the negative eigenvalue is tangential to it. This behavior along the homogeneous sectors of the particle boundary applies to all particles studied here, i.e., to the ones in Fig. \ref{fig02_01} and to the needle with a homogeneous boundary.  The boundary of the upper half $w$ plane with switching points also displays this behavior (see Refs. \cite{EB_16,BX, BX_91}).

Results for certain two-point correlation functions at large distance from a solitary particle in the bulk are derived in the Appendices \ref{correlations} and \ref{appendix_D}. In the corresponding discussion of the energy-energy two-point correlation function in the presence of a quadrupolar particle and its SPOE in Appendix \ref{appendix_D}, we identify, as a byproduct, the dimension 4 operators in the operator product expansion (OPE) (Eq. \ref{OPE_refined}) of two energy densities in the bulk.

\item[(ii)] \emph{Small particle operator expansion.} In Sec. \ref{SPOE} we have established SPOEs for each of the particles shown in Fig. \ref{fig02_01}. These are series of operators with increasing scaling dimensions. Their prefactors can be determined and the SPOE can be checked, because it must reproduce all aforementioned one- and multi-point correlation functions in the presence of a single particle at large distances from that particle. While the prefactors of low-dimensional operators follow already from the one-point correlation functions (i.e., profiles) in Secs \ref{symJanus}-\ref{quadrupoles}, knowledge of two-point correlation functions is necessary in order to disentangle (see Appendices \ref{correlations} and \ref{appendix_D}) the multitude of operators for a given high scaling dimension. The operators appear in three groups, corresponding to the three primary operators $\varepsilon$, $\Phi$, $\mathbb{I}$ of the Ising model and to descendants thereof. Each of the operators has to be consistent with the underlying symmetries of the particle. Due to the dipole symmetry of the symmetric Janus circle and of the Janus needle in Figs. \ref{fig02_01}(a) and \ref{fig02_01}(d), a multi-point correlation function of $\varepsilon$ and $\Phi$, in the presence of the Janus particle, keeps its value or merely changes its sign when all its arguments are mirror reflected about the center of the Janus particle according to $(z, \bar{z}) \rightarrow (-z, -\bar{z})$ --- depending on whether it contains an even or an odd number of factors of the order parameter $\Phi$. The same even or odd invariance holds for multi-point correlation functions in the presence of the quadrupolar particle in Fig. \ref{fig02_01}(c) when all their arguments are rotated by $90$ degrees according to $(z, \bar{z}) \rightarrow (iz, -i\bar{z})$. The operators in the SPOEs in Eqs. (\ref{J_spoe_en})-(\ref{J_spoe_id}), (\ref{nJ_spoe_en})-(\ref{nJ_spoe_id}), and (\ref{Q_spoe_en})-(\ref{Q_spoe_id}) reproduce these symmetries. It is instructive to compare the examples in Eqs. (\ref{cf_001}) and (\ref{cf_002}) for the Janus particles with Eqs. (\ref{sQ_02})-(\ref{sQ_04}) for the quadrupolar particle.

\item[(iii)] \emph{Interactions at close distances.} In Refs. \cite{BE_95, ER_95} it was shown that right at the critical point the Derjaguin approximation for the interaction between two close spheres with {\it homogeneous} boundary conditions becomes exact. Here we have considered the interaction between close particles in two dimensions with {\it inhomogeneous} surfaces such as the Janus and the quadrupolar particles.
\begin{itemize}
  \item[(iiia)] In Appendix \ref{appendix_F} we have considered two Janus circles I and II with $(\alpha_{\rm I} , \alpha_{\rm II}) = (0,0)$ or $(0, \pi)$, i.e., with their switching points facing each other, and a Janus circle (with $\alpha=0$) facing with its switching point a vertical $+$ wall. We have mapped the two-circle geometry of the two Janus particles, with the closest surface-to-surface distance $\mathcal{C}$ being much smaller than their radius $R$, onto an annulus which is thin and which near the switching points of the boundary condition resembles a strip with corresponding inhomogeneous boundary conditions. Determining the position-dependent disjoining pressure of the latter has allowed us to calculate the force between the close circular particles in leading and next-to-leading order. We have found that in leading order the force is given by a Derjaguin expression, which looks like being stitched together (i.e., the arithmetic mean of two Derjaguin expressions corresponding to the pairs of boundary conditions on the two sides of the switches; compare the remark below Eq. (\ref{nJanuswall6})). While this corresponds to a step-like spatial dependence of the disjoining pressure, for the next-to-leading order force contribution the actual smooth dependence of the pressure has to be taken into account. The explicit result for this contribution is, by the order $(\mathcal{C}/R)^{1/2}$, smaller than the leading one. We have also explicitly calculated the next-to-leading contribution for two circles with homogeneous boundary conditions and have found it to be smaller by the same order than the leading (Derjaguin) contribution.
  \item[(iiib)] A Janus needle of length $D$ in a half plane and oriented perpendicular to the boundary, as considered in Eq. (\ref{nJanuswall6}), can be replaced by a semi-infinite Janus needle if the distance of the close end from the boundary is much smaller than $D$. This reduces the geometry to a simply connected one for which the force has been calculated in Eq. (D12) of Ref. \cite{EB_16} in the case of the $+$ boundary. 
\end{itemize}

\item[(iv)] \emph{Janus and quadrupolar particles in confinement.} Transferring one of these particles from the bulk into a confinement, at large distances from the confining walls the cost of free energy $\delta\mathcal{F}$ is basically given by the confinement averages of the SPOE operators (see Eq. (\ref{int_01})). 
In Secs. \ref{sec_J_half} and \ref{wall_section2} we have applied this for the insertion into a half plane, a strip, and a wedge, with various boundary conditions. For a Janus particle in the half plane, in Subsec. \ref{JnJ_HP} we have compared the corresponding behavior at large distances with that at small distances from the boundary. In many cases the qualitative dependence of the free energy $\delta\mathcal{F}$ of insertion on the position and the orientation of the particle can be obtained by considering the ``degree of fitting'', i.e., to which extent the properties around the particle in the bulk (such as the distribution of order and disorder represented by the profiles $\langle \Phi \rangle_{\rm particle}$ and  $\langle \varepsilon \rangle_{\rm particle}$, respectively) fit to the corresponding properties of the confinement without the particle. For good or bad fitting at a given position and orientation, $\delta\mathcal{F}$ is low or high, respectively. For example, a Janus particle in the right half plane with its center at a given distance from an ''ordinary'' vertical half plane boundary exhibits the lowest $\delta\mathcal{F}$ for the orientation $\alpha =0$ for which the disorder regions of the boundary and the switching points are as close as possible. Likewise, for a $+$ boundary the orientation of the Janus particle with the lowest free energy is the one with $\alpha =\pi /2$ where the $+$ segment of the Janus particle is as close as possible to the boundary, because both induce order.

\item[(iv a)] The results in Subsec. \ref{JnJ_HP} for inserting a Janus particle into the half plane at large and at small distances from the boundary imply, in some cases, an interesting \emph{non-monotonic} dependence of $\delta\mathcal{F}$ on distance (see Figs. \ref{fig07_01} and \ref{fig07_02}). For Janus particles interacting with a $+$ boundary, in Fig. \ref{fig07_02} there are conflicting repulsion-attraction tendencies from the energy density and the order parameter at large distances. Therefore, in this case, our quantitative SPOE approach is indispensable in order to be able to decide in favor of repulsion at large distances, while at small distances the Derjaguin-type approach described in (iii) predicts attraction so that the insertion free energy $\delta\mathcal{F}$ must have a maximum in between which, however, might be rather flat.

\item[(iv b)] Inserting a small Janus particle into a strip or wedge, besides the energy and the order parameter profiles, the profile of the stress tensor $\langle T \rangle_{\textrm{P}}$ is also nonvanishing and has to be taken into account (see Subsecs. \ref{sec_J_strip}-\ref{sec_J_wedge}). For a strip with two ordinary boundaries it yields, via the SPOEs in Eqs. (\ref{J_spoe_en})-(\ref{J_spoe_id}) and (\ref{nJ_spoe_en})-(\ref{nJ_spoe_id}), even the leading dependence on orientation (see Eq. (\ref{Janus_ordinary})) predicting the lowest free energy to occur for $\alpha = \pm \pi /2$ for which the dipole vector is parallel to the strip. This can be understood qualitatively from a best fit between the stress tensor profiles of the isolated Janus particle and the empty strip. In Subsecs. \ref{sec_J_strip}-\ref{sec_J_wedge} we have also discussed the orientation dependence of $\delta \mathcal{F}$ for inserting a Janus particle into strips with other boundary conditions and into a wedge; we have checked the validity of our results for a general opening angle of the wedge if the wedge reduces to the right half plane and the horizontal 
strip.

\item[(iv c)] In Sec. \ref{wall_section2} we have investigated the orientation dependence of $\delta\mathcal{F}$ for inserting a small {\it quadrupolar} particle into a half plane, a strip, and a wedge with various boundary conditions by using the SPOE in Eqs. (\ref{Q_spoe_en})-(\ref{Q_spoe_id}). Besides the energy density operator and its derivatives and the quadrupolar descendant of the order parameter operator, $\delta\mathcal{F}$ contains the combination $s^{(\mathbb{I})}$ of fourth order descendants of the identity. Similar to the stress tensor in Eqs. (\ref{J_spoe_id}) and (\ref{nJ_spoe_id}) for the Janus particles, the combination $s^{(\mathbb{I})}$ in Eq. (\ref{Q_spoe_id}) for the quadrupolar particle contains anisotropic operators with nonvanishing profiles in strips and wedges and determines the leading orientation dependence of the quadrupolar particle if all boundaries are ordinary ones. Concerning the wedge in Fig. \ref{fig08_01}, with both boundaries being ordinary and with both boundaries being $+$, we have determined how $\delta\mathcal{F}$ depends on the orientation of the quadrupole and we have determined the orientations with lowest $\delta \mathcal{F}$ (see below Eqs. (\ref{Q_Owedge}) and (\ref{wedge_019}), respectively). These latter orientations depend on the opening angle $\pi / \Xi$ of the wedge and can be understood most easily for a quadrupole centered on the symmetry axis of the wedge (i.e., the real axis). In the case of an $OO$ wedge these are the orientations for which the two $+$ segments of the quadrupole are either parallel or perpendicular to the real axis (equal to the equilibrium orientations in the horizontal strip) if $\pi / \Xi < \pi$. For $\pi /\Xi > \pi$ the equilibrium orientations are rotated by 45 degrees. In the case of a $++$ wedge the equilibrium orientations change at an opening angle $0.686 \pi$ of the wedge. For smaller opening angles the two $+$ segments of the quadrupole are parallel to the real axis, as for the strip, while for larger opening angles they are perpendicular to the real axis, as for the right half plane. 

\item[(v)] \emph{Interaction between two small particles.} The SPOEs in Eqs. (\ref{J_spoe_en})-(\ref{J_spoe_id}) and (\ref{Q_spoe_en})-(\ref{Q_spoe_id}) can also be used to determine the interaction between two (or more) particles which are much smaller than the interparticle distances (compare Eq. (\ref{int_01})). In Sec. \ref{interaction_JJ} we have presented such results for two Janus circles, two Janus needles, two quadrupolar particles, and a Janus circle or Janus needle interacting with a needle where both sides induce disorder (i.e., ordinary boundary conditions). Finally our results for two distant Janus circles, together with the results obtained within the Derjaguin approximation when the Janus circles are close to each other, have been compared with the simulation results in Ref. \cite{BPK}.

These simulations were based on the Ising model on a square-lattice with embedded pieces of material, each of which is composed of neighboring elementary squares intended to approximate a particle of circular shape. Suitable couplings of the elementary squares of such a patch of material to the Ising spins generate the character of a circular Janus particle. We expect that this model enters the universal scaling region described by our theory, provided the size of the patch and the distances between such patches are large on the scale of the lattice constant. For patches with a diameter of 4 lattice constants, as used in the simulations, one can expect qualitative agreement at best. We expect that it is easier to find agreement between our continuum theory and lattice simulations in the case of Janus {\it needles}.

\end{itemize}

\section*{Acknowledgments}
We thank T. W. Burkhardt for useful discussions. A.S. acknowledges the hospitality of the Physikzentrum Bad Honnef during the 671 WE-Heraeus-Seminar \emph{``Fluctuation-induced Phenomena in Complex Systems''}, and the Galileo Galilei Institute for Theoretical Physics (Arcetri, Florence) during the winter school \emph{``SFT 2019: Lectures on Statistical Field Theories''}, where this work has been concluded.

\bibliographystyle{unsrt}
\bibliography{bibliography}


\begin{appendices}

\newpage
\section{Glossary and list of symbols}
\label{glossary}
In order to make the reading of the manuscript easier, here we summarize the mathematical symbols and abbreviations used throughout the main text of this work. Certain extra symbols used only in the Appendices are not listed here.  
\vspace{10mm}

\textbf{Mathematical symbols and acronyms}
\vspace{2mm}

\begin{tabular}{cp{0.7\textwidth}}
\hline
BCFT & boundary conformal field theory \\
CCF & critical Casimir force \\
CFT & conformal field theory \\
OPE & operator product expansion \\
SPOE & small particle operator expansion \\
$\partial\equiv\partial_{z}$, $\bar{\partial}\equiv\partial_{\bar{z}}$ 	& complex derivatives \\
$\rm P \in \{ \textrm{J}, \textrm{Jn}, \textrm{J}_{\chi}, \textrm{Q}, \textrm{N}\}$ & particle label: J for Janus particle, Jn for Janus needle, $\textrm{J}_{\chi}$ for generalized Janus particle, Q for quadrupolar particle and N for needle. \\
\hline
\end{tabular}

\vspace{10mm}

\textbf{Greek}
\vspace{2mm}

\begin{tabular}{cp{0.7\textwidth}}
\hline
$\alpha_{\rm P}$ & orientation angle of a particle with label $\rm P$; $\alpha_{\rm P}=0$ corresponds to Fig. \ref{fig02_01}; see Sec. \ref{SPOE} \\
$\delta\mathcal{F}$ & excess free energy in units of $k_{\rm B}T$ required to transfer a particle from the bulk to the configuration of interest; see Eqs. (\ref{int_01}) and (\ref{int_02}) \\
$\delta\mathcal{F}_{\textrm{P}_1 , \textrm{P}_2}$ & Casimir free energy in units of $k_{\rm B}T$ between two particles with labels $\textrm{P}_{1}$ and $\textrm{P}_{2}$; see Eq. (\ref{int_04}) \\
$\delta\mathcal{F}_{\rm P_1 , \dots , \rm P_N}$ & Casimir free energy in units of $k_{\rm B}T$ between $N$ particles with labels $\textrm{P}_{1}, \dots , \textrm{P}_N$; see Eq. (\ref{OPE_10_new}) \\
$\delta\mathcal{F}_{a-\textrm{wall},\rm P}$ & excess free energy in units of $k_{\rm B}T$ of a particle labelled by $\rm P$ in the half-plane domain with uniform boundary conditions $a$ along the boundary; see Sec. \ref{JnJ_HP}. \\
$\delta\mathcal{H}_{\rm P}$ & Hamiltonian in units of $k_{\rm B}T$ defining the local Boltzmann weight for the particle with label $\rm P$; see Eq. (\ref{def_SPOE}) \\
$\Delta$ & Laplacian operator in two dimensions: $\Delta=\partial_{x}^{2}+\partial_{y}^{2}=4\partial_{z}\partial_{\bar{z}}$ \\
$\Delta_{ab}$ & universal amplitude for the CCF in a strip with  boundary conditions $a$ and $b$; $\Delta_{ab} \equiv \Theta_{ab}(0)$ in $d=2$; see Eqs. (\ref{Casimir_plates}) and (\ref{Casimir_plates'})\\
$\varepsilon$ & energy density operator \\
$\theta$ & argument of the complex number $z=x+iy$, i.e., $z=|z|\textrm{e}^{i\theta}$ \\
$\Theta_{ab}$ & scaling function for the CCF between parallel plates with boundary conditions $a$ and $b$; see Eq. (\ref{Casimir_plates}). \\
\hline
\end{tabular}

\newpage
\textbf{Greek (continued)}
\vspace{2mm}

\begin{tabular}{cp{0.7\textwidth}}
\hline
$\lambda_{\rm P}$ & eigenvalue of the complex stress tensor corresponding to the particle labelled by $\rm P$; see Eqs. (\ref{03_14a}), (\ref{gen_Janus_eigenvalue}), and (\ref{Q_eigenvalue}) \\
$\xi$ & bulk correlation length; see Eq. (\ref{Casimir_plates}) \\
$\Xi$ & parameter defining the wedge opening angle as $\pi/\Xi$; see Eq. (\ref{map_wedge}) and Fig. \ref{fig08_01} \\
$\Phi$ & order parameter operator \\
$\Psi$ & conformal operator, not necessarily primary \\
$\chi$ & characteristic angle of the generalized Janus particle; see Fig. \ref{fig02_01}(b) and Eq. (\ref{composite}) \\
\hline
\end{tabular}

\vspace{10mm}

\textbf{Latin}
\vspace{2mm}

\begin{tabular}{cp{0.7\textwidth}}
\hline
$A$ & prefactor of the operator $L_{-3} \Phi$ in Appendix \ref{correlations} (see Eq. (\ref{cf_010})); complex position variable in the annulus geometry of Appendix \ref{appendix_F} \\ 
$a$ & label for boundary conditions, $a \in \{ +, -, O\}$; e.g., Eq. (\ref{Casimir_plates}) \\
$\mathbb{A}$ & annulus of Appendix \ref{appendix_F} \\
$\mathcal{A}_{\mathcal{O}}^{(\pm)}$, $\mathcal{A}_{\mathcal{O}}^{(O)}$ & universal amplitudes of one-point correlation functions of the operator $\mathcal{O}$ in the upper half-plane with fixed ($\pm$) and ordinary ($O$) boundary conditions, respectively; see Eqs. (\ref{03_03}) and (\ref{03_04_01})  \\
$B$ & prefactor of the operator $L_{-1}^{3} \Phi$ in Appendix \ref{correlations}; see Eq. (\ref{cf_010}) \\
$b$ & label for boundary conditions, $b \in \{ +, -, O\}$; e.g., Eq. (\ref{Casimir_plates}) \\
$C$ & prefactor of the operator $L_{-1}^{2}\bar{L}_{-1} \Phi$ in Appendix \ref{correlations}; see Eq. (\ref{cf_010}) \\
$c=\cos\alpha$ & shorthand notation used in Eq. (\ref{J_strip_average}) \\
$d$	 & spatial dimension; see Eq. (\ref{Casimir_plates}) \\ 
$D$ & needle length; see Fig. \ref{fig02_01}(d) \\
$\mathcal{D}=D/(2x_{0})$ & dimensionless ratio; see Eq. (\ref{wall_nJ_phi}) \\
$\mathscr{D}_{j}$ & linear combination of critical Casimir amplitudes $\Delta_{ab}$ corresponding to the configuration with label $j \in \{ \textrm{I}, \textrm{II}, \dots, \textrm{VII} \}$; see Eq. (\ref{short04}) \\
$\widehat{\1{e}}_{\pm}$ & unit eigenvector corresponding to the positive/negative eigenvalue of $T(z)$; see, e.g., Eq. (\ref{03_14c}) \\
$f_{x}$ & critical Casimir force in $x$-direction, in units of $k_{B}T_{c}$; see Eqs. (\ref{Januswall6}) and (\ref{nJanuswall6}) \\
$F_{ab}$ & disjoining CCF per area and $k_B T$ between two parallel walls with uniform boundary conditions $a$ and $b$; see Eq. (\ref{Casimir_plates}) \\
$\mathscr{G}$ & generic confining geometry; see Eq. (\ref{int_01}) \\
$\mathcal{J}=\pi R/\mathcal{W}$ & dimensionless parameter introduced in Eq. (\ref{J_strip_average}) \\
$\mathbb{H}_{a}$ & the upper half plane with uniform boundary conditions of type $a$ along its boundary; see for instance Eq. (\ref{03_03}) \\
\hline
\end{tabular}

\vspace{35mm}

\textbf{Latin (continued)}
\vspace{2mm}

\begin{tabular}{cp{0.7\textwidth}}
\hline
$\widetilde{\mathbb{H}}_{a}$ & the right half plane with uniform boundary conditions of type $a$ along its boundary; see Eq. (\ref{rotation}) \\
$\mathbb{H}_{-+}$ & the upper half plane with inhomogeneous boundary conditions $+/-$ merging at the origin; see Eqs. (\ref{03_05}) and (\ref{03_06}) \\
$\mathbb{H}_{-+-+}$ & upper half plane with three $+/-$ switches; image of the quadrupole; see Eqs. (\ref{07_01}) and (\ref{07_02}) \\
$\mathbb{I}(z,\bar{z})$ & identity operator \\
$L_{p}$, $\bar{L}_{p}$ & Virasoro generators; see Eq. (\ref{def_descendant}) \\
$\mathcal{N}=\pi D/(2\mathcal{W})$ & dimensionless parameter introduced in Eq. (\ref{nJ_strip_average}) \\
$\mathcal{O}(z,\bar{z})$ & primary operator in the complex plane with coordinates ($z,\bar{z}$) \\
$\langle \mathcal{O} \rangle$ & statistical average of $\mathcal{O}$ in the bulk \\
$\langle \mathcal{O}(w,\bar{w}) \rangle_{\mathbb{H}_{a}}$ & statistical average of $\mathcal{O}$ in the domain $\mathbb{H}_{a}$ with coordinates $(w,\bar{w})$; see Eq. (\ref{03_03}) \\
$\mathcal{P}$ & pattern of $+$ and $-$ segments on the $u$ axis (i.e., the boundary of the upper half $w$ plane), see Sec. \ref{single}  \\
$\1{r}=(x,y)$ & two-dimensional vector with Cartesian coordinates $x$, $y$ \\
$s_{\rm P}$ & operator series in the SPOE of a particle labelled by $\rm P$; see Eq. (\ref{def_SPOE}) \\
$s_{\rm P}^{(\mathcal{O})}$ & operator $\mathcal{O}$ and descendants thereof appearing in the SPOE of a particle labelled by $\rm P$; e.g., $s_{\textrm{J}}^{(\varepsilon)}$, see Eq. (\ref{def_SPOE}) \\
$R$ & radius of the circular particles shown in Figs. \ref{fig02_01}(a)-(c) \\
$S$ & Schwarzian derivative; see Eq. (\ref{stresstrafo}) \\
$\mathbb{S}_{ab}$ & strip geometry with boundary conditions $a$ and $b$ along the edges; see Sec. \ref{sec_J_half} \\
$T(z)$ & holomorphic part of the complex stress tensor operator \\
$\bar{T}(\bar{z})$ & anti-holomorphic part of the complex stress tensor operator \\
$T_{kl}(x,y)$, $k,l \in\{x,y\}$ & Cartesian components of the stress tensor in the Euclidean $(x,y)$-plane; see Eq. (\ref{cartcompstress}) \\
$T_{rr}(x,y)$ & radial component of the stress tensor in the Euclidean $(x,y)$-plane; see Eq. (\ref{radtang}) and the conclusive paragraphs of Sec. \ref{quadrupoles} \\
$t_{ab}$ & parameter characterizing the statistical average of the stress tensor; see Eq. (\ref{wall_04}) \\
$w=u+iv$	& points in the complex $w$-plane plane with coordinates $u$ and $v$ \\
$\mathcal{W}$ & strip width; see Sec. \ref{sec_J_half} \\
$\mathbb{W}$ & interior of the wedge with apex in the origin and centered about the positive real axis; see Fig. \ref{fig08_01} and the mapping in Eq. (\ref{map_wedge}) \\
\hline
\end{tabular}

\newpage
\textbf{Latin (continued)}
\vspace{2mm}

\begin{tabular}{cp{0.7\textwidth}}
\hline
$x_{\mathcal{O}}$ & scaling dimension of the primary operator $\mathcal{O}$; see Eq. (\ref{03_04bis}) \\
$x_{0}$ & distance of a Janus particle from a wall measured from the particle center; see Sec. \ref{JnJ_HP} \\
$\mathcal{X}_{0}=x_{0}/R$ & scaled distance of a circular particle of radius $R$ from a wall at a distance $x_{0}$ from the particle center; see Eqs. (\ref{Q_eps_rhp})-(\ref{Q_id_rhp}) \\
$Y=\pi y/\mathcal{W}$ & dimensionless parameter introduced in Sec. \ref{sec_J_strip} \\
$z=x+iy$ & a point in the complex plane with Cartesian coordinates $x$ and $y$ \\
$\bar{z}$ & complex conjugate of $z$ \\
$z_{12}=z_{1}-z_{2}$ & difference between two complex coordinates \\
\hline
\end{tabular}

\vspace{20mm}
\section{Two-point correlation functions for Janus particles ($\rm{J}$ and $\rm{Jn}$) and their SPOE}
\label{correlations}
\subsection{Mixed correlation functions $\langle\Phi\varepsilon\rangle_{\rm J}$, $\langle\Phi\varepsilon\rangle_{\rm Jn}$ and next-to-leading $\Phi$ descendants for Janus circles and Janus needles}
In this Appendix we treat the Janus circle and Janus needle on the same footing. Therefore in this Appendix we often use the subscript P for both, i.e., ${\rm P}\in\{\textrm{J},\textrm{Jn}\}$.

The prefactors of the operators $[\mathbb{I}, L_{-1}\bar{L}_{-1}, L_{-1}^{2}, \bar{L}_{-1}^{2}]\varepsilon$, $[L_{-1}, \bar{L}_{-1}]\Phi$, and $[L_{-2}, \bar{L}_{-2}]\mathbb{I}$ in the SPOEs in Eqs. (\ref{J_spoe_en})-(\ref{J_spoe_id}) and (\ref{nJ_spoe_en})-(\ref{nJ_spoe_id}) for a Janus circle and a Janus needle follow from their profiles $\langle \varepsilon \rangle_{\textrm{P}}$, $\langle \Phi \rangle_{\textrm{P}}$, and $\langle T \rangle_{\textrm{P}}$ in Secs. \ref{symJanus} and \ref{App_needle}, respectively. Here we use the large distance behavior of the two-point correlation function $\langle \Phi\varepsilon \rangle_{\textrm{P}}$ in order to determine the prefactors $A, B, C$ in the linear combination of the third order\footnote{We do not have to consider the operators $L_{-1}L_{-2}\Phi$ and $L_{-2}L_{-1}\Phi$, because they can be expressed as $L_{-1}L_{-2}\Phi = (4/3) L_{-1}^{3}\Phi$ and as $L_{-2}L_{-1}\Phi = [L_{-3} + (4/3)L_{-1}^{3}]\Phi$. This follows from the level $2$ degeneracy of $\Phi$ (see Eq. (\ref{degeneracy})) and from the Virasoro algebra (Eq. (\ref{VAextended})) (for further details see, e.g., Ref. \cite{CFT_books_1}).} descendants $[L_{-3}, L_{-1}^{3}, L_{-1}^{2}\bar{L}_{-1}]\Phi$ of $\Phi$ and their conjugates, which represent the next-to-leading contribution to $s_{\textrm{P}}^{(\Phi)}$ in Eqs. (\ref{J_spoe_en})-(\ref{J_spoe_id}) and (\ref{nJ_spoe_en})-(\ref{nJ_spoe_id}), if $\alpha=0$ and $R=D/2=1$. In this case the next-to-leading order contribution to $\langle\Phi\rangle_{\textrm{P}}$ does not provide sufficient information. The reason for this is, that the first two terms in this contribution
\begin{align} \nonumber
\label{cf_001}
2 \textrm{Re} \bigl\langle \Phi(z_{1},\bar{z}_{1}) \left( [AL_{-3}+BL_{-1}^{3}+CL_{-1}^{2}\bar{L}_{-1}]\Phi(z_{0},\bar{z}_{0}) \right) \bigr\rangle \big\vert_{z_{0}=0} = & \\
= 2 \textrm{Re} \biggl[ \left( \frac{A}{4} + \frac{153}{512} B \right) z_{1}^{-3} + \frac{9C}{512} z_{1}^{-2}\bar{z}_{1} \biggr] |z_{1}|^{-1/4} &
\end{align}
to $\langle \Phi(z_{1},\bar{z}_{1}) \rangle_{\rm P}$ are identical apart from a numerical prefactor which does not allow one to disentangle $A$ from $B$. However, the analytic structure of the corresponding terms in the next-to-leading contribution to the two-point correlation function $\langle \Phi(z_{1},\bar{z}_{1}) \varepsilon(z_{2},\bar{z}_{2}) \rangle_{\rm P}$
\begin{align}
\label{cf_002}
2 \textrm{Re} \bigl\langle \Phi(z_{1},\bar{z}_{1}) \varepsilon(z_{2},\bar{z}_{2}) \left( [AL_{-3}+BL_{-1}^{3}+CL_{-1}^{2}\bar{L}_{-1}]\Phi(z_{0},\bar{z}_{0}) \right) \bigr\rangle \big\vert_{z_{0}=0} = 2 \Theta \textrm{Re}(\mathcal{T}) \, ,
\end{align}
with
\begin{eqnarray} \nonumber
\label{cf_003}
\mathcal{T} & = & \frac{A}{4} \left( z_{1}^{-3} - 6 z_{2}^{-3} +2z_{1}^{-1}z_{2}^{-2}+2z_{1}^{-2}z_{2}^{-1} \right) + \\ \nonumber
& + & \frac{3B}{512} \left( 65 z_{1}^{-3} - 320 z_{2}^{-3} + 144 z_{1}^{-1}z_{2}^{-2} + 60 z_{1}^{-2}z_{2}^{-1} \right) + \\ \nonumber
& + & \frac{3C}{512} \left( -15 z_{1}^{-2}\bar{z}_{1}^{-1} - 64 z_{2}^{-2}\bar{z}_{2}^{-1} + 20 z_{1}^{-2}\bar{z}_{2}^{-1} + 48 z_{2}^{-2}\bar{z}_{1}^{-1} - 24 |z_{1}|^{-2} z_{2}^{-1} + 32 |z_{2}|^{-2} z_{1}^{-1} \right) \, , \\
\end{eqnarray}
does provide sufficient information to allow the identification of $A$ and $B$ separately. In Eq. (\ref{cf_002}), $\Theta$ is the bulk three-point correlation function:
\begin{equation}
\label{cf_004}
\Theta \equiv \langle \Phi(z_{1},\bar{z}_{1}) \varepsilon(z_{2},\bar{z}_{2})\Phi(0,0) \rangle = -\frac{1}{2} |z_{1}|^{3/4}|z_{2}|^{-1}|z_{1}-z_{2}|^{-1} \, .
\end{equation}
Concerning the necessary explicit determination of $\langle \Phi(z_{1},\bar{z}_{1}) \varepsilon(z_{1},\bar{z}_{1}) \rangle_{\textrm{P}}$, we proceed similarly to the case of the profiles in Secs. \ref{symJanus} and \ref{App_needle} by transforming the corresponding two-point correlation function $\langle \Phi(w_{1},\bar{w}_{1}) \varepsilon(w_{1},\bar{w}_{1}) \rangle_{\mathbb{H}_{-+}}$ (in the upper half plane $w = u + iv$ with $+$ and $-$ boundary conditions on the positive and negative real axis, respectively) to the geometry of the Janus circle and Janus needle in the $z$ plane upon using the transformations $w(z)$ in Eqs. (\ref{03_02}) and (\ref{App_needle_02}), respectively. Using the expression for the half plane function given in the last but one equation (4.3) in Ref. \cite{BX_91} this reads
\begin{eqnarray}
\label{cf_005}
\langle \Phi(z_{1},\bar{z}_{1}) \varepsilon(z_{2},\bar{z}_{2}) \rangle_{\rm P} & = & \langle \Phi(z_{1},\bar{z}_{1}) \rangle_{+} \langle \varepsilon(z_{2},\bar{z}_{2}) \rangle_{+} \, K_{\Phi\varepsilon}(z_{1},\bar{z}_{1};z_{2},\bar{z}_{2})
\end{eqnarray}
with
\begin{equation}
\label{cf_006}
K_{\Phi\varepsilon}(z_{1},\bar{z}_{1};z_{2},\bar{z}_{2}) = \frac{1}{2p\sqrt{1+p^{2}}} \biggl[ -3U_{1}\left( 1-(4/3)U_{2}^{2} \right)(1+2p^{2}) + 4U_{2}V_{1}V_{2} \biggr]
\end{equation}
and
\begin{eqnarray} \nonumber
\label{cf_007}
\frac{1}{2p} & \equiv & \left( \frac{v_{1}v_{2}}{|w_{1}-w_{2}|^{2}} \right)^{1/2} \rightarrow \frac{|z_{1}z_{2}|}{|z_{1}-z_{2}|} \biggl\{ \frac{1}{2} - \frac{1}{4}\left( |z_{1}|^{-2} + |z_{2}|^{-2} \right) ; 1 - \frac{1}{8} \bigl( z_{1}^{-2} + z_{2}^{-2} + \\ \nonumber
&& + \bar{z}_{1}^{-2} + \bar{z}_{2}^{-2}  + |\bar{z}_{1}|^{-2} + |z_{2}|^{-2} +z_{1}^{-1}z_{2}^{-1} + \bar{z}_{1}^{-1}\bar{z}_{2}^{-1} \bigr) \biggr\} \, \\ \nonumber
U_{j} & \equiv & \frac{u_{j}}{|w_{j}|} \rightarrow i\left( z_{j}^{-1} - \bar{z}_{j}^{-1} \right) \biggl\{ 1 + \frac{1}{2}\left( z_{j}^{-2} + \bar{z}_{j}^{-2} \right) ; 1 + \frac{3}{8}\left( z_{j}^{-2} + \bar{z}_{j}^{-2} \right) + \frac{1}{4}z_{j}^{-1}\bar{z}_{j}^{-1} \biggr\} \, , \\
V_{j} & \equiv & \frac{v_{j}}{|w_{j}|} \rightarrow i \left( z_{j}^{-1} - \bar{z}_{j}^{-1} \right) \biggl\{ 1 + \frac{1}{2}\left( z_{j}^{-1} - \bar{z}_{j}^{-1} \right)^{2} ; 1 + \frac{1}{8}\left( z_{j}^{-1} - \bar{z}_{j}^{-1} \right)^{2} \biggr\}
\end{eqnarray}
for $\textrm{P}$ corresponding to the Janus \{circle; needle\}. The symbol $\rightarrow$ indicates that on the right hand side only the leading order and next-to-leading order terms in the expansion for large $|z_{j}|$ are retained. The differences between the two cases arise from the different behaviors at large $z$:
\begin{equation}
\label{cf_008}
w(z) \rightarrow \biggl\{ i \Bigl[ 1+2\left(z^{-1}+z^{-2}+z^{-3}\right) \Bigr] ; i \Bigl[ 1+z^{-1}+\left(z^{-2}+z^{-3}\right)/2 \Bigr] \biggr\} \, ,
\end{equation}
where $w(z)$ is the mapping given in Eqs. (\ref{03_02}) and (\ref{App_needle_map}) for the Janus circle and Janus needle, respectively. The lower index $j$ in the last two equations (\ref{cf_007}) is either $1$ or $2$. The first two factors on the right hand side of Eq. (\ref{cf_005}) are the profiles of the order parameter and of the energy densities around two particles with the same (circular and needle) shapes as above but with homogeneous boundary conditions $+$. These particles have been denoted by ``circle,+'' and ``needle,+'' in Eqs. (\ref{03_09}) and (\ref{App_needle_04}), respectively. To the order of interest the profiles are
\begin{equation}
\begin{aligned}
\label{cf_009}
\langle \Phi(z_{1},\bar{z}_{1}) \rangle_{+} & \rightarrow 2^{1/8} |z_{1}|^{-1/4} \biggl\{ 2^{1/8} \biggl[ 1+\frac{1}{8}|z_{1}|^{-2}\biggr] ; 1+\frac{1}{8}\biggl[ \frac{3}{8} \left( z_{1}^{-2} + \bar{z}_{1}^{-2} \right) + \frac{1}{4}|z_{1}|^{-2}\biggr] \biggr\} \, , \\
\langle \varepsilon(z_{2},\bar{z}_{2}) \rangle_{+} & \rightarrow -\frac{1}{2} |z_{2}|^{-2} \biggl\{ 2 \biggl[ 1+|z_{2}|^{-2}\biggr] ; 1+\biggl[ \frac{3}{8} \left( z_{2}^{-2} + \bar{z}_{2}^{-2} \right) + \frac{1}{4}|z_{2}|^{-2}\biggr] \biggr\} \, .
\end{aligned}
\end{equation}
By expanding Eq. (\ref{cf_005}) for large distances upon using Eqs. (\ref{cf_007}) and (\ref{cf_009}), one realizes that the next-to-leading contribution is indeed identical to the expression in Eq. (\ref{cf_002}) if one chooses
\begin{equation}
\label{cf_010}
(A,B,C) = i \biggl\{ 2^{13/4} \left(-\frac{11}{7},\frac{32}{21},-\frac{8}{3}\right) ; 2^{9/8} \left(-\frac{9}{7},\frac{10}{7},-2\right) \biggr\} \, .
\end{equation}
We note that establishing the agreement of the next-to-leading contribution to $\langle \Phi\varepsilon \rangle_{\textrm{P}}$ in Eq.~(\ref{cf_005}) with the complicated expression in Eq. (\ref{cf_002}) by adapting only three numbers is nontrivial and a welcome opportunity to see the SPOE machinery at work. Equation (\ref{cf_001}), with the values of $A$, $B$, and $C$ from Eq. (\ref{cf_010}) inserted, reproduces the next-to-leading order contributions of the order parameter profiles $\langle \Phi(z_{1},\bar{z}_{1}) \rangle_{\textrm{J}}$ and $\langle \Phi(z_{1},\bar{z}_{1}) \rangle_{\textrm{Jn}}$ for the Janus circle and the Janus needle given in Eqs. (\ref{03_09}) and (\ref{App_needle_06bis}), respectively. In Secs. \ref{wall} and \ref{interaction_JJ} we use these three numbers in order to determine corresponding properties of the free energy of interaction involving the Janus particles.

The present prefactors $A,B,C$ of the operators $[L_{-3},L_{-1}^{3},L_{-1}^{2}\bar{L}_{-1}]\Phi$ and $-A$, $-B$, $-C$ of their conjugates in the SPOE apply to a Janus circle and a Janus needle with radius $1$ and length $2$, respectively, both in the standard orientation $\alpha=0$, i.e., they apply to the particle configurations studied in Secs. \ref{symJanus} and \ref{App_needle}. For a Janus circle and a Janus needle with radius $R$ and length $D$, respectively, both rotated counterclockwise by an angle $\alpha$ from their standard orientation $\alpha=0$, the expressions in front of $A$, $B$, and $C$ must be multiplied by $R^{3+1/8}\exp(3i\alpha)$ for the Janus circle, and by $(D/2)^{3+1/8}\exp(3i\alpha)$ for the Janus needle. Analogously, the expressions in front of the complex conjugated operators are multiplied by $R^{3+1/8}\exp(-3i\alpha)$ and $(D/2)^{3+1/8}\exp(-3i\alpha)$, for the Janus circle and Janus needle, respectively. This is due to the dilatation and rotation properties of the operators determined by their scaling dimension and spin as discussed below Eq. (\ref{VAextended}) and it leads to the expressions for $s_{\rm P}^{(\Phi)}$ in Eqs. (\ref{J_spoe_phi}) and (\ref{nJ_spoe_phi}).

\subsubsection{Next-to-leading $\Phi$ descendants and their averages in a confined geometry}
\label{appb11}
For the circular or needle-like Janus particle $\textrm{P}$ embedded in a confined system ${\mathscr{G}}$, here we determine the next-to-leading contribution to the average $\langle s_P^{(\Phi)} \rangle_{\mathscr{G}}$, which is formed by the corresponding averages of the $\Phi$ descendants of third order, as considered above, and which is required for the free energy of transfer $\delta\mathcal{F}$ in Eq. (\ref{int_01}). As mentioned in Sec. \ref{wall}, for $\mathscr{G}$ we consider a half plane, a strip, and a wedge. The average $\langle L_{-3}\Phi(z_{0},\bar{z}_{0})\rangle_{\mathscr{G}}$ requires knowledge of the corresponding average $\langle T(z) \Phi(z_{0},\bar{z}_{0})\rangle_{\mathscr{G}}$ (see Eq. (\ref{def_descendant})). If ${\mathscr{G}}$ has a homogeneous boundary $a$, the latter average can be obtained (via the conformal transformations studied in Sec. \ref{wall}) from the upper half $w$ plane $\mathbb{H}_{a}$ for which the conformal Ward identity yields
\begin{align} \nonumber
\label{cf_011}
\langle T(w) \mathcal{O}(w_{0},\bar{w}_{0}) \rangle_{\mathbb{H}_{a}} & = \biggl[ \frac{x_{\mathcal{O}}/2}{(w-w_{0})^{2}} + \frac{1}{w-w_{0}}\partial_{w_{0}} + \frac{x_{\mathcal{O}}/2}{(w-\bar{w}_{0})^{2}} + \frac{1}{w-\bar{w}_{0}}\partial_{\bar{w}_{0}} \biggr] \langle \mathcal{O}(w_{0},\bar{w}_{0}) \rangle_{\mathbb{H}_{a}} \\
& \equiv (x_{\mathcal{O}}/2) (w-w_{0})^{-2}(w-\bar{w}_{0})^{-2}(w_{0}-\bar{w}_{0})^{2} \langle \mathcal{O}(w_{0},\bar{w}_{0}) \rangle_{\mathbb{H}_{a}}
\end{align}
(compare Ref. \cite{Cardy_BCFT}). Here $\mathcal{O}$, with scaling dimension $x_{\mathcal{O}}$, is a (scalar) primary operator such as $\Phi$ or $\varepsilon$. The simplest case is the right half plane $\widetilde{\mathbb{H}}_{a}$  for which the rotation in Eq. (\ref{rotation}) yields
\begin{equation}
\label{cf_012}
\langle T(z) \mathcal{O}(z_{0},\bar{z}_{0}) \rangle_{\widetilde{\mathbb{H}}_{a}} = (x_{\mathcal{O}}/2) (z-z_{0})^{-2}(z+\bar{z}_{0})^{-2}(z_{0}+\bar{z}_{0})^{2} \langle \mathcal{O}(z_{0},\bar{z}_{0}) \rangle_{\widetilde{\mathbb{H}}_{a}}
\end{equation}
and via Eq. (\ref{def_descendant})
\begin{equation}
\label{cf_013}
\langle L_{-3} \mathcal{O}(z_{0},\bar{z}_{0}) \rangle_{\widetilde{\mathbb{H}}_{a}} = -2 x_{\mathcal{O}} (z_{0}+\bar{z}_{0})^{-3} \langle \mathcal{O}(z_{0},\bar{z}_{0}) \rangle_{\widetilde{\mathbb{H}}_{a}} \, .
\end{equation}
For $\mathcal{O}=\Phi$ and for the half plane with its boundary belonging to the universality class $a=+$ or $a=-$, this implies that for $R=D/2=1$ and arbitrary $\alpha$ the next-to-leading (nl) contribution is given by
\begin{align} \nonumber
\label{cf_014}
\langle s_{\rm P}^{(\Phi)} \rangle_{\widetilde{\mathbb{H}}_{a}}^{(\rm nl)}  & = 2 \textrm{Re} \bigl\langle \bigl[ \textrm{e}^{3i\alpha} \left( A L_{-3} + B L_{-1}^{3} \right) + \textrm{e}^{i\alpha} C L_{-1}^{2}\bar{L}_{-1} \bigr] \Phi(z_{0},\bar{z}_{0}) \bigr\rangle_{\widetilde{\mathbb{H}}_{a}} \\
& = - 2 \textrm{Re} \biggl[ \textrm{e}^{3i\alpha} \left( \frac{A}{4} + \frac{153B}{512} \right) + \textrm{e}^{i\alpha} \frac{153C}{512} \biggr] (2x_{0})^{-3} \mathcal{A}_{\Phi}^{(a)} x_{0}^{-1/8}
\end{align}
and yields, with $A, B, C$ from Eq. (\ref{cf_010}), the results for the next-to-leading averages of $s_{\textrm{P}}^{(\Phi)}$ in the right half plane as provided in Eqs. (\ref{wall_J_eps})-(\ref{wall_nJ_phi}). 

For the strip and the wedge the same procedure applies upon combining Eq. (\ref{cf_011}) with the mappings in Eqs. (\ref{map_strip}) and (\ref{map_wedge}), respectively. For the strip with equal boundary conditions $aa$ and $a \in \{ +, -, O\}$ at both edges, we find\footnote{The Schwarzian derivative $S(w(z))$ appearing in the transformation formula leading from the half plane to the strip is independent of $z$ and does not contribute to the integral in Eq. (\ref{def_descendant}) which defines $\langle L_{-3} \mathcal{O} \rangle_{\mathbb{S}}$.}
\begin{equation}
\langle L_{-3} \mathcal{O}(z_{0},\bar{z}_{0}) \rangle_{\mathbb{S}_{aa}} = \frac{i x_{\mathcal{O}}}{16} \left( \frac{\pi}{\mathcal{W}} \right)^{3} \left( 3t_{0} + 4t_{0}^{3} \right) \langle \mathcal{O}(z_{0},\bar{z}_{0}) \rangle_{\mathbb{S}_{aa}}
\end{equation}
and therefore 
\begin{equation}
\begin{aligned}
\label{cf_014_nl}
\langle s_{\textrm{P}}^{(\Phi)} \rangle_{\mathbb{S}_{aa}}^{(\textrm{nl})} & \equiv 2 \textrm{Re} \bigl\langle \bigl[ \textrm{e}^{3i\alpha} \left( A L_{-3} + B L_{-1}^{3} \right) + \textrm{e}^{i\alpha} C L_{-1}^{2}\bar{L}_{-1} \bigr] \Phi(z_{0},\bar{z}_{0})\bigr\rangle_{\mathbb{S}_{aa}} \, , \\
& = \biggl[ \frac{i A \cos(3\alpha)}{64} \left( 3 t_{0} + 4 t_{0}^{3} \right) + \frac{i B \cos(3\alpha) - iC \cos\alpha }{2048} \left( 152 t_{0} + 153 t_{0}^{3} \right) \biggr] \left( \frac{\pi}{\mathcal{W}} \right)^{3} \times \\
& \times \langle \Phi(z_{0},\bar{z}_{0}) \rangle_{\mathbb{S}_{aa}} \, ,
\end{aligned}
\end{equation}
where
\begin{equation}
t_{0} \equiv \tan Y_{0} \, , \qquad Y_{0} \equiv \pi y_{0}/ \mathcal{W}
\end{equation}
and
\begin{equation}
\langle \Phi(z_{0},\bar{z}_{0}) \rangle_{\mathbb{S}_{aa}} = \mathcal{A}_{\Phi}^{(a)} \left( \frac{\pi}{\mathcal{W}\cos Y_{0}} \right)^{1/8}
 \, .
\end{equation}
Inserting the values of $A,B,C$ (Eq. (\ref{cf_010})) for a circular and needle-like Janus particle yields the results in Eq. (\ref{nl_averages}).

\subsubsection{Order parameter two-point correlation function $\langle\Phi\Phi\rangle_{\rm{J}}$ for symmetric Janus particles}
Here and in the following subsection we check that the operators $\varepsilon$ and $T$ in $s_{\textrm{J}}^{(\varepsilon)}$ and $s_{\textrm{J}}^{(\mathbb{I})}$ of the SPOE reproduce the behavior of $\langle\Phi\Phi\rangle_{\textrm{J}}$ and $\langle\varepsilon\varepsilon\rangle_{\textrm{J}}$ at large distances. Since the corresponding prefactors have been determined already from the profiles $\langle\varepsilon\rangle_{\textrm{J}}$ and $\langle T\rangle_{\textrm{J}}$, we can be very brief and limit the discussion to the symmetric circular Janus particle. The relations
\begin{equation}
\label{cf_015}
\frac{1}{|z_{12}|^{2}} = \vert w^{\prime}(z_{1}) \vert \vert w^{\prime}(z_{2}) \vert \frac{1}{|w_{12}|^{2}}
\end{equation}
and
\begin{equation}
\label{cf_016}
|w_{12}| = \frac{2|z_{12}|}{|z_{1}-1| |z_{2}-1|} \, ,
\end{equation}
where $z_{12}=z_{1}-z_{2}$ and $w_{12}=w_{1}-w_{2}$, will be useful in the following. While the first relation applies to any M$\ddot{\textrm{o}}$bius transformation $w(z)$, the second one follows from our specific choice for the transformation in Eq. (\ref{03_02}).

As above, here we start from the order parameter two-point correlation function in the upper half $w$ plane with boundary conditions $+$ and $-$ on the positive and negative real axis, respectively. This correlation function can be inferred from the first equation in Eq. (4.3) in Ref. \cite{BX_91}. For the present purpose it is useful to rewrite it as\footnote{Within our normalization (Eqs. (\ref{03_03}), (\ref{03_04bis}), and (\ref{03_04_01})), the amplitudes $\mathcal{A}$ and $\mathcal{B}$ appearing in Eq. (4.3) of Ref. \cite{BX_91} have the values $\mathcal{A}=2^{1/8}$ and $\mathcal{B}=-1/2$; these amplitudes correspond to the ones in Eq. (\ref{03_04_01}).}
\begin{equation}
\label{cf_017}
\langle \Phi(w_{1},\bar{w}_{1}) \Phi(w_{2},\bar{w}_{2}) \rangle_{\mathbb{H}_{-+}} = |w_{12}|^{-1/4} \left( 1+p^{2} \right)^{1/8} K_{\Phi\Phi}
\end{equation}
with
\begin{equation}
\label{cf_018}
K_{\Phi\Phi} = \left( 1+\mathcal{U}^{-2} \right)^{1/2} U_{1}U_{2} + \frac{1-\mathcal{U}^{-2}}{\left( 1+\mathcal{U}^{-2} \right)^{1/2}} V_{1}V_{2} \, ,
\end{equation}
and
\begin{equation}
\label{cf_019}
\mathcal{U} = p^{-1/2} \left( 1+p^{2} \right)^{1/4} \, ,
\end{equation}
where $p$, $U_{j}$, and $V_{j}$ are defined in Eq. (\ref{cf_007}). (We note that the quantity called $\mathcal{U}$ here corresponds to the quantity called $u$ in Ref. \cite{BX_91}.) We now proceed by calculating $\langle \Phi(z_{1},\bar{z}_{1})\rangle_{\textrm{J}}$ which equals $\langle \Phi(w_{1},\bar{w}_{1})\rangle_{\mathbb{H}_{-+}}$ in Eq. (\ref{cf_017}) with the prefactor $|w_{12}|^{-1/4}$ replaced by $|z_{12}|^{-1/4}$. This follows from the transformation formula in Eq. (\ref{03_08}), applied to two-point correlation functions and by using Eq. (\ref{cf_015}). With this, the large distance expansions in Eq. (\ref{cf_007}) imply
\begin{equation}
\label{cf_020}
K_{\Phi\Phi} = 1 - \frac{3}{2}p + \frac{7}{8}p^{2} - 2 \left( \frac{y_{1}}{|z_{1}|^{2}} - \frac{y_{2}}{|z_{2}|^{2}} \right)^{2} + O(r^{-3}) \, .
\end{equation}
Hence the large-distance behavior of the correlation function $\langle \Phi(z_{1},\bar{z}_{1}) \Phi(z_{2},\bar{z}_{2}) \rangle_{\textrm{J}}$ is given by
\begin{equation}
\label{corr_13}
\langle \Phi(z_{1},\bar{z}_{1}) \Phi(z_{2},\bar{z}_{2}) \rangle_{\textrm{J}} = |z_{12}|^{-1/4} \left( 1 - \frac{3}{2} \frac{|z_{12}|}{|z_{1}||z_{2}|} + 2G + O(r^{-3}) \right) \, ,
\end{equation}
where
\begin{equation}
\label{corr_14}
G = \frac{1}{2} \frac{|z_{12}|^{2}}{|z_{1}|^{2}|z_{2}|^{2}} - \left( \frac{y_{1}}{|z_{1}|^{2}} - \frac{y_{2}}{|z_{2}|^{2}} \right)^{2} \, .
\end{equation}
This is in agreement with the SPOE in Eqs. (\ref{J_spoe_en})-(\ref{J_spoe_id}), which predicts
\begin{eqnarray} \nonumber
\label{corr_15}
\langle \Phi(z_{1},\bar{z}_{1}) \Phi(z_{2},\bar{z}_{2}) \rangle_{\textrm{J}} & = & \langle \Phi(z_{1},\bar{z}_{1}) \Phi(z_{2},\bar{z}_{2}) \rangle + 3 \langle \Phi(z_{1},\bar{z}_{1}) \Phi(z_{2},\bar{z}_{2}) \varepsilon(0,0) \rangle + \\
& + & 16\pi \langle \Phi(z_{1},\bar{z}_{1}) \Phi(z_{2},\bar{z}_{2}) T_{yy}(0,0) \rangle + \dots \, ,
\end{eqnarray}
because within the normalization adopted in Eq. (\ref{03_04bis}), the three-point correlation functions appearing above are given by
\begin{equation}
\label{corr_16}
\langle \Phi(z_{a},\bar{z}_{a}) \Phi(z_{b},\bar{z}_{b}) \varepsilon(z_{c},\bar{z}_{c}) \rangle = C_{\Phi\Phi\varepsilon} |z_{ab}|^{3/4} |z_{ac}|^{-1}|z_{bc}|^{-1}
\end{equation}
with the structure constant $C_{\Phi\Phi\varepsilon}=-1/2$ and by
\begin{equation}
\label{corr_17}
\langle \mathcal{O}(z_{1},\bar{z}_{1}) \mathcal{O}(z_{2},\bar{z}_{2}) T_{yy}(0,0) \rangle = \frac{x_{\mathcal{O}}}{\pi|z_{12}|^{2x_{\mathcal{O}}}} G
\end{equation}
with scaling dimension $x_{\mathcal{O}}$ for $\mathcal{O}=\Phi$ or $\mathcal{O}=\varepsilon$.

\subsubsection{Energy density two-point correlation function $\langle\varepsilon\varepsilon\rangle_{\rm{J}}$ for symmetric Janus particles}
Writing the corresponding half plane expression, given by Eq. (4.3) in Ref. \cite{BX_91}, in the form
\begin{equation}
\label{corr_18}
\langle \varepsilon(w_{1},\bar{w}_{1})\varepsilon(w_{2},\bar{w}_{2}) \rangle_{\mathbb{H}_{-+}} = |w_{12}|^{-2} K_{\varepsilon\varepsilon}
\end{equation}
with
\begin{eqnarray} \nonumber
\label{corr_19}
K_{\varepsilon\varepsilon} & = & p^{2} \left( 1-4V_{1}^{2} \right) \left( 1-4V_{2}^{2} \right) + 1 - \frac{4}{|w_{1}||w_{2}|}\left( u_{1}^{2}v_{2}^{2} + u_{2}^{2}v_{1}^{2} - 2 u_{1}u_{2}v_{1}v_{2} \right) + \\
& - & \frac{p^{2}}{1+p^{2}} \biggl[ 1 - \frac{4}{|w_{1}||w_{2}|}\left( u_{1}^{2}v_{2}^{2} + u_{2}^{2}v_{1}^{2} + 2 u_{1}u_{2}v_{1}v_{2} \right) \biggr]
\end{eqnarray}
yields
\begin{equation}
\label{corr_20}
\langle \varepsilon(z_{1},\bar{z}_{1})\varepsilon(z_{2},\bar{z}_{2}) \rangle_{\textrm{J}} = |z_{12}|^{-2} K_{\varepsilon\varepsilon} \, .
\end{equation}
Expanding this expression for large distance by using Eq. (\ref{cf_007}) and
\begin{equation}
\begin{aligned}
u_{j} & = \frac{2y_{j}}{|z_{j}|^{2}} + O\left( |z_{j}|^{-2} \right) \, , \\
v_{j} & = 1 + \frac{2x_{j}}{|z_{j}|^{2}} + O\left( |z_{j}|^{-2} \right) \, ,
\end{aligned}
\end{equation}
leads to
\begin{equation}
\label{corr_21}
\langle \varepsilon(z_{1},\bar{z}_{1})\varepsilon(z_{2},\bar{z}_{2}) \rangle_{\textrm{J}} = |z_{12}|^{-2} \left( 1 + 16G + O(r^{-3}) \right) \, ,
\end{equation}
where $G$ is given in Eq. (\ref{corr_14}). This expression is in agreement with the SPOE in Eqs. (\ref{J_spoe_en})-(\ref{J_spoe_id}) upon taking into account the bulk three-point correlation function in Eq. (\ref{corr_17}) with $\mathcal{O}=\varepsilon$.

\section{SPOE for generalized Janus particles}
\label{appendix_C}
In this section we derive the SPOE for generalized Janus particles. First we expand the profiles $\langle \Phi \rangle_{\rm J_{\chi}}$ and $\langle \varepsilon \rangle_{\rm J_{\chi}}$ given in Eq. (\ref{04_04}) and Eq. (\ref{04_05}), respectively, for large $|z| = \sqrt{x^2 +y^2}$. This requires expanding $\langle \mathcal{O} \rangle_{\rm{circle},+}$ defined below Eq. (\ref{03_10}) for $\mathcal{O}=\Phi$ and $\varepsilon$ and expanding the expression $\mathcal{C}$ in Eq. (\ref{04_06}) for which we find
\begin{equation}
\label{Janus_SPOE_01}
\mathcal{C}(\chi;x,y) = \sum_{k=0}^{\infty} \mathcal{C}_{k} |z|^{-k} \, ,
\end{equation}
where the coefficients $\mathcal{C}_{k}$ are dimensionless functions of $\chi$. The first few of them are
\begin{eqnarray}
\label{Janus_SPOE_02}
\mathcal{C}_{0} & = & \sin\chi \, , \\
\label{Janus_SPOE_03}
\mathcal{C}_{1} & = & 2\cos^{2}\chi\sin\theta \, , \\
\label{Janus_SPOE_04}
\mathcal{C}_{2} & = & (3\cos2\theta-1)\cos^{2}\chi\sin\chi \, ,
\end{eqnarray}
and
\begin{equation}
\label{Janus_SPOE_04bis}
\mathcal{C}_{3} = \frac{1}{2} \cos^2\chi \Bigl[ (1-3 \cos2\chi )\sin\theta + (5 \cos2\chi-3)\sin3\theta \Bigr]
\end{equation}
with $\theta=\textrm{arg}(z)$. Together with the corresponding expansion of the expression for the stress tensor in Eq. (\ref{03_10}) this yields
\begin{eqnarray} \nonumber
\label{Janus_SPOE_05}
\langle \Phi(z,\bar{z}) \rangle_{\textrm{J}_{\chi}} & = & \frac{2^{1/4}}{|z|^{1/4}} \biggl[ \mathcal{C}_{0} + \frac{\mathcal{C}_{1}}{|z|} + \left( \mathcal{C}_{2} + \frac{\mathcal{C}_{0}}{8} \right)\frac{1}{|z|^{2}} \biggr] + O\left( |z|^{-3-1/4} \right) \, , \\ \nonumber
\label{Janus_SPOE_06}
\langle \varepsilon(z,\bar{z}) \rangle_{\textrm{J}_{\chi}} & = & -\frac{1}{|z|^2} \biggl[ \left( 4\mathcal{C}_{0}^{2}-3 \right) + \frac{8\mathcal{C}_{0}\mathcal{C}_{1}}{|z|} + \frac{4\mathcal{C}_{0}^{2}-3+4\mathcal{C}_{1}^{2}+8\mathcal{C}_{0}\mathcal{C}_{2}}{|z|^{2}} \biggr] + O\left( |z|^{-5} \right) \, , \\
\langle T(z) \rangle_{\textrm{J}_{\chi}} & = & \frac{2\cos^{2}\chi}{z^{4}} - \frac{8i\sin\chi\cos^{2}\chi}{z^{5}} + O\left( z^{-6} \right) \, .
\end{eqnarray}
The contributions to the SPOE given in Eqs. (\ref{genJ_spoe_en})-(\ref{genJ_spoe_id}) follow from the requirement to reproduce the profiles in Eq. (\ref{Janus_SPOE_06}) on using Eq. (\ref{Psidotdot}). 

\section{Descendants of the identity operator in the SPOE for quadrupolar particles}
\label{appendix_D}
Here we consider the leading operators in the SPOE of quadrupolar particles which are descendants of unity. Due to the symmetries of these particles the leading descendants are the five operators
\begin{equation}
\label{sQ_01}
L_{-2}\bar{L}_{-2} \mathbb{I}(z,\bar{z}) = T(z) \bar{T}(\bar{z}) \, , \quad L_{-2}^{2} \mathbb{I}(z,\bar{z}) \, , \quad \bar{L}_{-2}^{2} \mathbb{I}(z,\bar{z}) \, , \quad L_{-4} \mathbb{I}(z,\bar{z}) \, , \quad \bar{L}_{-4} \mathbb{I}(z,\bar{z}) \, ,
\end{equation}
with scaling dimension 4. We note that $L_{-3}L_{-1}\mathbb{I}$ vanishes due to $\partial_z \mathbb{I} =0$, and that both $L_{-1}^2 L_{-2}\mathbb{I}$ and $L_{-1}L_{-3}\mathbb{I}$, being equal to $2L_{-4}\mathbb{I}$, are not independent. This follows from the Virasoro algebra (see Eq. (\ref{VAextended})). Accordingly, these three operators and their conjugates do not appear in Eq. (\ref{sQ_01}). 

\subsection{Prefactors of the operators in the SPOE}
In order to determine their prefactors we apply the SPOE to the energy density two-point correlation function $\langle \varepsilon_1 \, \varepsilon_2 \rangle_{\rm Q}$ because in this case --- due to the $+/-$ and the duality symmetry of the Ising model --- it is only the descendants of the unit operator which via Eq. (\ref{Psidotdot}) contribute to the correlation function. Thus beyond the bulk contribution $\langle \varepsilon_{1} \varepsilon_{2} \rangle = |z_{12}|^{-2}$ the leading large distance behavior of $\langle \varepsilon_{1} \varepsilon_{2} \rangle_{\textrm{Q}}$ must be a linear combination of the three expressions
\begin{eqnarray}
\label{sQ_02}
\langle \varepsilon_{1}\varepsilon_{2} T(0)\bar{T}(0) \rangle & = & \frac{|z_{12}|^{2}}{4|z_{1}|^{4}|z_{2}|^{4}} \equiv \mathfrak{a} \, , \\
\label{sQ_03}
\langle \varepsilon_{1}\varepsilon_{2} L_{-2}^{2} \mathbb{I}(0,0) \rangle & = & \frac{z_{12}^{2}}{4|z_{12}|^{2}z_{1}^{4}z_{2}^{4}} \left( 5z_{1}^{2}+2z_{1}z_{2}+5z_{2}^{2} \right) \equiv \mathfrak{b} \, ,
\end{eqnarray}
and
\begin{eqnarray}
\label{sQ_04}
\langle \varepsilon_{1}\varepsilon_{2} L_{-4} \mathbb{I}(0,0) \rangle & = & \frac{z_{12}^{2}}{2|z_{12}|^{2}z_{1}^{4}z_{2}^{4}} \left( 3z_{1}^{2}+4z_{1}z_{2}+3z_{2}^{2} \right) \equiv \mathfrak{c} \, ,
\end{eqnarray}
and of their complex conjugates. For simplicity, here we assume that the particle center is located at the origin and we use the short notations $z_{12}\equiv z_{1}-z_{2}$ and $\varepsilon_{1} \equiv \varepsilon(z_{1},\bar{z}_{1})$, $\varepsilon_{2} \equiv \varepsilon(z_{2},\bar{z}_{2})$. These expressions are of the order $O\left( r^{-6} \right)$, where $|z_{j}|=O(r)$, $j=1,2$, is the distance from the quadrupole. They explicitly show that $T\bar{T}$ is an isotropic operator and that the other operators in Eq. (\ref{sQ_01}) are anisotropic ones respecting the (even) quadrupole symmetry.

For the quadrupole in standard orientation $\alpha =0$, the explicit calculation of $\langle \varepsilon_{1} \varepsilon_{2} \rangle_{\textrm{Q}}$ is explained below Eq. (\ref{sQ_09}) and yields
\begin{eqnarray} \nonumber
\label{sQ_05}
\langle \varepsilon_{1} \varepsilon_{2} \rangle_{\textrm{Q}} - \frac{1}{|z_{12}|^{2}} & = & \frac{35}{3} \frac{|z_{12}|^{2}}{|z_{1}|^{4}|z_{2}|^{4}} + \frac{8}{3 |z_{12}|^{2}} \biggl[ \frac{z_{12}^{2}}{z_{1}^{4}z_{2}^{4}}\left( z_{1}^{2} + 4z_{1}z_{2} + z_{2}^{4} \right) + \textrm{cc} \biggr] + O\left(r^{-8}\right) \, , \\
& = & \frac{140}{3} \mathfrak{a} + \frac{16}{21} \Bigl[ \left( -8 \mathfrak{b} + 9 \mathfrak{c} \right) + \textrm{cc} \Bigr] + O\left(r^{-8}\right) \, .
\end{eqnarray}
This displays indeed the linear combination mentioned above and determines the prefactors of the operators with scaling dimension $4$ in the SPOE as given in Eqs. (\ref{Q_spoe_en}) and (\ref{Q_spoe_id}). We note that $\langle \varepsilon_{1} \varepsilon_{2} \rangle_{\rm Q}$ does not contain terms of the order $-3$, $-4$, and $-5$ in the distance.

Equations (\ref{Q_spoe_en}) and (\ref{Q_spoe_id}) must describe the large distance behavior of $\langle T(z) \rangle_{\textrm{Q}}$ as well. By using the bulk two-point correlation functions
\begin{eqnarray}
\label{sQ_06}
\langle T(z) T(0) \bar{T}(0) \rangle & = & 0 \, , \\
\label{sQ_07}
\langle T(z) L_{-2}^{2} \mathbb{I}(0,0) \rangle = \frac{3}{2z^{6}} & \equiv & \mathfrak{B} \, ,
\end{eqnarray}
and
\begin{eqnarray}
\label{sQ_08}
\langle T(z) L_{-4} \mathbb{I}(0,0) \rangle = \frac{5}{2z^{6}} & \equiv & \mathfrak{C} \, ,
\end{eqnarray}
the SPOE predicts
\begin{equation}
\label{sQ_09}
\langle T(z) \rangle_{\textrm{Q}} \stackrel{z\rightarrow\infty}{\longrightarrow} \frac{16}{21} \left( -8\mathfrak{B} + 9\mathfrak{C} \right) = \frac{8}{z^{6}} \, ,
\end{equation}
in agreement with the result in Eq. (\ref{07_09}) obtained by direct calculation.

In order to obtain the result in Eq. (\ref{sQ_05}), one follows the same route as before, relating --- via Eqs. (\ref{03_08}) and (\ref{03_02}) --- the energy density two-point correlation function to its known\footnote{See the remarks ahead of Eq. (\ref{07_01}).} counterpart in the half plane. This is given by
\begin{equation}
\label{sQ_10}
\langle \varepsilon(z_{1},\bar{z}_{1}) \varepsilon(z_{2},\bar{z}_{2}) \rangle_{\textrm{Q}} = 4 |(z_{1}-1)(z_{2}-1)|^{-2} \langle \varepsilon(w_{1},\bar{w}_{1}) \varepsilon(w_{2},\bar{w}_{2}) \rangle_{\mathbb{H}_{-+-+}} \, .
\end{equation}
In terms of the notations used in Ref. \cite{BG_93}, the correlation function on the right hand side of Eq. (\ref{sQ_10}) can be expressed as
\begin{equation}
\label{sQ_11}
\langle \varepsilon(w_{1},\bar{w}_{1}) \varepsilon(w_{2},\bar{w}_{2}) \rangle_{\mathbb{H}_{-+-+}} = \lim_{\zeta_{4}\rightarrow-\infty}\langle \varepsilon(w_{1},\bar{w}_{1}) \varepsilon(w_{2},\bar{w}_{2}) \rangle_{\zeta_{1}=1, \zeta_{2}=0, \zeta_{3}=-1, \zeta_{4}} = -\frac{2}{3} \textrm{Pf}^{(8)} M_{ab} \, ,
\end{equation}
with the Pfaffian of the $8\times8$ antisymmetric matrix with the elements
\begin{equation}
\label{sQ_12}
M_{ab}=
\begin{cases}
(\omega_{a}-\omega_{b})^{-1}      	& \text{for} \quad 1 \leqslant a < b \leqslant 7 , \\
M_{a8}=1      					& \text{for} \quad 1 \leqslant a \leqslant 7 ,
\end{cases}
\end{equation}
where $(\omega_{1}, \dots, \omega_{7}) = (w_{1}, \bar{w}_{1}, w_{2}, \bar{w}_{2}, 1, 0, -1)$. One has $w_{1}\equiv w(z_{1})$, $w_{2}\equiv w(z_{2})$, where the mapping $w(z)$ is the one given in Eq. (\ref{03_02}). We have used that $\zeta_{4}$, in the limit $\zeta_{4} \rightarrow -\infty$, drops out of the corresponding Eq. (16a) of Ref. \cite{BG_93}, because it enters into both the numerator and the denominator only as an overall factor $1/\zeta_{4}$. The matrix elements in Eq. (\ref{sQ_11}) are given by
\begin{align} \nonumber
\label{sQ_13}
& M_{12} = \frac{|z_{1}-1|^{2}}{2i\left( |z_{1}|^{2}-1 \right)} \, , & &  M_{13} = - \frac{(z_{1}-1)(z_{2}-1)}{2i (z_{1}-z_{2})} \, , & &  M_{14} = - \frac{(z_{1}-1)(\bar{z}_{2}-1)}{2i (z_{1}\bar{z}_{2}-1)} \, , \\
& M_{15} = \left( i\frac{z_{1}+1}{z_{1}-1} - 1 \right)^{-1} \, , & &  M_{16} = \frac{z_{1}-1}{i (z_{1}+1)} \, , & &  M_{17} = \left( i\frac{z_{1}+1}{z_{1}-1} + 1 \right)^{-1} \, ,
\end{align}
and by
\begin{align} \nonumber
\label{sQ_15}
& (M_{23}, M_{24}, M_{25}, M_{26}, M_{27}) = (\overline{M}_{14}, \overline{M}_{13}, \overline{M}_{15}, \overline{M}_{16}, \overline{M}_{17}) \, , & \\ \nonumber
& (M_{34}, M_{35}, M_{36}, M_{37}) = (M_{12}, M_{15}, M_{16}, M_{17})_{z_{1} \mapsto z_{2}} \, , & \\ \nonumber
& (M_{35}, M_{36}, M_{37}) = (\overline{M}_{45}, \overline{M}_{46}, \overline{M}_{47}) \, , & \\
& M_{56}=1 \, , \quad M_{57}=1/2 \, , \quad M_{67}=1 \, ,
\end{align}
where the overbar denotes complex conjugation. We note that upon expanding for large $z_{1}$ and $z_{2}$, the leading distance dimension of $M_{13}$ and $M_{24}$ is $1$ while it is $0$ for all the other matrix elements.

Within the recursive representation of the Pfaffian $\textrm{Pf}^{(8)}M_{ab} = (12345678)$ it is the term
\begin{equation}
\label{Pf}
-(13)(24)(5678) = \frac{3}{2}|M_{13}|^{2} = - \frac{3}{2} \frac{|(z_{1}-1)(z_{2}-1)|^{2}}{4|z_{12}^{2}|}
\end{equation}
which yields --- via Eqs. (\ref{sQ_10}) and (\ref{sQ_11}) --- the bulk contribution $|z_{12}|^{-2}$ in Eq. (\ref{sQ_05}). Here one uses the notation $(kl) \equiv M_{kl}$ and the bracket with four entries is expressed by means of the recursion formula in terms of the sum of three products of two matrix elements. We refer to Ref. \cite{Biggs} for further details concerning the recursive definition of Pfaffians. The expansion of $\langle \varepsilon(z_{1},\bar{z}_{1}) \varepsilon(z_{2},\bar{z}_{2}) \rangle_{\textrm{Q}}$, given by Eqs. (\ref{sQ_10})-(\ref{sQ_15}) up to the order $-6$ in the distance, has been carried out by using \texttt{Mathematica} and has yielded the result in Eq. (\ref{sQ_05}).

\subsection{Descendant averages in a half plane, a strip, and a wedge}
\label{appendix_E}
In order to calculate via Eq. (\ref{int_01}) the cost of free energy to transfer a small quadrupolar particle from the bulk into a confined geometry $\mathscr{G}$ such as a half plane, strips, or wedges, our approach requires knowledge of the averages of the operators in the SPOE (Eqs. (\ref{Q_spoe_en})-(\ref{Q_spoe_id})) for these geometries. As for the operators in Eq. (\ref{sQ_01}), these averages are given by
\begin{equation}
\begin{aligned}
\label{desc_01}
\langle L_{-2}\bar{L}_{-2} \mathbb{I}(z_{0},\bar{z}_{0}) \rangle_{\mathscr{G}} & = \langle T(z_{0}) \bar{T}(\bar{z}_{0}) \rangle_{\mathscr{G}} = \bigl[ \langle T(z_{0}) \bar{T}(\bar{z}_{1}) \rangle_{\mathscr{G}} \bigr]_{\bar{z}_{1}\rightarrow \bar{z}_{0}} \, , \\
\langle L_{-2}^{2} \mathbb{I}(z_{0},\bar{z}_{0}) \rangle_{\mathscr{G}} & = \int_{\mathcal{C}_{z_{0}}} \frac{\textrm{d}z_{1}}{2\pi i} \frac{1}{z_{1}-z_{0}} \langle T(z_{0})T(z_{1}) \rangle_{\mathscr{G}} \, , \\
\langle L_{-4}\mathbb{I}(z_{0},\bar{z}_{0}) \rangle_{\mathscr{G}} & = \frac{1}{2}\frac{\textrm{d}^{2}}{\textrm{d}z_{0}^{2}} \langle T(z_{0}) \rangle_{\mathscr{G}} \, ,
\end{aligned}
\end{equation}
where $\mathcal{C}_{z_0}$ is an integration path which encloses counterclockwise the point $z_0$ (see Eq. (\ref{def_descendant}) and the remarks following it).

\subsection*{Half plane}
In the upper half $w$ plane with homogeneous boundary condition $a \in \{ +, O \}$ on the real axis the two-point correlation functions of the stress tensor are given by (see Ref. \cite{Cardy_BCFT})\footnote{Eqs. (2.29) and (3.35) in Refs. \cite{BPZ_1} and \cite{BPZ_2} contain a misprint: there $c$ should be replaced by $c/2$.}
\begin{equation}
\label{desc_02}
\langle T(w_{0}) T(w_{1}) \rangle_{\mathbb{H}_{a}} = \frac{1/4}{(w_{0}-w_{1})^{4}} \, , \qquad \langle T(w_{0}) \bar{T}(\bar{w}_{1}) \rangle_{\mathbb{H}_{a}} = \frac{1/4}{(w_{0}-\bar{w}_{1})^{4}} \, ,
\end{equation}
so that\footnote{Equation (3.60) in Ref. \cite{Eisenriegler_04} contains a misprint. In the second of these two equations the number $1/16$ should be replaced by $1/64$ so that $\mathcal{A}_{\lambda_{2}}^{(+)}=\mathcal{A}_{\lambda_{2}}^{(O)}=1/64$.}
\begin{equation}
\label{desc_03}
\langle T(w_{0}) \bar{T}(\bar{w}_{0}) \rangle_{\mathbb{H}_{a}} = \frac{1/4}{(w_{0}-\bar{w}_{0})^4} = \frac{1}{64 v_{0}^{4}} \, ,
\end{equation}
where $v_{0}=\textrm{Im}(w_{0})$ is the distance of $w_{0}$ from the boundary. The averages in ${\mathbb{H}_{a}}$ of the other two operators $L_{-2}^{2}\mathbb{I}$ and $L_{-4}\mathbb{I}$ vanish, because due to the functional form of the stress two-point correlation function (see Eq. (\ref{desc_02})) the contour integral corresponding to Eq. (\ref{desc_01}) vanishes and the average $\langle T(w_{0}) \rangle_{\mathbb{H}_{a}}$ vanishes as well.

\subsection*{Strip}
For the horizontal strip ${\mathbb{S}_{ab}}$ of width $\mathcal{W}$ in the $z$ plane, as considered in Sec. \ref{sec_J_half}, the averages can be calculated from those in the upper half $w$ plane $\mathbb{H}_{ab}$ given above. Here one uses the general transformation formulae\footnote{It is instructive to re-derive the transformation formulae given in Eq.~(\ref{desc_05}) for products of the \emph{non-primary} operators $T(z)$ and $\bar{T}(\bar{z})$ from the common formula $\langle \varepsilon(z_{A},\bar{z}_{A}) \varepsilon(z_{B},\bar{z}_{B}) \varepsilon(z_{C},\bar{z}_{C}) \varepsilon(z_{D},\bar{z}_{D}) \rangle_{\mathscr{G}_{z}} = \vert w^{\prime}(z_{A}) w^{\prime}(z_{B}) w^{\prime}(z_{C}) w^{\prime}(z_{D}) \vert \langle \varepsilon(w_{A},\bar{w}_{A}) \varepsilon(w_{B},\bar{w}_{B}) \varepsilon(w_{C},\bar{w}_{C}) \varepsilon(w_{D},\bar{w}_{D}) \rangle_{\mathscr{G}_{w}}$ for the product of the four \emph{primary} operators $\varepsilon$, in the geometries $\mathscr{G}_{z}$ and $\mathscr{G}_{w}$, by considering $z_{AB}\equiv z_{A} - z_{B}\rightarrow0$, $z_{CD}\equiv z_{C} - z_{D}\rightarrow0$ while keeping $(z_{A}+z_{B})/2$ and $(z_{C}+z_{D})/2$ fixed. Here one uses the OPE $\varepsilon(z_{A},\bar{z}_{A}) \varepsilon(z_{B},\bar{z}_{B}) \rightarrow |z_{AB}|^{-2}\bigl\{ 1 + 2 z_{AB}^{2} T((z_{A}+z_{B})/2) + 2 \bar{z}_{AB}^{2} \bar{T}((\bar{z}_{A}+\bar{z}_{B})/2) \bigr\}$ and the relation $\vert (z_{AB}/w_{AB})^{2} w^{\prime}(z_{A}) w^{\prime}(z_{B}) \vert \rightarrow 1+\bigl[ z_{AB}^{2} \mathcal{S}\left( w((z_{A}+z_{B})/2) \right) + \textrm{cc} \bigr]/12$ together with their correspondences for the pair $CD$. Equating the terms $\propto z_{AB}^{2}z_{CD}^{2}$ on both sides of the primary transformation and identifying $(z_{A}+z_{B})/2 \equiv z_{0}$ and $(z_{C}+z_{D})/2 \equiv z_{1}$ yields the first equation in Eq.~(\ref{desc_05}) while the terms $\propto z_{AB}^{2}\bar{z}_{CD}^{2}$ yield the second equation in Eq.~(\ref{desc_05}). The same type of argument serves to confirm the  validity of the transformation formula $\langle \mathcal{O}(z_{0},\bar{z}_{0}) T(z_{1}) \rangle_{\mathscr{G}_{z}} = \langle |w^{\prime}(z_{0})|^{x_{\mathcal{O}}} \mathcal{O}(w_{0},\bar{w}_{0}) [ (w^{\prime}(z_{1}))^{2} T(w_{1}) + (1/24)S(w(z_{1})) ] \rangle_{\mathscr{G}_{w}}$ for the product of a primary operator and the stress tensor as well as the validity of the transformation formula given in Eq. (\ref{stresstrafo}) for the stress tensor.}
\begin{equation}
\begin{aligned}
\label{desc_05}
\langle T(z_{0}) T(z_{1}) \rangle_{\mathscr{G}} & = \biggl\langle \Bigl[ \left( w^{\prime}(z_{0}) \right)^{2} T(w_{0}) + \frac{1}{24} S (w(z_{0})) \Bigr] \Bigl[ \left( w^{\prime}(z_{1}) \right)^{2} T(w_{1}) + \frac{1}{24} S (w(z_{1})) \Bigr] \biggr\rangle_{\mathbb{H}_{ab}} \, , \\
\langle T(z_{0}) \bar{T}(\bar{z}_{1}) \rangle_{\mathscr{G}} & = \biggl\langle \Bigl[ \left( w^{\prime}(z_{0}) \right)^{2} T(w_{0}) + \frac{1}{24} S (w(z_{0})) \Bigr] \Bigl[ \left( \bar{w}^{\prime}(\bar{z}_{1}) \right)^{2} \bar{T}(\bar{w}_{1}) + \frac{1}{24} S (\bar{w}(\bar{z}_{1})) \Bigr] \biggr\rangle_{\mathbb{H}_{ab}} \, ,
\end{aligned}
\end{equation}
relating the stress tensor two-point correlation functions in the geometries $\mathscr{G}$ and $\mathbb{H}_{ab}$, which are conformally connected via $w(z)$. Here $S$ is the Schwarzian derivative (see Eq. (\ref{stresstrafo})). In the present case we consider the strip ${\mathscr{G}} = \mathbb{S}_{ab}$ and $w(z)$ taken from Eq. (\ref{map_strip}).
 
For the strip with equal boundary conditions $a=b \in\{O,+\}$, this yields
\begin{equation}
\begin{aligned}
\label{strip_01}
\langle T(z_{0})T(z_{1}) \rangle_{\mathbb{S}_{aa}} & = \frac{1}{64} \left( \frac{\pi}{\mathcal{W}} \right)^{4} \biggl[ \frac{1}{\sinh^{4}\left(\pi (z_{1}-z_{0})/(2\mathcal{W}) \right)} + \frac{1}{36} \biggr] \, , \\
\langle T(z_{0})\bar{T}(\bar{z}_{0}) \rangle_{\mathbb{S}_{aa}} & = \frac{1}{64} \left( \frac{\pi}{\mathcal{W}} \right)^{4} \biggl[ \frac{1}{\cosh^{4}\left(\pi y_{0}/\mathcal{W} \right)} + \frac{1}{36} \biggr] \, ,
\end{aligned}
\end{equation}
where $z_{0}=x_{0}+iy_{0}$. The above relations imply
\begin{equation}
\begin{aligned}
\label{des_strip}
\langle L_{-2}^{2}\mathbb{I}(z,\bar{z}) \rangle_{\mathbb{S}_{aa}} = \frac{49}{11520} \left( \frac{\pi}{\mathcal{W}} \right)^{4} \, , \\
\langle L_{-2}\bar{L}_{-2}\mathbb{I}(z,\bar{z}) \rangle_{\mathbb{S}_{aa}} = \frac{37}{2304} \left( \frac{\pi}{\mathcal{W}} \right)^{4} \, ,
\end{aligned}
\end{equation}
where the second equation in Eq. (\ref{desc_01}) has been used. The average of $L_{-4}\mathbb{I}(z,\bar{z})$ vanishes because the average of the stress tensor in the strip is spatially constant.

\subsection*{Wedge}
For the wedge ${\mathbb{W}}_{ab}$ considered in Sec. \ref{sec_J_half} the averages of all operators appearing in Eq. (\ref{sQ_01}) are nonzero for a generic opening angle $\pi/\Xi$. Like for the strip, we use the transformation in Eq. (\ref{desc_05}) with ${\mathscr{G}} = \mathbb{W}_{ab}$ and $w(z)$ from Eq. (\ref{map_wedge}) in order to derive the present stress tensor two-point correlation functions from the corresponding results in Eqs. (\ref{desc_02}) and (\ref{desc_03}) in the half plane and then apply the general relations in Eq. (\ref{desc_01}) in order to determine the averages under consideration. For a wedge with equal boundary conditions $a=b$ this yields
\begin{equation}
\begin{aligned}
\label{wedge_desc_01}
\langle T(z_{0}) \bar{T}(\bar{z}_{0}) \rangle_{\mathbb{W}_{aa}} & = \frac{1}{64|z_{0}|^{4}} \biggl[ \frac{\Xi^{4}}{\cos^{4}(\Xi \theta_{0})} + \frac{ \left( 1-\Xi^{2} \right)^{2} }{ 36 }\biggr] \, , \quad z_{0}=|z_{0}|\textrm{e}^{i\theta_{0}} \, , \\
\langle L_{-4} \mathbb{I}(z_{0},\bar{z}_{0}) \rangle_{\mathbb{W}_{aa}}    &  = \frac{ \left( 1-\Xi^{2} \right)^{2} }{ 16 z_{0}^{4} } \, ,
\end{aligned}
\end{equation}
as well as
\begin{equation}
\langle T(z_{0})T(z_{1}) \rangle_{\mathbb{W}_{aa}} = \frac{1}{4(z_{0} z_{1})^{2}} \biggl[ \Xi^{4} \frac{ (z_{0} z_{1})^{2\Xi} }{ \left( z_{0}^{\Xi}-z_{1}^{\Xi} \right)^{4} } + \frac{ \left( 1-\Xi^{2} \right)^{2} }{ 576 } \biggr] \, ,
\end{equation}
which implies
\begin{eqnarray}
\label{wedge_desc_02}
\langle L_{-2}^{2} \mathbb{I}(z_{0},\bar{z}_{0}) \rangle_{\mathbb{W}_{aa}} & = & \frac{\left( \Xi^{2}-1 \right)\left( 49\Xi^{2}-481 \right)}{11520 z_{0}^{4}} \, .
\end{eqnarray}
For $\Xi \rightarrow \infty$ and $|z_{0}| \rightarrow \infty$ with the ratio $\Xi/|z_{0}|$ fixed at the finite expression $\pi/\mathcal{W}$, the wedge reduces to the strip considered above, $\Xi\theta_{0}$ tends to $\pi y_{0}/\mathcal{W}$, and one may check that the wedge results for $\langle T\bar{T} \rangle_{\mathbb{W}}$ and $\langle L_{-2}^{2}\mathbb{I} \rangle_{\mathbb{W}}$ reduce to those of the strip. This also applies to $\langle L_{-4}\mathbb{I} \rangle_{\mathbb{W}}$ which tends to zero for $z_{0}\rightarrow\infty$.

\subsection*{Operator product expansion for two energy densities}
We have determined the two-point correlation functions $\langle \varepsilon_{1} \varepsilon_{2} \rangle_{\mathscr{G}}$ for energy densities $\varepsilon_{1} \equiv \varepsilon(z_{1},\bar{z}_{1})$ and $\varepsilon_{2} \equiv \varepsilon(z_{2},\bar{z}_{2})$ close to each other in ${\mathscr{G}}$, which is a half plane, a strip, or a wedge with homogeneous boundaries. Upon comparison with the descendant averages we have found the form
\begin{eqnarray} \nonumber
\label{OPE_refined}
\varepsilon_{1}\varepsilon_{2} & = & \frac{1}{|z_{12}|^{2}} + \frac{2}{|z_{12}|^{2}} \left( z_{12}^{2}T(\hat{z}) + \bar{z}_{12}^{2}\bar{T}(\bar{\hat{z}}) \right) + 4 |z_{12}|^{2} T(\hat{z}) \bar{T}(\bar{\hat{z}}) + \\
& + & \frac{1}{7|z_{12}|^{2}} \biggl[ z_{12}^{4} \left( 2L_{-2}^{2}-\frac{1}{2}L_{-4} \right) + \bar{z}_{12}^{4} \left( 2\bar{L}_{-2}^{2}-\frac{1}{2}\bar{L}_{-4} \right) \biggr] \mathbb{I}(\hat{z},\bar{\hat{z}})
\end{eqnarray}
of the OPE of two energy density operators. Here $z_{12} \equiv z_{1}-z_{2}$ and $\hat{z}_{12} \equiv (z_{1}+z_{2})/2$ and the terms neglected on the right hand side of Eq. (\ref{OPE_refined}) are at least two orders higher in $z_{12}$. The operator equation (\ref{OPE_refined}) provides a linear relation between the averages of $\varepsilon_{1}\varepsilon_{2}$ in ${\mathscr{G}}$ and those of the descendant operators with prefactors which are {\it independent} of ${\mathscr{G}}$. Our results for these quantities in the half plane, strip, and wedge geometries fulfill these necessary conditions for the validity of the operator equation (\ref{OPE_refined}). As further checks of Eq. (\ref{OPE_refined}) we apply it to the bulk three-point correlations functions $\langle \varepsilon_{1}\varepsilon_{2} T(z) \rangle$ and $\langle \varepsilon_{1}\varepsilon_{2} T(z)\bar{T}(\bar{z}) \rangle$. Apart from the importance in its own right, the OPE in Eq. (\ref{OPE_refined}) is very useful in the present context by enabling us to read off the averages of $T$, $\bar{T}$, and the combination $( 4L_{-2}^2 - L_{-4}) {\mathbb I}$ for {\it any} geometry $\mathscr{G}$, once the corresponding two-point correlation function $\langle \varepsilon_{1}\varepsilon_{2} \rangle_{\mathscr{G}}$ is known up to second order in $z_{12}$. This encompasses geometries both with homogeneous and inhomogeneous boundaries.

Here we provide certain details of our calculations. For a discussion of the half plane case, in which only $T\bar{T}$ contributes, see the paragraph\footnote{See the footnote below Eq. (\ref{desc_02}).} containing Eqs. (3.59)-(3.63) in Ref. \cite{Eisenriegler_04}. For the horizontal strip we use the well known form
\begin{eqnarray} \nonumber
\label{strip}
\langle \varepsilon_{1}\varepsilon_{2} \rangle_{\mathbb{S}_{aa}} & = & \left( \frac{\pi}{\mathcal{W}} \right)^{2} \biggl[ \frac{1/4}{\sinh\zeta\sinh\bar{\zeta}} + \frac{1/2}{\cos(2\pi\hat{y}/\mathcal{W}) + \cosh(\zeta-\bar{\zeta})} - \frac{1/2}{\cos(2\pi\hat{y}/\mathcal{W}) + \cosh(\zeta+\bar{\zeta})} \biggr] \\
\zeta & \equiv & \pi z_{12}/(2\mathcal{W}) \, , \qquad \bar{\zeta} \equiv \pi \bar{z}_{12}/(2\mathcal{W}) \, , \qquad \hat{y}=(y_{1}+y_{2})/2  
\end{eqnarray}
of the two-point correlation function in the strip, which follows from its counterpart in the half plane given in Eq. (4.2) in Ref. \cite{BX_91}; the corresponding conformal transformation is taken from Eq. (\ref{map_strip}). For fixed $\mathcal{W}$ and $\hat{y}$ we expand the right hand side of Eq. (\ref{strip}) up to fourth order in $\zeta$, $\bar{\zeta}$. The result can indeed be identified with the strip-average of the right hand side of Eq. (\ref{OPE_refined}) by taking into account the form of the strip average $\langle T(z) \rangle_{\mathbb{S}_{aa}}=-\pi^{2}/(48\mathcal{W}^{2})$ of $T$ (with $a\in\{O,+\}$) as well as the averages of $T\bar{T}$ and $L_{-2}^{2}\mathbb{I}$ given in Eqs. (\ref{strip_01}) and (\ref{des_strip}) and the vanishing of $L_{-4}\mathbb{I}$.

For the wedge we have followed the same route to verify that the corresponding averages of the left and right hand sides of Eq. (\ref{OPE_refined}) are identical up to order $z_{12}^2$. Here we use the transformation in Eq. (\ref{map_wedge}), the wedge average $\langle T(z) \rangle_{\mathbb{W}_{aa}} = \left( 1-\Xi^{2} \right)/(48z^{2})$ of $T$, which follows from Eq. (\ref{wall_04}), and the averages in Eqs. (\ref{wedge_desc_01}) and (\ref{wedge_desc_02}) of $T\bar{T}$, $L_{-2}^{2}\mathbb{I}$, and $L_{-4}\mathbb{I}$.

Further confirmation of the expression in Eq. (\ref{OPE_refined}) stems from the bulk three-point correlation function
\begin{equation}
\label{bulkthree}
\langle \varepsilon_{1}\varepsilon_{2}T(z) \rangle = \frac{1}{|z_{12}|^{2}} \biggl[ \frac{z_{12}^{2}}{2\left( \hat{z}-z \right)^{4}} + \frac{z_{12}^{4}}{4\left( \hat{z}-z \right)^{6}} + \dots \biggr] \, ,
\end{equation}
where the ellipses contain terms which are by two orders higher in $z_{12}$. Equation (\ref{bulkthree}) together with the bulk two-point correlation functions
\begin{equation}
\label{ }
\langle L_{-2}^{2}\mathbb{I}(\hat{z},\bar{\hat{z}}) T(z) \rangle = \frac{3/2}{ \left( \hat{z}-z \right)^{6} } \, , \qquad \langle L_{-4}\mathbb{I}(\hat{z},\bar{\hat{z}}) T(z) \rangle = \frac{5/2}{ \left( \hat{z}-z \right)^{6} } \, ,
\end{equation}
and the vanishing of $\langle T(\hat{z})\bar{T}(\bar{\hat{z}})T(z) \rangle$ is again consistent with Eq. (\ref{OPE_refined}). The consistency of Eq. (\ref{OPE_refined}) with the bulk three-point correlation function $\langle \varepsilon_{1}\varepsilon_{2}T(z)\bar{T}(\bar{z}) \rangle$ has been shown in Ref. \cite{Eisenriegler_04}.

\section{Proximal interaction for symmetric Janus circles}
\label{appendix_F}
In order to investigate the disjoining force between two close Janus particles or between a Janus particle and a close ``wall'' (i.e., the boundary of its enclosing half plane) it is advantageous to conformally transform the system to the annulus geometry. The reason is that the corresponding annulus is \emph{thin}, i.e., its width $\mathcal{W}$ is much smaller than its diameter, so that its local properties such as the average of the stress tensor are asymptotically equal to those of an infinitely long \emph{strip} of width $\mathcal{B}$. This has been exploited in Refs. \cite{BE_95} and \cite{ER_95} in order to show that for two circles (or two spheres in $d = 3$) with \emph{homogeneous} boundary conditions the Derjaguin approximation \cite{Derjaguin} actually is \emph{exact} right at the critical point. Here we extend this reasoning to close circles with \emph{inhomogeneous} boundary conditions, such as the circular Janus particles considered here, realizing that the stress tensor in the corresponding strip can be calculated rather easily.
The transformation
\begin{equation}
\label{D1}
w(c) = i \exp\left( \frac{\pi c}{\mathcal{B}} \right)
\end{equation}
maps the upper half $w$ plane $\mathbb{H}$ to a vertical strip $\widetilde{\mathbb{S}}$ in the $c$ plane of width $\mathcal{B}$ with its midline coinciding with the imaginary axis. Here the four segments $[-\infty < w < -1, -1 < w < 0 ]$ and $[0<w<1, 1<w<\infty]$ of the half plane boundary are mapped onto the [lower right, upper right] and [upper left, lower left] boundary segments of the vertical strip, which are separated by $c=\mathcal{B}/2$ and $c=-\mathcal{B}/2$, respectively.

The M$\ddot{\textrm{o}}$bius transformation\footnote{The present complex position variable $A$ in the annulus geometry should not be confused with the prefactor $A$ of the descendant operator $L_{-3} \Phi$ in Appendix \ref{correlations} (see Eq. (\ref{cf_001})).}
\begin{equation}
\label{D2}
\frac{z(A)}{R_{0}} = \frac{R_{0}+A}{R_{0}-A} \, ,
\end{equation}
which is similar to Eq. (\ref{03_02}), maps the annulus onto two non-overlapping circles $1$ and $2$. It maps the interior and exterior of the circle of radius $R_{0}$, centered at the origin of the entire $A$-plane, to the right and to the left half of the $z$-plane, separated by the imaginary axis, which is the image of the above circle. The interior of the annulus $\mathbb{A}$, centered at the origin of the $A$-plane and bounded by circles with radii $R_{<} \leqslant R_{0}$ and $R_{>} \geqslant R_{0}$, is mapped onto the exterior of the circles $1$ and $2$. The latter are the images of the smaller and wider boundary circles of $\mathbb{A}$ and are located in the right and the left half of the $z$-plane, respectively, with their centers on the real axis. Denoting their radii by $R_{1}$ and $R_{2}$ and the closest distance between them by $\mathcal{C}$, the invariance of cross ratios yields the relation\footnote{A corresponding discussion concerning two spheres in arbitrary dimensions is given in Refs. \cite{BE_95} and \cite{ER_95}.}
\begin{equation}
\label{D3}
\frac{R_{<}}{R_{>}} + \frac{R_{>}}{R_{<}} = \frac{(\mathcal{C} + R_{1} + R_{2})^{2} - R_{1}^{2} - R_{2}^{2}}{R_{1}R_{2}}
\end{equation}
between the two geometries. In the following we consider two special cases.
\begin{itemize}
\item[(i)] The two circles are of equal size $R_{1}=R_{2}=R$ and their centers are located at $z=\pm[R + \mathcal{C}/2]$ as generated by Eq. (\ref{D2}) if $R_{<}R_{>}=R_{0}^{2}$. In this case Eqs. (\ref{D3}) and (\ref{D2}) imply
\begin{equation}
\label{D4}
\frac{\mathcal{C}}{R} = \biggl[ \left( \frac{R_{>}}{R_{<}} \right)^{1/4} - \left( \frac{R_{<}}{R_{>}} \right)^{1/4} \biggr]^{2} \, , \qquad \frac{2}{R} = \frac{1}{R_{<}} - \frac{1}{R_{>}} \, .
\end{equation}
\item[(ii)] For $R_{>}=R_{0}$, circle $2$ reduces to the imaginary axis, forming the boundary of the right half $z$ plane, which contains circle $1$ with radius $R_{1}\equiv R$ and its center at $z=\mathcal{C}+R$. In this case one has
\begin{equation}
\label{D5}
\frac{\mathcal{C}}{R} = \frac{1}{2} \left( \sqrt{\frac{R_{0}}{R_{<}}} - \sqrt{\frac{R_{<}}{R_{0}}} \right)^{2} \, , \qquad \frac{2R_{0}}{R} = \frac{1}{2} \left( \frac{R_{0}}{R_{<}} - \frac{R_{<}}{R_{0}} \right)  \, .
\end{equation}
\end{itemize}
Equations (\ref{D4}) and (\ref{D5}) apply to arbitrary size-to-distance ratios and imply for close distances $\mathcal{C} \ll R$ the following relations for the width of the annulus $\mathcal{B}=R_{>}-R_{<}$
\begin{equation}
\label{D6}
\mathcal{B} \rightarrow 2\mathcal{C} \, , \qquad R_{0} \rightarrow \sqrt{\mathcal{C}R} \, , \qquad \textrm{case (i),}  
\end{equation}
and
\begin{equation}
\label{D7}
\mathcal{B} \rightarrow 2\mathcal{C} \, , \qquad R_{0} \rightarrow \sqrt{2\mathcal{C}R} \, , \qquad \textrm{case (ii),}
\end{equation}
which will be needed in the following.

We now consider the two circles of case (i) or the circle of case (ii) endowed with inhomogeneous boundary conditions, representing particles such as the circular Janus ones or quadrupolar particles. Our goal is to determine the disjoining force $f_x$, i.e., the force necessary to shift the two particles or the particle and the boundary apart in $x$ direction with the orientations kept fixed. Using the transformation given by Eq. (\ref{D2}), this force can be written as \cite{VED,EB_16}
\begin{eqnarray} \nonumber
\label{D8}
f_{x} & \equiv & -\frac{\partial}{\partial \mathcal{C}} {\cal F}_{1,2} = \int_{-\infty}^{\infty}\textrm{d}y \, \langle T_{xx}(0,y) \rangle_{1,2} = -\frac{1}{\pi} \textrm{Re} \frac{1}{i} \int_{-i\infty}^{i\infty}\textrm{d}z \, \langle T(z) \rangle_{1,2} \, \\
& = & -\frac{1}{2\pi} \textrm{Re} \frac{1}{i} \oint_{\mathfrak{C}} \textrm{d}A \, \langle T(A) \rangle_{\mathbb{A}} (A/R_{0}-1)^{2} \, ,
\end{eqnarray}
i.e., in terms of the average of the Cartesian stress tensor component $T_{xx}(x,y)$ or of the complex stress tensor $T(z)$ in the system of the two circular particles (i) or of the system of the particle and the boundary (ii). The force can also be expressed in terms of the average of $T(A)$ (see Eq. (\ref{D2})) in the annulus geometry $\mathbb{A}$ with corresponding boundary conditions. Here, ${\cal F}_{1,2}$ is the free energy of the 1,2 system. The closed integration path $\mathfrak{C}$ runs clockwise along the circle with radius $R_{0}$ and the factor accompanying $\langle T(A) \rangle_{\mathbb{A}}$ is $1/(\textrm{d}z/\textrm{d}A)$.

As a first step, we use Eq. (\ref{D8}) in order to re-derive the known disjoining force $f_{x}$ between two circles of equal size [case (i)] or between a circle and a boundary [case (ii)] with {\it homogeneous} boundaries $a$ and $b$. For the corresponding (rotationally invariant) annulus $\mathbb{A} \equiv \mathbb{A}_{ab}$ with homogeneous boundary conditions $a$ and $b$ on the inner and outer boundary, respectively, one has
\begin{equation}
\label{D9}
\langle T(A) \rangle_{\mathbb{A}_{ab}} A^{2} = -\pi \langle T_{nn} \rangle_{\mathbb{A}_{ab}} |A|^{2} \equiv \tau_{ab}(R_{<}/R_{>}) \, ,
\end{equation}
which solely depends on $(a,b)$ and the ratio $R_{<}/R_{>}$; $T_{nn}$ is the component of the Cartesian stress tensor normal to the annulus boundaries. Equation (\ref{D8}) implies
\begin{equation}
\label{D10}
f_{x} = -\frac{2\tau_{ab}}{R_{0}} \, .
\end{equation}
For a \emph{thin} annulus, $\langle T_{nn} \rangle_{\mathbb{A}_{ab}}$ equals the perpendicular stress tensor component of an \emph{infinite strip} of width $R_{>}-R_{<} \equiv \mathcal{B}$ so that $\langle T_{nn} \rangle_{\mathbb{A}_{ab}} \rightarrow \Delta_{ab}/\mathcal{B}^{2}$, $\tau_{ab}(R_{<}/R_{>}) \rightarrow -\pi \Delta_{ab} (R_{0}/\mathcal{B})^{2}$ due to $|A| \rightarrow R_{0}$, and $\langle T(A) \rangle_{\mathbb{A}_{ab}} \to -A^{-2} \pi \Delta_{ab}(R_{0}/\mathcal{B})^{2}$ in leading order of an expansion for small annulus width $\mathcal{B}$. Upon inserting $\mathcal{B}$ and $R_{0}$ from Eqs. (\ref{D6}) and (\ref{D7}), Eqs. (\ref{D8}) and (\ref{D9}) yield the leading order contribution to the force:
\begin{equation}
\label{D11}
f_{x} \stackrel{\mathcal{C}\rightarrow0}{\longrightarrow} \pi \Delta_{ab} \frac{R^{1/2}}{2\mathcal{C}^{3/2}} [1,\sqrt{2}] - \frac{1}{48\mathcal{C}} \equiv \Bigl[ \Upsilon_{ab}^{(\textrm{i})} , \Upsilon_{ab}^{(\textrm{ii})} \Bigr] \, , \qquad \mathcal{C} \ll R \, ,
\end{equation}
in the cases [(i), (ii)] with homogeneous boundary conditions $ab$. As discussed in Refs. \cite{EB_16, ER_95}, the leading contribution to the exact expansion in Eq. (\ref{D11}) is identical to the result of the Derjaguin approximation (see Ref. \cite{Derjaguin}). For later comparison, in Eq. (\ref{D11}) we have included the next-to-leading contribution $-1/(48\mathcal{C})$, which is independent of $ab$ and the same for both cases (i) and (ii). This expression follows from the next-to-leading contribution of $\tau_{ab}(R_{<}/R_{>})$ in the limit $R_{<}/R_{>} \nearrow 1$ 
\footnote{For homogeneous boundary conditions, the free energy of interaction and the disjoining force between two circles and between a circle and a wall in $d = 2$ can be obtained \cite{BE_95,MVS} for \emph{arbitrary} values of $R/\mathcal{C}$, because $\tau_{ab}$ in Eq. (\ref{D10}) is known \cite{Cardy_BCFT} for an arbitrary ratio $R_{<}/R_{>}$ of the radii of the boundary circles of the annulus. A convenient representation is $\tau_{ab}(R_{<}/R_{>}) = (1/48) - \delta^{2}\Delta_{ab}(1/\delta)/(4\pi)$, where $\delta=2\pi/|\ln(R_{<}/R_{>})|$, in terms of the amplitude $\Delta_{ab}(1/\delta)=W^{2}\langle T_{\bot\bot} \rangle_{W \times W\delta}$ of the spatially constant average of the stress tensor component $T_{\bot\bot}$ perpendicular to the parallel boundaries of a semi-periodic rectangular system. In this system the distance between the boundaries is $W$ and their length is the periodicity length $W\delta$. It is related to the annulus via the conformal transformation $\sigma(A) = W\delta(2\pi)^{-1}\ln(A/R_{0})$ if the rectangle is in the $\sigma$ plane with one of its boundaries extending from $W\delta(2\pi)^{-1}\ln(R_{<}/R_{0})$ to $W\delta[i+(2\pi)^{-1}\ln(R_{<}/R_{0})]$ and the other from $W\delta(2\pi)^{-1}\ln(R_{>}/R_{0})$ to $W\delta[i+(2\pi)^{-1}\ln(R_{>}/R_{0})]$. The first term, $1/48$, in $\tau_{ab}$ arises from the contribution due to the Schwarzian derivative $[\sigma^{\prime\prime\prime}/\sigma^{\prime} - (3/2)(\sigma^{\prime\prime}/\sigma^{\prime})^{2}]/24 = 1/(48A^{2})$ to $\langle T(A) \rangle$. The explicit form of $\Delta_{ab}(1/\delta)$ can be found, e.g., in Eqs. (A7)-(A12) in Ref. \cite{VED}. }.

As intuitively expected, the result in Eq. (\ref{D11}) also applies if the circles exhibit Janus boundary conditions, provided they face each other or the wall with the homogeneous sections of their boundaries which is realized, in particular, for the orientations $\alpha=\pi/2$ or $\alpha=-\pi/2$. Within the present approach, in, say, case (i), the reason for this behavior is, that the four switching points $z=\pm (\mathcal{C}/2+ R) \pm i R$ correspond, via the mapping in Eq. (\ref{D2}) for close circular particles, to the points $A \rightarrow R_{0} [1+\sqrt{\mathcal{C}/R} (\pm1\pm i)]$ on the annulus boundaries, which are very close to $A = R_{0}$. Thus in the last integral in Eq. (\ref{D8}) the contributions from the two non-facing homogeneous sections of the two Janus boundaries are suppressed due to the factor accompanying $\langle T(A) \rangle_{\mathbb{A}}$.

\subsection{Two close Janus circles with their switching points facing each other}
Here we consider again two circular particles in configuration (i) (Eq. (\ref{D4})) but now both are taken with Janus boundary conditions with orientation $\alpha=0$ so that the $+$ and $-$ boundary conditions are located at the upper and lower half of their surfaces, respectively. By the mapping according to Eq. (\ref{D2}), the upper and lower half of the two boundary circles of the corresponding annulus $\mathbb{A}$ also exhibit the boundary conditions $+$ and $-$, respectively. The deviation $\delta f_{x}$ of the disjoining force $f_{x}$ from the plausible expectation $(\Upsilon_{++}^{(\textrm{i})}+\Upsilon_{--}^{(\textrm{i})})/2 = \Upsilon_{++}^{(\textrm{i})}$ (for $\Upsilon$ see Eq. (\ref{D11})) is determined via Eq. (\ref{D8}) by the corresponding change $\delta\langle T(A) \rangle_{\mathbb{A}} \equiv \langle T(A) \rangle_{\mathbb{A}} - \langle T(A) \rangle_{\mathbb{A}_{++}}$ in the expression for $\langle T(A) \rangle_{\mathbb{A}}$. For close particles, i.e., a thin annulus, a non-vanishing contribution $\delta\langle T(A) \rangle_{\mathbb{A}}$ potentially appears only in the two small regions around $A=\pm R_{0}$ with a size of the order of $\mathcal{B}$. We note, however, that the region around $A=+R_{0}$ does not contribute due to the accompanying factor in Eq. (\ref{D8}) and that outside these two regions one has $\langle T_{nn} \rangle=-\pi/(48\mathcal{B}^{2})$ because $\Delta_{++}=\Delta_{--}=-\pi/48$.

In the region around $A=-R_{0}$ the annulus appears like a vertical strip $\widetilde{\mathbb{S}}$ of width $\mathcal{W}=\mathcal{B}$ for which $\delta \langle T(c) \rangle_{\widetilde{\mathbb{S}}} \equiv \langle T(c) \rangle_{\widetilde{\mathbb{S}}} - \langle T(c) \rangle_{\widetilde{\mathbb{S}}_{++}}$ can be calculated from $\langle T(w) \rangle_{\mathbb{H}}$ in the upper half plane by using Eq. (\ref{D1}). Since the upper sections of the left and right boundaries of the vertical strip $\widetilde{\mathbb{S}}$ have boundary conditions $+$ and the two lower ones have boundary conditions $-$, this corresponds to the upper half $w$ plane $\mathbb{H}_{-+-}$ with two switching points $w=\pm1$ and with the stress tensor average $\langle T(w) \rangle_{\mathbb{H}_{-+-}} = 2/(w^{2}-1)^{2}$ so that Eq. (\ref{D1}) yields, along the midline $c=ib$ of the vertical strip, in the $c = a + ib$ plane
\begin{equation}
\label{D12}
\langle T(c=ib) \rangle_{\tilde{\mathbb{S}}} = \frac{\pi^2}{B^{2}} \biggl[ \frac{1}{2\cosh^{2}(\pi b/\mathcal{B})} + \frac{1}{48} \biggr] \equiv \delta \langle T(c=ib) \rangle_{\tilde{\mathbb{S}}} + \frac{\pi^{2}}{48\mathcal{B}^{2}} \, .
\end{equation}
Since $\delta f_{x}$ is given by Eq. (\ref{D8}) with $\langle T(A) \rangle_{\mathbb{A}}$ replaced by $\delta\langle T(A) \rangle_{\mathbb{A}}$, the integration differential $\textrm{d}A$ near $A=-R_{0}$ can be replaced by $\textrm{d}(ib)$, and the factor $(A/R_{0}-1)^{2}$ can be replaced by $4$. This yields
\begin{eqnarray} \nonumber
\label{D13}
\delta f_{x} & = & -\frac{1}{\pi} \textrm{Re} \frac{2}{i} \int_{-\infty}^{\infty}\textrm{d}(ib) \, \delta \langle T(c=ib) \rangle_{\widetilde{\mathbb{S}}} \\
& = & -\frac{1}{\mathcal{B}} \int_{-\infty}^{\infty}\textrm{d}\beta \, \cosh^{-2}\beta = -\frac{2}{\mathcal{B}} = - \frac{1}{\mathcal{C}} \, ,
\end{eqnarray}
i.e., a disjoining force $f_{x}=\Upsilon_{++}^{(\textrm{i})}-1/\mathcal{C}$, where we have used Eq. (\ref{D6}) in the last step. Thus $\delta f_{x}$ is attractive. The two present circular Janus particles at close distances are attracted more strongly than in the reference case of two circular particles with the same homogeneous boundary. This is plausible because the switching points face each other and ``fit together'', in line with their common property of an increased energy density. However, compared with this reference case of two circular particles with the same homogeneous boundary (both $+$ or both $-$) (see Eq. (\ref{D11})), the magnitude of $\delta f_{x}$ is only of the order of the next-to-leading contribution, i.e., by a factor of order $(\mathcal{C}/R)^{1/2}$ weaker than the leading contribution of the Derjaguin form.

The other case of the switching points facing each other occurs for one Janus particle oriented with $\alpha=0$ and the other one with $\alpha=\pi$. In this case one has $\langle T_{nn} \rangle_{\mathbb{A}}=23\pi/(48\mathcal{B}^{2})$ outside the two regions $A=\pm R_{0}$ because $\Delta_{+-}=\Delta_{-+}=23\pi/48$. The vertical strip has $+$ boundary conditions on its upper left and lower right boundary sections and $-$ boundary conditions on the two remaining sections so that the corresponding half plane boundary has three points, where the boundary conditions switch: at $w=-1$, $w=0$, and $w=1$ with the stress tensor $\langle T(w) \rangle_{\mathbb{H}_{+-+-}}$ given by the right hand side of Eq. (\ref{07_08}). This leads to the form
\begin{equation}
\label{D14}
\langle T(c=ib) \rangle_{\widetilde{\mathbb{S}}} = \delta \langle T(c=ib) \rangle_{\widetilde{\mathbb{S}}} - \frac{23\pi^{2}}{48\mathcal{B}^{2}}
\end{equation}
of the midline tensor of the strip where $\delta \langle T(c=ib) \rangle_{\tilde{\mathbb{S}}}$ is \emph{identical} to the one in Eq. (\ref{D12}) so that, upon using Eq. (\ref{D13}), $\delta f_{x}$ is again given by $\delta f_{x} = -1/\mathcal{C}$ so that the force is $f_{x}=\Upsilon_{+-}^{(\textrm{i})}-1/\mathcal{C}$.

\subsection{Janus circles with their switching points facing a $+$ wall}
We consider a Janus circle with orientation $\alpha=0$ near a vertical wall with a homogeneous $+$ boundary so that the Janus circle faces the wall with its switching point, which specifies the wall-circle and the corresponding annulus geometry, as in case (ii) associated with Eq. (\ref{D5}). A plausible expectation for the disjoining force is $f_{x} \approx (\Upsilon_{++}^{(\textrm{ii})} + \Upsilon_{+-}^{(\textrm{ii})})/2$, which is the mean value $(11\pi^{2}/48)(R/2)^{1/2}\mathcal{C}^{-3/2}-1/(48\mathcal{C})$ of the forces between the wall and a homogeneous $+$ circle and the wall and a homogeneous $-$ circle (see Eq. (\ref{D11})). The leading order for small distances of this expression is equivalent to the Derjaguin approximation and leads to Eq. (\ref{Januswall6}). Here we show that in this case the disjoining force is given by
\begin{equation}
\label{D15}
f_{x} = (\Upsilon_{++}^{(\textrm{ii})} + \Upsilon_{+-}^{(\textrm{ii})})/2 +1/(4\mathcal{C}) \, ,
\end{equation}
so that the deviation $\delta f_{x} = 1/(4\mathcal{C})$ of the force $f_{x}$ from the expected expression is a repulsive contribution. This is in accordance with intuition because the increased energy density (i.e., increased disorder) near the switching point on the Janus circle does ``not fit'' with the decreased energy density (i.e., increased order) near the homogeneous $+$ wall. Like in the cases of the previous subsection, the magnitude of $\delta f_{x}$ is only of the order of the next-to-leading term in the expression expected above for $f_{x}$, i.e., it is by a factor of the order of $(\mathcal{C}/R)^{1/2}$ weaker than its leading term.

In order to derive Eq. (\ref{D15}) we write the deviation $\delta f_{x}$ from the expected expression as
\begin{eqnarray} \nonumber
\label{D16}
\delta f_{x} & = & -\frac{1}{2\pi} \textrm{Re} \frac{1}{i} \oint_{\mathfrak{C}}\textrm{d}A \, \biggl\{ \Theta(\textrm{Im} A) \bigl[ \langle T(A) \rangle_{\mathbb{A}} - \langle T(A) \rangle_{\mathbb{A}_{++}} \bigr] + \\
& + & \Theta(-\textrm{Im} A) \bigl[ \langle T(A) \rangle_{\mathbb{A}} - \langle T(A) \rangle_{\mathbb{A}_{+-}} \bigr] \biggr\} (A/R_{0}-1)^{2} \, ,
\end{eqnarray}
where $\Theta$ is the Heaviside step function ($1$ and $0$ for positive and negative argument, respectively) and the boundary conditions of $\mathbb{A}$ are those of a Janus particle with orientation $\alpha =0$ at the inner circle while the outer circle has a homogeneous $+$ boundary. Here we have used Eq. (\ref{D8}) which also tells that for a homogeneous annulus the upper and the lower halves of the integration path $\mathfrak{C}$ separately yield the same contributions to the force, i.e., one half of it. The contributions to the integral in Eq. (\ref{D16}) stem --- like in the previous subsection --- from a small region around $A=-R_{0}$, where the annulus looks like a vertical strip with inhomogeneous boundary conditions on its right boundary $c=(\mathcal{B}/2)+ib$ ($+$ for $b>0$ and $-$ for $b<0$) while its left boundary is a homogeneous $+$ boundary. By using Eq. (\ref{D1}) the corresponding stress tensor along the left homogeneous strip boundary is given by
\begin{equation}
\label{D17}
\langle T(c=-(\mathcal{B}/2)+ib) \rangle_{\tilde{\mathbb{S}}} = - \frac{\pi^{2}}{2\mathcal{B}^{2}} \Biggl[ \frac{1}{ \left( \textrm{e}^{\pi b/\mathcal{B}}+1 \right)^{2} } - \frac{1}{24} \Biggr] \, ,
\end{equation}
so that the strip integral, corresponding to Eq. (\ref{D16}), converges. It yields $\delta f_{x} \rightarrow 1/(2\mathcal{B})$ which together with Eq. (\ref{D7}) leads to the result given in Eq. (\ref{D15}).

It is interesting to extend this investigation to other cases such as a Janus particle facing with its switching point an ordinary wall, which involves three boundary conditions: $+$, $-$, and $O$ \cite{EB_unpublished}. For such a system there is competition between the disorder effect near the switching point fitting well to the disorder induced by the ordinary wall and thus tending to attraction and the effect of the $+$ and $-$ segments tending to repulsion. It turns out that for a Janus \emph{circle} repulsion dominates at short distances and has the form of a Derjaguin expression which is stitched together. The particle, which according to Eq. (\ref{Januswall1}) is attracted at large distances, finds a stable position (i.e., the free energy minimum) at a finite and nonzero separation from the ordinary wall. Besides the Janus circle, the behavior of the corresponding Janus \emph{needle} oriented perpendicular to the ordinary wall and with one end close to it, is also of interest. In the present context it can be considered to be of semi-infinite extent and the disorder effect from the tip dominates and leads to attraction. These properties can be derived from the stress tensor average $\langle T(w) \rangle_{\mathbb{H}_{+O-}}$ in the upper half plane with boundary conditions $+$, $O$, $-$ in the intervals $-\infty < u < -1$, $-1<u<1$, and $1<u<\infty$, respectively.

\section{Degeneracy on level two}
\label{discussion_degeneracy}
In order to understand the notion of degeneracy it is useful to consider a larger class of conformal models, including the Ising model, the three-state Potts model etc., each of which is characterized by a different value for the central charge $c$ in its Virasoro algebra given in Eq. (\ref{VAextended}). Primary operators $\cal O$ are characterized by the property $L_{p} {\cal O} =0$ for all $p=1,2,3,\dots$ so that the primary operator has the lowest scaling dimension in its conformal class. Using the Virasoro algebra it is easy to see that the combination $\mu_{\cal O} \equiv [L_{-1}^2 -(2/3)(x_{\cal O} +1) L_{-2}] {\cal O}$ of level two descendants, which is proportional to the difference of the right hand side and left hand side of Eq. (\ref{VAextended}), also has this property, provided $x_{\cal O}$ is related to $c$ via
\begin{eqnarray}
\label{L2O}
c =c(x_{\cal O}) \equiv x_{\cal O}(5-4x_{\cal O})/(x_{\cal O} +1) \, .
\end{eqnarray}
While the condition $L_{1} \mu_{\cal O} =0$ is satisfied for {\it any} primary operator ${\cal O}$, the condition $L_{2} \mu_{\cal O} =0$ represents the above relation between $x_{\cal O}$ and $c$. Belavin, Polyakov, and Zamolodchikov \cite{BPZ_1,BPZ_2} (BPZ) have shown that in this case any correlation function of the form
\begin{equation}
\label{ }
\langle  \Psi_{\rm I} (z_1 , \bar{z}_1) \dots \Psi_{\rm N} (z_N , \bar{z}_N) \mu_{\cal O} (z_0, \bar{z}_0) \rangle
\end{equation}
vanishes so that it is consistent to put $\mu_{\cal O} =0$ which implies Eq. (\ref{degeneracy}). This vanishing holds not only in the bulk but also at internal points in a system with boundaries. Correspondingly $\mu_{\cal O}$ is called a ``null-field'' or ``null-operator'' and ${\cal O}$ is called to be ``degenerate on level two''. It is instructive to check this vanishing for the simple case of three-point correlation functions $\langle {\cal O}_{\rm I} {\cal O}_{\rm II} \mu_{\cal O} \rangle$ in the bulk made up of two primary operators which are multiplied by $\mu_{\cal O}$. Like the three-point correlation function of three primary operators (cp. the expression in footnote 12) it has a model-independent structure and for ${\cal O}_{\rm I} = {\cal O}_{\rm II}$ it is given by 
\begin{eqnarray}
\label{three}
&&\langle {\cal O}_{\rm I} (1) {\cal O}_{\rm I} (2) \mu_{\cal O} (0) \rangle /  \langle {\cal O}_{\rm I} (1) {\cal O}_{\rm I} (2) {\cal O} (0) \rangle = \Delta \Bigl[ (\Delta +1)( z_{20}^{-2} + z_{01}^{-2} ) -2 \Delta ( z_{20}z_{01} )^{-1} \Bigr] \nonumber \\
&& \quad \qquad \qquad \qquad \quad - (2/3) (x_{\cal O} +1) \Bigl[ (\Delta_{\rm I}+\Delta) ( z_{20}^{-2} + z_{01}^{-2} ) + (2 \Delta_{\rm I} - \Delta) ( z_{20}z_{01} )^{-1} \Bigr]  \, 
\end{eqnarray}
where $2 \Delta_{\rm I} \equiv x_{{\cal O}_{\rm I}}$ and $2 \Delta \equiv x_{\cal O}$ are the scaling dimensions of ${\cal O}_{\rm I}$ and ${\cal O}$ and the notation $(m) \equiv (z_{m}, \bar{z}_{m})$ where $m=0,1,2$ as well as the notation $z_{mn} \equiv z_{m} -z_{n}$ have been used. Here the first and the second line of the right hand side arises from the first and second terms in $\mu_{\cal O}$, respectively, and we assume the three-point correlation function of primary operators in the denominator to be non-vanishing. Like in Appendix \ref{appb11} the derivation is obtained by combining Eq. (\ref{def_descendant}) with the appropriate conformal Ward identity (cp. Eq. (\ref{cf_011})). In the following we show, for three examples of a non-vanishing three-point correlation function $\langle {\cal O}_{\rm I} {\cal O}_{\rm I} {\cal O} \rangle$ with $x_{\cal O}$ obeying Eq. (\ref{L2O}), that --- in agreement with BPZ --- the right hand side of Eq. (\ref{three}) indeed vanishes.
\begin{itemize}
  \item[(i)] In the Ising model, which corresponds to $c=1/2$ we consider $\langle \Phi (1) \Phi (2) \varepsilon (0) \rangle$. Here the scaling dimension $x_{\cal O} = 2 \Delta = x_{\varepsilon}  =1$ satisfies Eq. (\ref{L2O}) and --- with $2 \Delta_{\rm I} = x_{\Phi} =1/8$ --- the right hand side of Eq. (\ref{three}) and thus $\langle \Phi (1) \Phi (2) \mu_{\varepsilon} (0) \rangle$ vanishes. Since the scaling dimension of $\Phi$ also satisfies Eq. (\ref{L2O}), the correlation function $\langle \Phi \varepsilon \mu_{\Phi} \rangle$ must also vanish. This can be shown along similar lines.
  \item[(ii)] For the three-state Potts model one has $c=x_{\varepsilon} =4/5$ and $x_{\Phi} = 2/15$, and $\langle \Phi (1) \Phi (2) \varepsilon (0) \rangle$ is non-vanishing. Thus Eq. (\ref{L2O}) is fulfilled for ${\cal O}=\varepsilon$ and $\langle \Phi (1) \Phi (2) \mu_{\varepsilon} (0) \rangle$ is expected to vanish. Indeed, the right hand side of Eq. (\ref{three}) vanishes for the corresponding values $2 \Delta = 4/5$ and $2\Delta _{\rm I} = 2/15$.
  \item[(iii)] Finally, we consider the Yang-Lee edge singularity model (i.e., the critical point of an Ising model in a purely imaginary magnetic field) which has only one primary operator ${\cal O}$ the three-point correlation function of which is non-vanishing (see Ref. \cite{Cardy_review}). Here $c=-22/5$ and $x_{\cal O}=-2/5$ satisfy Eq. (\ref{L2O}) so that $\langle {\cal O} (1) {\cal O} (2) \mu_{{\cal O}} (0) \rangle$ is expected to vanish. This is corroborated by the vanishing of the right hand side of Eq. (\ref{three}) for the corresponding values $\Delta_{\rm I} =\Delta = -1/5$.
\end{itemize}
As an example for a system with a boundary we consider in the upper half $z$ plane ${\mathbb H}_a$ with a homogeneous boundary $a=+$ or $O$ the correlation function with the stress tensor, $\langle T \mu_{\cal O} \rangle_{{\mathbb H}_a}$,   which has a model-independent form. If $T$ is located at the origin $z=0$, i.e., right at the boundary, this form is particularly simple and reads
\begin{eqnarray}
\label{Tmu}
\nonumber
\langle T(0) \mu_{\cal O} (z, \bar{z}) \rangle_{{\mathbb H}_a} /  \langle {\cal O} (z, \bar{z}) \rangle_{{\mathbb H}_a} = [c(x_{\cal O}) - c] z^{-4} (x_{\cal O}+1)/3 
\end{eqnarray}
with $c(x_{\cal O})$ as given above. The expression on the right hand side is independent of $a$ and indeed vanishes for $c=c(x_{\cal O})$.

\end{appendices}

\end{document}